\documentclass[preprint,12pt,3p]{elsarticle}

\usepackage{graphicx}
\usepackage{caption}
\usepackage{subcaption}
\usepackage{psfrag}
\usepackage{array}
\usepackage{multirow}
\usepackage{comment}

\usepackage{amsmath}%
\usepackage{amsfonts}%
\usepackage{amssymb}%%% The amsthm package provides extended theorem environments
\usepackage{algorithm}
\usepackage{algpseudocode}
\usepackage{setspace}
\usepackage{gensymb}

\journal{Journal of Computational Physics}

\DeclareMathOperator*{\argmin}{arg\,min}

\begin{document}

\begin{frontmatter}

\title{Deep-learning-based reduced-order modeling for subsurface flow simulation}
% \tnotetext[label0]{This is only an example}

\author[label1]{Zhaoyang Larry Jin\corref{cor1}}%\fnref{label3}}
\address[label1]{Department of Energy Resources Engineering, Stanford University, Stanford, CA, 94305}
% \address[label2]{Address Two\fnref{label4}}
% \address[label5]{}

\cortext[cor1]{Corresponding author}
% \fntext[label3]{I also want to inform about\ldots}
% \fntext[label4]{Small city}

\ead{zjin@stanford.edu}
% \ead[url]{author-one-homepage.com}

\author[label1]{Yimin Liu}

\ead{yiminliu@stanford.edu}

\author[label1]{Louis J. Durlofsky}
\ead{lou@stanford.edu}

\begin{abstract}

A new deep-learning-based reduced-order modeling (ROM) framework is proposed for application in subsurface flow simulation. The reduced-order model is based on an existing embed-to-control (E2C) framework and includes an auto-encoder, which projects the system to a low-dimensional subspace, and a linear transition model, which approximates the evolution of the system states in low dimension. In addition to the loss function for data mismatch considered in the original E2C framework, we introduce a physics-based loss function that penalizes predictions that are inconsistent with the governing flow equations. The loss function is also modified to emphasize accuracy in key well quantities of interest (e.g., fluid production rates). The E2C ROM is shown to be very analogous to an existing ROM, POD-TPWL, which has been extensively developed for subsurface flow simulation. The new ROM is applied to oil-water flow in a heterogeneous reservoir, with flow driven by nine wells operating under time-varying control specifications. A total of 300 high-fidelity training simulations are performed in the offline stage, and the network training requires 10-12~minutes on a Tesla V100 GPU node. Online (runtime) computations achieve speedups of $\mathcal{O}$(1000) relative to full-order simulations. Extensive test case results, with well controls varied over large ranges, are presented. Accurate ROM predictions are achieved for global saturation and pressure fields at particular times, and for injection and production well responses as a function of time. Error is shown to increase when 100 or 200 (rather than 300) training runs are used to construct the E2C ROM. 

\end{abstract}

\begin{keyword}
reservoir simulation \sep reduced-order model \sep deep learning \sep physics-informed neural network \sep auto-encoder \sep embed-to-control \sep E2C
\end{keyword}

\end{frontmatter}

% \linenumbers

\section{Introduction}\label{intro}
% -------------------------------------------------------------
% -------------------------------------------------------------

Reservoir simulation is widely applied to model and manage subsurface flow operations. However, due to the nonlinear nature of the governing equations and the multiscale character of the geological description, computational costs can be high, especially when highly resolved models are used. Computational demands can become prohibitive when simulation tools are applied for optimization, uncertainty quantification, and data assimilation, in which case thousands of simulation runs may be required.

Reduced-order models (ROMs) have been developed and applied to accelerate flow predictions in a variety of settings. Our goal in this work is to develop a new deep-learning-based reduced-order modeling procedure. Following the embed-to-control framework, the approach introduced here is comprised of a linear transition model and an auto-encoder (AE, also referred to as encoder-decoder). An encoder-decoder architecture is used to achieve dimension reduction by constructing the mapping to and from the low-dimensional representation. The AE component is a stack of multiple convolutional neural network (CNN) layers and dense feed-forward layers. The linear transition model represents the step-wise evolution of the system states with multiple linear feed-forward layers. %The method is applied to simulate oil production with water injection in heterogeneous reservoirs. 
The E2C procedure is constructed to predict key well quantities, such as time-varying production and injection rates and/or bottom-hole pressure (BHPs), as well as global pressure and saturation fields, in oil-water reservoir simulation problems.

ROM methodologies have received a large amount of attention in recent years. These procedures typically involve an offline (train-time) component, where training runs are performed and relevant solution information is processed and saved, and an online (test-time) component, where new (test) runs are performed. A popular category of methods is proper-orthogonal-decomposition-based (POD-based) ROMs, in which POD is applied to enable the low-dimensional representation of solution unknowns in the online computations. These approaches also require the projection of the system of equations to low dimension (this projection is also referred to as constraint reduction). Galerkin projection and least-squares Petrov-Galerkin projection are the two approaches typically used for this step. 

A treatment of solution nonlinearity is also required, and there have been a number of treatments for this within the context of POD-based ROMs. One effective approach is Gauss-Newton with approximated tensors or GNAT, which also uses POD for state reduction and least-squares Petrov-Galerkin projection. GNAT was developed by \citet{carlberg2011efficient}, and has since been used for structural and solid mechanics \citep{zahr2017multilevel}, electromechanics \citep{amsallem2012nonlinear}, and computational fluid dynamics \citep{carlberg2013gnat}. GNAT represents a generalization of the discrete empirical interpolation method (DEIM) \citep{chaturantabut2010nonlinear}), and the two methods (GNAT and POD-DEIM) have been applied in a number of studies involving subsurface flow simulation \citep{yoon2016hyper, yang2016fast, efendiev2016online, tan2019trajectory, jiang2019implementation, florez2019model}. A radial basis function (RBF) multidimensional interpolation method has also been used to treat nonlinearity in the low-dimensional space represented by POD, and the resulting procedure is referred to as POD-RBF method \citep{xiao2015non, kostorz2019non}. Trajectory piecewise linearization, originally introduced by \citet{rewienski2003trajectory}, entails linearization around `nearby' training solutions.  %which is considered a different treatment of nonlinearity compared to GNAT, POD-DEIM or POD-RBF. 
POD-TPWL has been widely applied for subsurface flow simulations involving oil-water, oil-gas compositional, CO$_2$ storage, and coupled flow-geomechanics systems \citep{cardoso2010linearized, he2011enhanced, he2014reduced, he2015constraint, jin2018reduced, jin2019reduced}. \citet{trehan2016trajectory} extended POD-TPWL to include a quadratic term, which gives a trajectory piecewise quadratic (POD-TPWQ) procedure.

The recent success of deep learning in image processing has inspired the rapid development of algorithms for subsurface modeling that make use of deep neural networks. These methods have been applied for geological parameterization, uncertainty quantification, and surrogate/reduced-order modeling. For geological parameterization and uncertainty quantification, \citet{canchumuni2018history} generated new geological realizations from randomized low-dimensional latent variables using a variational auto-encoder (VAE). A VAE entails a convolutional encoder-decoder neural network architecture similar to the AE, where the encoder component projects a high-dimensional distribution into a low-dimensional random vector, with each element following an independent Gaussian distribution. The decoder acts as the inverse of the encoder and projects the sampled Gaussian-distributed random variables back to the high dimension. \citet{laloy2018training} achieved a similar goal using a generative adversarial network (GAN), where the projection to high dimension is determined by training two adversarial neural networks (known as the generator and the discriminator). \citet{liu2018deep} and \citet{liu2019multilevel} extended principal component analysis (PCA) based representations to a CNN-PCA procedure. This approach applied the `fast neural style transfer' algorithm \citep{johnson2016perceptual} to represent complex geological models characterized by multipoint spatial statistics, and was shown to enable more efficient data assimilation. \citet{zhu2018bayesian} formulated surrogate modeling as an image-to-image regression, and constructed a Bayesian deep convolutional neural network for geological uncertainty quantification. Subsequently, \citet{mo2019deep} extended this model to handle multiphase flow problems, and further improved performance by introducing additional physical constraints. 
%(i.e., binary cross entropy loss for saturation fields) to the system. 

Recent developments involving the use of deep-learning techniques in ROMs indicate great potential for such approaches. \citet{lee2018model} introduced an improved GNAT procedure by replacing POD with AE. The resulting method was applied to a one-dimensional dynamic Burgers' equation and a two-dimensional quasi-static chemically reacting flow problem, with the boundary conditions in the test runs different from those in the training runs. \citet{kani2019reduced} developed a deep residual recurrent neural network (DR-RNN) procedure, which employed RNN to approximate the low-dimensional residual functions for the governing equations in a POD-DEIM procedure. The resulting ROM was then applied to a one-dimensional oil-water problem with the distribution of porosity in the test runs perturbed from that of the training runs. \citet{zhang2019deep} used a fully-connected network to replace the Newton iterations in a POD-DEIM procedure. The method was used to predict well responses in a two-dimensional oil-water problem, in which combinations of well controls and permeability fields for test runs were different from those of the training simulations. Though improvements in accuracy were achieved by all of the above approaches relative to the `standard' implementations, all of these developments were within existing ROM settings; i.e., none adopted an end-to-end deep-learning framework.

Other researchers have developed ROM methodologies that represent more of a departure from existing approaches. \citet{wang2018model}, for example, used the long-short-term-memory (LSTM) RNN \citep{gers1999learning} to approximate flow dynamics in a low-dimensional subspace constructed by POD. Subsequently, \citet{gonzalez2018deep} replaced the POD step with VAE \citep{kingma2013auto} for the low-dimensional representation. Both of these approaches, however, were applied on relatively simple problems, where the only differences between online and offline simulation runs were the initial conditions of the systems (boundary conditions were identical). In the subsurface flow equations, wells appear as localized source/sink terms, which essentially act as `internal' boundary conditions. The ability to vary well settings between offline and online computations is an essential feature for ROMs used in oil production optimization and related areas. Thus the above implementations may not be directly applicable for these problems. Another potential limitation is that these procedures are purely data driven and do not take the underlying governing equations into consideration. This could result in solutions that are visually appealing but physically unrealistic.

%It appears to us that there is a need for a deep-learning-based ROM for subsurface flow that includes physical constraints and can handle varying well-control settings. 
A number of methods have been applied to incorporate physical constraints into deep neural networks. These procedures have different names but often share the same key ideas. \citet{raissi2019physics} introduced a physics-informed deep learning framework (later referred to as physics-informed neural network or PINN) that used densely connected feed-forward neural networks. In PINN, the residual functions associated with the governing partial differential equations (PDEs) are introduced into the loss function of the neural network. \citet{zhu2019physics} extended this PDE-constraint concept to a deep flow-based generative model (GLOW~\citep{kingma2018glow}), and constructed a surrogate model for uncertainty quantification using residuals of the governing equations rather than simulation outputs. \citet{watter2015embed} proposed an embed-to-control (E2C) framework, in the context of robotic planning systems, to predict the evolution of system states using direct sensory data (images) and time-varying controls as inputs. The E2C framework combines a VAE, which is used as both an inference model to project the system states to a low-dimensional subspace, and a generative model to reconstruct the prediction results at full order, with a linear transition model. The latter approximates the evolution of low-dimensional states based on the time-varying control inputs.

%The preliminary application of the E2C model in oil production problems is introduced by \citet{temirchev2019deep}. However, without physical constraints from PDEs, the resulting ROM performed poorly with relative error on the target quantity over 25\%.

In this paper, we develop a deep-learning framework for reduced-order modeling of subsurface flow systems based on the E2C model \citep{watter2015embed} and the aforementioned physics-informed treatments \citep{raissi2019physics, zhu2019physics}. Two key modifications of the existing E2C model are introduced. Specifically, we simplify the VAE to an AE to achieve better accuracy for deterministic test cases, and we incorporate a comprehensive loss function that introduces both PDE-based physical constraints and enforces consistency in well data. The latter treatment is important for improving the accuracy of well quantities of interest, which are essential in oil production optimization procedures. Since we are considering a supervised learning problem with labeled data (input and output pairs), the way we introduce physical constraints distinguishes our model from those of \citet{raissi2019physics} and \citet{ zhu2019physics}, where access to the PDE residuals is required to compute the loss function during the training process. Our treatment may be more appropriate in many practical settings, where the residual values of the underlying PDEs may not be accessible. This could be the case, for example, if a commercial simulator is used. Interestingly, our E2C procedure is quite analogous to existing POD-TPWL methodologies, and we discuss the relationships between the two approaches in some detail.

This paper proceeds as follows. In Section~\ref{sec::method}, we present the governing equations for subsurface oil-water flow and then briefly describe the POD-TPWL ROM. In Section~\ref{sec::e2c}, the E2C formulation is presented, and the correspondences between E2C and POD-TPWL are highlighted. We present results for a two-dimensional oil-water problem in Section~\ref{results}. Test cases involve the specification of different time-varying well settings, as would be encountered in an optimization problem. We also present a detailed error assessment for several key quantities. We conclude with a summary and suggestions for future work in Section~\ref{conclusion}. The detailed architectures for the encoder and decoder used in the E2C model are provided in \ref{appendix-e2c}. Additional simulation results with the deep-learning-based ROM are presented in \ref{appendix-case}.

% -------------------------------------------------------------
% -------------------------------------------------------------
\section{Governing equations and POD-TPWL ROM}\label{sec::method}
% -------------------------------------------------------------
% -------------------------------------------------------------

In this section, we present the equations for oil-water flow. We then provide an overview of the POD-TPWL ROM for this problem, which will allow us to draw analogies with the E2C ROM.

%Following the ones in \citep{he2011enhanced, he2015constraint, jin2018reduced, jin2019reduced}, a quick overview of the formulation for POD-TPWL is presented here to draw analogies with embed-to-control, which will be presented in the following section. 

% -------------------------------------------------------------
\subsection{Governing equations}\label{subsec::gov_equ}
% -------------------------------------------------------------

%We simulate oil-water flow using the existing Stanford's Automatic Differentiation-based General Purpose Research Simulator, AD-GPRS \citep{zhou2012parallel}. 

The governing equations for immiscible oil-water flow derive from mass conservation for each component combined with Darcy's law for each phase. The resulting equations, 
%for two-dimensional horizontal systems 
with capillary pressure effects neglected, are
\begin{equation}\label{equ::gov}
    \frac{\partial}{\partial{t}}(\phi S_{j}\rho_{j})\ - \nabla\cdot(\lambda_{j}\rho_{j}\mathbf{k}\nabla p) + \sum_{w}\rho_{j}q^{w}_{j} = 0,
\end{equation}
where subscript $j$ ($j = o, w$ for oil and water) denotes fluid phase. The geological characterization is represented in Eq.~\ref{equ::gov} through porosity $\phi$ and the permeability tensor $\mathbf{k}$, while the interactions between rock and fluids are specified by the phase mobilities $\lambda_{j}$, where $\lambda_{j}=k_{rj}/\mu_j$, with $k_{rj}$ the relative permeability of phase $j$ and $\mu_j$ the viscosity of phase $j$. Other variables are pressure $p$ and phase saturation $S_j$ (these are the primary solution variables), time $t$, and phase density $\rho_{j}$. The $q^{w}_{j}$ term denotes the phase source/sink term for well $w$. This oil-water model is completed by enforcing the saturation constraint $S_{o}+S_{w} =1$. Because the system considered in this work is horizontal (in the $x-y$ plane), gravity effects are neglected.

The oil and water flow equations are discretized using a standard finite-volume formulation, and their solutions are computed for each grid block. In this work, we use Stanford's Automatic Differentiation-based General Purpose Research Simulator, AD-GPRS \citep{zhou2012parallel}, for all flow simulations. Let $n_b$ denote the number of grid blocks in the model. The flow system is fully defined through the use of two primary variables, $p$ and $S_{w}$, in each grid block, so the total number of variables in the system is $2n_b$. We define $\mathbf{x}_{t} = [\mathbf{p}_{t}^{\rm T}, \mathbf{S}_{t}^{\rm T}]^{\rm T}\in\mathbb{R}^{2n_b}$ to be the state vector for the flow variables at a specific time step $t$, where $\mathbf{p}_{t}\in\mathbb{R}^{n_b}$ and $\mathbf{S}_{t}\in\mathbb{R}^{n_b}$ denote the pressure and saturation in every grid block at time step $t$. 

The set of nonlinear algebraic equations representing the discretized fully-implicit system can be expressed as:
\begin{equation}\label{equ::disc_gov}
    \mathbf{g}\big(\mathbf{x}_{t+1},\mathbf{x}_{t},\mathbf{u}_{t+1}\big) = \mathbf{0},
\end{equation}
where $\mathbf{g} \in \mathbb{R}^{2n_b}$ is the residual vector (set of nonlinear algebraic equations) we seek to drive to zero, the subscript $t$ indicates the current time level and $t+1$ the next time level, and $\mathbf{u}_{t+1}\in\mathbb{R}^{n_{w}}$ designates the well control variables, which can be any combination of bottom-hole pressures (BHPs) or well rates. Here $n_w$ denotes the number of wells in the system. In this work we operate production wells under BHP specifications and injection wells under rate specifications. Our treatments are general in this regard, and other control settings could also be applied.

Newton's method is typically used to solve the full-order discretized nonlinear system defined by Eq.~\ref{equ::disc_gov}. This requires constructing the sparse Jacobian matrix of dimension $2n_b \times 2n_b$, and then solving a linear system of dimension $2n_b$, at each iteration for every time step. Solution of the linear system is often the most time-consuming part of the simulation. As will be explained later, both POD-TPWL and the deep-learning-based E2C ROM avoid the test-time construction and solution of this high-dimensional system.

% -------------------------------------------------------------
\subsection{POD-TPWL formulation}
% -------------------------------------------------------------
%%%%%%%%%%%%%%%%%%%%%%%%%%%%%%%%%%%%%%%%%%%%%%%%%%%%%%%%%%
% 3 paragraphs making the analogy between POD-TPWL and E2C
%%%%%%%%%%%%%%%%%%%%%%%%%%%%%%%%%%%%%%%%%%%%%%%%%%%%%%%%%%

Many deep-learning-based models involve treatments that are not directly analogous to those used in existing ROMs. Rather, they entail machine-learning methods that derive from image classification, language recognition, or other non-PDE-based applications. Our E2C ROM is somewhat different in this sense, as its three main components are directly analogous to those used in an existing ROM, POD-TPWL. Because POD-TPWL has been extensively developed for subsurface flow applications, we believe it is worthwhile to discuss the correspondences between the POD-TPWL and E2C ROMs. To enable this discussion, we first provide a high-level overview of POD-TPWL for reservoir simulation. For full details on recent POD-TPWL implementations, please consult \citep{he2014reduced, he2015constraint, jin2018reduced, jin2019reduced}.

As discussed earlier, POD-TPWL and other POD-based ROMs involve an offline (train-time) stage and an online (test-time) stage. During the offline stage, a number of training simulation runs are performed using a full-order simulator (AD-GPRS in this work). The goal here is to predict test-time results with varying well control sequences. Therefore, during training runs, we apply different well control sequences $\mathbf{U} = [\mathbf{u}_{1}, \dots, \mathbf{u}_{N_{\text{ctrl}}}]\in\mathbb{R}^{n_{w}\times N_{\text{ctrl}}}$, where $\mathbf{u}_{k}\in\mathbb{R}^{n_w}$, $k=1, \dots, N_{\text{ctrl}}$, contains the settings for all wells at control step $k$, and $N_{\text{ctrl}}$ denotes the total number of control steps in a training run. There are many fewer control steps than time steps in a typical simulation (in our examples we have 20 control steps and around 100 time steps).
%are differs from (usually less than) that of the time steps for each training simulation ($N_{\text{tr}}$), since controls may remain constant for multiple time steps. For instance, a simulation of 2000 days with 100 time steps (20 days per time step) may have 20 control steps (100 day per control step). Therefore a control will remain unchanged for 5 time steps. 
State variables in all grid blocks (referred to as snapshots) and derivative matrices are saved at each time step in the training runs. At test-time, simulations with control sequences that are different from those of the training runs are performed. Information saved from the training runs is used to (very efficiently) approximate test solutions.

POD-TPWL entails (1) projection from a high-dimensional space to a low-dimensional subspace, (2) linear approximation of the dynamics in the low-dimensional subspace, and (3) projection back to the high-dimensional space. A projection matrix $\boldsymbol{\Phi}\in\mathbb{R}^{2n_b\times l_{\xi}}$ is constructed based on the singular value decomposition (SVD) of the solution snapshot matrices (these snapshot matrices contain full-order solutions at all time steps in all training runs). Given $\boldsymbol{\Phi}$, the high-dimensional states $\mathbf{x}\in\mathbb{R}^{2n_b}$ can be represented in terms of the low-dimensional variable $\boldsymbol{\xi}\in\mathbb{R}^{l_{\xi}}$ using
\begin{equation}\label{equ::pod}
    \mathbf{x} \approx \boldsymbol{\Phi}\boldsymbol{\xi},
\end{equation}
where $l_{\xi}$ is the dimension of the reduced space, with $l_{\xi} \ll n_b$. Note that in practice, the SVD and subsequent projections are performed separately for the pressure and saturation variables. Because $\boldsymbol{\Phi}$ is orthonormal, we also have $\boldsymbol{\xi}=\boldsymbol{\Phi}^T\mathbf{x}$.

Before discussing the POD-TPWL approximation in low-dimensional space, we first show the linearization in high dimension. Following \citep{he2015constraint}, the TPWL formulation (with the POD representation for states, $\mathbf{x}=\boldsymbol{\Phi}\boldsymbol{\xi}$, applied to the right-hand side) can be expressed as
\begin{equation}\label{equ::tpwl_full}
    \mathbf{J}_{i+1}\hat{\mathbf{x}}_{t+1} = \mathbf{J}_{i+1}\boldsymbol{\Phi}\boldsymbol{\xi}_{i+1} - [\mathbf{A}_{i+1}\boldsymbol{\Phi}(\boldsymbol{\xi}_{t} - \boldsymbol{\xi}_{i}) + \mathbf{B}_{i+1}(\mathbf{u}_{t+1} - \mathbf{u}_{i+1})],
\end{equation}
where
\begin{equation}\label{eq:jab}
\mathbf{J}_{i+1} = \frac{\partial \mathbf{g}^{i+1}}{\partial \mathbf{x}^{i+1}} \in \mathbb{R}^{2n_b \times 2n_b}, \quad
\mathbf{A}_{i+1} = \frac{\partial \mathbf{g}^{i+1}}{\partial \mathbf{x}^{i}} \in \mathbb{R}^{2n_b \times 2n_b}, \quad
\mathbf{B}_{i+1} = \frac{\partial \mathbf{g}^{i+1}}{\partial \mathbf{u}^{i+1}} \in \mathbb{R}^{2n_b \times n_w}.
\end{equation}
Here the subscripts $t$ and $t+1$ denote time steps in the test run, while the subscripts $i$ and $i+1$ designate time steps in the training simulations. Note that Eq.~\ref{equ::tpwl_full} differs slightly from the expressions in \citep{he2015constraint} since the time step designations are now subscripted, for consistency with the embed-to-control equations shown later. The variable $\boldsymbol{\xi}_{t}$ is the projection of true (high-order) solution of Eq.~\ref{equ::disc_gov} at time step $t$. The variable $\hat{\mathbf{x}}_{t+1}\in\mathbb{R}^{2n_b}$ is distinct from $\mathbf{x}_{t+1}$, in that it represents the full-order variable at time step $t+1$ approximated through linearization instead of via solution of the full-order system (Eq.~\ref{equ::disc_gov}). From here on, we will use variables without `hats' to denote the true high-order solution (e.g., $\mathbf{x}$) or the true solution projected with matrix $\boldsymbol{\Phi}$ (e.g., $\boldsymbol{\xi}=\boldsymbol{\Phi}^T\mathbf{x}$). And, we will use variables with `hats' ($\hat{\mathbf{x}}$ and $\hat{\boldsymbol{\xi}}$) to designate solutions approximated (either reconstructed or predicted, as will be explained in detail later) by the ROM. The variables $\mathbf{u}_{t+1}, \mathbf{u}_{i+1} \in \mathbb{R}^{n_w}$ are the well settings at time step $t+1$ and $i+1$ --- these are prescribed by the user or specified by an optimization algorithm.

Applying the POD representation on the left-hand side and constraint reduction (projection) on both sides of Eq.~\ref{equ::tpwl_full}, the solution approximation in low-dimensional space, after some rearrangement, is given by
\begin{equation}\label{equ::tpwl_reduced}
    \hat{\boldsymbol{\xi}}_{t+1} = \boldsymbol{\xi}_{i+1} - (\mathbf{J}^{r}_{i+1})^{-1}[\mathbf{A}^{r}_{i+1}(\boldsymbol{\xi}_{t} - \boldsymbol{\xi}_{i}) + \mathbf{B}^{r}_{i+1}(\mathbf{u}_{t+1} - \mathbf{u}_{i+1})],
\end{equation}
with the reduced derivative matrices defined as
\begin{equation}\label{equ:reduced_matrices}
\mathbf{J}_{i+1}^{r} = (\mathbf{\Psi}_{i+1})^{T}\mathbf{J}_{i+1}\mathbf{\Phi}, \quad \mathbf{A}_{i+1}^{r} = (\mathbf{\Psi}_{i+1})^{T}\mathbf{A}_{i+1}\mathbf{\Phi}, \quad \mathbf{B}_{i+1}^{r} = (\mathbf{\Psi}_{i+1})^{T}\mathbf{B}_{i+1}.
\end{equation}
Here $\mathbf{J}_{i+1}^{r}\in\mathbb{R}^{l_{\xi}\times l_{\xi}}$, $\mathbf{A}_{i+1}^{r}\in\mathbb{R}^{l_{\xi}\times l_{\xi}}$ and $\mathbf{B}_{i+1}^{r}\in\mathbb{R}^{l_{\xi}\times n_w}$. The matrix $\boldsymbol{\Psi}_{i+1}$ denotes the constraint reduction matrix at time step $i+1$. The variable $\hat{\boldsymbol{\xi}}_{t+1}\in\mathbb{R}^{l_{\xi}}$ represents the reduced variable approximated through linearization at time step $t+1$.

%is distinct from $\boldsymbol{\xi}_{t+1}$ to  instead of the projection of the true solution.

During the online stage (test-time), we do not know $\boldsymbol{\xi}_{t}$ (the projected true solution of Eq.~\ref{equ::disc_gov} at time step $t$). Rather, we have $\hat{\boldsymbol{\xi}}_{t}$, the reduced variable approximated through linearization at time step $t$ (computed from Eq.~\ref{equ::tpwl_reduced} at the previous time step). Therefore, at test-time, Eq.~\ref{equ::tpwl_reduced} becomes
\begin{equation}\label{equ::tpwl}
    \hat{\boldsymbol{\xi}}_{t+1} = \boldsymbol{\xi}_{i+1} - (\mathbf{J}^{r}_{i+1})^{-1}[\mathbf{A}^{r}_{i+1}(\hat{\boldsymbol{\xi}}_{t} - \boldsymbol{\xi}_{i}) + \mathbf{B}^{r}_{i+1}(\mathbf{u}_{t+1} - \mathbf{u}_{i+1})].
\end{equation}
Note that $\hat{\boldsymbol{\xi}}_{t}$ now appears on the right-hand side instead of $\boldsymbol{\xi}_{t}$. At test-time, the training `point,' around which linearization is performed (this point defines $i$ and $i+1$), is determined using a `point-selection' procedure. This point selection depends on $\hat{\boldsymbol{\xi}}_{t}$ (see \citep{he2011enhanced, jin2018reduced} for details), so the reduced derivative matrices $\mathbf{J}^{r}_{i+1}$, $\mathbf{A}^{r}_{i+1}$ and $\mathbf{B}^{r}_{i+1}$ can all be considered to be functions of $\hat{\boldsymbol{\xi}}_{t}$. In the last step of POD-TPWL, the approximated solutions are projected back to the full-order space through application of $\hat{\mathbf{x}} = \boldsymbol{\Phi}\hat{\boldsymbol{\xi}}$, where $\hat{\boldsymbol{\xi}}\in\mathbb{R}^{l_{\xi}}$ is the approximated test state in the subspace, and the variable $\hat{\mathbf{x}}\in\mathbb{R}^{2n_{b}}$ is the corresponding approximated state in the full-dimensional space. 

%More details on POD-TPWL workflow should be consulted in \citep{he2011enhanced, he2015constraint, jin2018reduced, jin2019reduced}.

Each of the above-mentioned steps in POD-TPWL can be viewed as an optimization (in some cases heuristic), as we now consider. The projection matrix $\boldsymbol{\Phi}$ is constructed using the POD procedure. This has the property that the resulting basis matrix minimizes a projection error $e_{\text{proj}}$, defined as
\begin{equation}\label{equ::proj_error}
    e_{\text{proj}} = \|\mathbf{x}-\boldsymbol{\Phi}\boldsymbol{\Phi}^{\rm T}\mathbf{x}\|_{2}^{2},
\end{equation}
where $\mathbf{x}\in\mathbb{R}^{2n_b}$ is the full-order state variable.

In addition, as discussed by \citet{he2015constraint}, the constraint reduction error can be defined as
\begin{equation}\label{equ::cons_red_error}
    e_{\text{cr}} = \|\hat{\mathbf{x}}-\boldsymbol{\Phi}\hat{\boldsymbol{\xi}}\|_{\boldsymbol{\Theta}}^{2},
\end{equation}
where $\hat{\mathbf{x}}$ corresponds to the solution $\hat{\mathbf{x}}_{t+1}$ in Eq.~\ref{equ::tpwl_full} (before constraint reduction is applied); this variable was denoted as $\mathbf{x}_2$ in \citep{he2015constraint}. The variable $\hat{\boldsymbol{\xi}}$ corresponds to the solution $\hat{\boldsymbol{\xi}}_{t+1}$ in Eq.~\ref{equ::tpwl_reduced} (after constraint reduction is applied) and was expressed as $\boldsymbol{\xi}_{3}$ in \citep{he2015constraint}. The notation $\|\cdot\|_{\boldsymbol{\Theta}}$ is a norm defined as $\|\mathbf{e}\|_{\boldsymbol{\Theta}}=\sqrt{\mathbf{e}^{\rm T}\boldsymbol{\Theta}\mathbf{e}}$, with $\mathbf{e}\in\mathbb{R}^{2n_b}$ and $\boldsymbol{\Theta}\in\mathbb{R}^{2n_b\times 2n_b}$, where $\mathbf{\Theta}$ is a symmetric positive definite matrix. The optimal constraint reduction matrix $\boldsymbol{\Psi}$ can be determined by minimizing the constraint reduction error, i.e.,
\begin{equation}\label{equ::opt_e_cr}
    \boldsymbol{\Psi} = \argmin_{\boldsymbol{\Psi}}e_{\text{cr}}.
\end{equation}
If matrix $\boldsymbol{\Theta}$ is defined as $\mathbf{J}^{\rm T}\mathbf{J}$ then, following Eqs.~21 through 27 in \citep{he2015constraint}, we arrive at the least-squares Petrov-Galerkin projection, i.e.,
\begin{equation}\label{equ::petrov-galerkin}
    \boldsymbol{\Psi}^ = \mathbf{J}\boldsymbol{\Phi}.
\end{equation}
This treatment, which as we see is optimal in a particular norm, is now routinely used in POD-TPWL. 

Thus it is evident that, in the POD-TPWL procedure, the low-dimensional state representation and the constraint reduction procedure are based on two distinct optimizations, with (separate) objective functions $e_{\text{proj}}$ and $e_{\text{cr}}$. The remaining aspect of POD-TPWL to be considered is point selection. Different  point-selection strategies have been used for different applications, and these typically include a heuristic component. These procedures do, however, entail the minimization of a `distance' metric, which quantifies the distance (in an application-specific sense) between the current test point and a large set of training-run points. Thus, this step also entails an optimization. These POD-TPWL component optimizations correspond directly to the loss function minimization that will be applied in the embed-to-control framework. A key difference, however, is that in the E2C framework all of the steps are optimized together, rather than separately as in POD-TPWL.

%Therefore, in POD-TPWL procedure, POD and constraint reduction can be viewed as two separated optimization problem with objective functions $e_{\text{proj}}$ and $e_{\text{cr}}$, respectively. The objective (loss) function defined analogous to projection error ($e_{\text{proj}}$) and constraint reduction error ($e_{\text{cr}}$) for the optimization in the embed-to-control framework will be discussed in the following section. 

% -------------------------------------------------------------
% -------------------------------------------------------------
\section{Embed-to-control formulation}\label{sec::e2c}
% -------------------------------------------------------------
% -------------------------------------------------------------

In this section, we develop an embed-to-control ROM that includes physical constraints. Analogies to POD-TPWL are established for the various E2C components. The E2C model presented here generally follows that developed by \citet{watter2015embed}, though several important modifications are introduced, as will be discussed below.

% -------------------------------------------------------------
\subsection{E2C overview}
% -------------------------------------------------------------

The embed-to-control framework entails three processing steps: an encoder or inference model that projects the system variables from a high-dimensional space to a low-dimensional subspace (referred to here as the latent space), a linear transition model that approximates system dynamics in low-dimension, and a decoder or generative model that projects solutions back to high-dimensional (full-order) space. The E2C framework originally proposed by \citet{watter2015embed} used a VAE architecture for both the encoder and decoder procedures, which allowed them to account for uncertainty in  predictions. In the formulation here, the VAE architecture is reduced to an auto-encoder (AE) architecture, since we are considering deterministic systems. We note that the auto-encoder (AE) architecture is commonly used for semantic segmentation \citep{ronneberger2015u}, where each pixel of the image is associated with a class label, and for depth prediction \citep{eigen2014depth}, where the 3D geometry of a scene is inferred from a 2D image. In the context of subsurface flow simulation, AE architectures have been used to construct surrogate simulation models as an image-to-image regression, where the input images are reservoir properties (e.g., permeability fields) and the outputs are state variables  \citep{zhu2018bayesian, mo2019deep}.

\begin{figure}[htbp]
    \centering
    \includegraphics[width=0.9\textwidth]{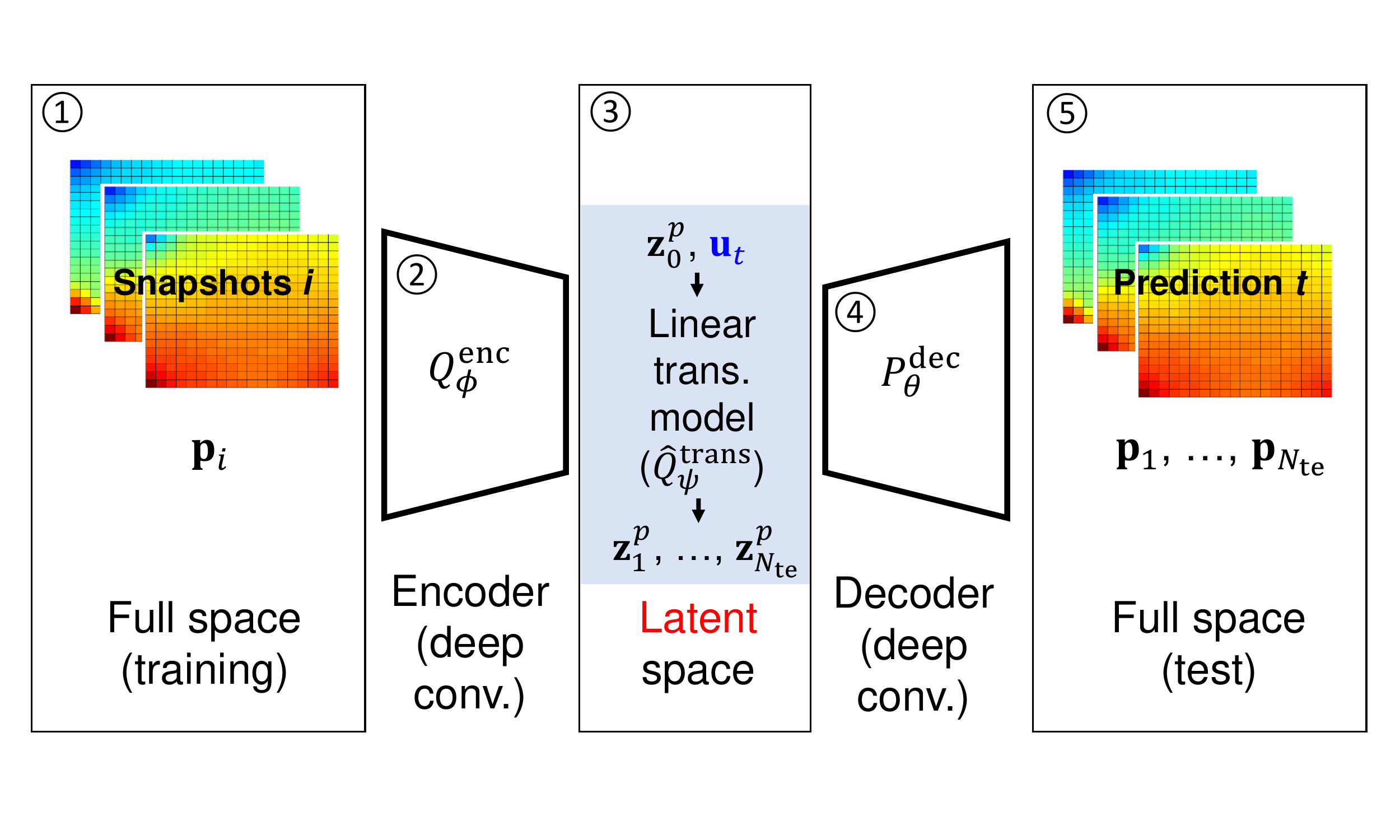}
    \caption{Embed-to-control (E2C) overview}
    \label{fig::e2c_overview}
\end{figure}

Figure~\ref{fig::e2c_overview} displays the overall workflow for our embed-to-control model. The pressure field $\mathbf{p}_{i}\in\mathbb{R}^{n_b}$ is the only state variable shown in this illustration (the subscript $i$, distinct from $t$, denotes the time steps in a training run), though our actual problem also includes the saturation field $\mathbf{S}_{i}\in\mathbb{R}^{n_b}$. Additional state variables would appear in more general settings (e.g., displacements if a coupled flow-geomechanics model is considered).

Box~1 in Fig.~\ref{fig::e2c_overview} displays pressure snapshots $\mathbf{p}_{i}\in\mathbb{R}^{n_b}, i=1,\dots, N_{s}$ in the full-order space, where $N_s$ is the total number of snapshots. 
%Note that $\mathbf{p}_{i}\in\mathbb{R}^{n_b}$ here is a subset of the full set of state variables within $\mathbf{x}_{i}=[\mathbf{p}_{i}^{\rm T}, \mathbf{S}_{i}^{\rm T}]^{\rm T}\in\mathbb{R}^{2n_b}$. 
The notation $Q_{\phi}^{\text{enc}}$ in Funnel~2 denotes the encoder, which projects the full space into a latent space, with $\phi$ representing all of the `learnable' parameters in the encoder. The variable $\mathbf{z}^{p}_{i}\in\mathbb{R}^{l_z}$ in Box~3 is the latent variable for pressure, with $l_z$ the dimension of the latent space. 
%This is analogous to the dimension of the reduced space, $l_{\xi}$, in POD-TPWL framework. 

In Box~3, the test simulation results are approximated in the latent space with a linear transition model. The variable $\mathbf{z}^{p}_{0}\in\mathbb{R}^{l_z}$ denotes the initial latent state for a test run, and $\mathbf{u}_{t} \in\mathbb{R}^{n_w}, t = 1, \dots, N_{\text{ctrl}}$ designates the control sequence for a test run, with $n_w$ the number of wells (as noted previously), the subscript $t$ indicates time step in the test run, and $N_{\text{ctrl}}$ is the number of control steps in the test run. The linear transition model $\hat{Q}_{\psi}^{\text{trans}}$ ($\psi$ denotes the learnable parameters) takes $\mathbf{z}^{p}_{0}\in\mathbb{R}^{l_z}$ and $\mathbf{u}_{t}\in\mathbb{R}^{n_w}$ as input, and outputs $\mathbf{z}^{p}_{t}\in\mathbb{R}^{l_z}, t=1,\dots,N_{\text{te}}$ sequentially, where $N_{\text{te}}$ is the total number of time steps in a test run. The decoder $P_{\theta}^{\text{dec}}$ (indicated by Funnel~4, with $\theta$ representing all of the learnable parameters in the decoder) then projects the variable $\mathbf{z}^{p}_{t}$ back to the full-order state $\mathbf{p}_{t}\in\mathbb{R}^{n_b}$, as shown in Box~5.

The embed-to-control ROM incorporates the control variable $\mathbf{u}_{t}\in\mathbb{R}^{n_w}$ naturally in the framework. This is an important distinction relative to the VAE-LSTM-based ROM developed by \citet{gonzalez2018deep}, where system controls were not included in the model. In the following subsections, the three main components of the embed-to-control framework, the encoder, the linear transition model, and the decoder, will be discussed in detail. A loss function with physical constraints, along with E2C implementation details, will also be presented.

% -------------------------------------------------------------
\subsection{Encoder component}
% -------------------------------------------------------------

The encoder provides a low-dimensional representation of the full-order state variables. In contrast to the original embed-to-control implementation by \citet{watter2015embed}, here we adopt an AE instead of a VAE architecture. With this treatment only the mean values of the latent variables are estimated, not the variances. Also, we do not require a sampling process in the latent space. Therefore, at train-time, the encoder can be simply expressed as
\begin{equation}\label{equ::enc}
    \mathbf{z}_{t} = Q_{\phi}^{\text{enc}}(\mathbf{x}_{t}),
\end{equation}
where $Q_{\phi}^{\text{enc}}$ represents the encoder, as explained previously. The variable $\mathbf{x}_{t}\in\mathbb{R}^{2n_b}$ is the full-order state variable at time step $t$, and $\mathbf{z}_{t}\in\mathbb{R}^{l_z}$ is the corresponding latent variable, with $l_z$ the dimension of the latent space. In the examples presented later, we consider a two-dimensional $60 \times 60$ oil-water model (which means the full-order system is of dimension 7200), and we set $l_z=50$. Note that Eq.~\ref{equ::enc} is analogous to Eq.~\ref{equ::pod} in the POD-TPWL procedure, except the linear projection in POD is replaced by a nonlinear projection $Q^{\text{enc}}_{\phi}$ in the encoder. Following the convention described earlier, we use variables without a `hat' to denote (projected) true solutions of Eq.~\ref{equ::disc_gov}, which are available from training runs. Variables with a hat designate approximate solutions provided by the test-time ROM. 
%Thus, during test-time, the input to Eq.~\ref{equ::enc} becomes $\hat{\mathbf{x}}_{t}\in\mathbb{R}^{2n_b}$, the predicted state variable from the E2C model at the previous time step, and the output becomes $\hat{\mathbf{z}}_{t}\in\mathbb{R}^{l_z}$ (defined correspondingly). 

\begin{figure}[htbp]
    \centering
    \includegraphics[width=0.9\textwidth]{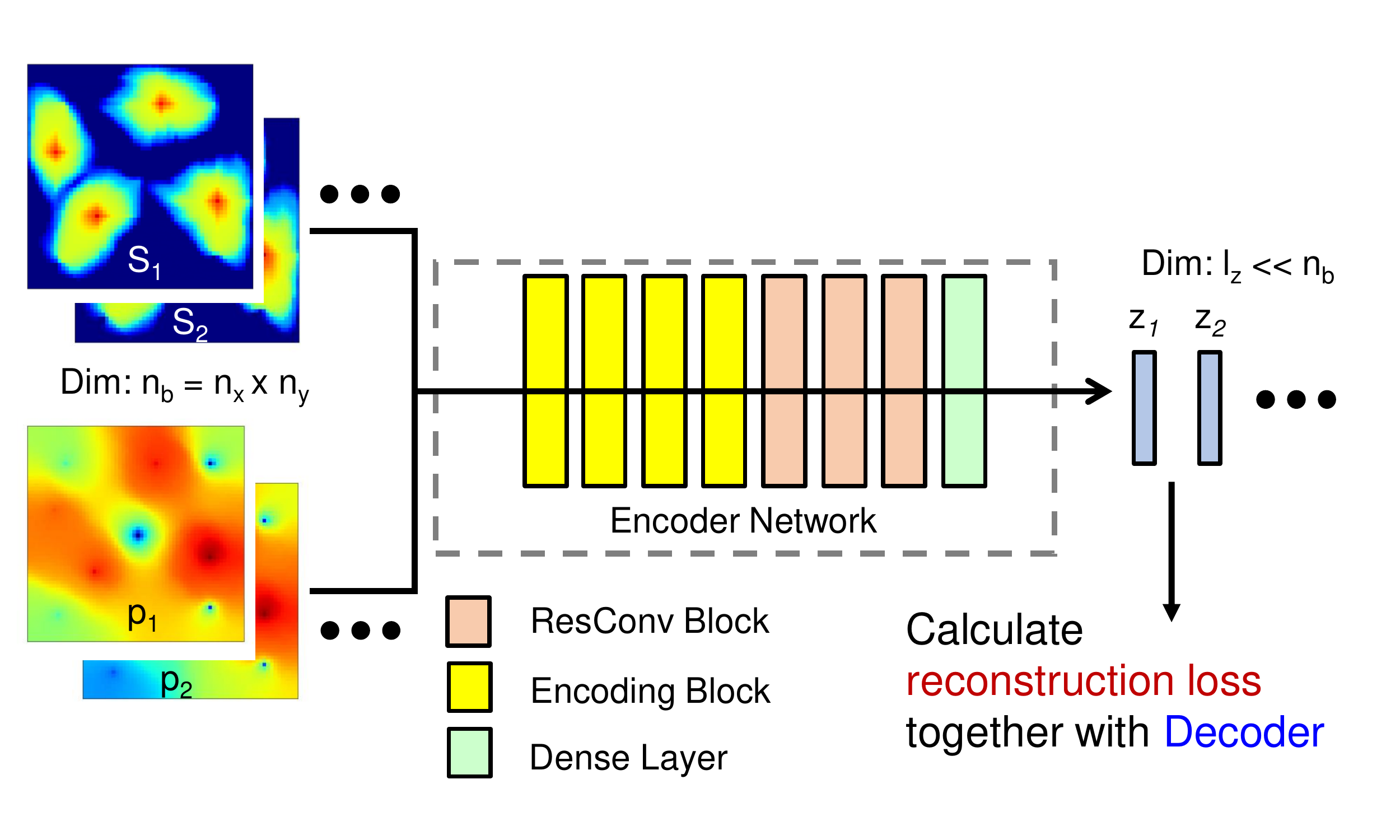}
    \caption{Encoder layout}
    \label{fig::encoder}
\end{figure}

The detailed layout of the encoder in the E2C model is presented in Fig.~\ref{fig::encoder}. During training, sequences of pressure and saturation snapshots are fed through the encoder network, and sequences of latent state variables $\mathbf{z}_t\in\mathbb{R}^{l_z}$ are generated. The encoder network used here is comprised of a stack of four encoding blocks, a stack of three residual convolutional (resConv) blocks, and a dense layer. The encoder in Fig.~\ref{fig::encoder} is more complicated (i.e., it contains resConv blocks and has more convolutional layers) compared to those used in \citep{watter2015embed}. A more complicated structure may be needed here because, compared to the prototype planning tasks addressed in \citep{watter2015embed} (e.g, cart-pole balancing, and three-link robotic arm planning), proper representation of PDE-based pressure and saturation fields requires feature maps from a deeper network.

Similar to the CNN-PCA proposed by \citet{liu2018deep}, which uses the filter operations in CNN to capture the spatial correlations that characterize geological features, the embed-to-control framework uses stacks of convolutional filters to represent the spatial distribution of the pressure and saturation fields determined by the underlying governing equations. Earlier implementations with AE/VAE-based ROMs \citep{lee2018model, gonzalez2018deep} have demonstrated the potential of convolutional filters to capture such fields in fluid dynamics problems. Thus, our encoder network is mostly comprised of these convolutional filters (in the form of two-dimensional convolutional layers, i.e., conv2D layer \citep{lecun1998gradient}). More detail on the encoder network is provided in Table~\ref{tab::enc} of \ref{appendix-e2c}.

The input to an encoding block is first fed through a convolution operation, which can also be viewed as a linear filter. Following the expression in \citep{liu2018deep}, the mathematical formulation of linear filtering is
\begin{equation}\label{equ::filter}
    F_{i,j}(\mathbf{x}) = \sum_{p=-n}^{n}\sum_{q=-n}^{n}\mathbf{w}_{p,q}\mathbf{x}_{i+p,j+q} + b,
\end{equation}
where $\mathbf{x}$ is the input state map, subscripts $i$ and $j$ denote $x$ and $y$ coordinate direction indices, $\mathbf{w}$ represents the weights of a linear filter (template) of size $(2n+1)\times(2n+1)$, $F_{i,j}(\mathbf{x})$ designates the filter response map (i.e., feature map) for $\mathbf{x}$ at spatial location $(i,j)$, and $b$ is a scalar parameter referred to as bias. Note that there are typically many filters associated with a conv2D layer, and the filter response map, which collects all of these operations, is thus a third-order tensor. The output filter response maps are then passed through a batch normalization (batchNorm) layer \citep{ioffe2015batch}, which applies normalization operations (shifts the mean to zero and rescales by the standard deviation) for each subset of training data. A batchNorm operation is a crucial step in the efficient training of deep neural networks, since it renders the learning process less sensitive to parameter initialization, which means a larger initial learning rate can be used. The nonlinear activation function ReLU (rectified linear unit, max(0,$x$)) \citep{glorot2011deep} is applied on the normalized filter response maps to give a final response (output) of the encoder block. This nonlinear response is referred to as the `activation' of the encoding block. The conv2D-batchNorm-ReLU architecture (with variation in ordering) is a standard processing step in CNNs. An illustration of the encoding block structure can be found in Fig.~\ref{fig::e2c_blocks}(a) of \ref{appendix-e2c}.

To properly incorporate feature maps capable of representing the spatial pressure and saturation distributions, as determined by the underlying governing equations, a deep neural network with many stacks of convolutional layers is required. Deep neural networks are, however, difficult to train, mostly due to the gradient vanishing issue \citep{glorot2010understanding}. By this we mean that gradients of the loss function with respect to the model parameters (weights of the filters) become vanishingly small, which negatively impacts training. \citet{he2016deep} addressed this issue by creating an additional identity mapping, referred to as resNet, that bypasses the nonlinear layer. Following the idea of resNet, we add a stack of resConv blocks to the encoder network to deepen the network while mitigating the vanishing-gradient issue. The nonlinear layer in the resConv block still generally follows the conv2D-batchNorm-ReLu architecture. See Fig.~\ref{fig::e2c_blocks}(c) in \ref{appendix-e2c} for a visual representation of the resConv block.

Similar to that of the encoding block, the output of resConv blocks is a stack of low-dimension feature maps. This stack of feature maps is `flattened' to a vector (which is still a relatively high-dimensional vector due to the large number of feature maps), and then input to a dense layer. A dense (fully-connected) layer is simply a linear projection that maps a high-dimensional vector to a low-dimensional vector. 

The overall architecture of the encoder network used here differs from that constructed by \citet{zhu2018bayesian} in three key aspects. First, resNet is used in our encoder while they used denseNet \citep{huang2017densely} to mitigate the vanishing-gradient issue. Another key distinction is that the encoder (and the decoder) in \citep{zhu2018bayesian} do not include the dense layer at the end, which means the encoder outputs a stack of feature maps at the end. A large number of feature maps (i.e., a tall but relatively thin third-order tensor) would be too high-dimensional for the sequential linear operations subsequently performed by the linear transition model. Finally, \citet{zhu2018bayesian} adopted a U-Net \citep{ronneberger2015u} architecture, which is reasonable when the output of the encoder-decoder (e.g., pressure field) differs from the input (e.g., permeability map), as was the case in their setting. However, the U-Net architecture is inappropriate and may lead to over-fitting when the inputs and outputs are of the same type (i.e., both are pressure and saturation fields), as is the case here.

The encoder (and decoder) in the embed-to-control ROM is analogous to the POD representation used in POD-TPWL. As noted earlier, the basis matrix $\boldsymbol{\Phi}$ constructed via SVD of the snapshot matrices has the feature that it minimizes $e_{\text{proj}}$ in Eq.~\ref{equ::proj_error}. In the context of the encoder, a reconstruction loss $\mathcal{L}_{\text{R}}$, which is similar to $e_{\text{proj}}$ for POD, is computed. Conceptually, the `best' $Q^{\text{enc}}_{\phi}$ is found by minimizing $\mathcal{L}_{\text{R}}$. However, as mentioned earlier, the optimization applied for the embed-to-control model involves all three processing steps considered together, so $\mathcal{L}_{\text{R}}$ is not minimized separately.

% -------------------------------------------------------------
\subsection{Linear transition model}
% -------------------------------------------------------------

\begin{figure}[htbp]
    \centering
    \includegraphics[width=0.9\textwidth]{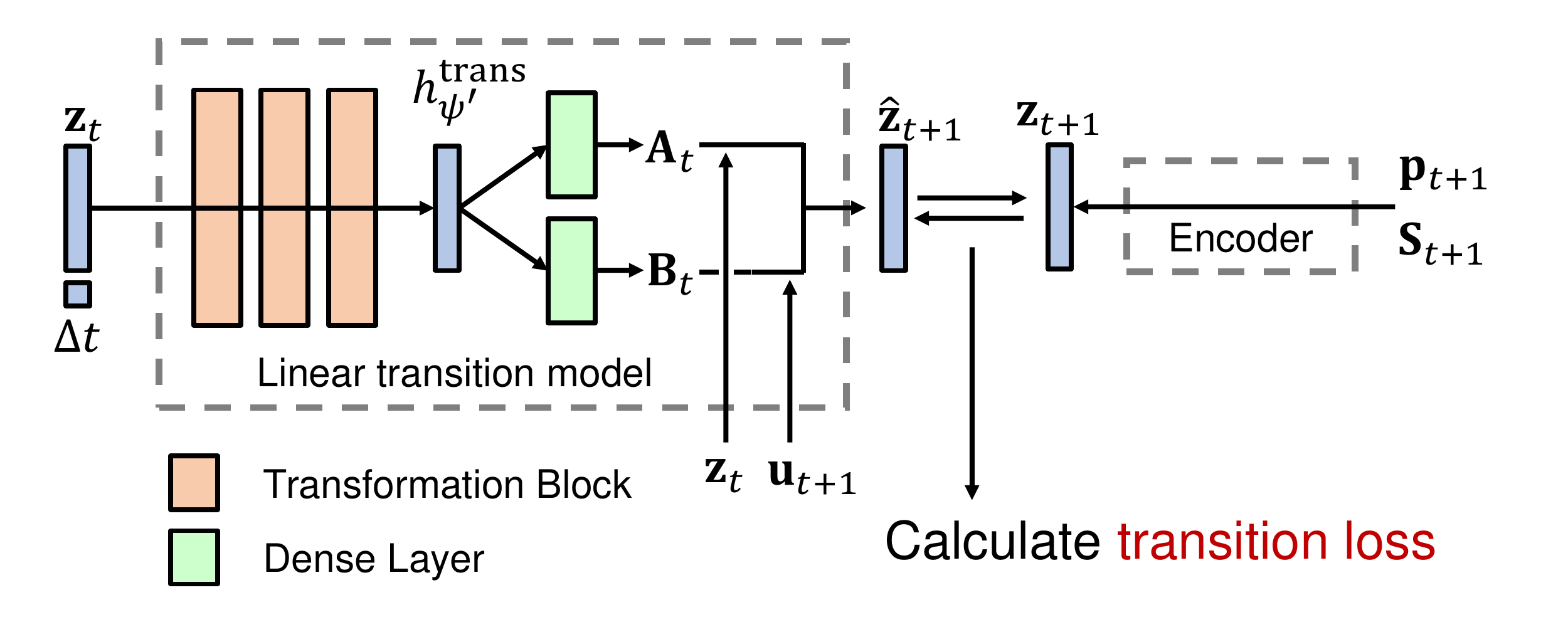}
    \caption{Linear transition model layout}
    \label{fig::linear_trains}
\end{figure}

The linear transition model evolves the latent variable from one time step to the next, given the controls. Fig.~\ref{fig::linear_trains} shows how the linear transition model is constructed and evaluated during the offline stage (train-time). The inputs to the linear transition model include the latent variable for the current state $\mathbf{z}_{t}\in\mathbb{R}^{l_z}$, the current step control $\mathbf{u}_{t+1}\in\mathbb{R}^{n_w}$, and time step size $\Delta t$. The model outputs the predicted latent state for the next time step $\hat{\mathbf{z}}_{t+1}\in\mathbb{R}^{l_z}$. We reiterate that $\hat{\mathbf{z}}_{t+1}$ represents the output of the linear transition model.
%is distinguished from $\mathbf{z}_{t+1}$ to (rather than the state projected by the encoder at time step $t+1$). 
The structure of the linear transition model, which generally follows that in \citep{watter2015embed}, is comprised of a stack of three transformation (trans) blocks and two dense layers. The trans block follows a dense-batchNorm-ReLU architecture (dense represents a dense layer), which is considered a standard processing step for fully-connected networks. The trans block architecture is shown in Fig.~\ref{fig::e2c_blocks}(c) in \ref{appendix-e2c}. The variables $\mathbf{z}_{t}$ and $\Delta t$ are first fed into the trans blocks. The final activation vector of the trans blocks, $h_{\psi_{\prime}}^{\text{trans}}$, is then used to construct the linearization matrices $\mathbf{A}_{t}\in\mathbb{R}^{l_z\times l_z}$ and $\mathbf{B}_{t}\in\mathbb{R}^{l_z\times n_w}$ through two separate dense layers. Matrices $\mathbf{A}_{t}$ and $\mathbf{B}_{t}$ are then combined with the latent variable for the current state $\mathbf{z}_{t}$ and current step control $\mathbf{u}_{t+1}$ to predict the latent variable at the next time step $\hat{\mathbf{z}}_{t+1}$. 

The optimization applied to determine the parameters for the linear transition model is again analogous to a key step in POD-TPWL. In POD-TPWL, the goal is to minimize he difference between the predicted reduced state $\hat{\boldsymbol{\xi}}^{t+1}$ and the projected true state $\boldsymbol{\xi}^{t+1}$. This is achieved, in part, by determining the optimal constraint reduction matrix $\boldsymbol{\Psi}$, as described in Eqs.~\ref{equ::cons_red_error} and \ref{equ::opt_e_cr}. Given this optimal $\boldsymbol{\Psi}$ matrix, the matrices appearing in the POD-TPWL evolution equation (Eq.~\ref{equ::tpwl_reduced}), namely $\mathbf{J}^{r}_{i+1}$, $\mathbf{A}^{r}_{i+1}$ and $\mathbf{B}^{r}_{i+1}$, are all fully defined. As discussed earlier, point-selection represents another (heuristic) optimization that appears in POD-TPWL. Similarly, in the embed-to-control formulation, a transition loss $\mathcal{L}_{\text{T}}$ is computed by comparing $\hat{\mathbf{z}}_{t+1}$ with $\mathbf{z}_{t+1}$, where $\hat{\mathbf{z}}_{t+1}$ is the output from the linear transition model, and $\mathbf{z}_{t+1}$ is the state projected by the encoder at time step $t+1$. The transition loss contributes to the total loss function, which is minimized during the offline stage.

The linear transition model at train-time can also be represented as
\begin{equation}\label{equ::trans1}
    \hat{\mathbf{z}}_{t+1} = \hat{Q}_{\psi}^{\text{trans}}(\mathbf{z}_{t}, \mathbf{u}_{t+1}, \Delta t),
\end{equation}
where $\Delta t$ is the time step size, the function $\hat{Q}_{\psi}^{\text{trans}}$ is the linear transition model as previously defined ($\psi$ denotes all the learnable parameters within the model), and $\hat{\mathbf{z}}_{t+1}\in\mathbb{R}^{l_z}$ denotes the latent variable at $t+1$ predicted by the linear transition model. To be more specific, Eq.~\ref{equ::trans1} can be expressed as
\begin{equation}\label{equ::trans_train}
    \hat{\mathbf{z}}_{t+1} = \mathbf{A}_{t}(\mathbf{z}_{t}, \Delta t)\mathbf{z}_{t} + \mathbf{B}_{t}(\mathbf{z}_{t}, \Delta t)\mathbf{u}_{t+1},
\end{equation}
where $\mathbf{A}_{t}\in\mathbb{R}^{l_z\times l_z}$ and $\mathbf{B}_{t}\in\mathbb{R}^{l_z\times n_w}$ are matrices. Consistent with the expressions in \citep{watter2015embed}, these matrices are given by
\begin{equation}\label{equ::trans_a}
    \text{vec}[\mathbf{A}_{t}] = \mathbf{W}_{A}h_{\psi^{\prime}}^{\text{trans}}(\mathbf{z}_{t}, \Delta t) + \mathbf{b}_{A},
\end{equation}
\begin{equation}\label{equ::trans_b}
    \text{vec}[\mathbf{B}_{t}] = \mathbf{W}_{B}h_{\psi^{\prime}}^{\text{trans}}(\mathbf{z}_{t}, \Delta t) + \mathbf{b}_{B},
\end{equation}
where $\text{vec}$ denotes vectorization, so $\text{vec}[\mathbf{A}_{t}]\in\mathbb{R}^{(l_z^2)\times 1}$ and $\text{vec}[\mathbf{B}_{t}]\in\mathbb{R}^{(l_z n_w)\times 1}$. The variable $h_{\psi^{\prime}}^{\text{trans}}\in\mathbb{R}^{n_{\text{trans}}}$ represents the final activation output after three transformation blocks (which altogether are referred to as the transformation network). The $\psi^{\prime}$ in Eqs.~\ref{equ::trans_a} and \ref{equ::trans_b} is a subset of $\psi$ in Eq.~\ref{equ::trans1}, since the latter also includes parameters outside the transformation network. Here $\mathbf{W}_{A}\in\mathbb{R}^{l_z^2\times n_{\text{trans}}}$, $\mathbf{W}_{B}\in\mathbb{R}^{(l_z n_w) \times n_{\text{trans}}}$, $\mathbf{b}_{A}\in\mathbb{R}^{(l_z^2)\times 1}$, and $\mathbf{b}_{B}\in\mathbb{R}^{(l_z n_w)\times 1}$, where $n_{\text{trans}}$ denotes the dimension of the transformation network. We set $n_{\text{trans}}=200$ in the model tested here.

During the online stage (test-time) the linear transition model is slightly different, since the latent variable fed into the model ($\hat{\mathbf{z}}_{t}\in\mathbb{R}^{l_{z}}$) is predicted from the last time step. Therefore, at test-time, Eq.~\ref{equ::trans_train} becomes
\begin{equation}\label{equ::trans_test}
    \hat{\mathbf{z}}_{t+1} = \mathbf{A}_{t}(\hat{\mathbf{z}}_{t}, \Delta t)\hat{\mathbf{z}}_{t} + \mathbf{B}_{t}(\hat{\mathbf{z}}_{t}, \Delta t)\mathbf{u}_{t+1}.
\end{equation}
Note the only difference is that $\mathbf{z}_{t}$ on the right-hand side of Eq.~\ref{equ::trans_train} is replaced by $\hat{\mathbf{z}}_{t}$ in Eq.~\ref{equ::trans_test}. 

The test-time formulation of the linear transition model is directly analogous to the linear representation step in POD-TPWL. In POD-TPWL, since the training step $i$ (and thus $i+1$) is determined based on the point-selection calculation involving $\hat{\boldsymbol{\xi}}_{t}$, the matrices appearing in the online expression (Eq.~\ref{equ::tpwl}) can be considered to be functions of $\hat{\boldsymbol{\xi}}_{t}$. After some reorganization, Eq.~\ref{equ::tpwl} can then be written as
\begin{equation}\label{equ::tpwl_reorg}
    \hat{\boldsymbol{\xi}}^{t+1} = \mathbf{A}_{t}^{\text{TPWL}}(\hat{\boldsymbol{\xi}}_{t})\hat{\boldsymbol{\xi}}_{t} + \mathbf{B}_{t}^{\text{TPWL}}(\hat{\boldsymbol{\xi}}_{t})\mathbf{u}_{t+1} + \mathbf{c}_{t}^{\text{TPWL}},
\end{equation}
where
\begin{equation}\label{equ::tpwl_reorg_elem}
\begin{aligned}
    &\mathbf{A}_{t}^{\text{TPWL}} =  - (\mathbf{J}_{r}^{i+1})^{-1}\mathbf{A}_{r}^{i+1}, \quad \mathbf{B}_{t}^{\text{TPWL}} =  - (\mathbf{J}_{r}^{i+1})^{-1}\mathbf{U}_{r}^{i+1}, \\ &\mathbf{c}_{t}^{\text{TPWL}} = -\mathbf{A}_{t}^{\text{TPWL}}\boldsymbol{\xi}^{i} - \mathbf{B}_{t}^{\text{TPWL}}\mathbf{u}^{i+1} + \boldsymbol{\xi}^{i+1}.
\end{aligned}
\end{equation}
Thus we see that Eq.~\ref{equ::trans_test} for the online stage of the embed-to-control formulation is of the same form as Eq.~\ref{equ::tpwl_reorg} for the online stage of POD-TPWL. The key difference is that matrices $\mathbf{A}_{t}$ and $\mathbf{B}_{t}$ in E2C are determined by a deep-learning model instead of being constructed from derivative matrices from training runs. Note also that $\mathbf{c}_{t}$ does not appear in the E2C formulation, since this representation does not entail expansion around nearby solutions.

% -------------------------------------------------------------
\subsection{Decoder component}
% -------------------------------------------------------------

The decoder is similar to the encoder and can be represented as
\begin{equation}\label{equ::dec}
    \hat{\mathbf{x}}_{t} = P_{\theta}^{\text{dec}}(\mathbf{z}_{t}),
\end{equation}
where $P_{\theta}^{\text{dec}}$ is the decoder as previously defined. The variable $\hat{\mathbf{x}}_{t}\in\mathbb{R}^{2n_b}$ denotes the reconstructed state variable at time step $t$ (which is distinct from the high-fidelity state variable $\mathbf{x}_{t}\in\mathbb{R}^{2n_b}$ from the training snapshots), though the input to the decoder $\mathbf{z}_{t}\in\mathbb{R}^{l_z}$ is the latent variable determined from the encoding of $\mathbf{x}_{t}$. If the input is instead $\hat{\mathbf{z}}_{t+1}\in\mathbb{R}^{l_z}$, which is the latent variable predicted at time step $t+1$ by the linear transition model, Eq.~\ref{equ::dec} becomes
\begin{equation}\label{equ::dec_t+1}
    \hat{\mathbf{x}}_{t+1} = P_{\theta}^{\text{dec}}(\hat{\mathbf{z}}_{t+1}),
\end{equation}
where $\hat{\mathbf{x}}_{t+1}$ is the predicted state variable at time step $t+1$. Note that Eq.~\ref{equ::dec} only appears in the train-time procedure (to compute reconstructed states), while Eq.~\ref{equ::dec_t+1} has the same form at both train-time and test-time.

\begin{figure}[htbp]
    \centering
    \includegraphics[width=0.9\textwidth]{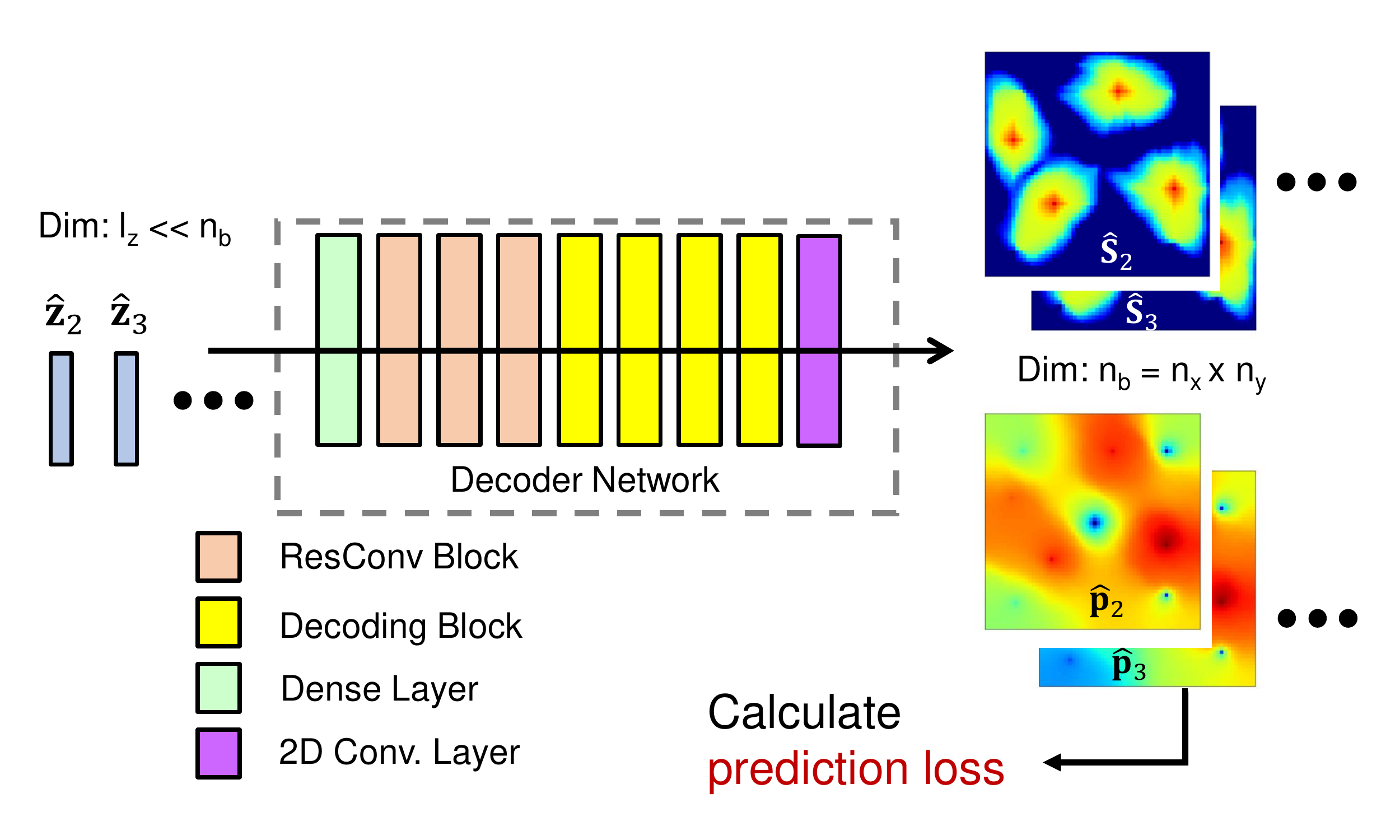}
    \caption{Decoder layout}
    \label{fig::decoder}
\end{figure}

The detailed structure of the decoder is shown in Fig.~\ref{fig::decoder}. Latent variables predicted by the linear transition model (at time step $t+1$) are fed to the decoder network as input, and the predicted high-dimensional states are output. The architecture of the decoder is analogous to that of the encoder except it is in reversed order (which is not surprising since the decoder is conducting the inverse operation). The decoder here is comprised of a dense layer, a stack of three resConv blocks, a stack of four decoding blocks, and a conv2D layer. The dense layer converts a low-dimensional latent vector to a stack of feature maps (after reshaping). The feature maps are expanded while going through stacks of resConv blocks and decoding blocks. The spatial distributions of the pressure and saturation fields are sequentially `extracted' from the feature maps as we proceed downstream in the decoder. The conv2D layer at the end converts the expanded feature maps to pressure and saturation fields as the final outputs. More detail on the encoder is provided in Table~\ref{tab::dec} in \ref{appendix-e2c}. The layout of the decoding block is shown in Fig.~\ref{fig::e2c_blocks}(b) of \ref{appendix-e2c}.

To determine the learnable parameters $\theta$ in the decoder, a prediction loss $\mathcal{L}_{\text{PD}}$ is minimized (along with the other losses) in the offline process. 
%The optimal decoder $P_{\theta}^{\text{dec}}$ is determined by minimizing the prediction loss (along with other losses in the lumped optimization process). 
More details on this optimization will be presented later.

% -------------------------------------------------------------
\subsection{Loss function with physical constraints}
% -------------------------------------------------------------

We have described each of the components of the embed-to-control framework. We now explain how the model parameters are determined during the offline stage. 
%In general, the goal at train-time (offline stage) of a deep-learning-based model is to determine all the parameters that are associated with the model through an optimization workflow that minimizes an objective function.
The parameters for the embed-to-control framework are $\phi$, $\psi$, and $\theta$ for the encoder, linear transition model, and decoder, respectively. The objective function to be minimized is the total loss function that quantifies the overall performance of the model in predicting the output state variables.

We have briefly introduced the reconstruction loss ($\mathcal{L}_{\text{R}}$), the linear transition loss ($\mathcal{L}_{\text{T}}$), and the prediction loss ($\mathcal{L}_{\text{PD}}$), which comprise major components of the total loss function. To be more specific, the reconstruction loss for a training data point $i$ can be expressed as
\begin{equation}\label{equ::recon_loss}
    (\mathcal{L}_{\text{R}})_{i} = \{\| \mathbf{x}_{t} - \hat{\mathbf{x}}_{t}\|_{2}^{2}\}_{i},
\end{equation}
where $i=1,\dots,N_t$, with $N_t$ denoting the total number of data points generated in the training runs. Note that $N_t = N_s - n_{\text{train}}$, where $N_s$ is the total number of snapshots in the training runs and $n_{\text{train}}$ is the number of training simulations performed. Here $N_t$ and $N_s$ differ because, for a training simulation containing $N_{\text{tr}}$ snapshots, only $N_{\text{tr}} - 1$ data points can be collected (since pairs of states, at sequential time steps, are required). The variable $\mathbf{x}_{t}$ is the state variable at time step $t$ from a training simulation, and $\hat{\mathbf{x}}_{t} = P_{\theta}^{\text{dec}}(\mathbf{z}_{t}) = P_{\theta}^{\text{dec}}(Q_{\phi}^{\text{enc}}(\mathbf{x}_{t}))$ denotes the states reconstructed by the encoder and decoder. 

The linear transition loss for training point $i$ is similarly defined as
\begin{equation}\label{equ::lt_loss}
    (\mathcal{L}_{\text{T}})_{i} = \{\| \mathbf{z}_{t+1} - \hat{\mathbf{z}}_{t+1}\|_{2}^{2}\}_{i},
\end{equation}
where $\mathbf{z}_{t+1} = Q_{\phi}^{\text{enc}}(\mathbf{x}_{t+1})$ is the latent variable encoded from the full-order state variable at $t+1$, and the variable $\hat{\mathbf{z}}_{t+1} = \hat{Q}_{\psi}^{\text{trans}}(\mathbf{z}_{t}, \mathbf{u}_{t+1}, \Delta t)$ denotes the latent variable predicted by the linear transition model. Finally, the prediction loss for training point $i$ is defined as
\begin{equation}\label{equ::pred_loss}
    (\mathcal{L}_{\text{PD}})_{i} = \{\| \mathbf{x}_{t+1} - \hat{\mathbf{x}}_{t+1}\|_{2}^{2}\}_{i},
\end{equation}
where $\mathbf{x}_{t+1}$ designates the state variable at time step $t+1$ from the training simulations, and $\hat{\mathbf{x}}_{t+1} = P_{\theta}^{\text{dec}}(\hat{\mathbf{z}}_{t+1})$ represents the full-order state variable predicted by the ROM. The data mismatch loss is the sum of these losses averaged over all training data points,
%
%We now consider the $\mathcal{L}_{2}$ norm regularized mean-square-error (MSE) training loss function. There are two parts in the loss function, a loss for data mismatch $\mathcal{L}_{d}$, and a loss for violation of physical constraints $\mathcal{L}_{p}$. Given the state variable at $t$, $t+1$ are $\mathbf{x}_{t}$, $\mathbf{x}_{t+1}$, and control variable during $[t, t+1]$ is $\mathbf{u}_{t+1}$, the loss for data mismatch can be expressed as
%
\begin{equation}\label{equ::loss_data}
    \mathcal{L}_{d} = \frac{1}{N_t}\sum_{i=1}^{N_t}(\mathcal{L}_{\text{R}})_{i} + (\mathcal{L}_{\text{PD}})_{i} + \lambda(\mathcal{L}_{\text{T}})_{i},
\end{equation}
where $\lambda$ is a weight term.

\begin{figure}[htbp]
  \centering
  \begin{subfigure}{.32\textwidth}
    \centering
    \includegraphics[width=\linewidth]{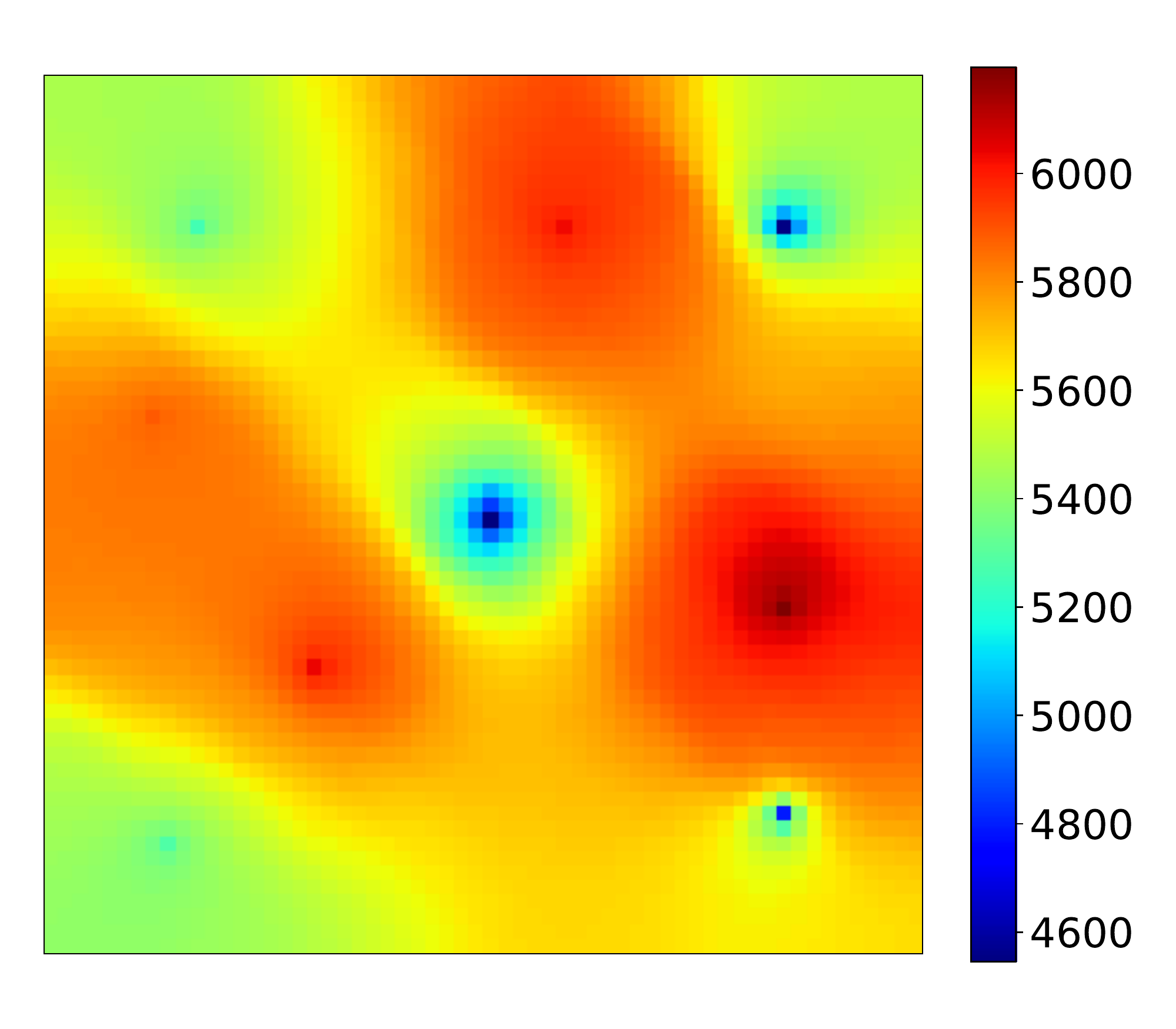}
    \caption{High-fidelity solution \\\hspace{\textwidth} (HFS$_{\text{test}}$)}
  \end{subfigure} \hfill
  \begin{subfigure}{.32\textwidth}
    \centering
    \includegraphics[width=\linewidth]{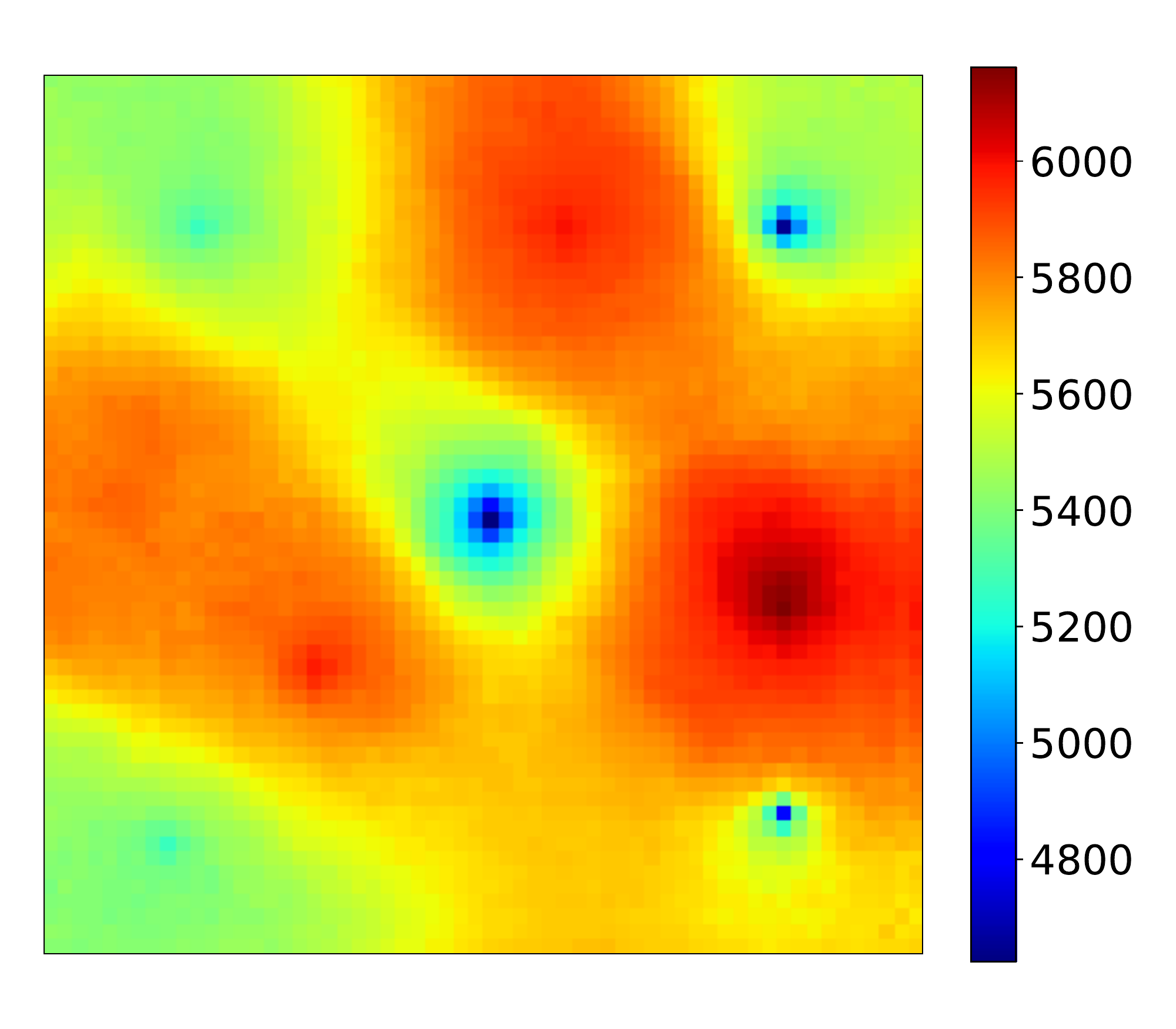}
    \caption{ROM solution \\\hspace{\textwidth} without $\mathcal{L}_{p}$}
  \end{subfigure} \hfill
  \begin{subfigure}{.32\textwidth}
    \centering
    \includegraphics[width=0.95\linewidth]{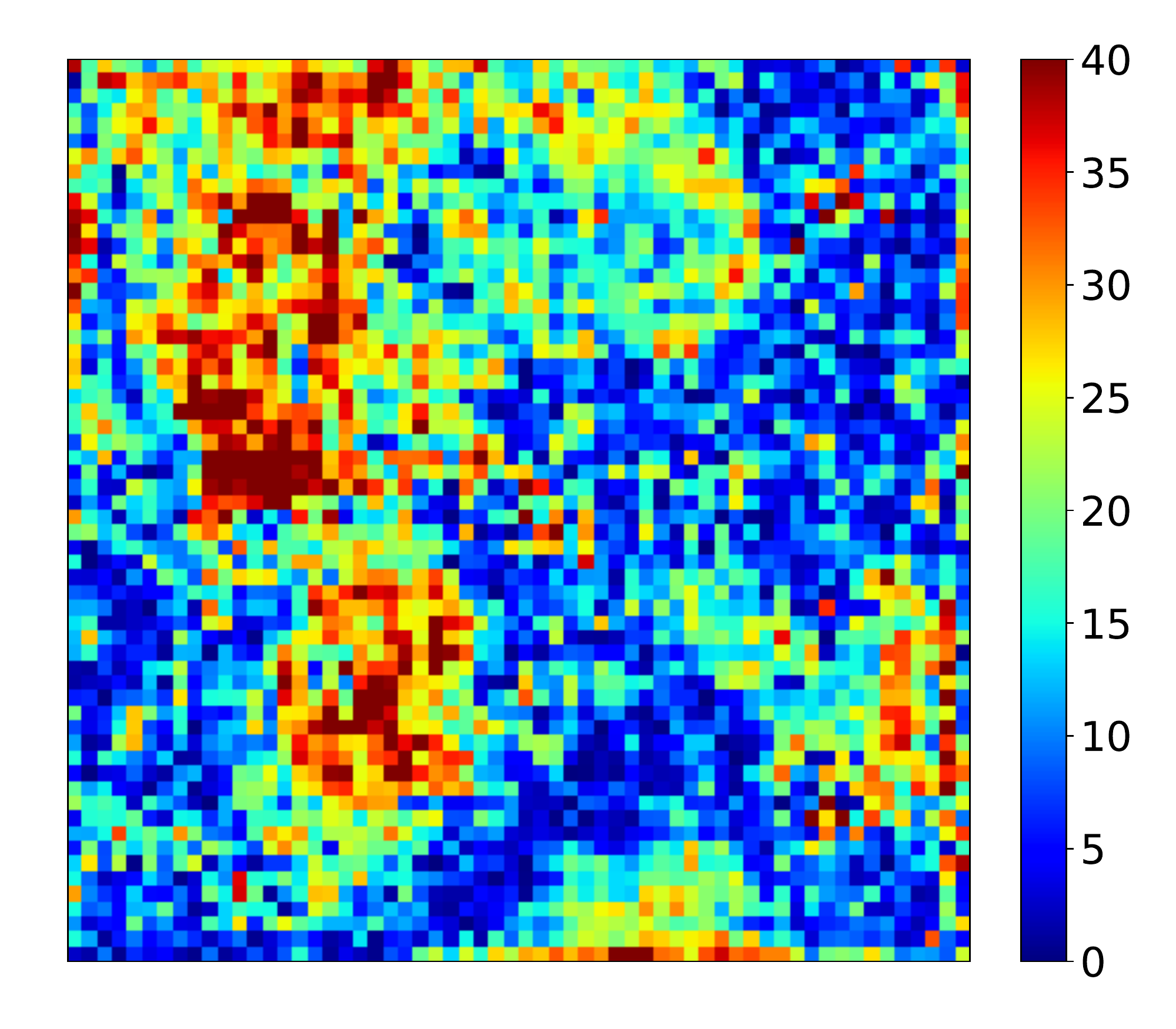}
    \caption{$|\text{HFS}_{\text{test}} - \text{ROM}_{\text{test}}|$ \\\hspace{\textwidth} without $\mathcal{L}_{p}$ (max error 97~psi)}
  \end{subfigure} \\
  \begin{subfigure}{.32\textwidth}
    \centering
    \includegraphics[width=\linewidth]{pres_eval_25_t_1000_true.pdf}
    \caption{High-fidelity solution \\\hspace{\textwidth} (HFS$_{\text{test}}$)}
  \end{subfigure} \hfill
  \begin{subfigure}{.32\textwidth}
    \centering
    \includegraphics[width=\linewidth]{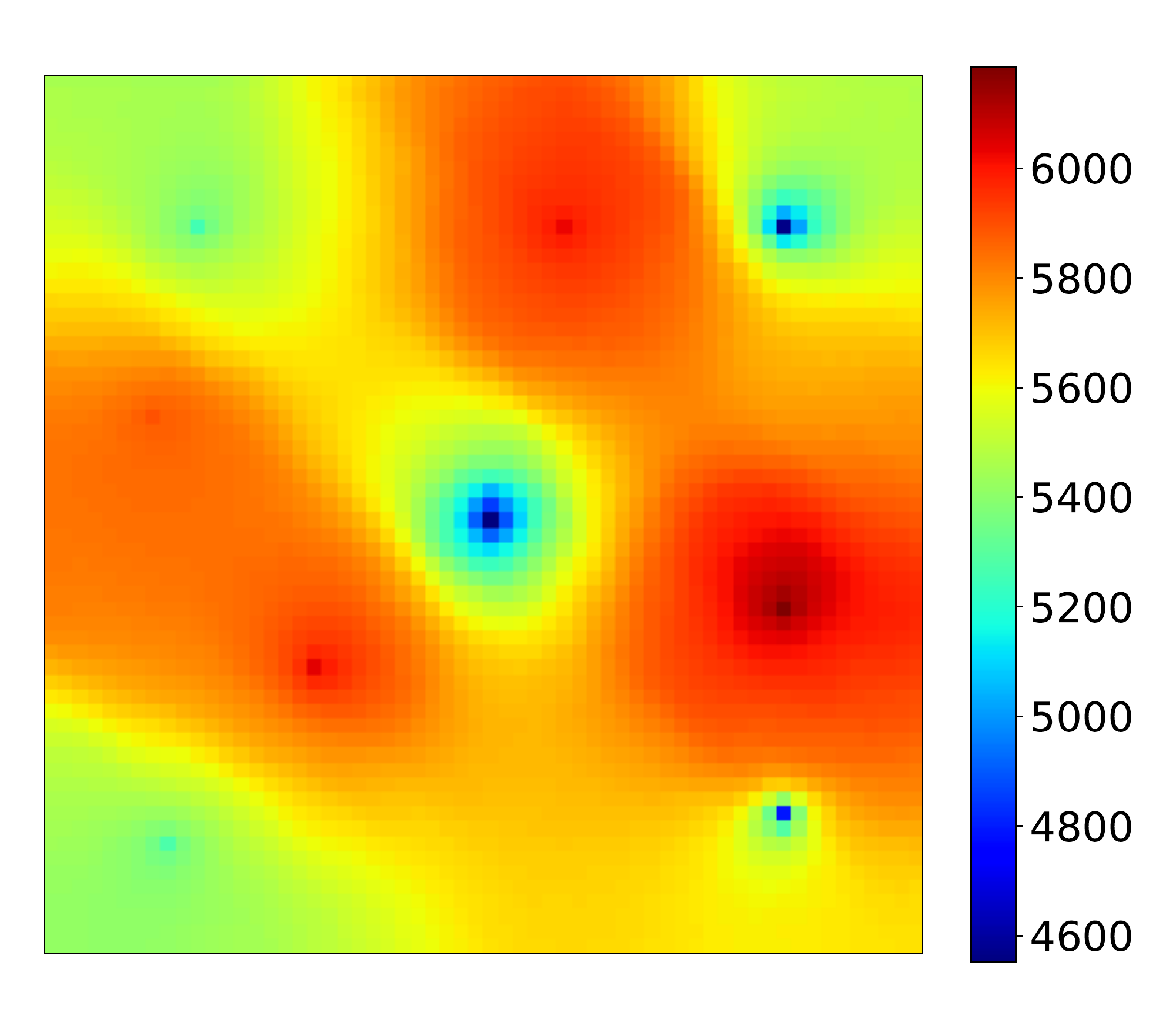}
    \caption{ROM solution \\\hspace{\textwidth} with $\mathcal{L}_{p}$}
  \end{subfigure} \hfill
  \begin{subfigure}{.32\textwidth}
    \centering
    \includegraphics[width=0.95\linewidth]{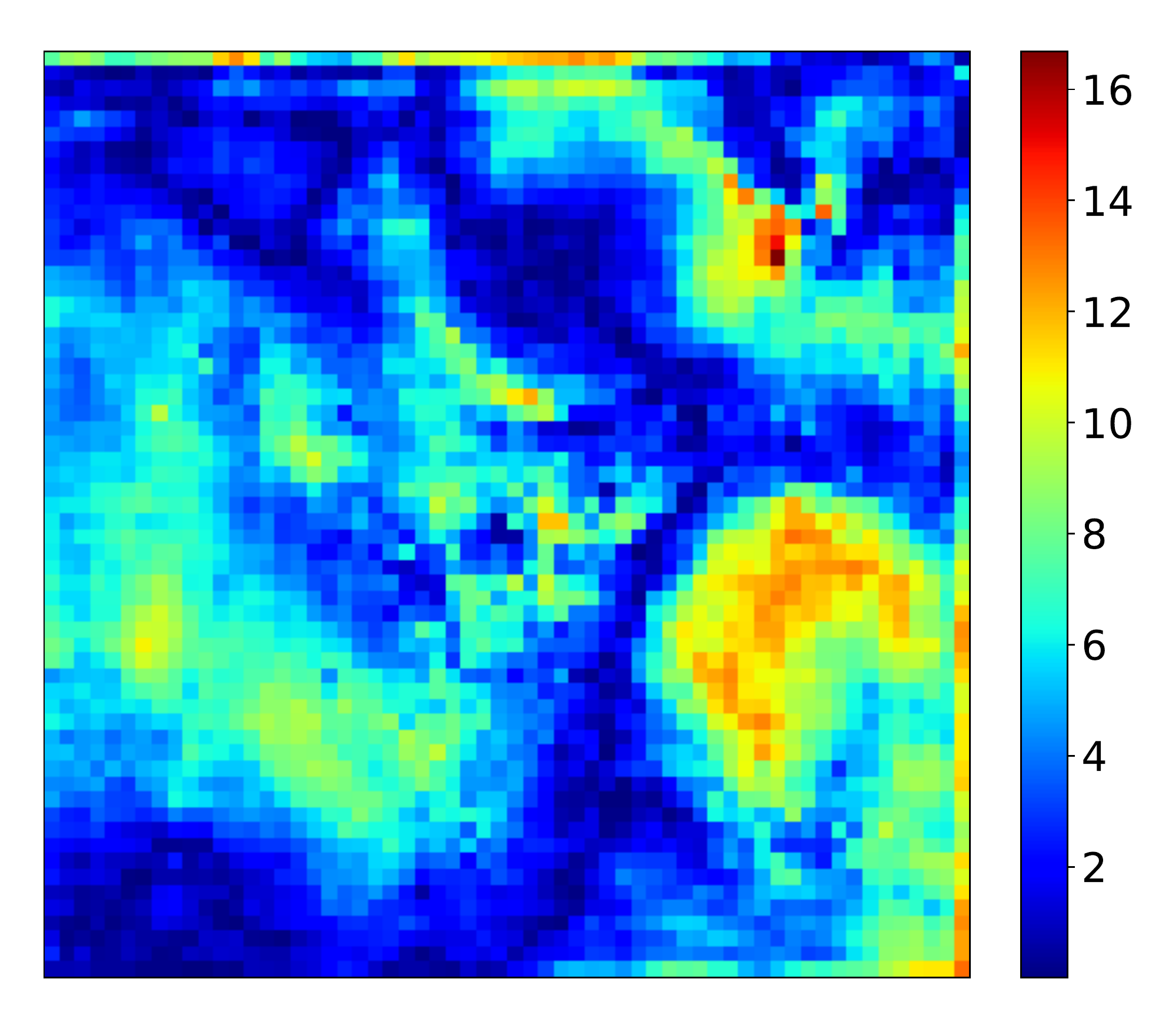}
    \caption{$|\text{HFS}_{\text{test}} - \text{ROM}_{\text{test}}|$ \\\hspace{\textwidth} with $\mathcal{L}_{p}$ (max error 16~psi)}
  \end{subfigure}
  \caption{Pressure field predictions with and without $\mathcal{L}_{p}$ (all colorbars in units of psi)}
  \label{fig::with_no_flux}
\end{figure}

The ROM as described up to this point is a purely data driven model, i.e., the goal of the model is to minimize the pixel-wise difference between the E2C output and the high-fidelity solution (the HFS is taken as the `true' reference solution). Physical behavior is, to some extent, inferred by E2C from the input pressure and saturation snapshots, but it is not explicitly enforced. If the ROM is trained using the loss function $\mathcal{L}_{d}$ given in Eq.~\ref{equ::loss_data}, unphysical effects can, however, be observed. This is illustrated in Fig.~\ref{fig::with_no_flux}, where we show predictions for the pressure field at a particular time (the problem setup will be described in detail in Section~\ref{results}). The  high-fidelity solution is shown in Fig.~\ref{fig::with_no_flux}(a), and the E2C pressure field based solely on $\mathcal{L}_{d}$ appears in Fig.~\ref{fig::with_no_flux}(b). Although the two results are visually similar, the difference map in Fig.~\ref{fig::with_no_flux}(c) indicates that the E2C result is not sufficiently smooth, and relatively large errors appear at some spatial locations. This could have a significant impact on well rate predictions, which are an essential ROM output.

To address this issue, we combine the loss for data mismatch with a loss function based on flow physics. Specifically, we seek to minimize the inconsistency in flux between each pair of adjacent grid blocks. Extra weight is also placed on key well quantities. We consider both reconstruction (at time step $t$) and prediction (at time step $t+1$). Thus we define the physics-based loss for each data point, $(\mathcal{L}_{p})_{i}$, as
\begin{equation}\label{equ::loss_physical}
\begin{aligned}
    (\mathcal{L}_{p})_{i} = 
    & \{\|\mathbf{k}\cdot[(\nabla\mathbf{p}_{t} - \nabla\hat{\mathbf{p}}_{t})_{\text{recon}} + (\nabla\mathbf{p}_{t+1} - \nabla\hat{\mathbf{p}}_{t+1})_{\text{pred}}]\|_{2}^2\}_{i} \\
    & + \gamma\{\|(\mathbf{q}_{t}^{w} - \hat{\mathbf{q}}_{t}^{w})_{\text{recon}} + (\mathbf{q}_{t+1}^{w} - \hat{\mathbf{q}}_{t+1}^{w})_{\text{pred}}\|_{2}^2\}_{i}.
\end{aligned}
\end{equation}
Here $\mathbf{p}_{t},\mathbf{p}_{t+1}\in\mathbb{R}^{n_b}$ are the pressure fields at time steps $t$ and $t+1$ from the training data, which are components of the state variables $\mathbf{x}_{t}$ and $\mathbf{x}_{t+1}$, and $\hat{\mathbf{p}}_{t},\hat{\mathbf{p}}_{t+1}\in\mathbb{R}^{n_b}$ represent the ROM pressure reconstruction (at time step $t$, defined after Eq.~\ref{equ::recon_loss}) and prediction (at time step $t+1$, defined after Eq.~\ref{equ::pred_loss}). The variables $\mathbf{q}^{w}_{t},\mathbf{q}^{w}_{t+1}\in\mathbb{R}^{n_w}$ are well quantities from the training data, and $\hat{\mathbf{q}}^{w}_{t}, \hat{\mathbf{q}}^{w}_{t+1}\in\mathbb{R}^{n_w}$ are well quantities reconstructed (at time step $t$) and predicted (at time step $t+1$) by the ROM. Recall that $n_w$ is the total number of wells. The variable $\gamma$ is a parameter that defines the weights for well-data loss in loss function $\mathcal{L}_{p}$.

The terms on the right hand side of Eq.~\ref{equ::loss_physical} correspond to the flux and source terms in Eq.~\ref{equ::gov}. In the examples in this paper, we specify rates for injection wells and BHPs for production wells. With this specification, the loss on injection rates is zero. The key quantity to track for production wells is the well-block pressure for each well. This is because production rate is proportional to the difference between wellbore pressure (BHP in this case, which is specified) and well-block pressure. The proportionality coefficient is the product of phase mobility $\lambda_j$ and the so-called well index \citep{peaceman1978interpretation}, which depends on permeability, block dimensions and wellbore radius. Because overall well rate in this case is largely impacted by well-block pressure, we set the second term on the right-hand side of Eq.~\ref{equ::loss_physical} to $\gamma^{\prime}\|\mathbf{p}^{w}_{j} - \hat{\mathbf{p}}^{w}_{j}\|_2^2$, where  $\mathbf{p}^{w}_{j}\in\mathbb{R}^{n_p}$ and $\hat{\mathbf{p}}^{w}_{j}\in\mathbb{R}^{n_p}$ ($j=t,t+1$) denote the true and ROM well-block pressures, and $n_p$ is the number of production wells. Here $\gamma^{\prime}$ is a modified weight that accounts for the well index.

%The second term on the right-hand-side of Eq.~\ref{equ::loss_physical} will become $\gamma^{\prime}\|\mathbf{p}^{w}_{j} - \hat{\mathbf{p}}^{w}_{j}\|_2^2$, where the variables $\mathbf{p}^{w}_{j}\in\mathbb{R}^{n_p}$ and $\hat{\mathbf{p}}^{w}_{j}\in\mathbb{R}^{n_p}$ ($j=t,t+1$) depict the true and ROM well-block pressure, and $n_p$ is the number of production wells. Note that the coefficient determines the relation between rate and well-block pressure (i.e., well index in \citep{peaceman1978interpretation}) is absorbed into $\gamma$ to give $\gamma^{\prime}$, and BHPs correspond to $\mathbf{p}^{w}_{j}$ and $\hat{\mathbf{p}}^{w}_{j}$ are the same and thus cancelled.

The physics-based loss function is computed by averaging $(\mathcal{L}_{p})_{i}$ over all data points, i.e.,
\begin{equation}\label{equ::loss_physical_avg}
\mathcal{L}_{p} = \frac{1}{N_t}\sum_{i=1}^{N_t}(\mathcal{L}_{p})_{i}.
\end{equation}
Combining the loss for data mismatch with this physics-based loss, the total loss function becomes
\begin{equation}\label{equ::loss_total}
    \mathcal{L} = \mathcal{L}_{d} + \alpha\mathcal{L}_{p},
\end{equation}
where $\alpha$ is a weight term. Through limited numerical experimentation, we found $\alpha = 0.033$ and $\gamma^{\prime} = 20$ to be appropriate values for these parameters. The E2C ROM prediction for the pressure field at a particular time, using the total loss function $\mathcal{L}$, is shown in Fig.~\ref{fig::with_no_flux}(e). Fig.~\ref{fig::with_no_flux}(d) is again the high-fidelity solution (identical to that in Fig.~\ref{fig::with_no_flux}(a)), and the difference map appears in Fig.~\ref{fig::with_no_flux}(f). We see that the ROM prediction is noticeably improved when $\mathcal{L}_{p}$ is included in the loss function. Specifically, the maximum pressure error is reduced from 97~psi to 16~psi, and the resulting field is smoother (and thus more physical). This demonstrates the benefit of incorporating physics-based losses into the E2C ROM.

% -------------------------------------------------------------
\subsection{E2C implementation and training details}
% -------------------------------------------------------------

To train the E2C model, we use a data set $\mathcal{D} = \{(\mathbf{x}_{t}, \mathbf{x}_{t+1}, \mathbf{u}_{t+1})_{i}\}, i=1, \dots, N_t$, containing full-order states and corresponding well controls, where $N_{t}$ is the total number of training run data points. In the examples in this paper, we simulate a total of 300 training runs. This is many more than are used with POD-TPWL (where we typically simulate three or five training runs), but we expect a much higher degree of robustness with E2C. By this we mean that the ROM is expected to provide accurate results over a large range of control specifications, rather than over a limited range as in POD-TPWL.

Rather than train over all snapshots, here we set $N_{\text{ctrl}} = N_{\text{tr}} = N_{\text{te}} = 20$. This accelerates training and focuses ROM predictions on quantities of interest at time steps when the controls are changing. This results in a total number of data points of $N_{t} = 300\times20=6000$. 

The gradient of the total loss function with respect to the model parameters $(\phi, \psi, \theta)$ is calculated via back-propagation through the embed-to-control framework. The adaptive moment estimation (ADAM) algorithm is used for this optimization, as it has been proven to be effective for optimizing deep neural networks \citep{kingma2014adam}. %ADAM represents one variation from stochastic gradient descent method, where the gradient of the objective function at each iteration is approximated with subsets of training data. 
The rate at which the model parameters are updated at each iteration is controlled by the learning rate $l_r$. Here we set $l_r=10^{-4}$.

Normalization is an important data preprocessing step, and its appropriate application can improve both the learning process and output quality. For saturation  we have $S\in[0, 1]$, so normalization is not required. Pressure and well data, including control variables, are normalized. Normalized rate $q^{0}$, and pressure (both grid-block pressure and BHP) $p^{0}$, are given by
\begin{equation}\label{equ::norm}
q^{0} = \frac{q -q_{\text{min}}}{q_{\text{max}}-q_{\text{min}}}, \quad p^{0} = \frac{p -p_{\text{min}}}{p_{\text{max}}-p_{\text{min}}}.
\end{equation}
Here $q$ denotes simulator rate output in units of m$^3$/day, $q_{\text{max}}$ and $q_{\text{min}}$ are the upper and lower injection-rate bounds, $p$ is either grid-block pressure or production-well BHP (units of psi), $p_{\text{min}}$ is the lower bound on BHP, and $p_{\text{max}}$ is 1.1 times the highest field pressure observed (the factor of 1.1 ensures essentially all data fall within the range).

%prespecified based on the highest field pressure observed and multiplied by $\sim$1.1 to create a safety buffer. Specifically, we set $p_{\text{max}}$ to 6162.5 psi (425 bar) and $p_{\text{min}}$ to 3770 psi (260 bar) for the examples shown in the following sections.

Each full-order training simulation requires about 60~seconds to run on dual Intel Xeon ES-2670 CPUs (24~cores). Our E2C ROM is implemented using Keras \citep{chollet2015keras} with TensorFlow \citep{tensorflow2015_whitepaper} backend. The offline training process (excluding training simulation runtime) takes around 10-12~minutes on a Tesla V100 GPU node (exact timings depend on the memory allocated, which can vary from 8-12~GB). The model is applied on 100 test runs, which will be discussed in detail in the following section. Nearly all of the test results presented are based on the use of 300 training runs, though we also present summary error statistics using 100 and 200 training runs. Offline training for these case requires about the same amount of time as for 300 training runs, except for the direct savings in the full-order training simulations.

% -------------------------------------------------------------
% -------------------------------------------------------------
\section{Results using embed-to-control ROM}\label{results}
% -------------------------------------------------------------
% -------------------------------------------------------------

In this section, we describe the model setup for the oil-water simulations and we present simulation results for the deep-learning-based ROM. One of the test cases is considered in detail in this section; results for two additional test cases are provided in \ref{appendix-case}. In this section we also present summary error results for all 100 test cases.

\subsection{Model setup}

\begin{figure}[htbp]
    \centering
    \includegraphics[width=.5\textwidth]{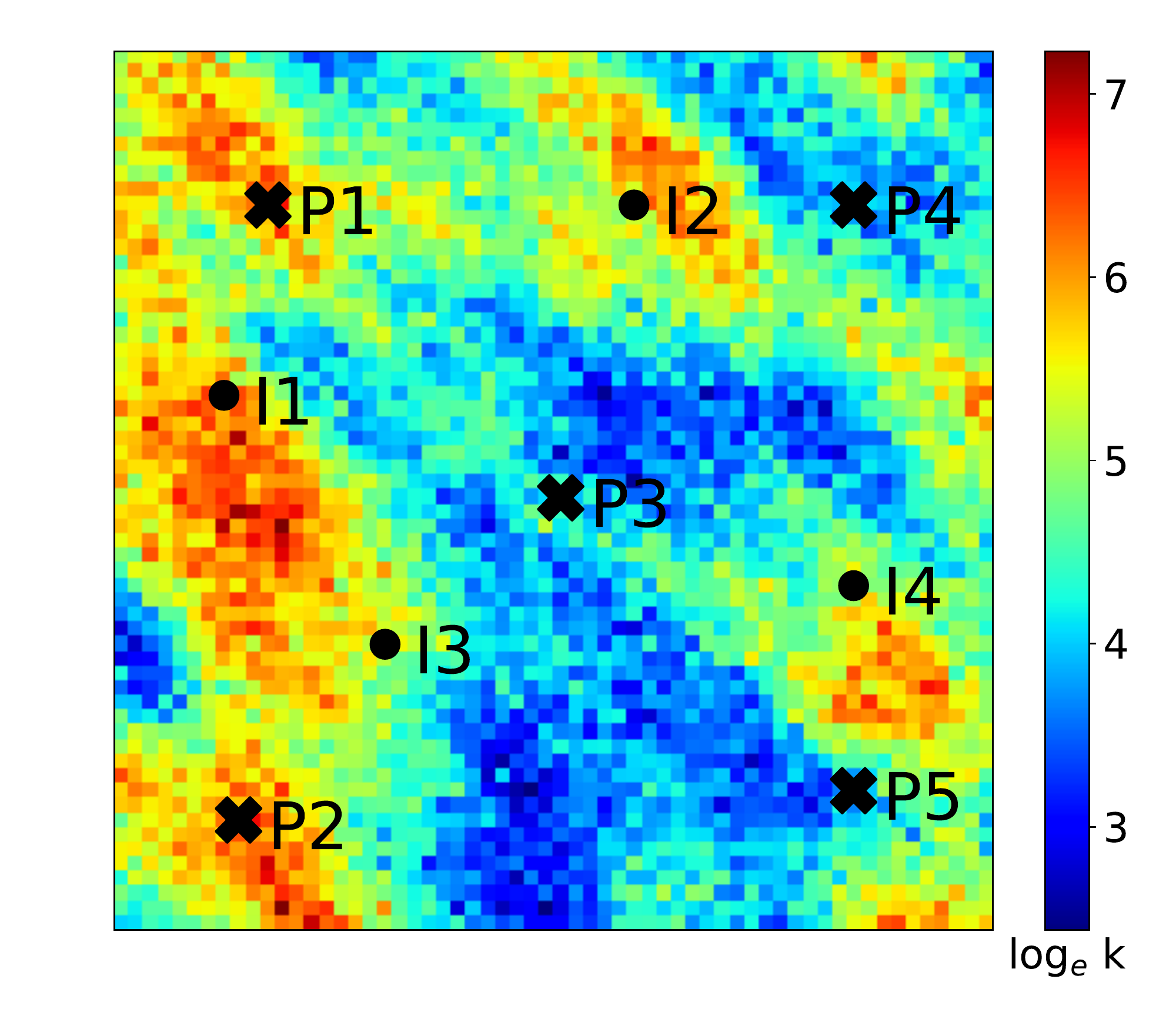}
    \caption{Log-permeability field and well locations}
    \label{fig::perm}
\end{figure}

The geological model, in terms of the log-permeability field, is shown in Fig.~\ref{fig::perm}. The locations of the four injection wells and five production wells are also displayed. The reservoir model contains $60 \times 60$ (total of 3600) grid blocks, with each block of dimensions $50~\text{m} \times 50~\text{m} \times 10~\text{m}$. The correlation structure of the log-permeability field is characterized by an exponential variogram model, with maximum and minimum correlation lengths of $\sim$1000~m and $\sim$500~m, and an azimuth of 45$^{\circ}$. The arithmetic mean permeability is 158~mD, and the standard deviation of log-permeability is 0.88. Permeability is taken to be isotropic, and porosity is set to a constant value of 0.2.

The relative permeability functions are given by
\begin{equation}\label{equ::rel_perm}
k_{ro}(S_w) = k_{ro}^{0}\bigg(\frac{1-S_w -S_{or}}{1-S_{wr}-S_{or}}\bigg)^{a}, \quad k_{rw}(S_w) = k_{rw}^{0}\bigg(\frac{S_w -S_{wr}}{1-S_{wr}-S_{or}}\bigg)^{b},
\end{equation}
where $k_{ro}^{0} = 1.0$, $k_{rw}^{0} = 0.7$, $S_{or} = 0.3$, $S_{wr} = 0.1$, $a =3.6$, and $b = 1.5$. 
Fluid densities are set to $\rho_{o}$ = 800~$\text{kg}/\text{m}^3$ and $\rho_{w}$ = 1000~$\text{kg}/\text{m}^3$, and viscosities are specified as $\mu_{o}$ = 0.91 cp and $\mu_{w}$ = 0.31 cp. Capillary pressure effects are neglected.

The initial pressure at the top of the reservoir is 4712~psi (325~bar), and the initial water saturation is 0.1. The total number of primary variables in the system is $3600 \times 2 = 7200$. The model is run for a total of 2000~days. The injection wells are controlled by specifying time-varying water rates, and the production wells are controlled by specifying time-varying BHPs. The controls for both production wells and injection wells are altered every 100~days, which means there are 20 control periods. Therefore, we have a total of $9 \times 20 = 180$ control parameters over the entire simulation time frame. The range for the injection rates is between 1500 and 6500~bbl/day (between 238 and 1033~m$^3$/day). This is a very large range for well operation compared with what is often considered with ROMs \citep{jin2018reduced, jin2019reduced}. The range for production BHPs is 3770 to 3988~psi (between 260 and 275~bar). 

The controls for the training and test runs are specified as follows. For each injection well, we randomly sample, from a uniform distribution between 2000 and 6000~bbl/day, a baseline injection rate $q_w^{\rm base}$. Then, at each control period, we sample uniformly a perturbation $q_w^\prime$ over the range $[-500, 500]$~bbl/day. The rate for the control period is then prescribed to be $q_w^{\rm base}+q_w^\prime$. Producer BHPs at each control step are sampled uniformly over the range $[3770, 3988]$~psi. For production wells there is not a baseline BHP, and the settings from control step to control step are uncorrelated. This approach for specifying injection rates results in a wide range of solution behaviors (e.g., saturation distributions), since well-by-well injection varies considerably from run to run. This treatment also avoids the averaging effect that can occur if injection rates are not referenced to a baseline value $q_w^{\rm base}$. Well specifications for a test case, generated using this procedure, are shown in Fig.~\ref{fig::test_1_well_ctrl}.

We perform 300 training simulations to construct the E2C ROM, except where otherwise indicated. As discussed in previous papers (e.g., \citep{jin2018reduced}), the types of well schedules shown in Fig.~\ref{fig::test_1_well_ctrl} are intended to represent the well control profiles evaluated during optimization procedures, where the goal is to maximize oil production or profitability, or to minimize environmental impact or some measure of risk. We note finally that the dimension of the E2C latent space, $l_z$, is set to 50. 

%%%%%%%%%%%%%%%%%%%%%%%%%%%%%%%%
%%%%%%%%%%% Case 1 %%%%%%%%%%%%%
%%%%%%%%%%%%%%%%%%%%%%%%%%%%%%%%
\subsection{Results for Test Case~1}

\begin{figure}[htbp]
  \centering
  \begin{subfigure}{.33\textwidth}
    \centering
    \includegraphics[width=\linewidth]{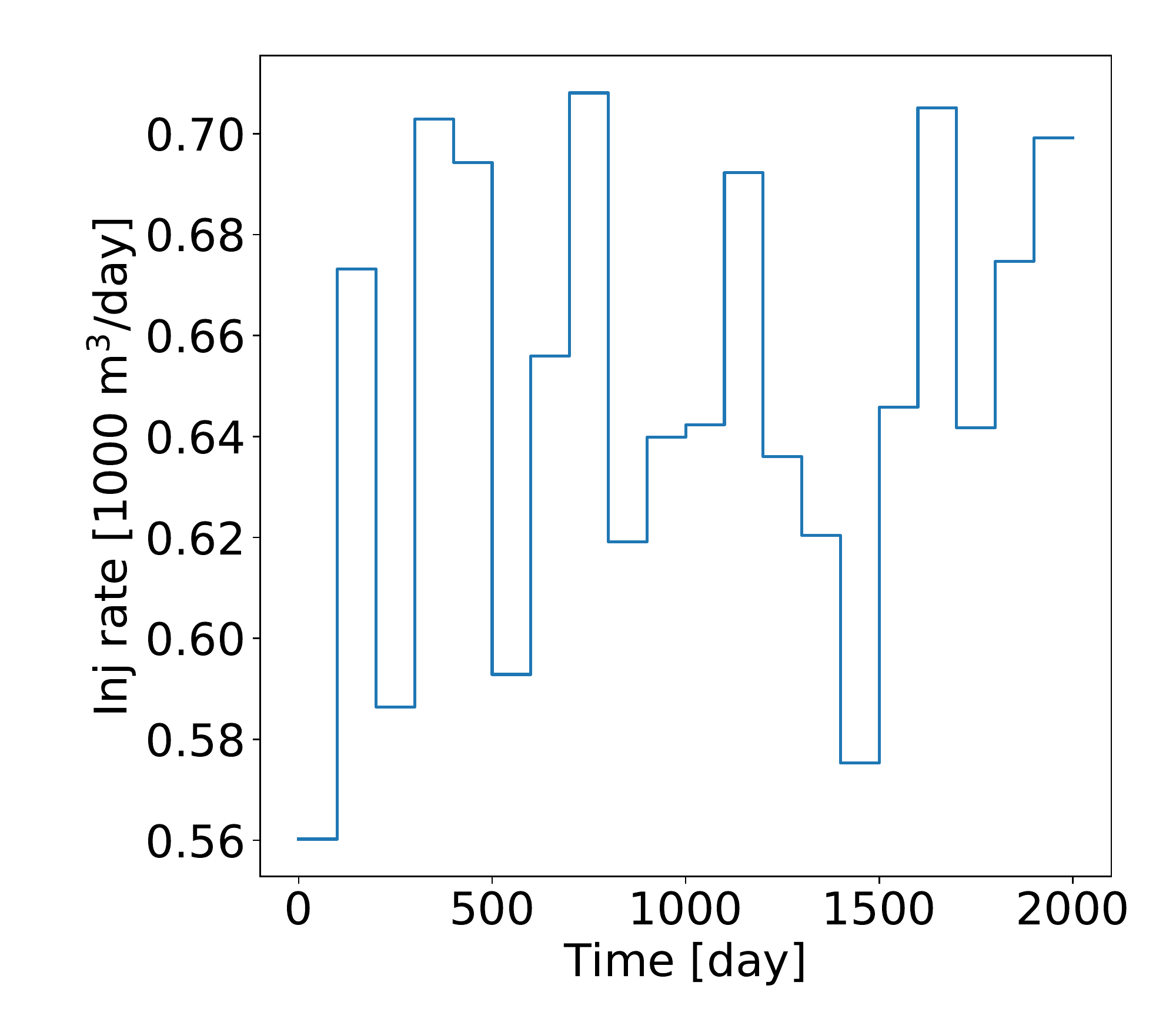}
    \caption{Well~I1}
  \end{subfigure}\hfill
  \begin{subfigure}{.33\textwidth}
    \centering
    \includegraphics[width=\linewidth]{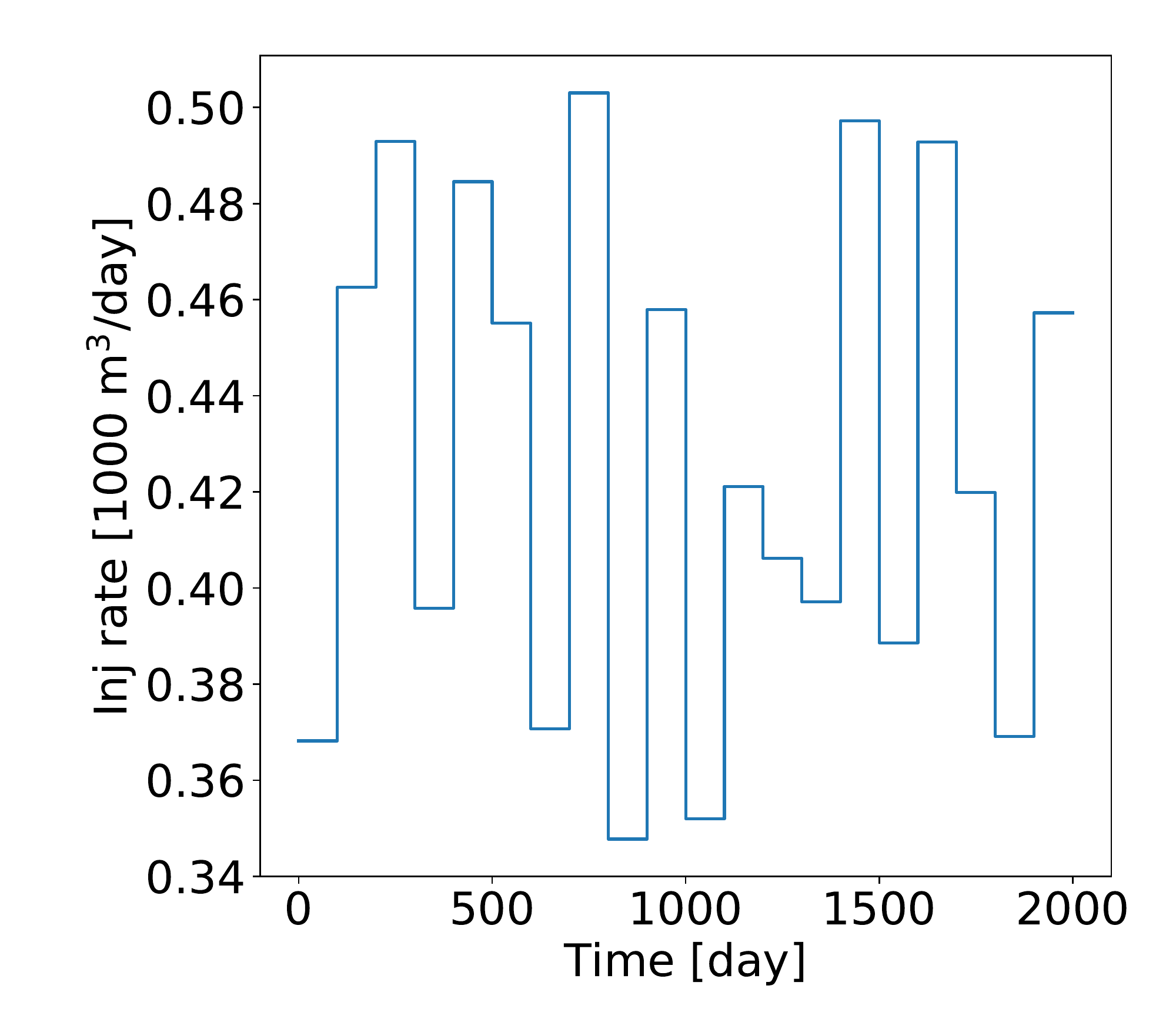}
    \caption{Well~I2}
  \end{subfigure} \hfill
  \begin{subfigure}{.33\textwidth}
    \centering
    \includegraphics[width=\linewidth]{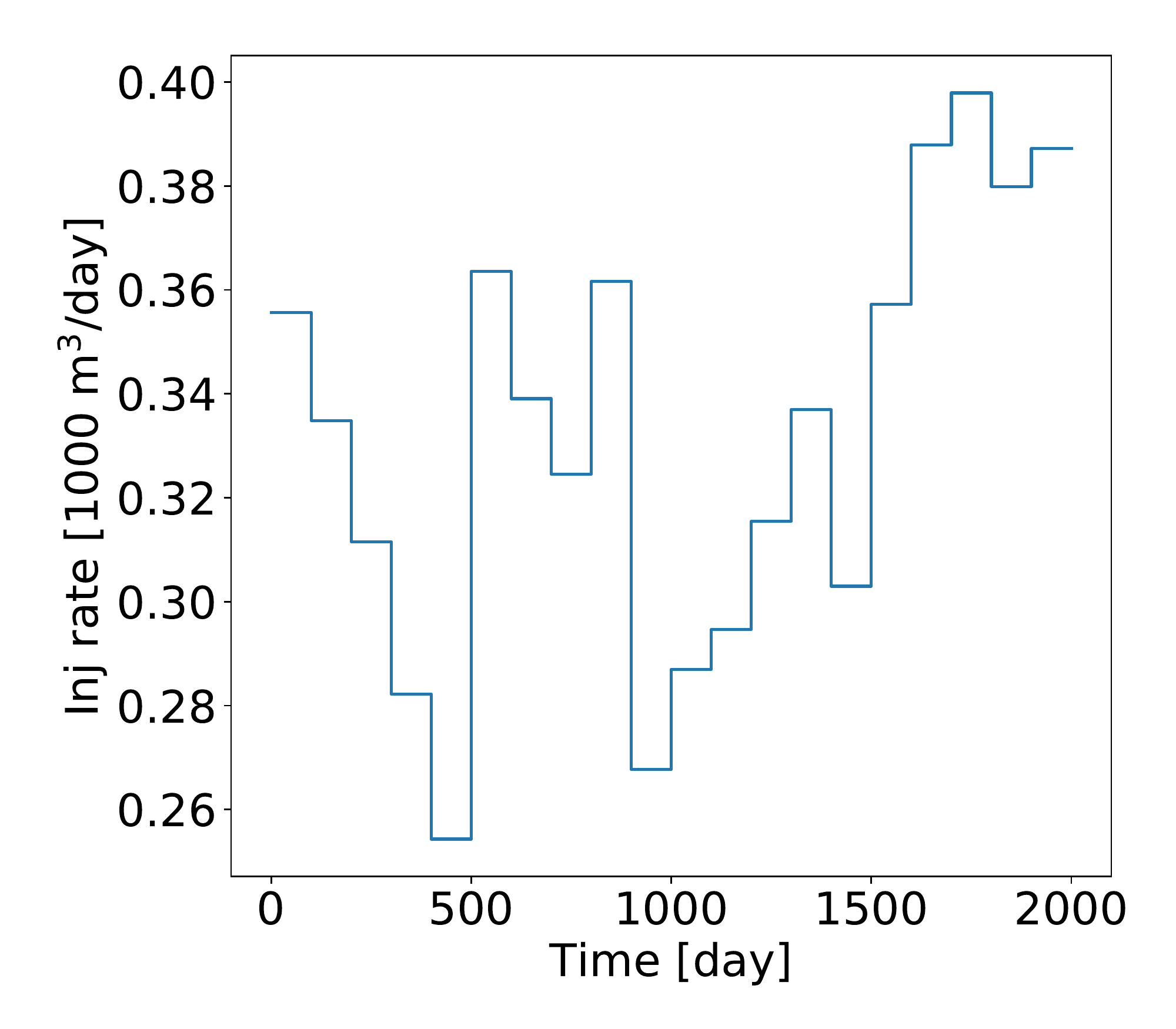}
    \caption{Well~I3}
  \end{subfigure} \\
  \begin{subfigure}{.33\textwidth}
    \centering
    \includegraphics[width=\linewidth]{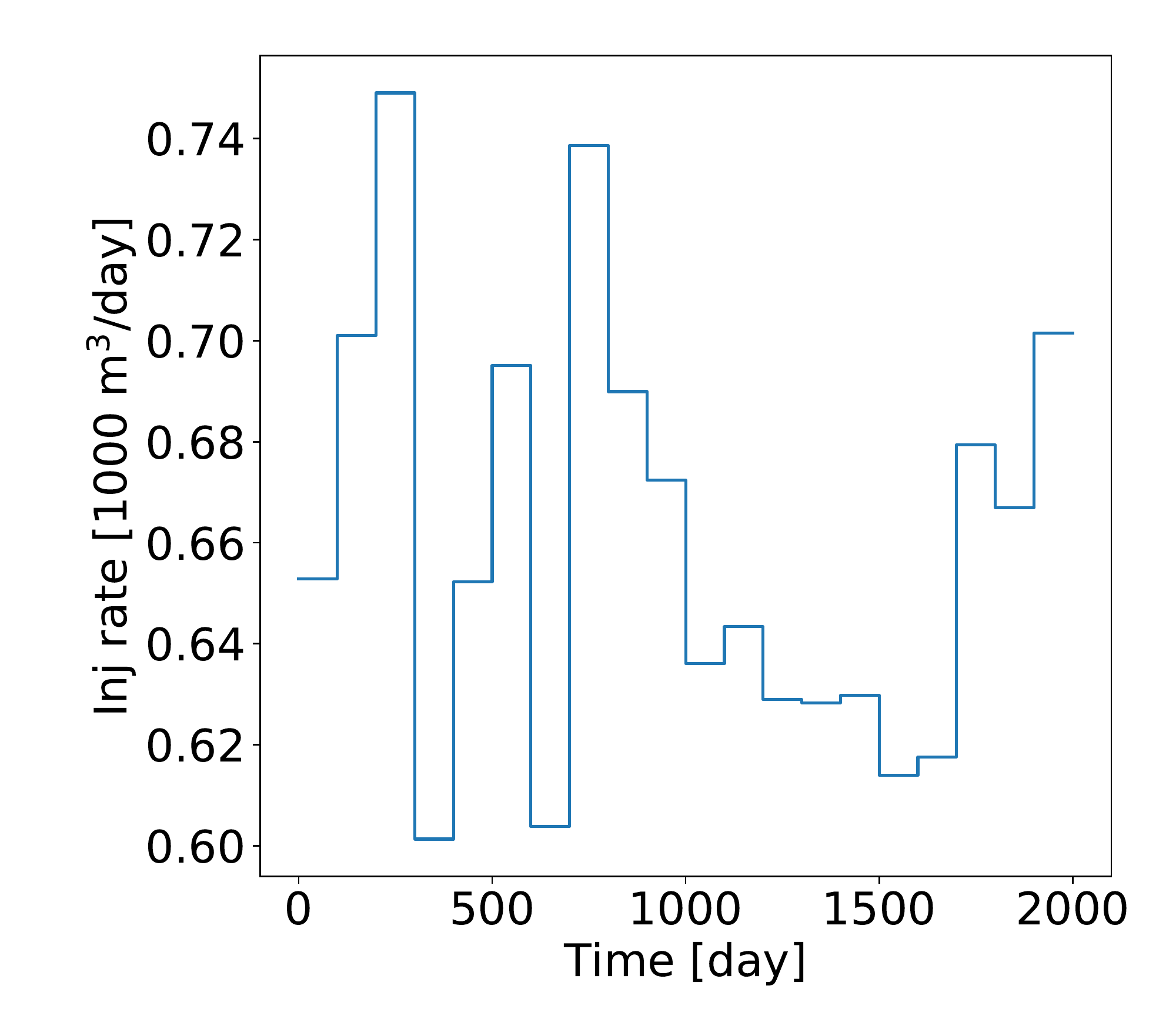}
    \caption{Well~I4}
  \end{subfigure}\hfill
  \begin{subfigure}{.33\textwidth}
    \centering
    \includegraphics[width=\linewidth]{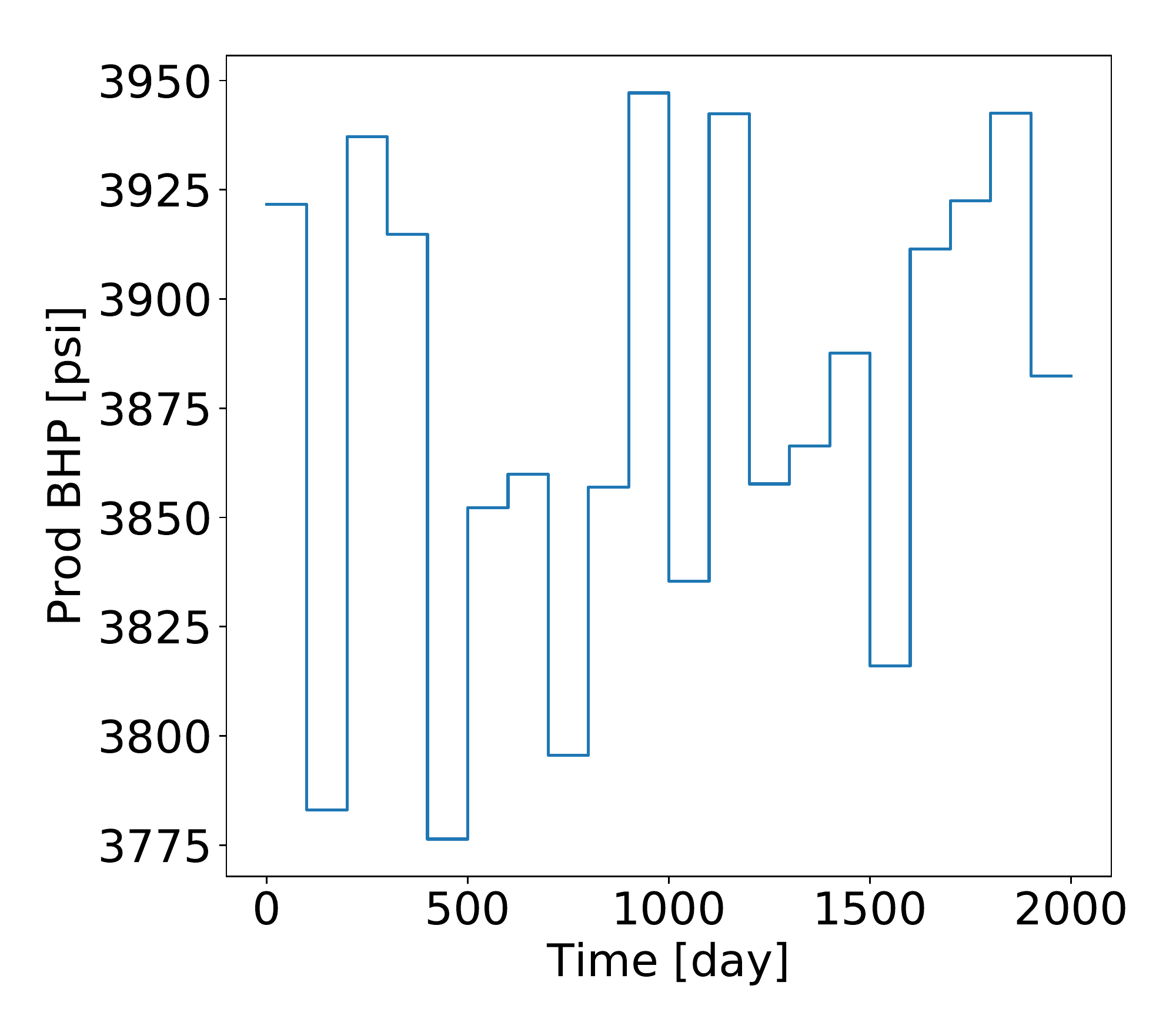}
    \caption{Well~P1}
  \end{subfigure}\hfill
    \begin{subfigure}{.33\textwidth}
    \centering
    \includegraphics[width=\linewidth]{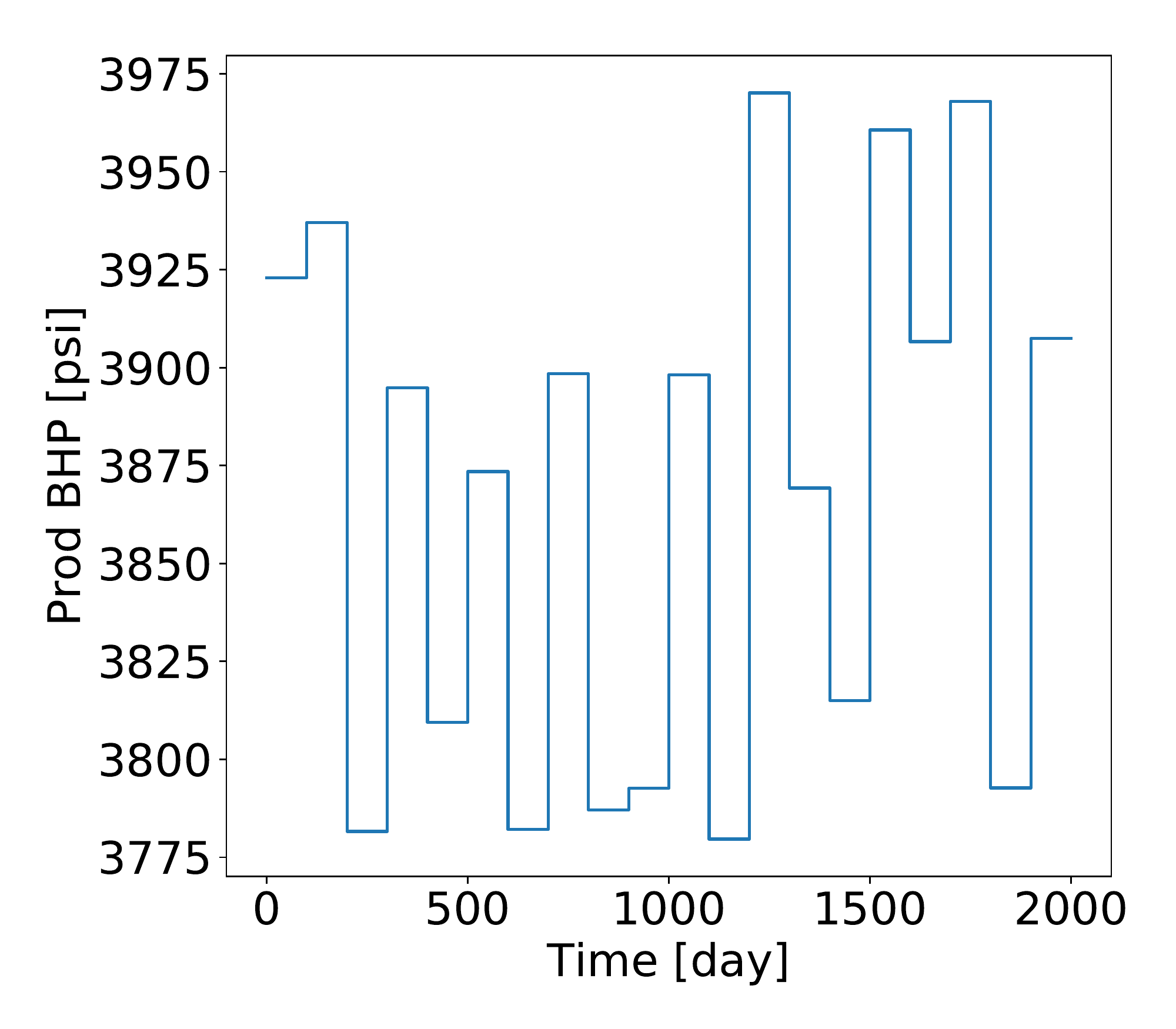}
    \caption{Well~P2}
  \end{subfigure} \\
    \begin{subfigure}{.33\textwidth}
    \centering
    \includegraphics[width=\linewidth]{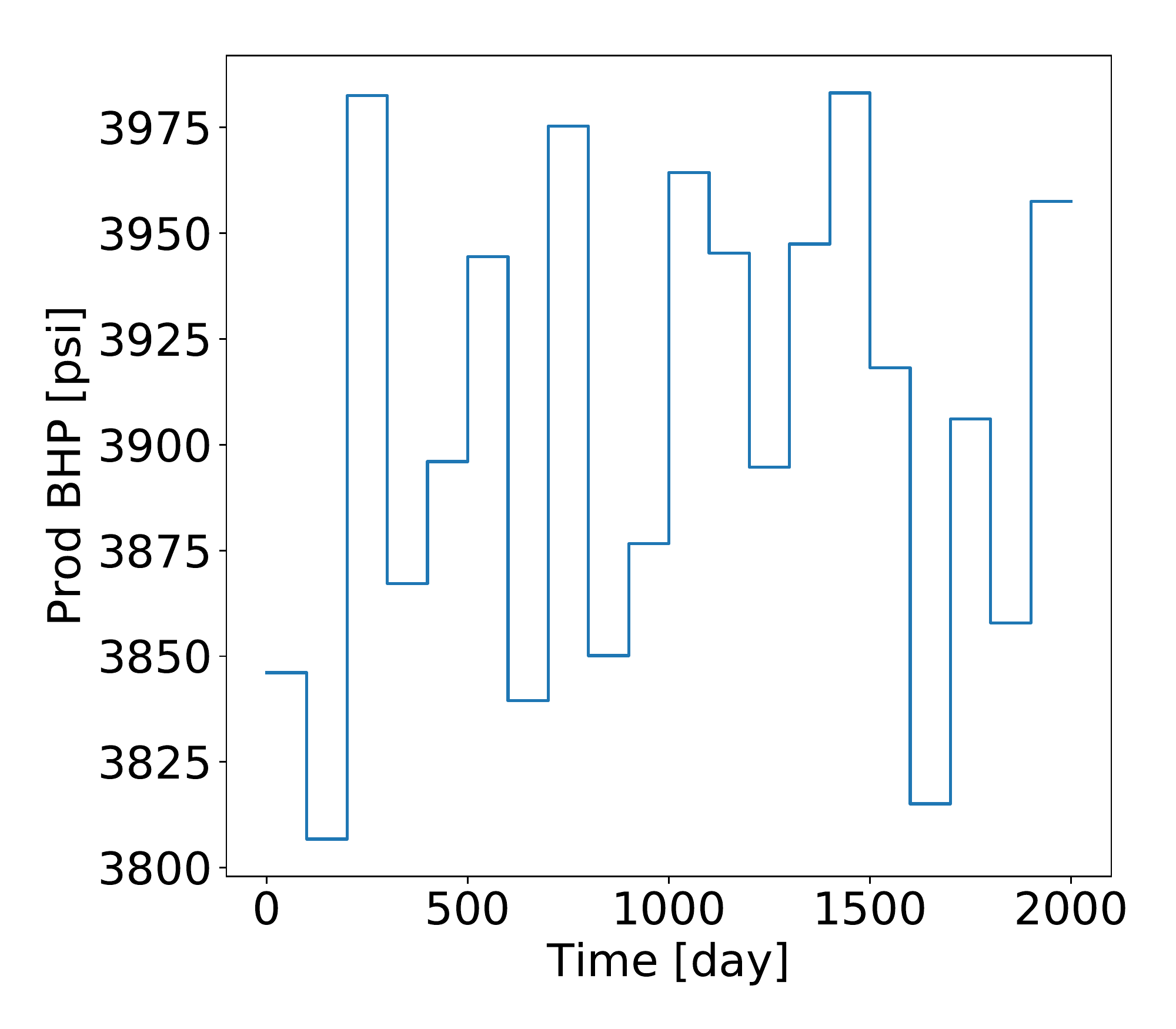}
    \caption{Well~P3}
  \end{subfigure}\hfill
  \begin{subfigure}{.33\textwidth}
    \centering
    \includegraphics[width=\linewidth]{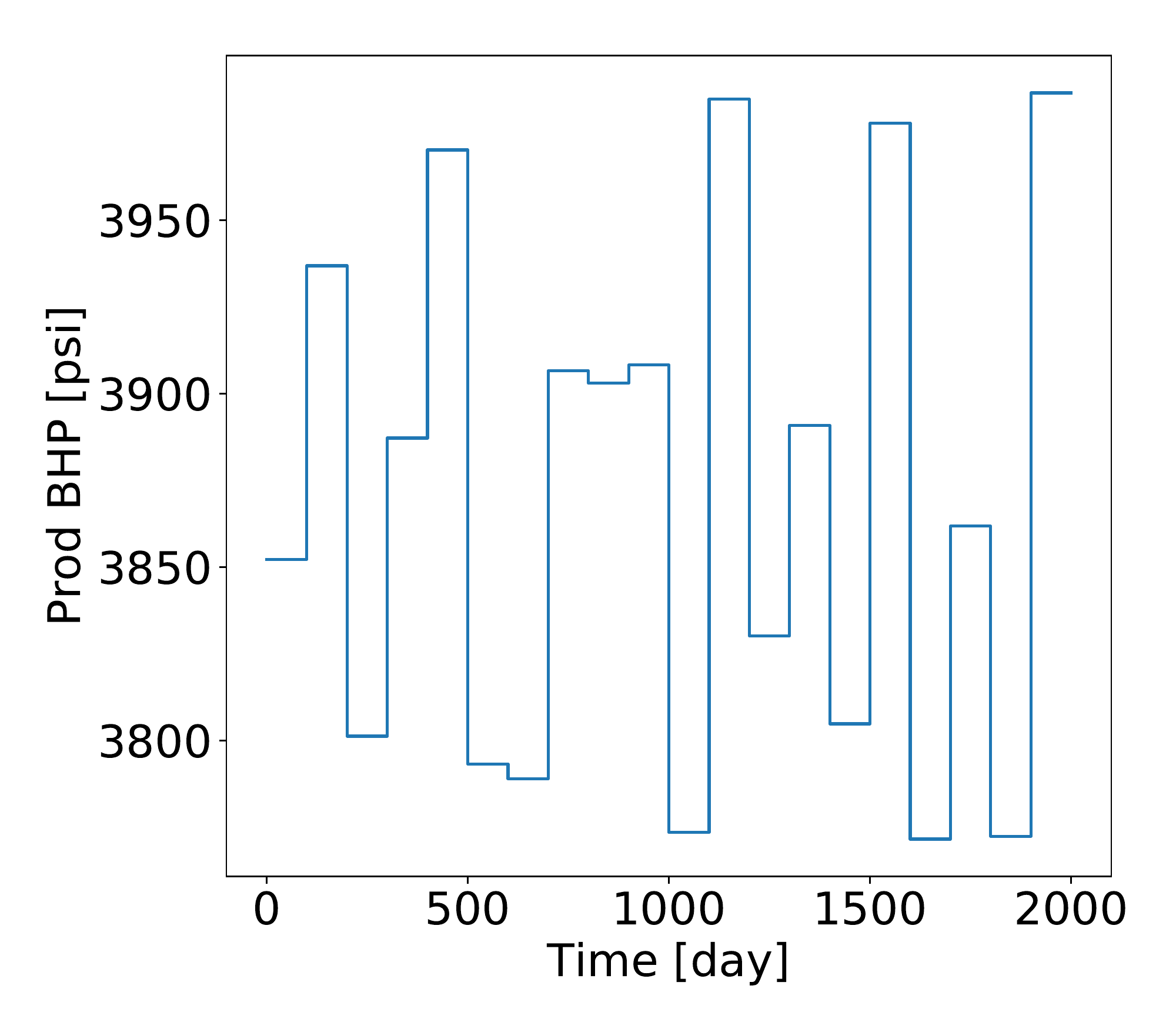}
    \caption{Well~P4}
  \end{subfigure}\hfill
    \begin{subfigure}{.33\textwidth}
    \centering
    \includegraphics[width=\linewidth]{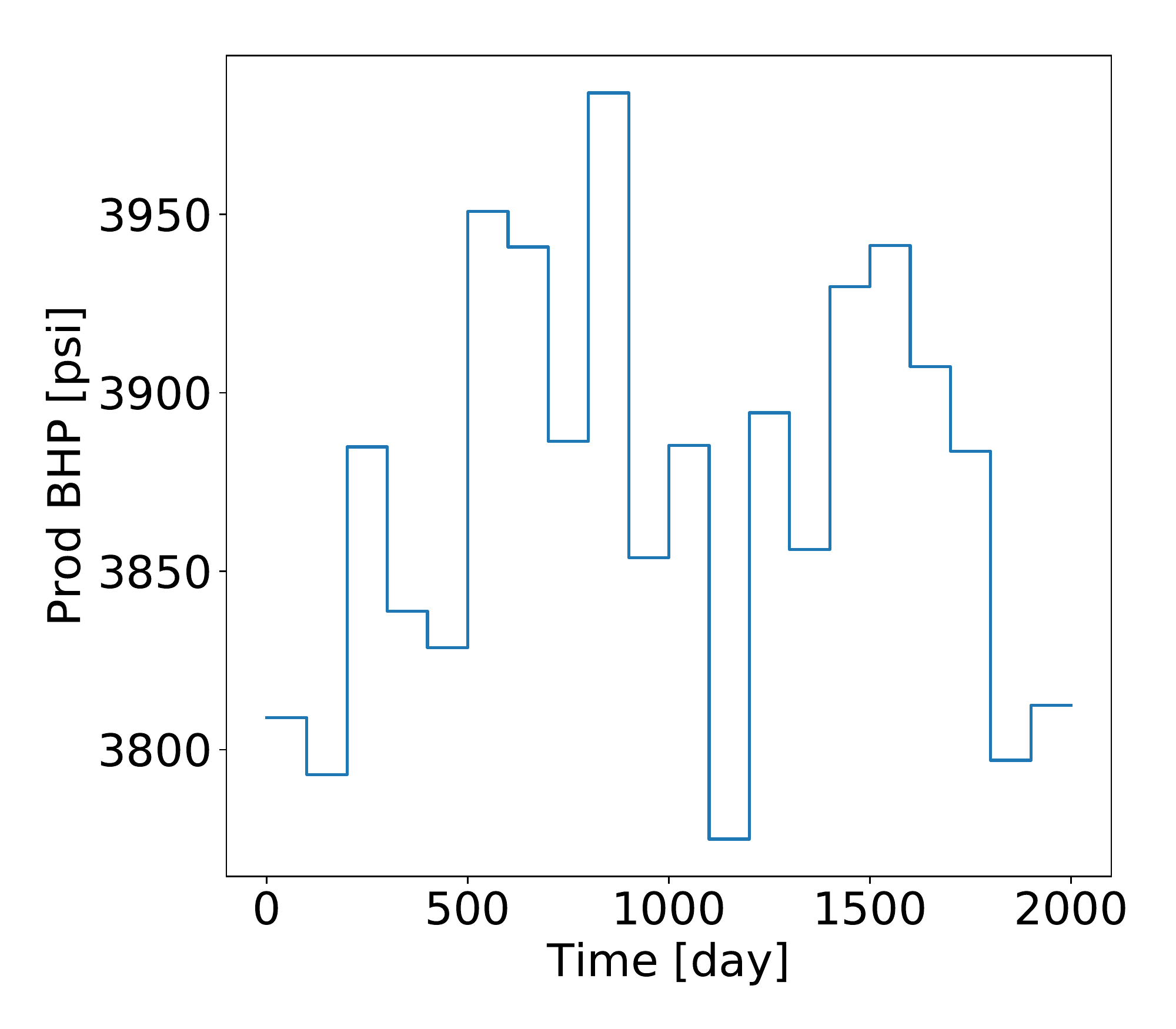}
    \caption{Well~P5}
  \end{subfigure}
  \caption{Test Case~1: well controls}
  \label{fig::test_1_well_ctrl}
\end{figure}

In this section we present detailed results for a particular test case. These include well quantities (injection BHPs and production rates) and global quantities (pressure and saturation fields). The injection rate and BHP profiles for Test Case~1 are displayed in Fig.~\ref{fig::test_1_well_ctrl}. Here we show the water rates for the four injection wells (Fig.~\ref{fig::test_1_well_ctrl}(a)-(d)), and the BHPs for the five production wells (Fig.~\ref{fig::test_1_well_ctrl}(e)-(i)). 
%These test-case controls are generated in the same way as the training controls described above. 
%Due to the high dimension of the control variables (dimension of 180 here), it is very difficult to find a training schedule resembles any of the test case control.

%%%%% case 1, well ctrl %%%%%%%%

\begin{figure}[htbp]
  \centering
  \begin{subfigure}{.45\textwidth}
    \centering
    \includegraphics[width=\linewidth]{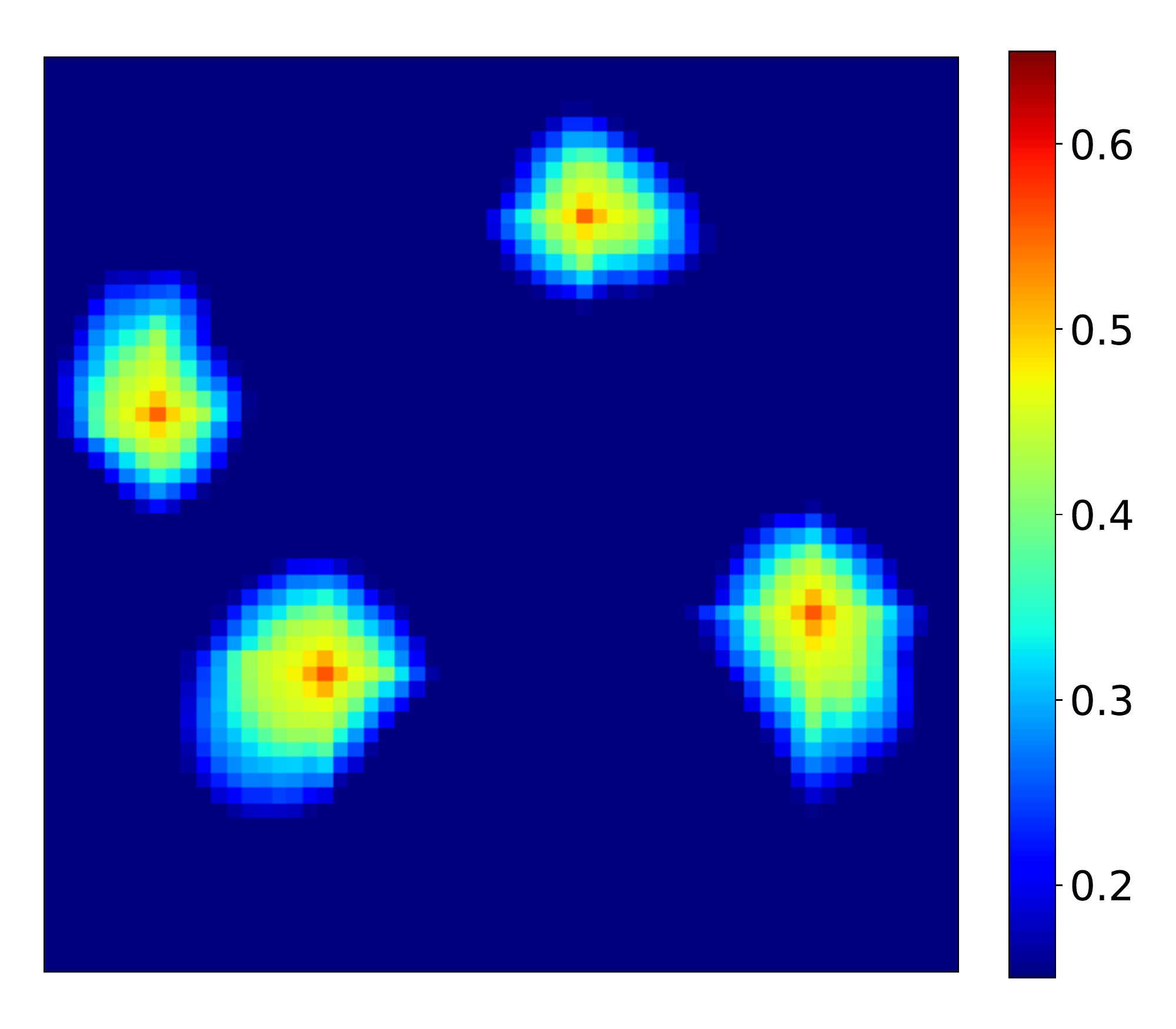}
    \caption{High-fidelity solution ($\text{HFS}_{\text{test}}$)}
  \end{subfigure}\hfill
  \begin{subfigure}{.45\textwidth}
    \centering
    \includegraphics[width=\linewidth]{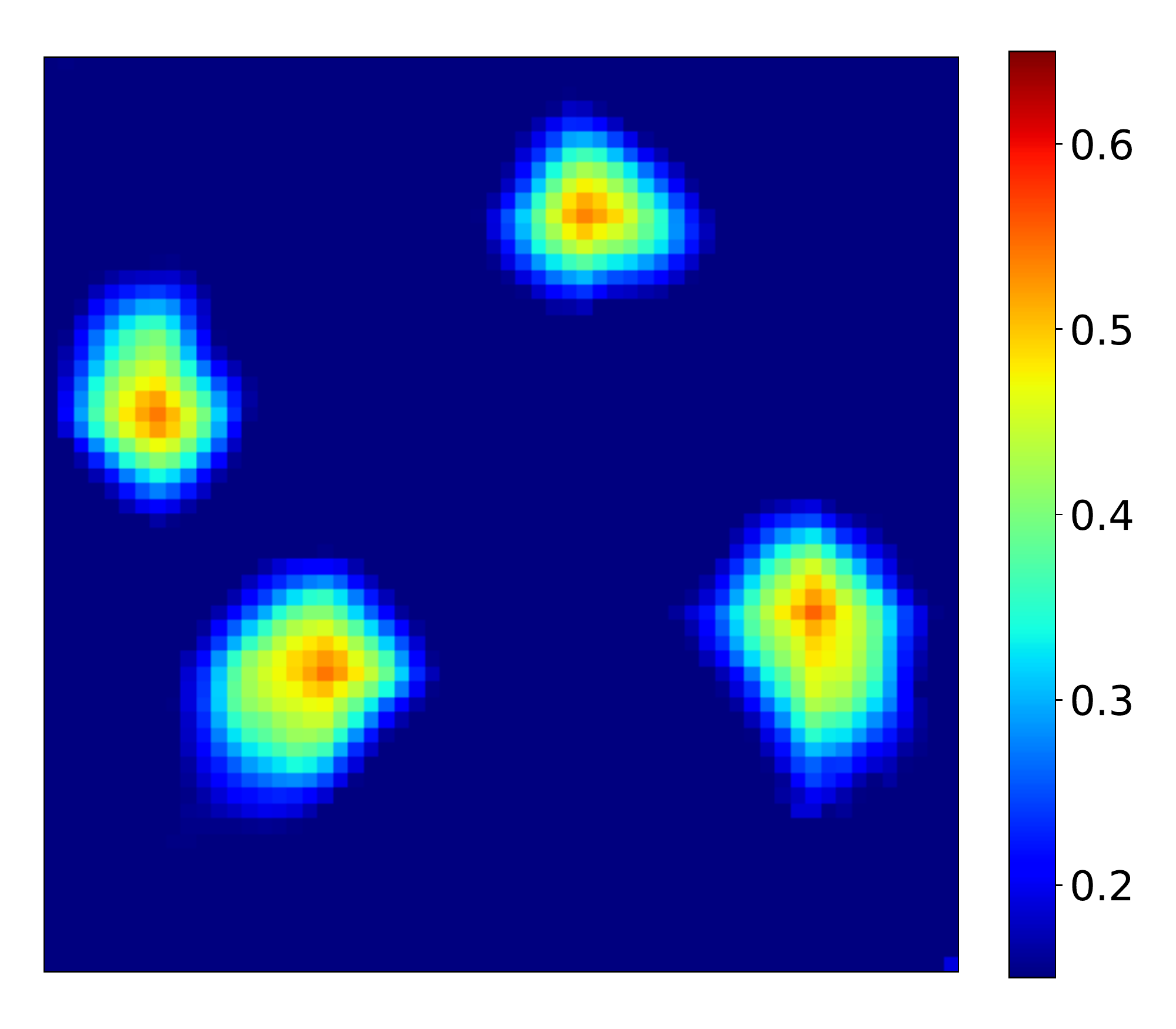}
    \caption{ROM solution ($\text{ROM}_{\text{test}}$)}
  \end{subfigure} \\
  \begin{subfigure}{.45\textwidth}
    \centering
    \includegraphics[width=\linewidth]{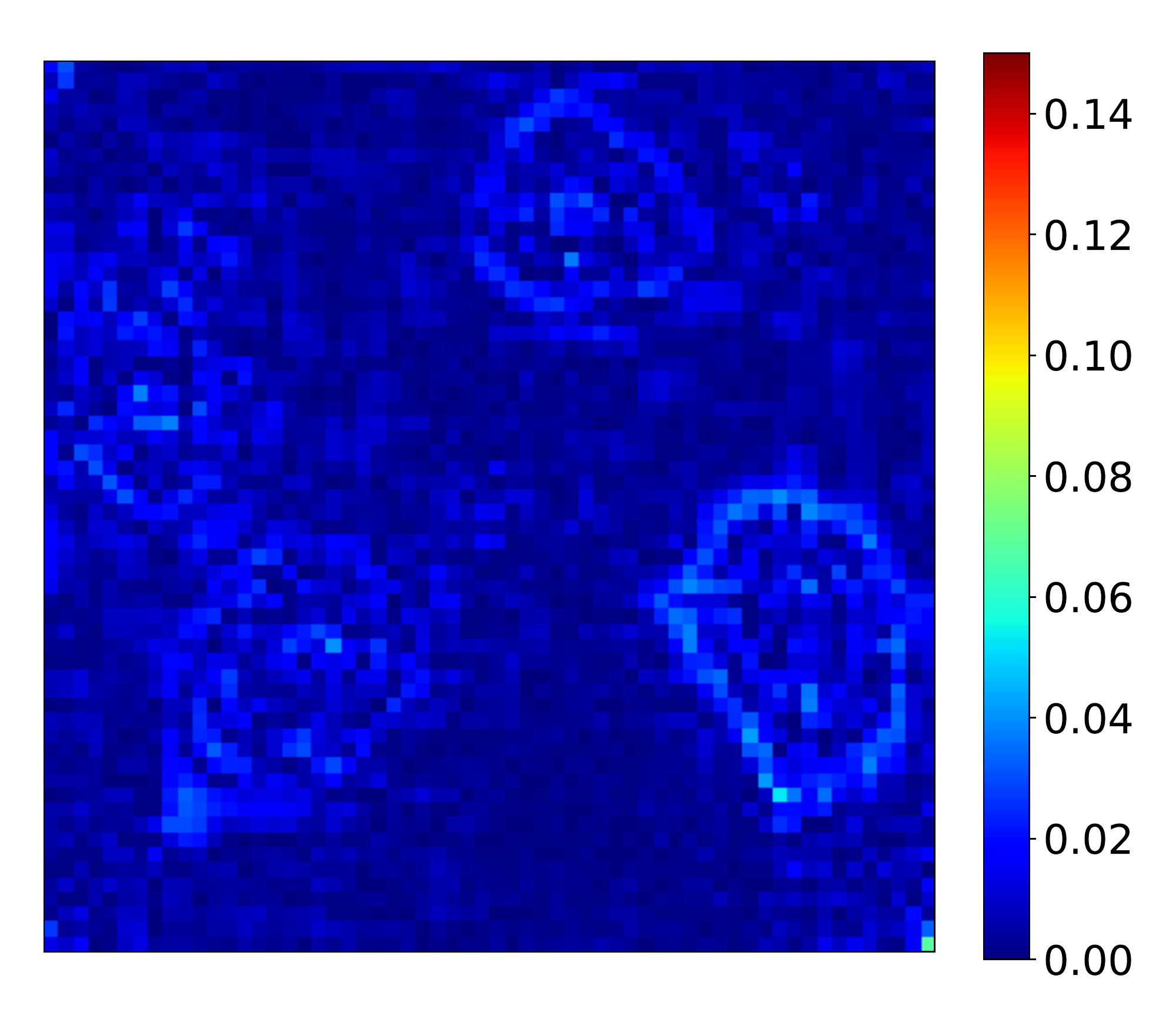}
    \caption{$|\text{HFS}_{\text{test}} - \text{ROM}_{\text{test}}|$}
  \end{subfigure}\hfill
    \begin{subfigure}{.45\textwidth}
    \centering
    \includegraphics[width=\linewidth]{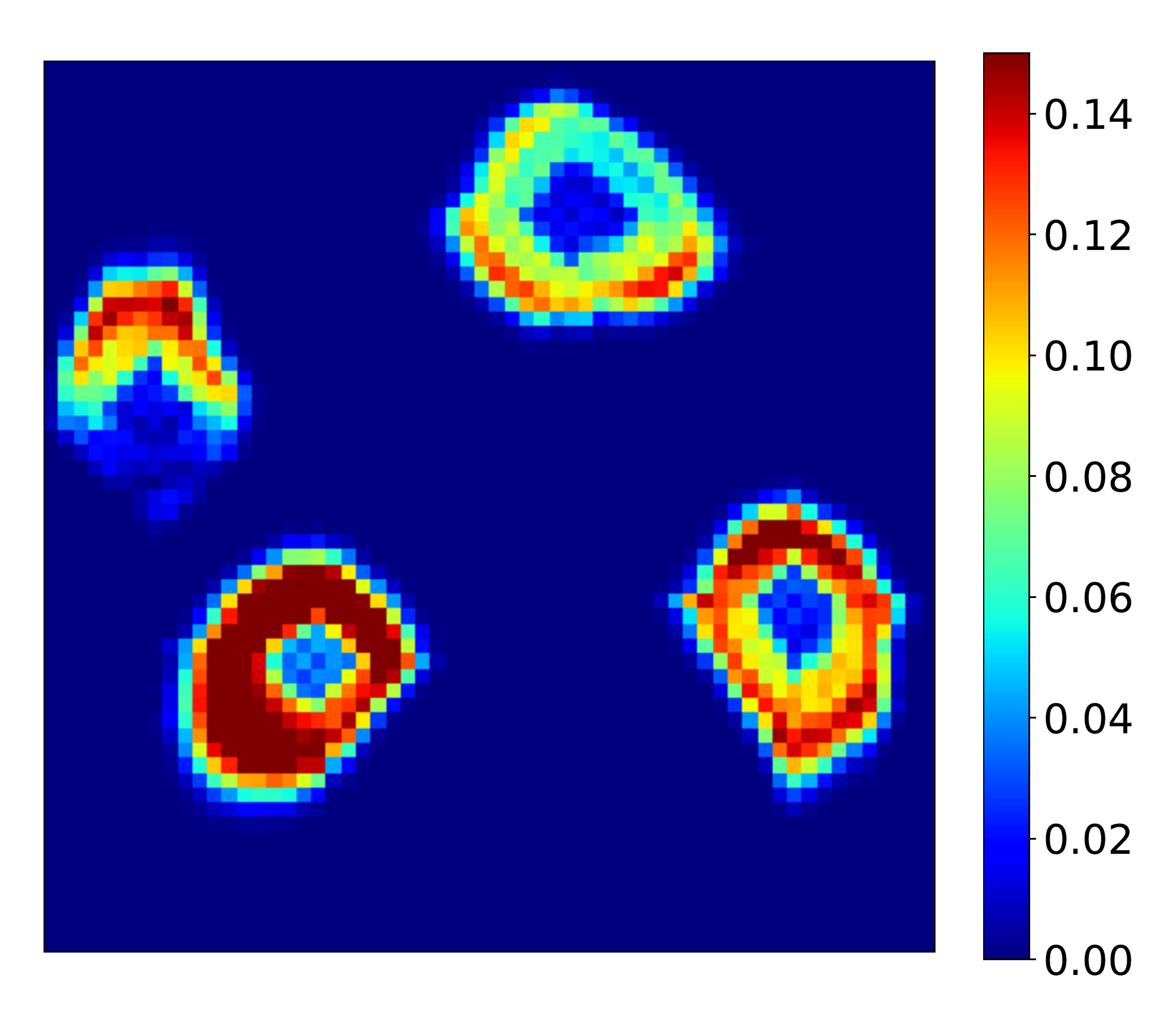}
    \caption{$|\text{HFS}_{\text{test}} - \text{HFS}_{\text{train}}|$}
  \end{subfigure}
  \caption{Test Case~1: saturation field at 200 days}
  \label{fig::test_1_sat_200}
\end{figure}

We now assess the performance of the deep-learning-based ROM for this test case. The time-evolution of the global saturation field is first considered. Fig.~\ref{fig::test_1_sat_200} displays the saturation field at 200~days. In Fig.~\ref{fig::test_1_sat_200}(a) the full-order saturation field (also referred to as the high-fidelity solution, HFS) is shown, and the corresponding E2C ROM result for saturation is presented in Fig.~\ref{fig::test_1_sat_200}(b). The color scale indicates water saturation value (thus red denotes water). The close visual agreement between Fig.~\ref{fig::test_1_sat_200}(a) and (b) suggests that the deep-learning-based ROM is able to provide accurate results for this quantity. The level of agreement between the two solutions is quantified in Fig.~\ref{fig::test_1_sat_200}(c), where the difference between the HFS and ROM solutions is displayed. Note that the colorbar scale here is very different than that in Fig.~\ref{fig::test_1_sat_200}(a) and (b). The error between the ROM and HFS results is clearly very small. 

In order to better quantify the predictive ability of the E2C ROM, we introduce the concept of the `closest training run.' We use this term to denote the specific training run, out of the 300 training runs performed, that most closely resembles (in a particular sense) the test case. The `distance' between the test run and each of the training runs is quantified in terms of the Euclidean distance between their vectors of normalized control parameters, and the `closest training run' ($k^{*}$) is the training run with the minimum distance. Specifically,
\begin{equation}\label{equ::dist}
    k^{*} = \argmin_{k} \|\mathbf{U}^{0}_{\text{te}} - \mathbf{U}^{0}_{\text{tr}, k}\|_2^2,
\end{equation}
where $k=1,\dots, 300$, denotes the index for the training runs,  $\mathbf{U}_{\text{te}}\in\mathbb{R}^{n_w\times N_{\text{ctrl}}}$ represents the control inputs for the test run, $\mathbf{U}_{\text{tr}, k}\in\mathbb{R}^{n_w\times N_{\text{ctrl}}}$ indicates the control inputs for training run $k$, and the superscript $0$ designates normalized pressures and rates in the controls, as per the normalizations in Eq.~\ref{equ::norm}. 

Eq.~\ref{equ::dist} provides a very approximate indicator of the `closest training run.' This definition has the advantage of simplicity, though more involved (and computationally demanding) assessments would be expected to provide closer training solutions. These would, however, require the application of an approach along the lines of the `point selection' procedure used in POD-TPWL \citep{jin2018reduced}. This would entail computing a measure of distance over many time steps for each training run. Since we have 300 training runs here (as opposed to three or five with POD-TPWL), this could become very time consuming. Thus we apply the simple approach defined in Eq.~\ref{equ::dist}, with the recognition that more sophisticated (and presumably more accurate) procedures could be devised.

%%%%% case 1, saturation %%%%%%%%

We now return to the global saturation results. Fig.~\ref{fig::test_1_sat_200}(d) shows the difference between the `closest training run' (determined as we just described and simulated at high fidelity), and the test-case saturation fields. The colorbar scale is the same as in Fig.~\ref{fig::test_1_sat_200}(c). The advantage of applying the deep-learning-based ROM is evident by comparing Fig.~\ref{fig::test_1_sat_200}(c) and (d). More specifically, the error in Fig.~\ref{fig::test_1_sat_200}(c) is about an order of magnitude less than the differences evident in Fig.~\ref{fig::test_1_sat_200}(d).

\begin{figure}[htbp]
  \centering
  \begin{subfigure}{.45\textwidth}
    \centering
    \includegraphics[width=\linewidth]{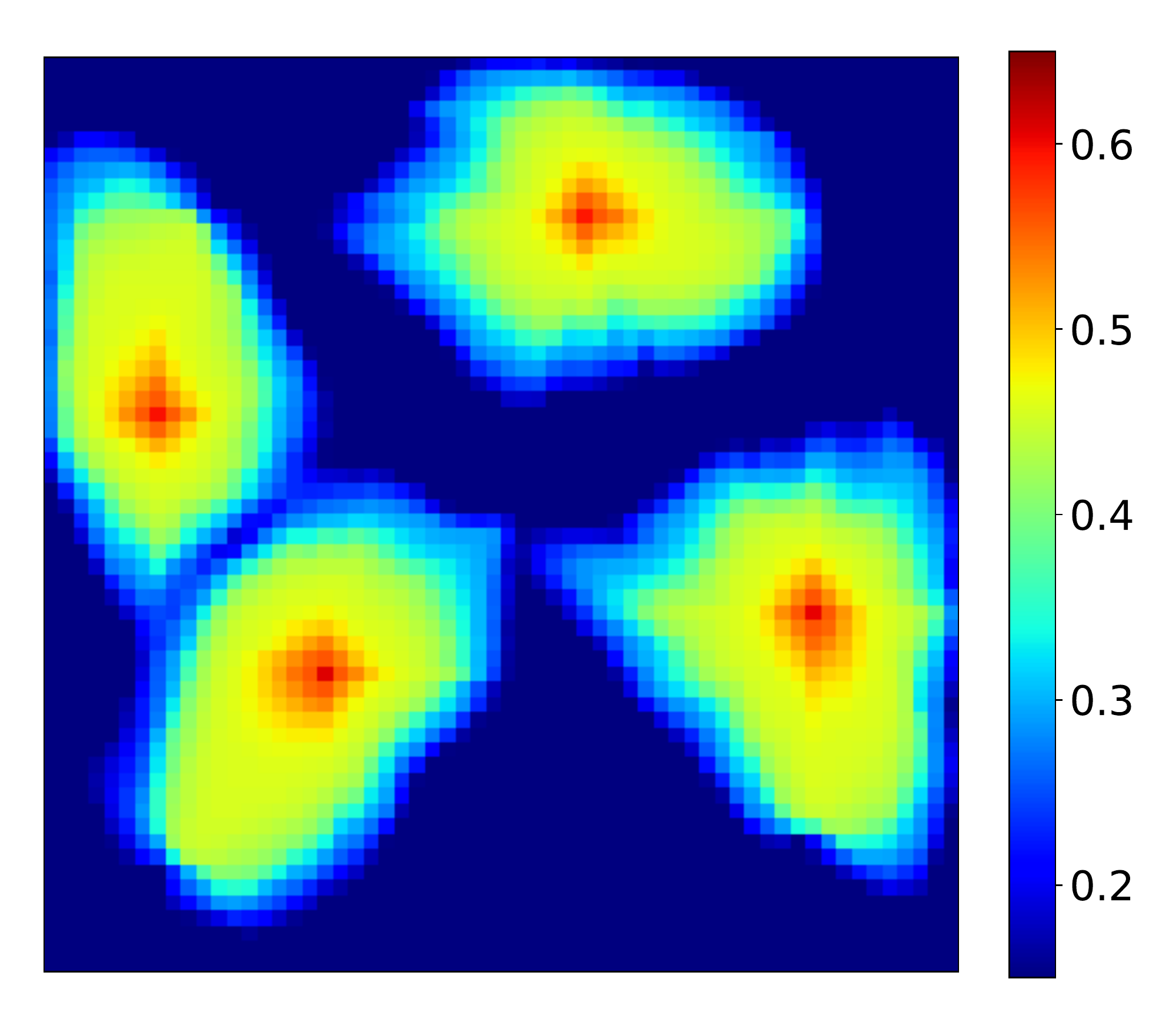}
    \caption{High-fidelity solution ($\text{HFS}_{\text{test}}$)}
  \end{subfigure}\hfill
  \begin{subfigure}{.45\textwidth}
    \centering
    \includegraphics[width=\linewidth]{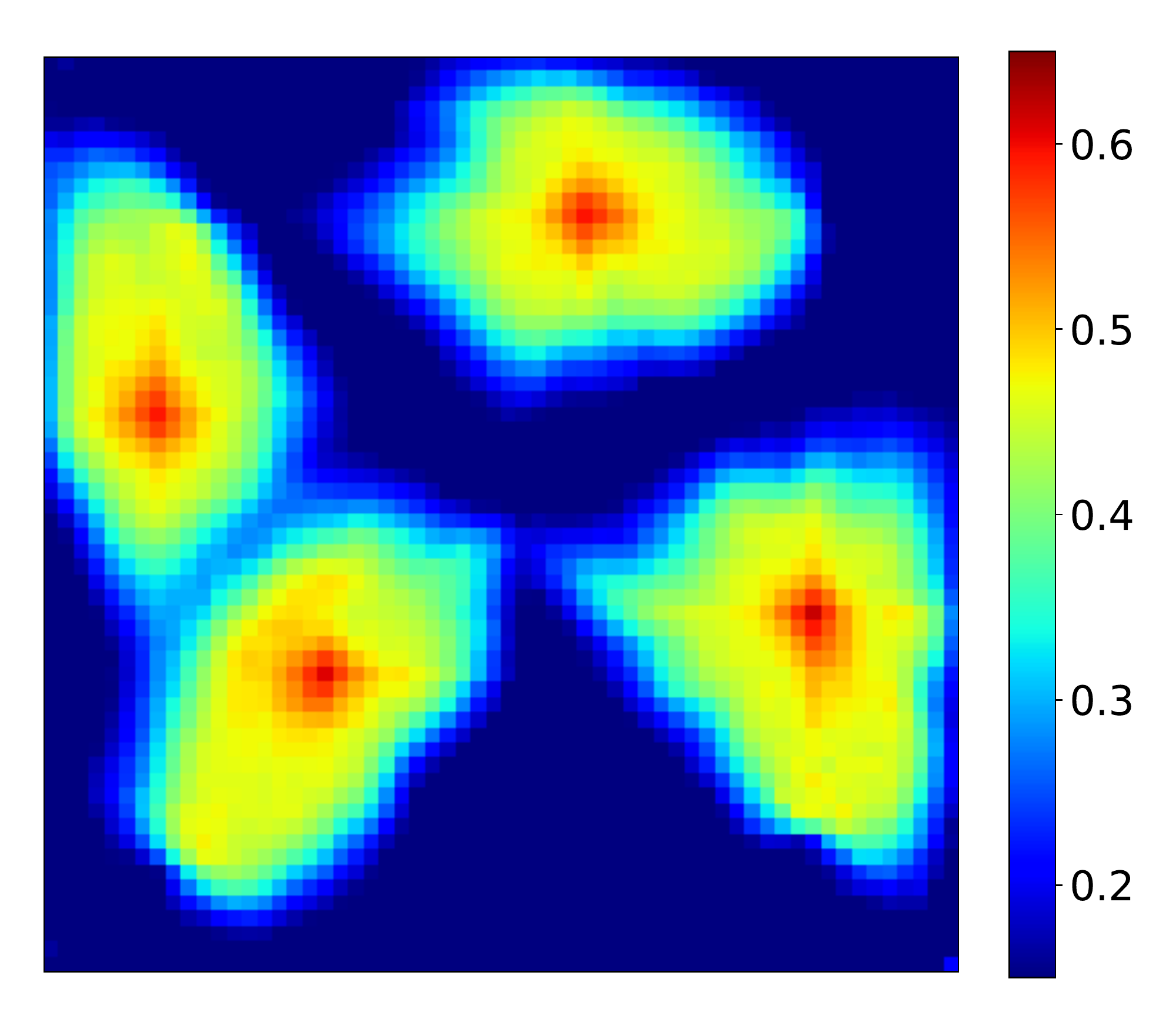}
    \caption{ROM solution ($\text{ROM}_{\text{test}}$)}
  \end{subfigure} \\
  \begin{subfigure}{.45\textwidth}
    \centering
    \includegraphics[width=\linewidth]{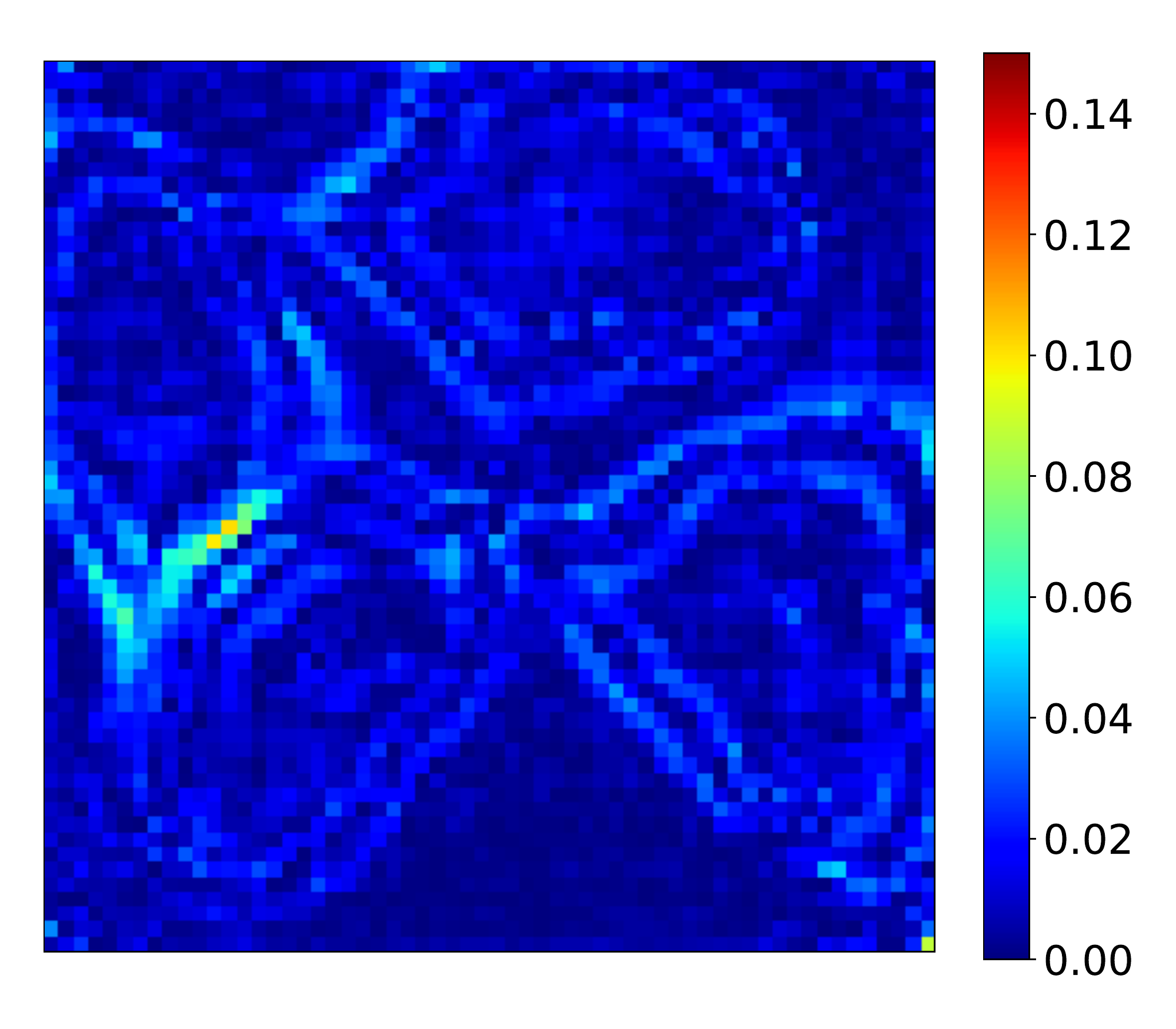}
    \caption{$|\text{HFS}_{\text{test}} - \text{ROM}_{\text{test}}|$}
  \end{subfigure}\hfill
    \begin{subfigure}{.45\textwidth}
    \centering
    \includegraphics[width=\linewidth]{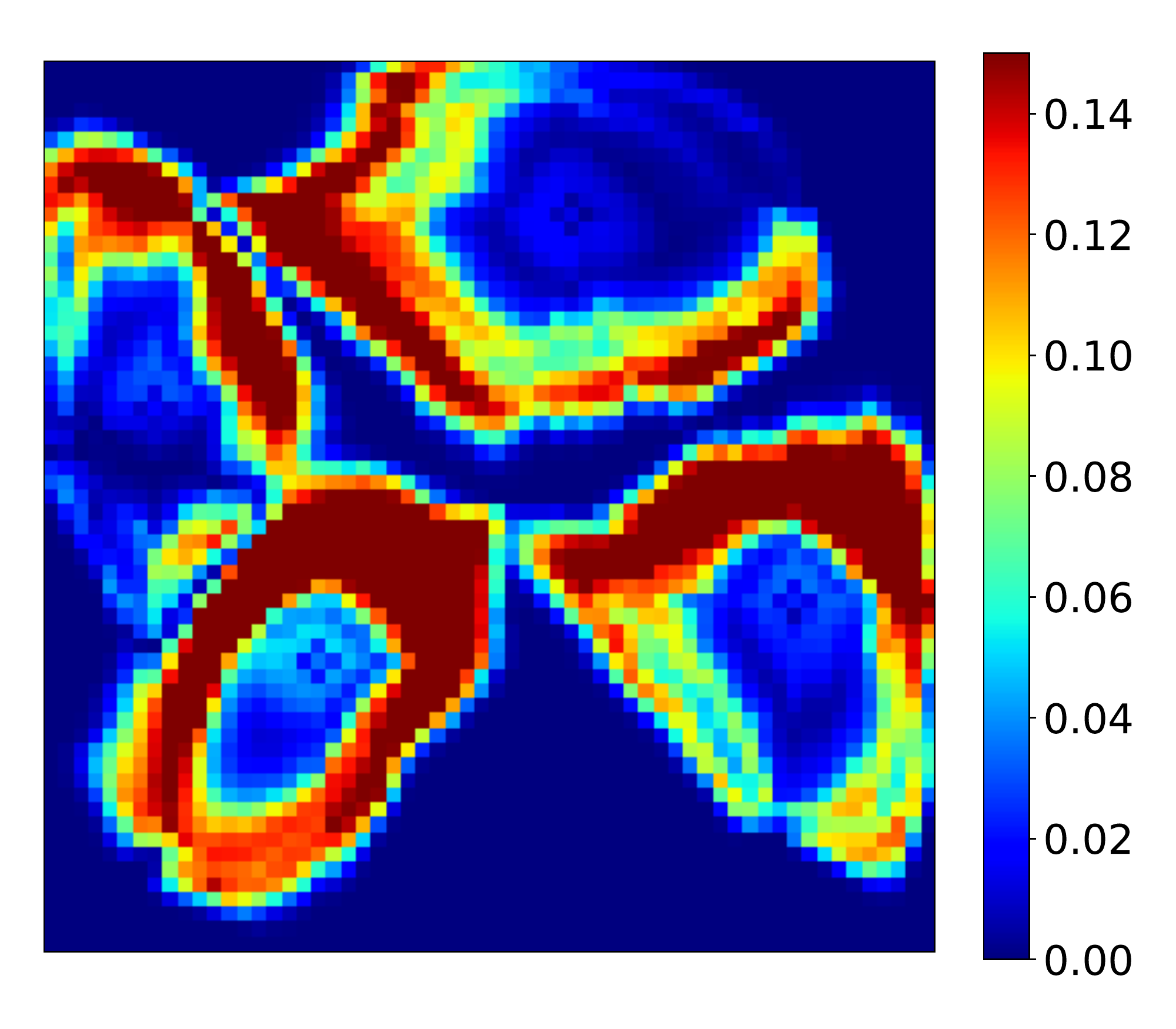}
    \caption{$|\text{HFS}_{\text{test}} - \text{HFS}_{\text{train}}|$}
  \end{subfigure}
  \caption{Test Case~1: saturation field at 1000 days}
  \label{fig::test_1_sat_1000}
\end{figure}

\begin{figure}[htbp]
  \centering
  \begin{subfigure}{.45\textwidth}
    \centering
    \includegraphics[width=\linewidth]{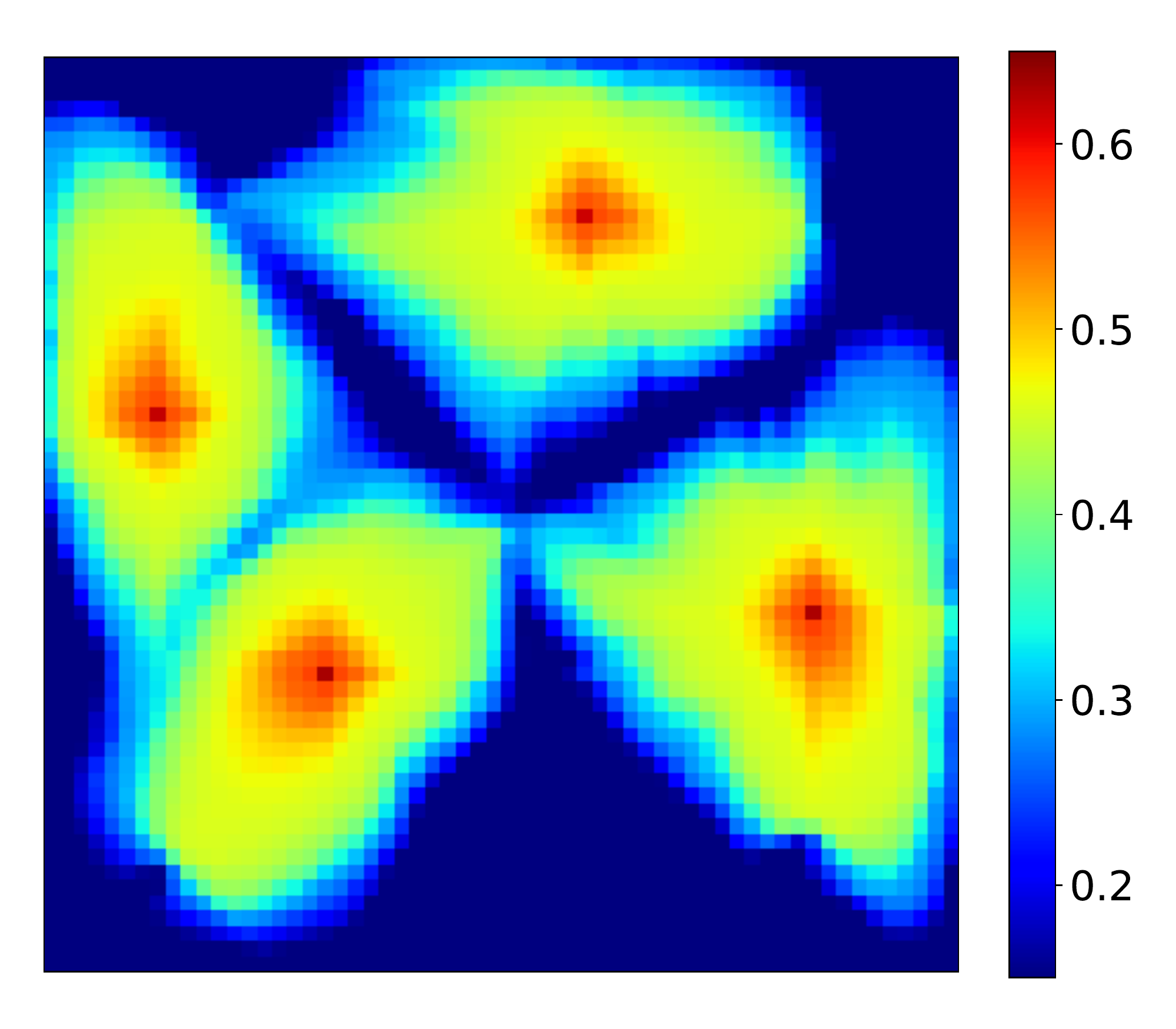}
    \caption{High-fidelity solution ($\text{HFS}_{\text{test}}$)}
  \end{subfigure}\hfill
  \begin{subfigure}{.45\textwidth}
    \centering
    \includegraphics[width=\linewidth]{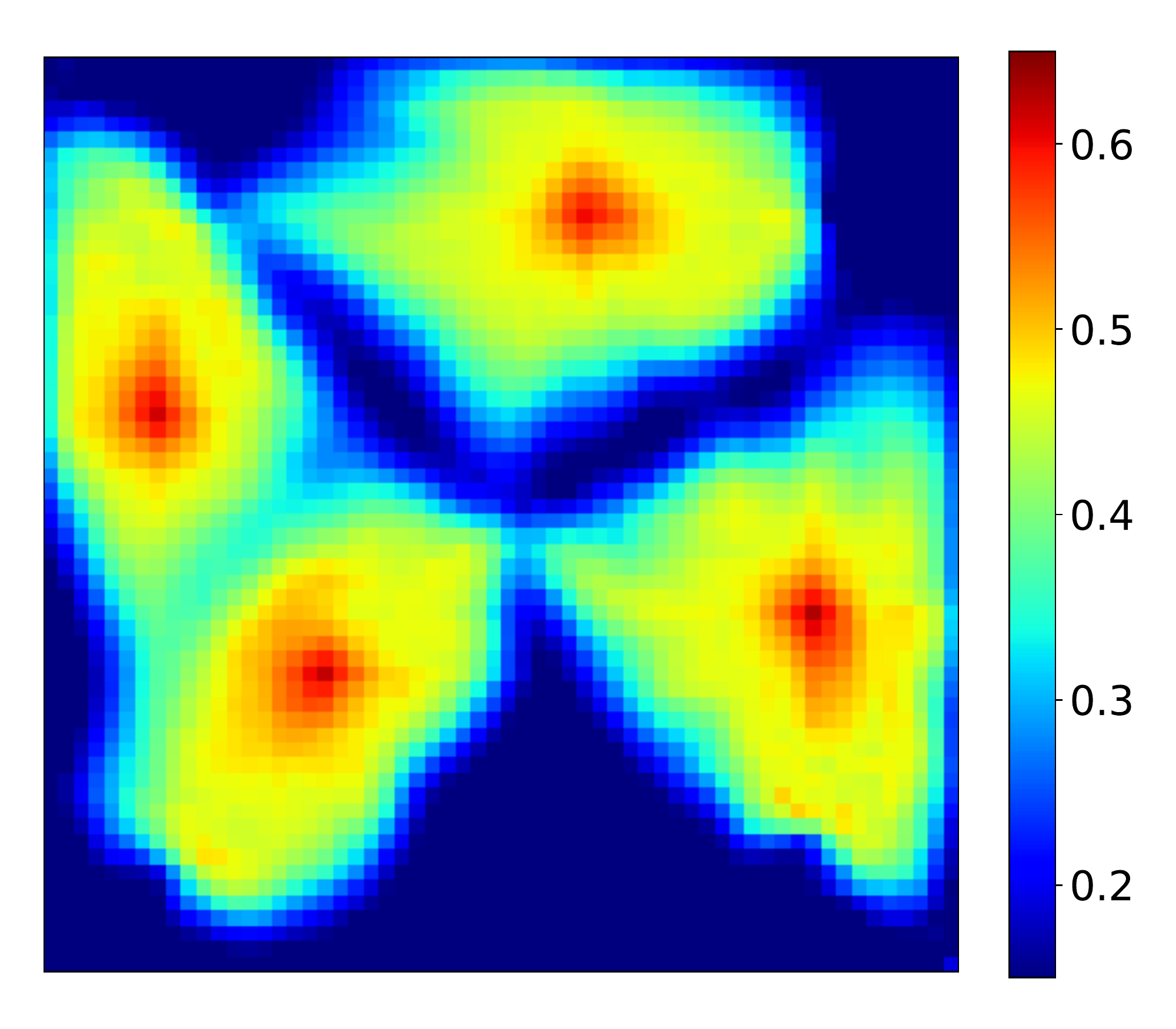}
    \caption{ROM solution ($\text{ROM}_{\text{test}}$)}
  \end{subfigure} \\
  \begin{subfigure}{.45\textwidth}
    \centering
    \includegraphics[width=\linewidth]{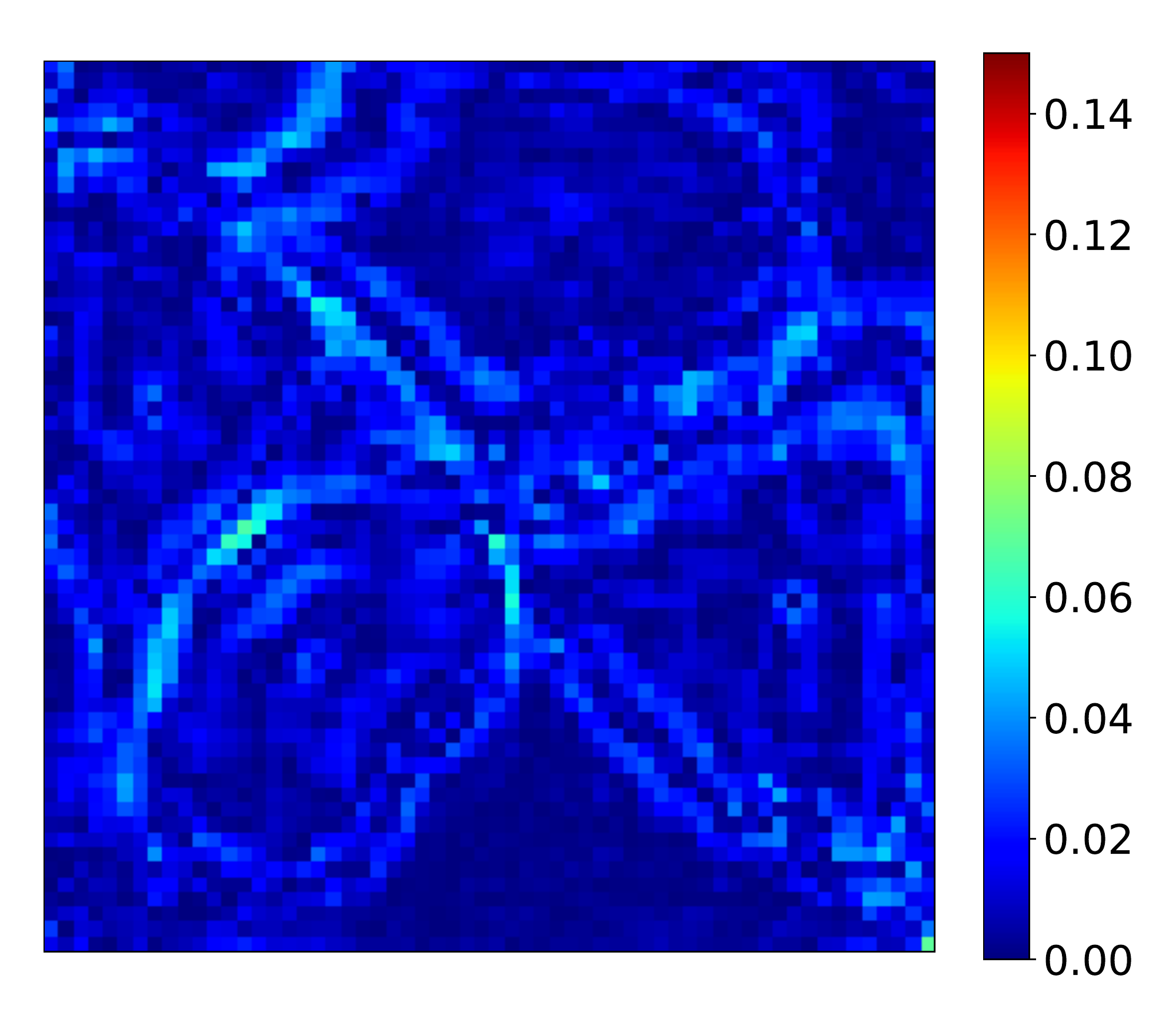}
    \caption{$|\text{HFS}_{\text{test}} - \text{ROM}_{\text{test}}|$}
  \end{subfigure}\hfill
    \begin{subfigure}{.45\textwidth}
    \centering
    \includegraphics[width=\linewidth]{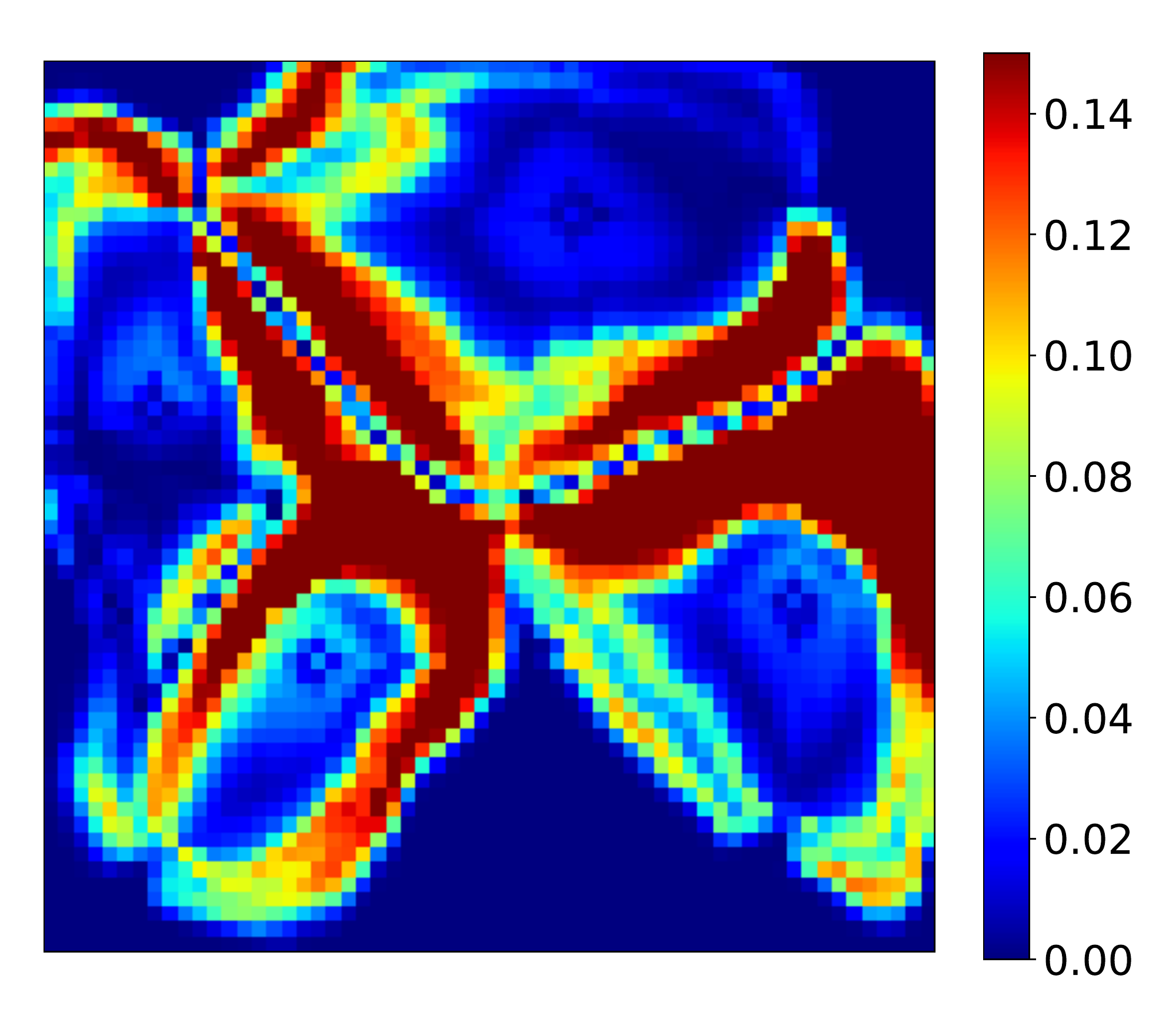}
    \caption{$|\text{HFS}_{\text{test}} - \text{HFS}_{\text{train}}|$}
  \end{subfigure}
  \caption{Test Case~1: saturation field at 1800 days}
  \label{fig::test_1_sat_1800}
\end{figure}

Figs.~\ref{fig::test_1_sat_1000} and \ref{fig::test_1_sat_1800} display analogous results for saturation at 1000 and 1800~days. The evolution of the saturation field with time is apparent, and the deep-learning-based ROM solutions (Figs.~\ref{fig::test_1_sat_1000}(b) and \ref{fig::test_1_sat_1800}(b)) are again seen to be in close visual agreement with the HFS (Figs.~\ref{fig::test_1_sat_1000}(a) and \ref{fig::test_1_sat_1800}(a)). The error maps in Figs.~\ref{fig::test_1_sat_1000}(c) and \ref{fig::test_1_sat_1800}(c) further quantify the accuracy of the deep-learning-based ROM. These errors are quite small compared with the difference maps between the `closest training run' and the HFS, shown in Figs.~\ref{fig::test_1_sat_1000}(d) and \ref{fig::test_1_sat_1800}(d), which further illustrates the effectiveness of the ROM.

Note that in the ROM solutions, we do observe some local (unphysical) extrema within the saturation plumes. This is a minor issue here since the difference maps show small overall discrepancies between the ROM and high-fidelity solutions. In some cases, however, this could be a cause for concern. A potential remedy for this would be to add a term to the physics-based loss function such that local extrema in saturation that are inconsistent with the governing flow equations are penalized.

\begin{figure}[htbp]
  \centering
  \begin{subfigure}{.45\textwidth}
    \centering
    \includegraphics[width=\linewidth]{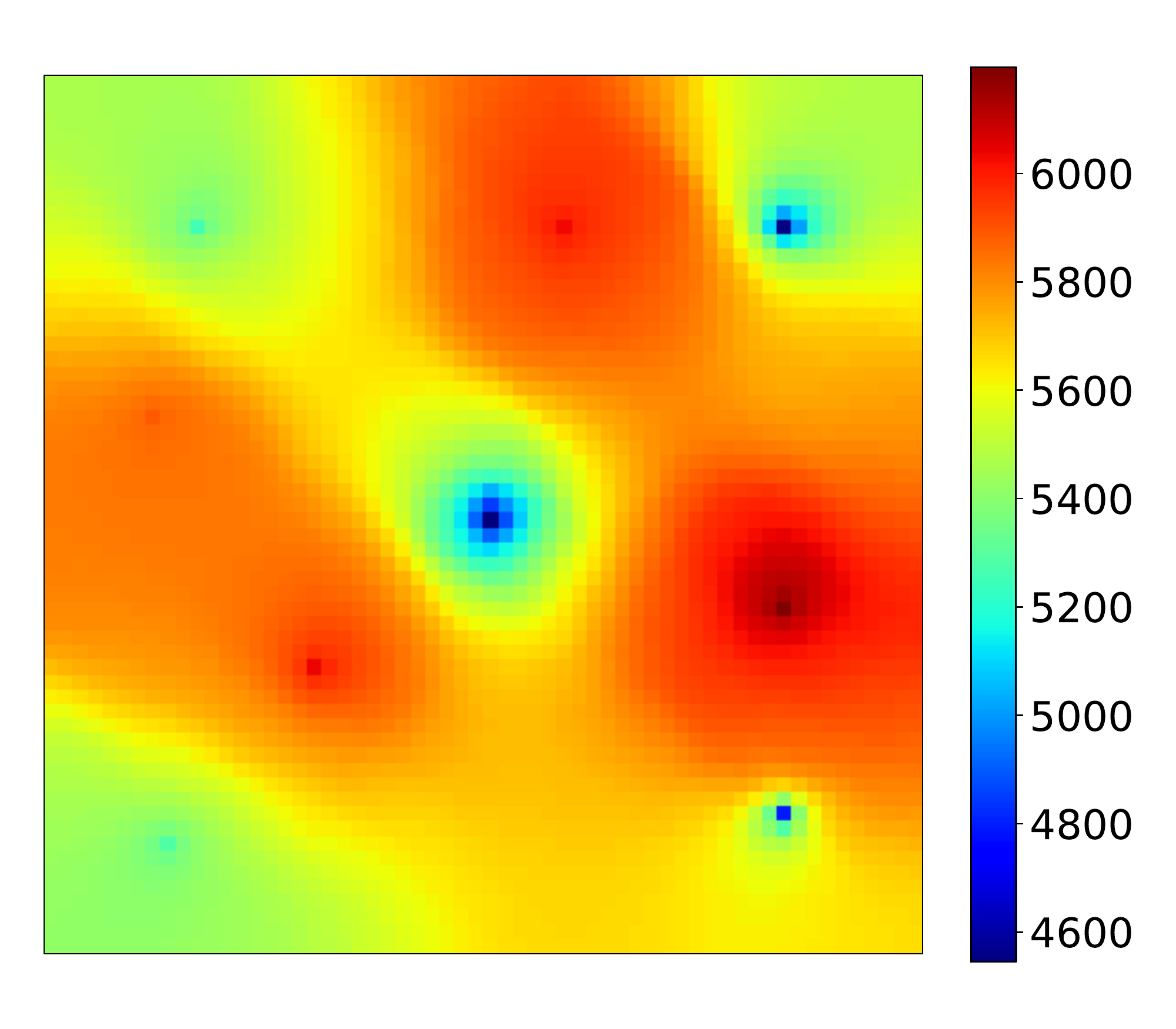}
    \caption{High-fidelity solution ($\text{HFS}_{\text{test}}$)}
  \end{subfigure}\hfill
  \begin{subfigure}{.45\textwidth}
    \centering
    \includegraphics[width=\linewidth]{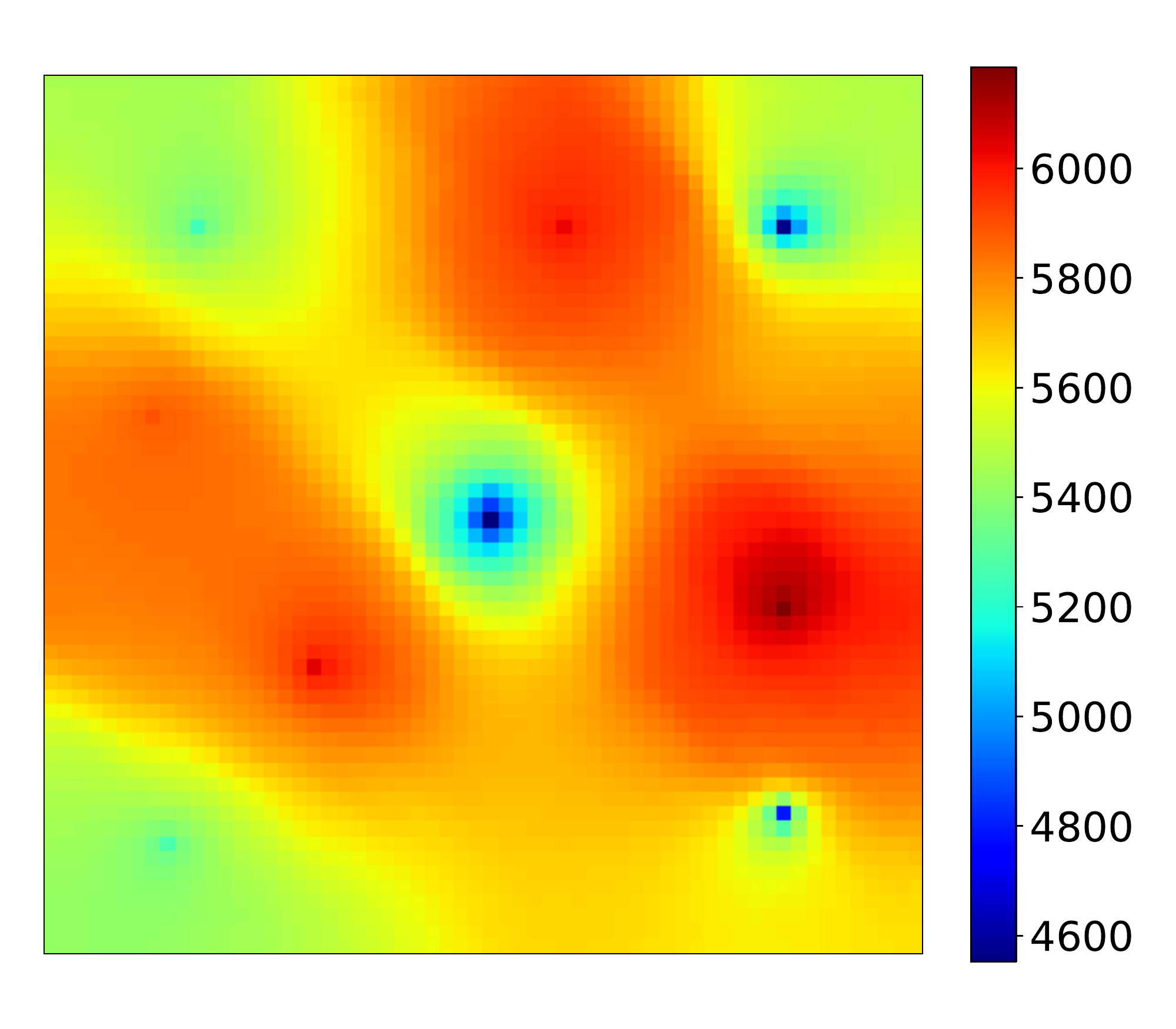}
    \caption{ROM solution ($\text{ROM}_{\text{test}}$)}
  \end{subfigure} \\
  \begin{subfigure}{.45\textwidth}
    \centering
    \includegraphics[width=\linewidth]{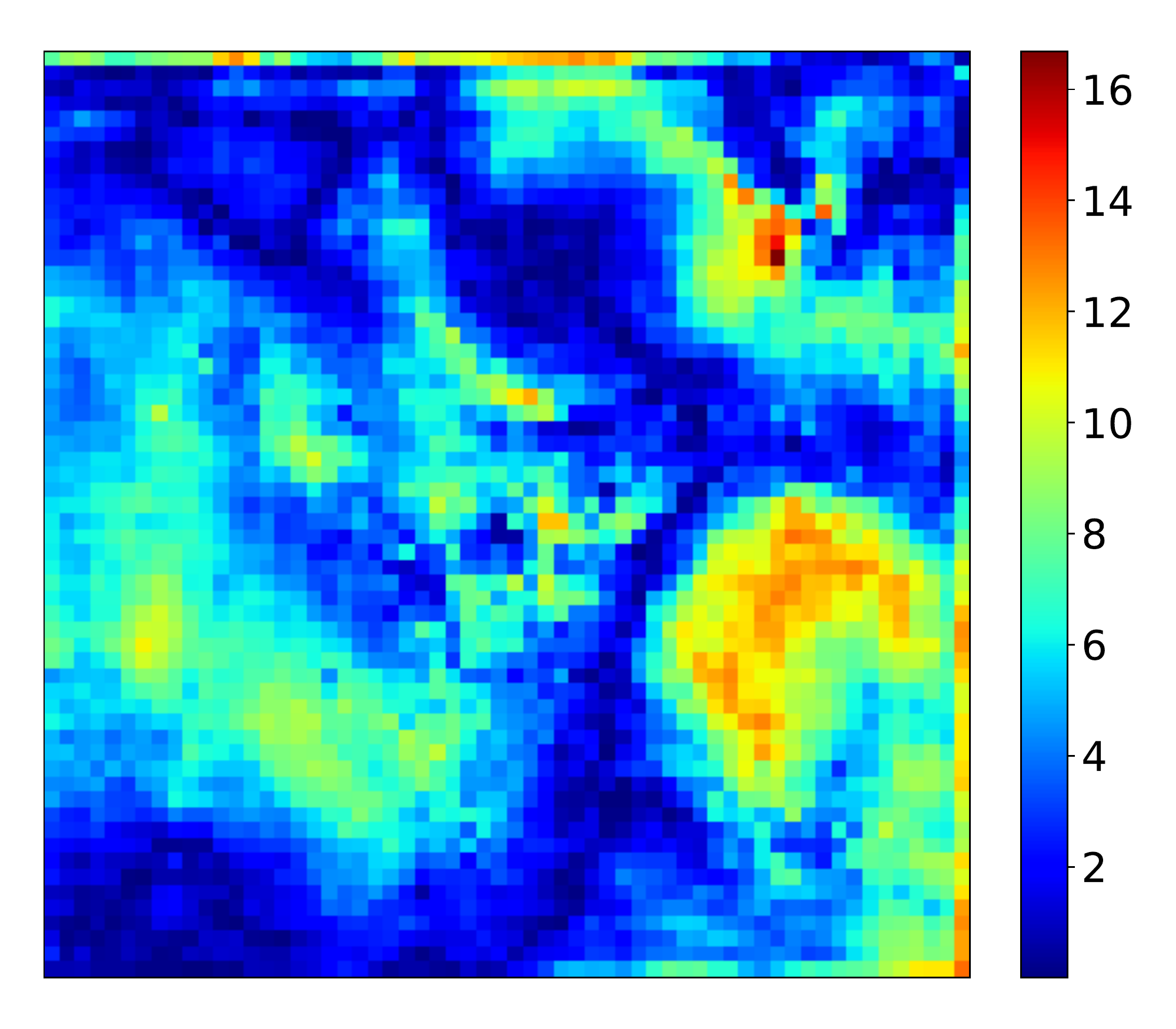}
    \caption{$|\text{HFS}_{\text{test}} - \text{ROM}_{\text{test}}|$}
  \end{subfigure}\hfill
    \begin{subfigure}{.45\textwidth}
    \centering
    \includegraphics[width=\linewidth]{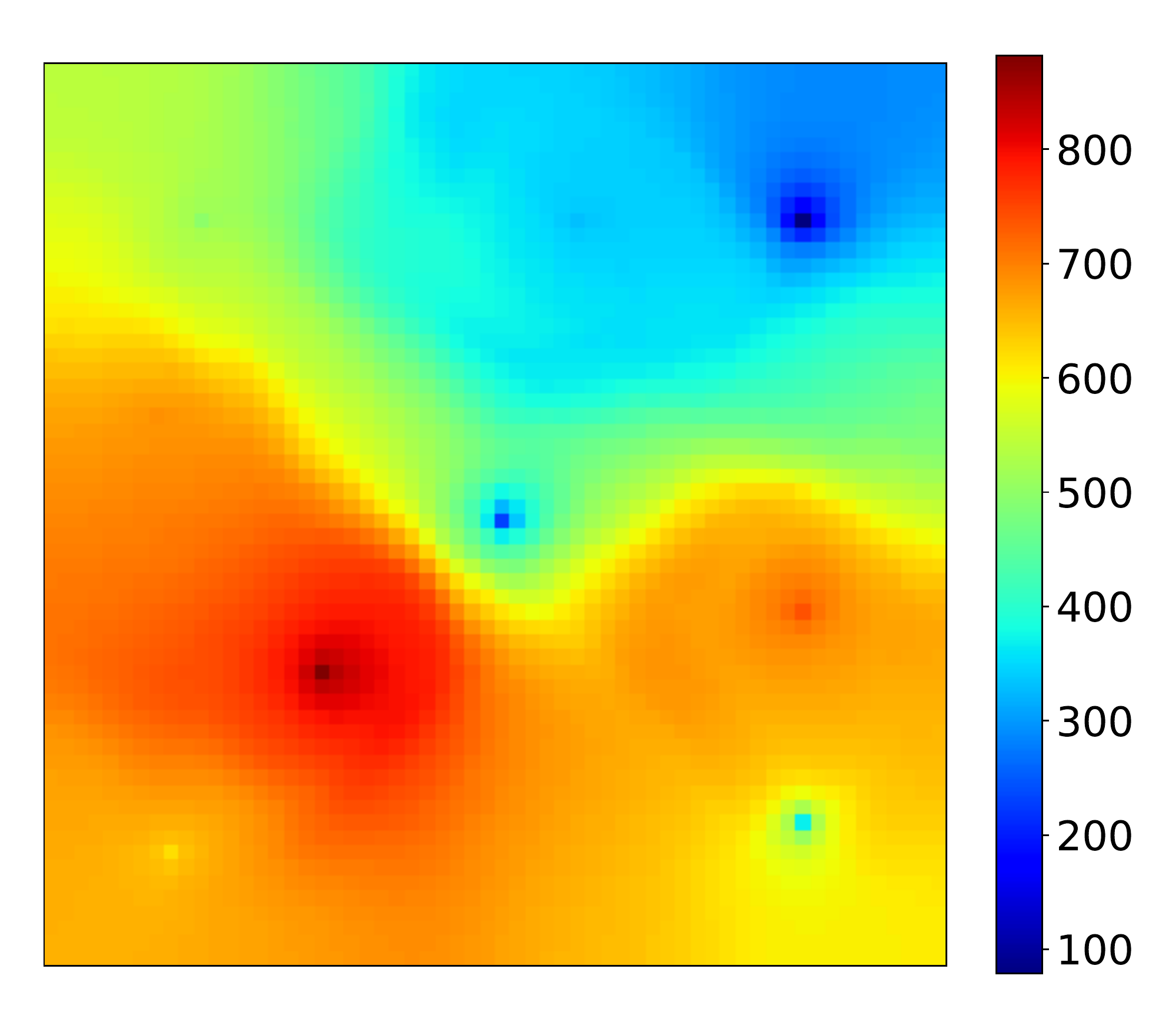}
    \caption{$|\text{HFS}_{\text{test}} - \text{HFS}_{\text{train}}|$}
  \end{subfigure}
  \caption{Test Case~1: pressure field at 1000 days (all colorbars in units of psi)}
  \label{fig::test_1_pres_1000}
\end{figure}

The global pressure field at particular times is also of interest. In Fig.~\ref{fig::test_1_pres_1000}(a) and (b) we display the HFS and ROM pressure solutions at 1000~days. The close visual agreement suggests that the deep-learning-based ROM is able to provide accurate (and smooth) pressure predictions. Fig.~\ref{fig::test_1_pres_1000}(c) shows the error map for the ROM solution, where we see that errors are indeed very small. These errors are much less than those for the `closest training run,' which are shown in Fig.~\ref{fig::test_1_pres_1000}(d).

%%%%% case 1, pressure %%%%%%%%

\begin{figure}[htbp]
  \centering
  \begin{subfigure}{.5\textwidth}
    \centering
    \includegraphics[width=\linewidth]{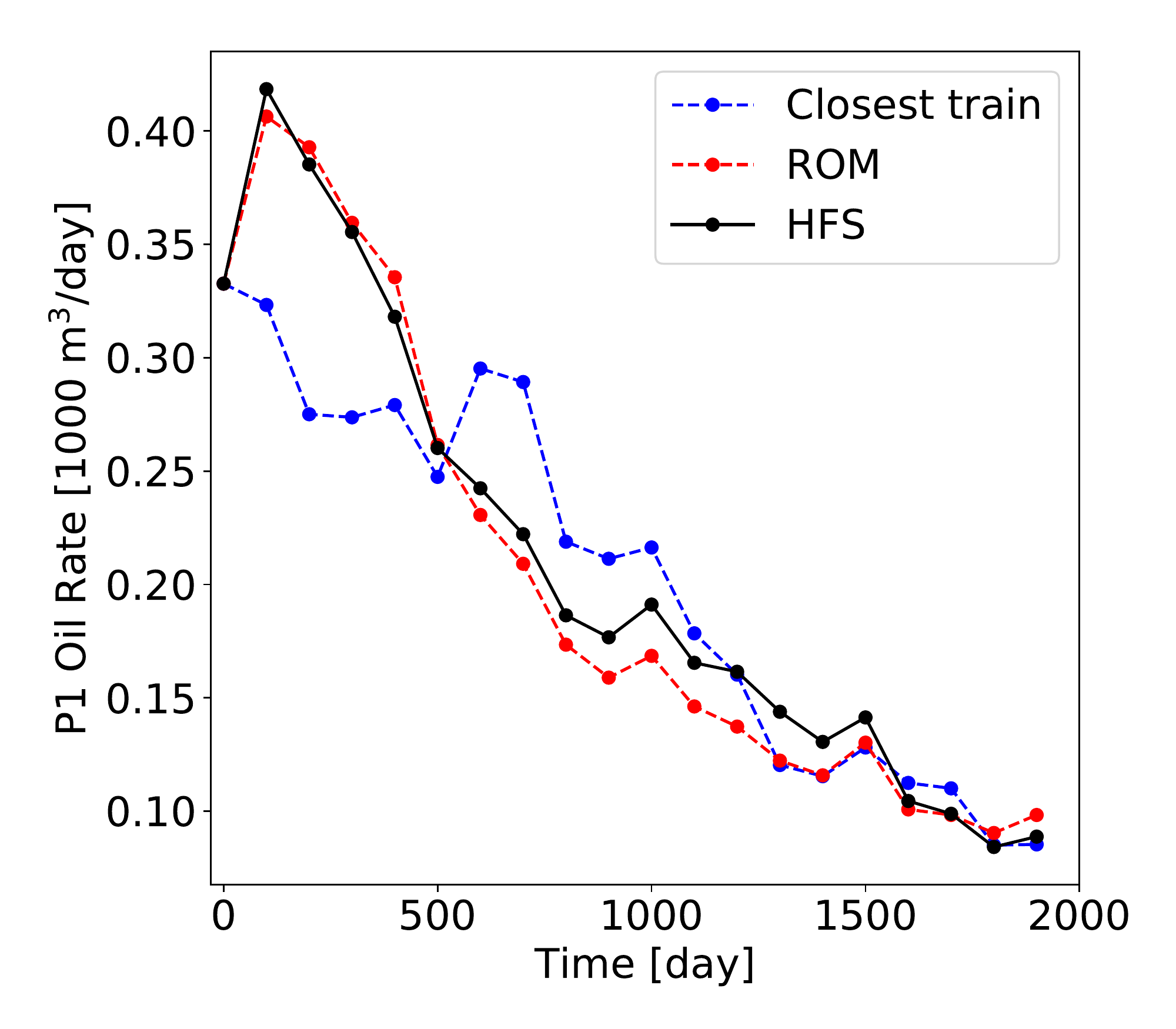}
    \caption{Oil rate}
  \end{subfigure}\hfill
  \begin{subfigure}{.5\textwidth}
    \centering
    \includegraphics[width=\linewidth]{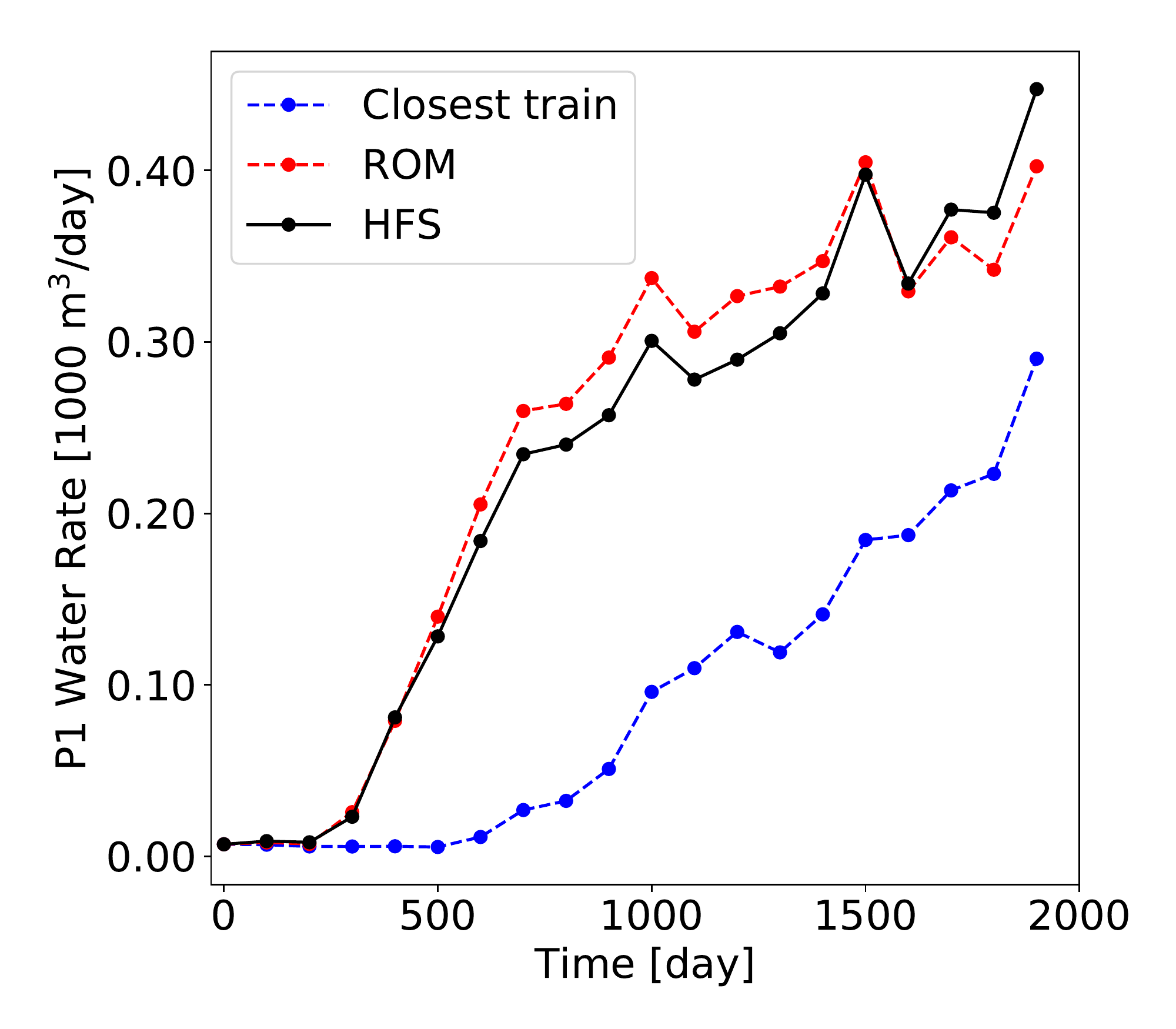}
    \caption{Water rate}
  \end{subfigure}
  \caption{Test Case~1: production rates for Well P1}
  \label{fig::test_1_rate_p1}
\end{figure}

\begin{figure}[htbp]
  \centering
  \begin{subfigure}{.5\textwidth}
    \centering
    \includegraphics[width=\linewidth]{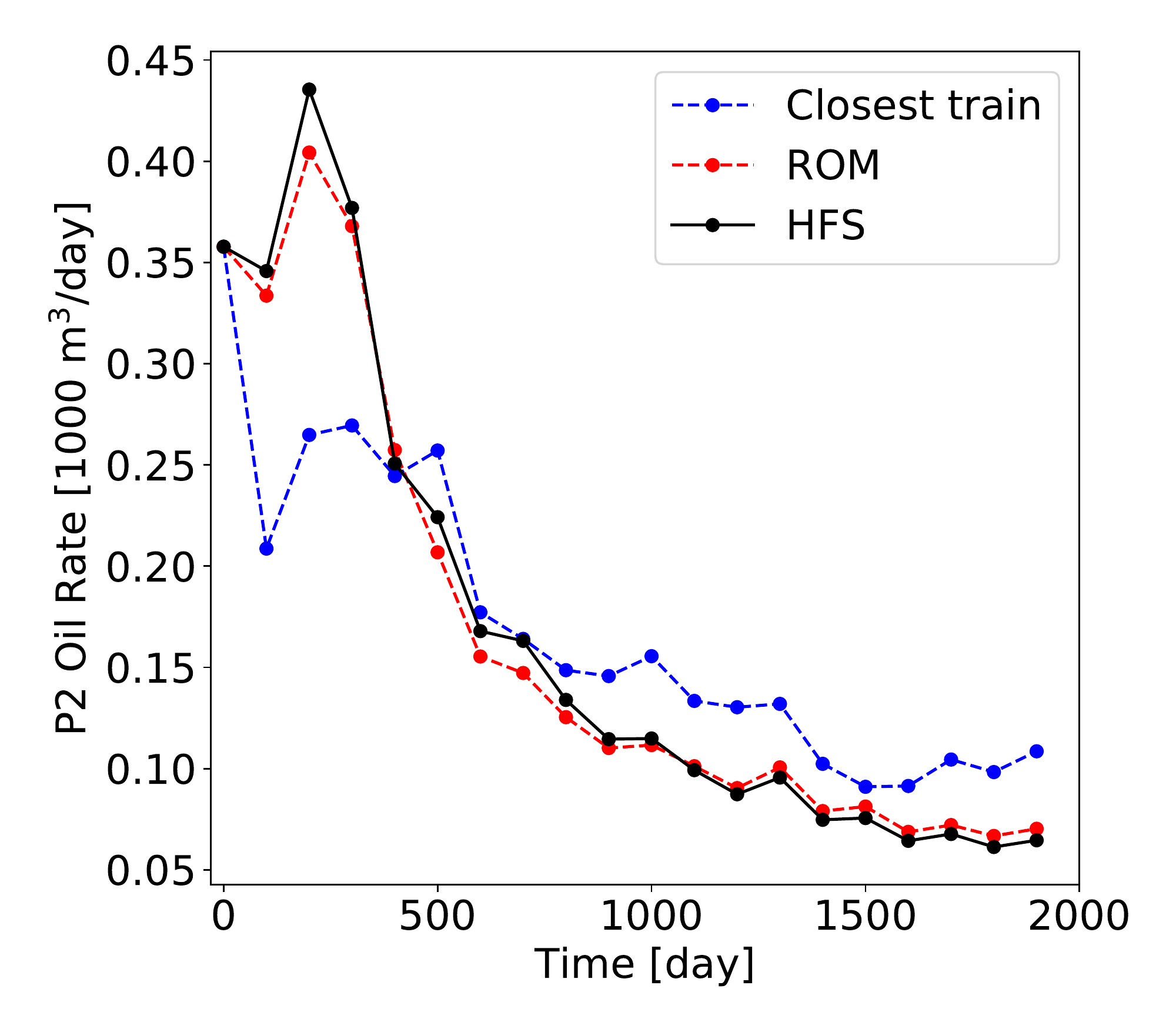}
    \caption{Oil rate}
  \end{subfigure}\hfill
  \begin{subfigure}{.5\textwidth}
    \centering
    \includegraphics[width=\linewidth]{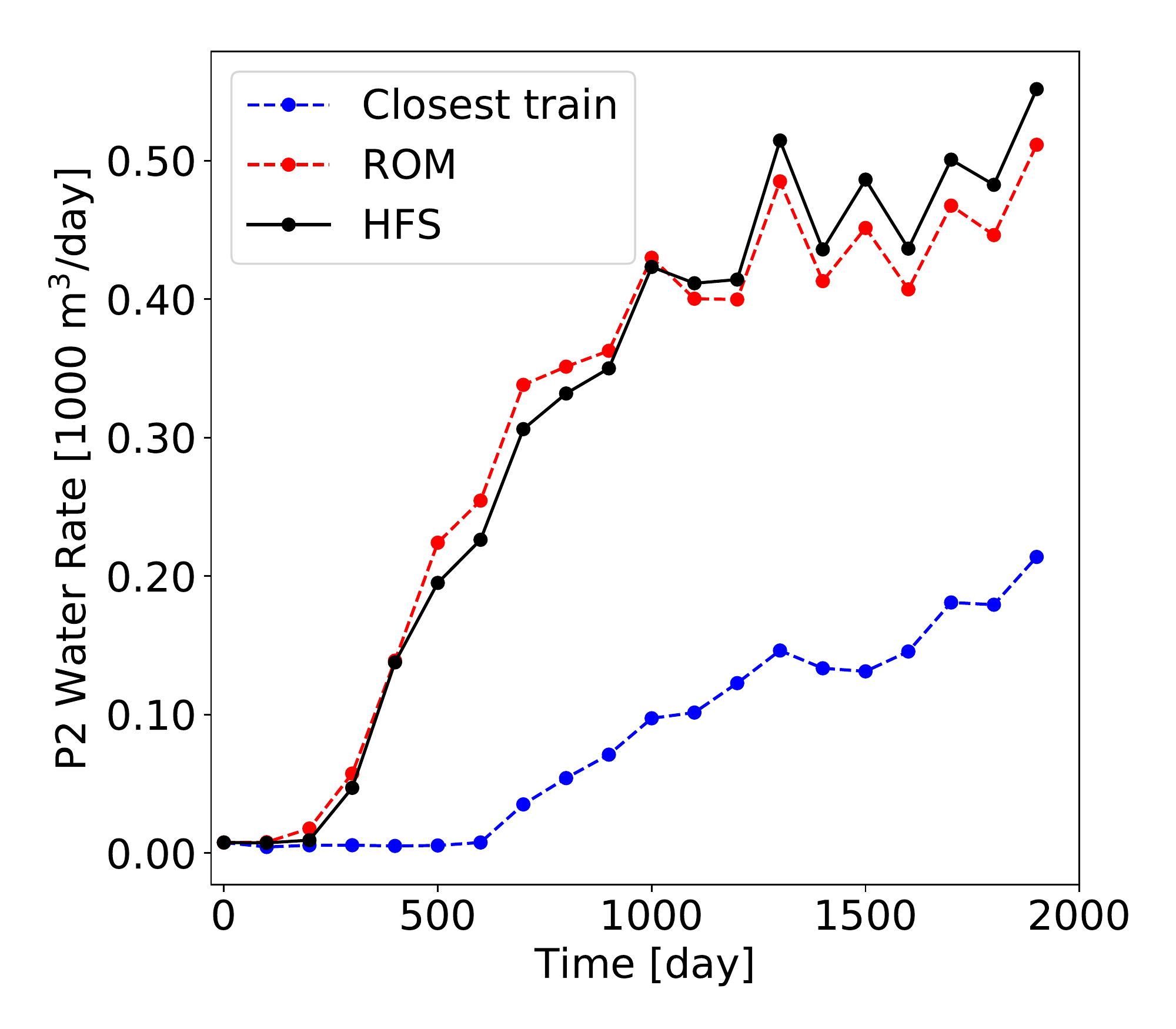}
    \caption{Water rate}
  \end{subfigure}
  \caption{Test Case~1: production rates for Well P2}
  \label{fig::test_1_rate_p2}
\end{figure}

\begin{figure}[htbp]
  \centering
  \begin{subfigure}{.45\textwidth}
    \centering
    \includegraphics[width=\linewidth]{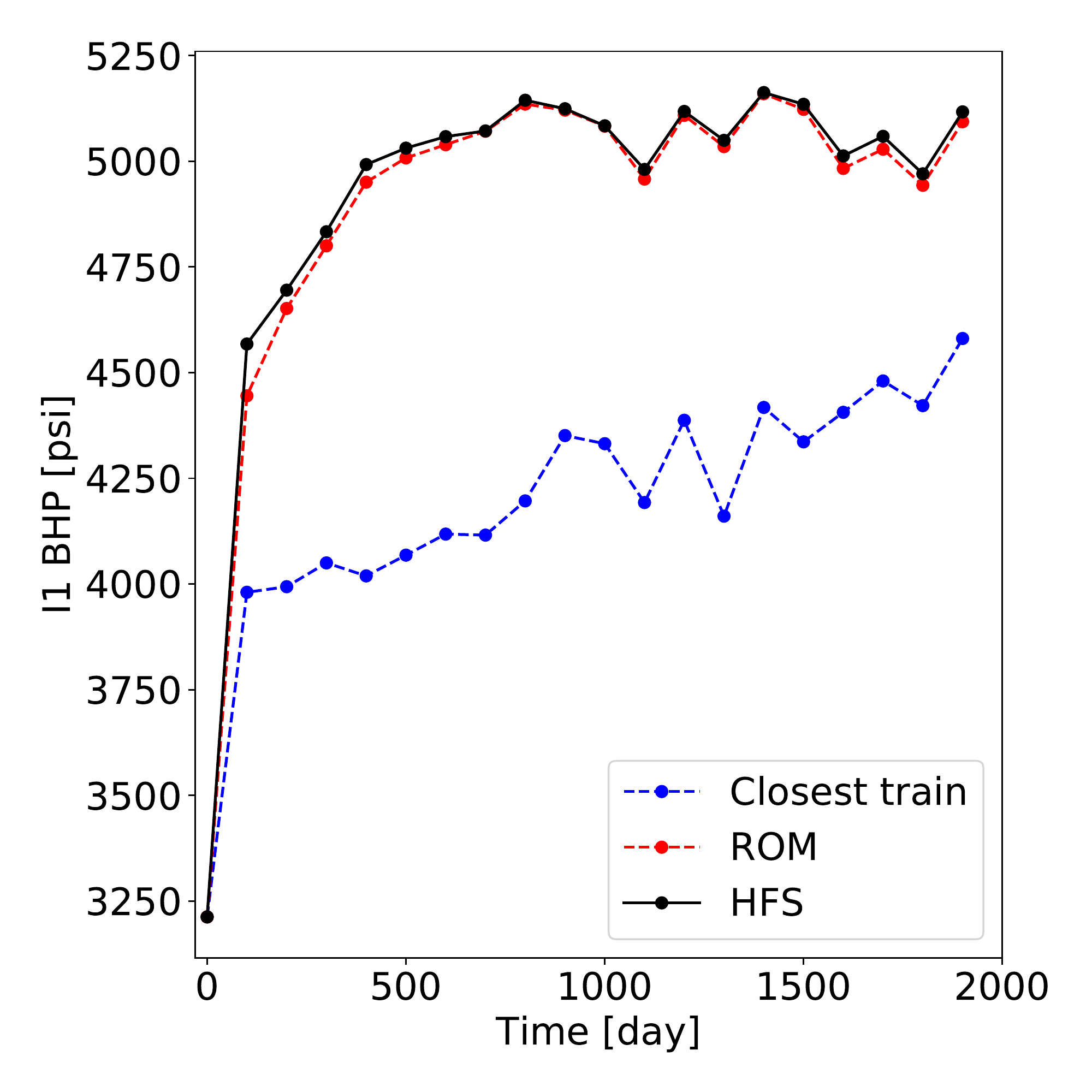}
    \caption{Well~I1}
  \end{subfigure}\hfill
  \begin{subfigure}{.45\textwidth}
    \centering
    \includegraphics[width=\linewidth]{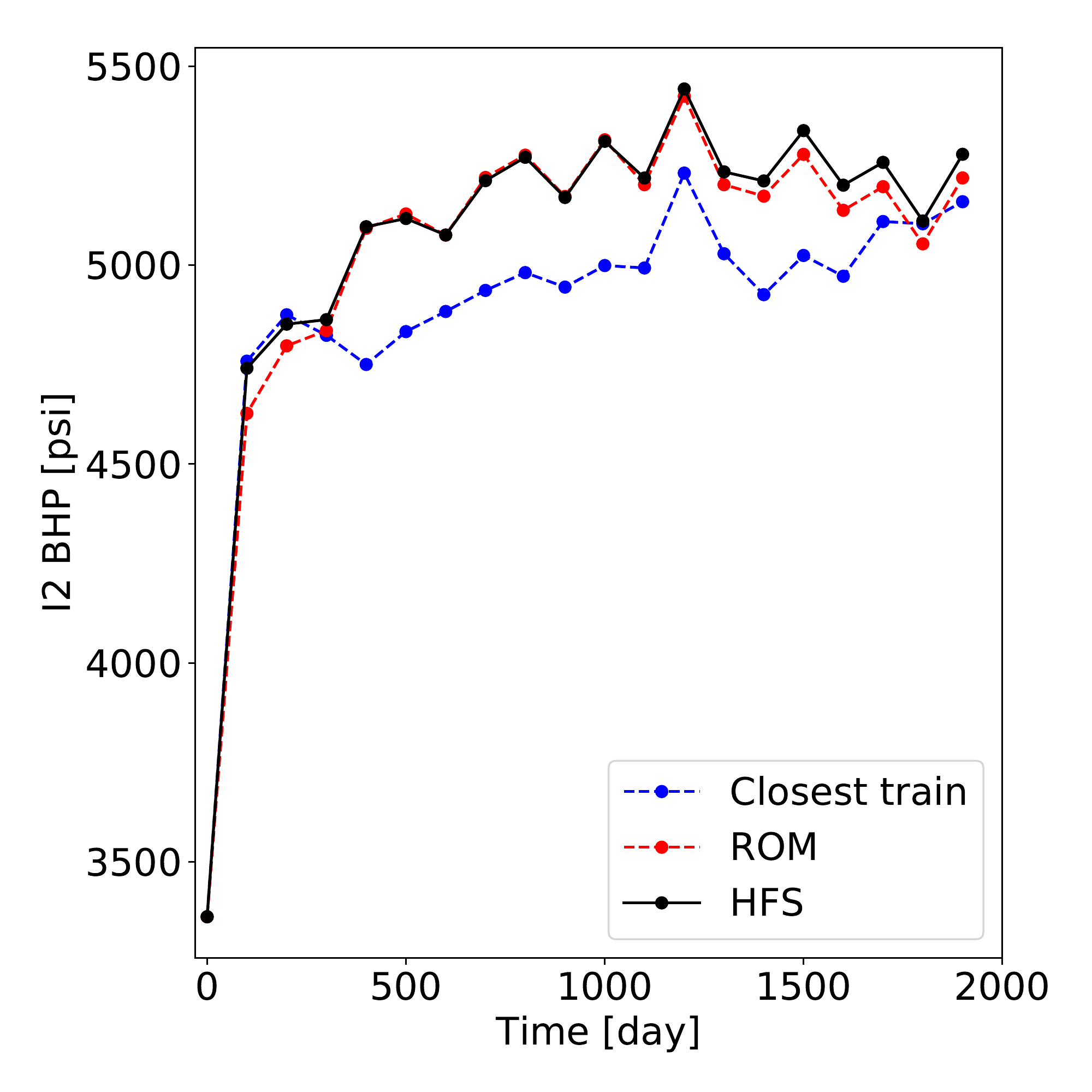}
    \caption{Well~I2}
  \end{subfigure} \\
  \begin{subfigure}{.45\textwidth}
    \centering
    \includegraphics[width=\linewidth]{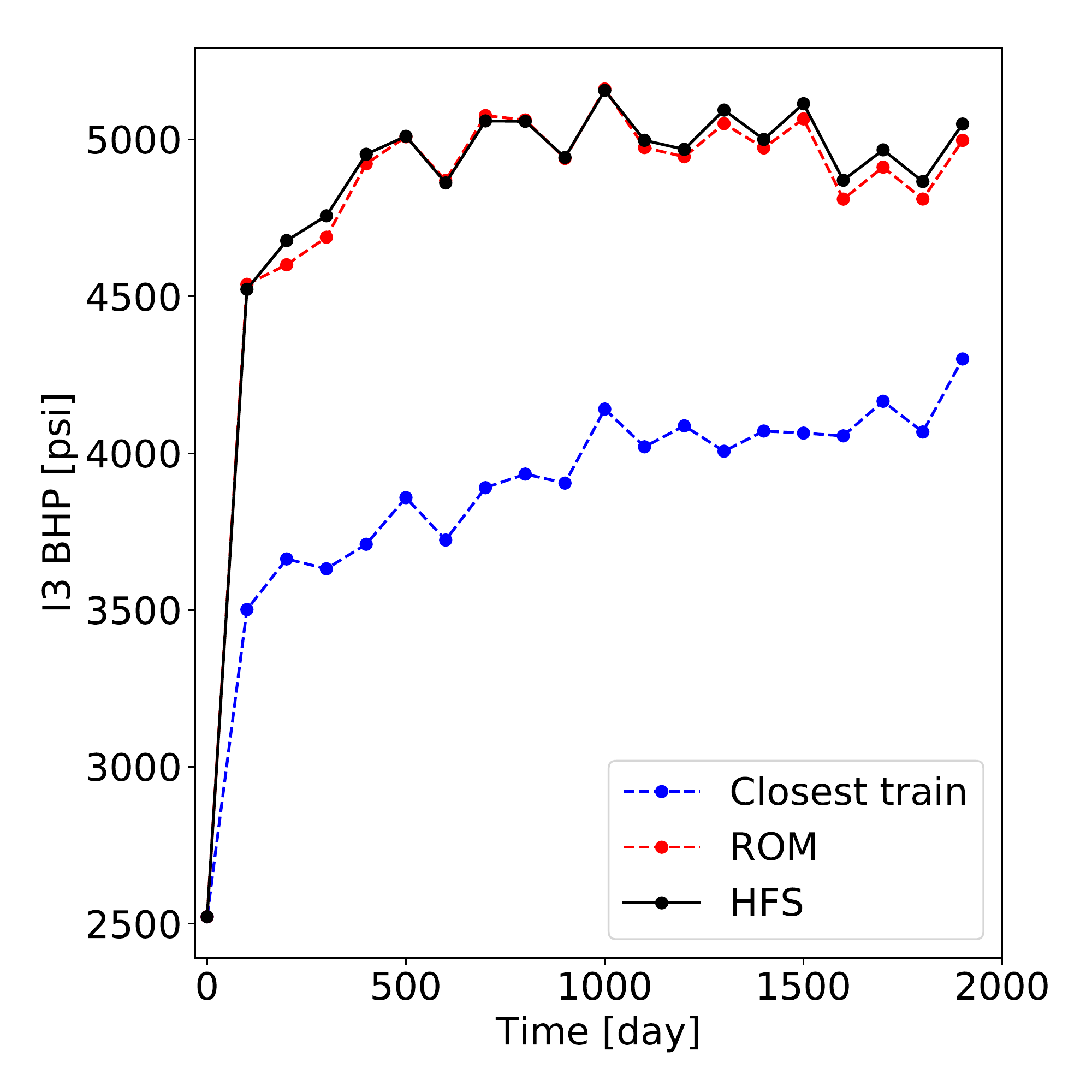}
    \caption{Well~I3}
  \end{subfigure}\hfill
    \begin{subfigure}{.45\textwidth}
    \centering
    \includegraphics[width=\linewidth]{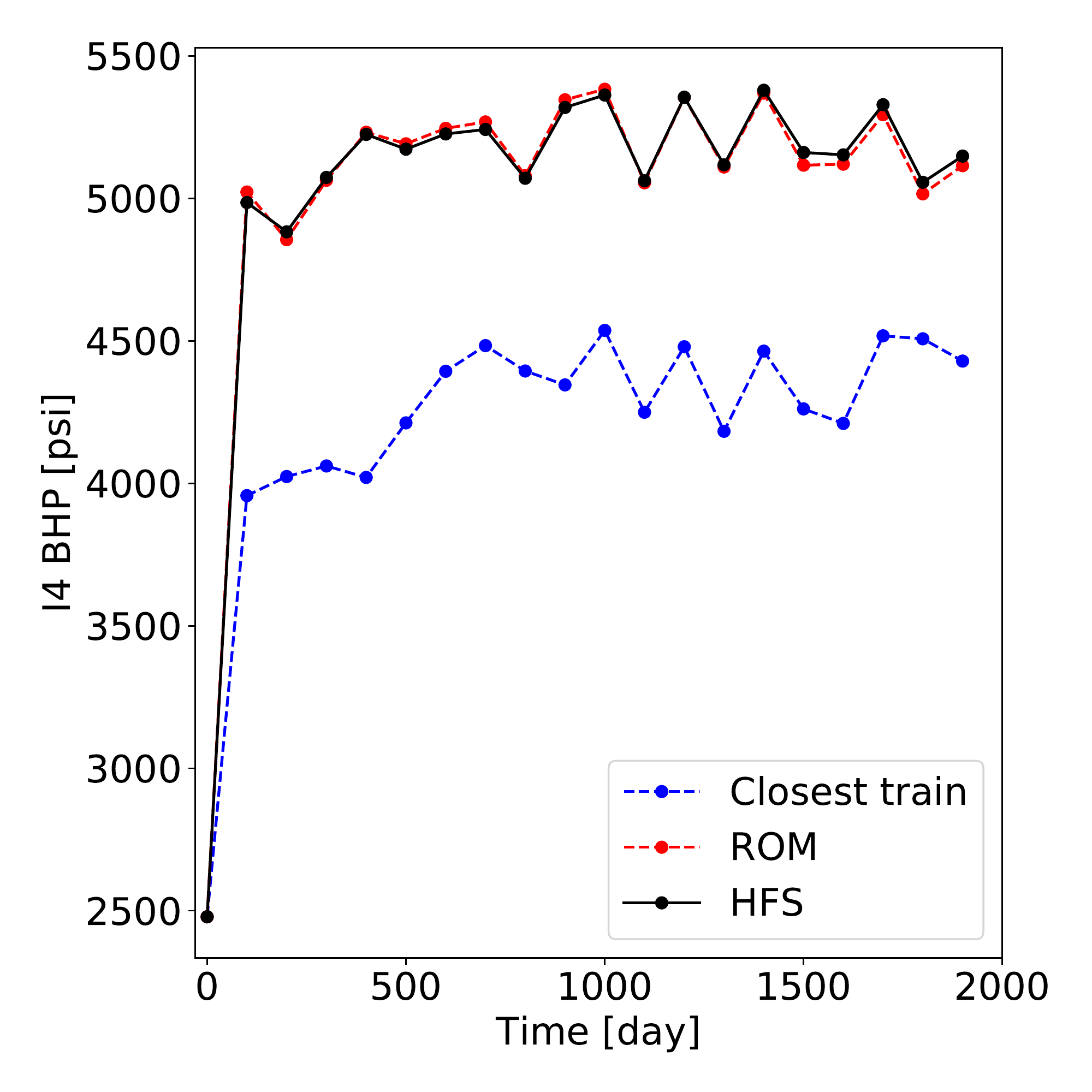}
    \caption{Well~I4}
  \end{subfigure}
  \caption{Test Case~1: injection BHPs}
  \label{fig::test_1_inj}
\end{figure}

In many subsurface flow applications the well responses are of primary interest. E2C ROM predictions for these quantities will now be assessed. Since in this problem we specify injection rates and production well BHPs, the quantities of interest are injection well BHPs and oil and water production rates. Figs.~\ref{fig::test_1_rate_p1} and \ref{fig::test_1_rate_p2} display the phase flow rates for Wells~P1 and P2, which are the wells contributing most to total field production. Fig.~\ref{fig::test_1_inj} shows the BHP responses for all four injection wells. In all figures the black curves represent the full-order (reference) HFS, the red curves are the deep-learning-based ROM results, and the blue curves are the results for the `closest training run.' A high degree of accuracy between the ROM and HFS results is consistently observed. The level of agreement in these essential quantities is enhanced through the additional weighting placed on well-block quantities in loss function $\mathcal{L}_{p}$ (see Eq.~\ref{equ::loss_physical}).

%%%%% case 1, prod rates %%%%%%%%

%%%%% case 1, inj bhps %%%%%%%%

We present results for two more examples (Test Cases~2 and 3) in \ref{appendix-case}. These results corroborate our observations here; namely, that the deep-learning-based ROM is able to accurately predict both global saturation and pressure distributions and well quantities of interest.

Finally, we discuss the timings for the high-fidelity and ROM runs. The high-fidelity test cases take 60 seconds each to simulate using AD-GPRS on a node with dual Intel Xeon CPUs (24~cores). The full batch of 100 test cases can be evaluated using the E2C ROM in about 2~seconds on a Tesla V100 GPU node with about 1~GB of memory allocated. A direct (though simplistic) comparison indicates a speedup factor of 3000.

\subsection{Results and error measurement for all test cases}

In this section we assess the accuracy of the ROM results for the full ensemble of 100 test cases. We first consider field cumulative oil and water production, which are given by
\begin{equation}\label{equ::cum_rates}
Q_{j} = \int_{0}^{T} \sum_{w=1}^{n_p} q_{j}^{w}(t) \text{d}t.
\end{equation}
Here $j = o, w$ denotes the phase, $n_p$ is the total number of production wells, $T$ designates the total simulation time, and $q_{j}^{w}(t)$ represents the fluid rate for phase $j$ at time step $t$.

\begin{figure}[htbp]
  \centering
  \begin{subfigure}{.45\textwidth}
  	\centering
  	\includegraphics[width=\linewidth]{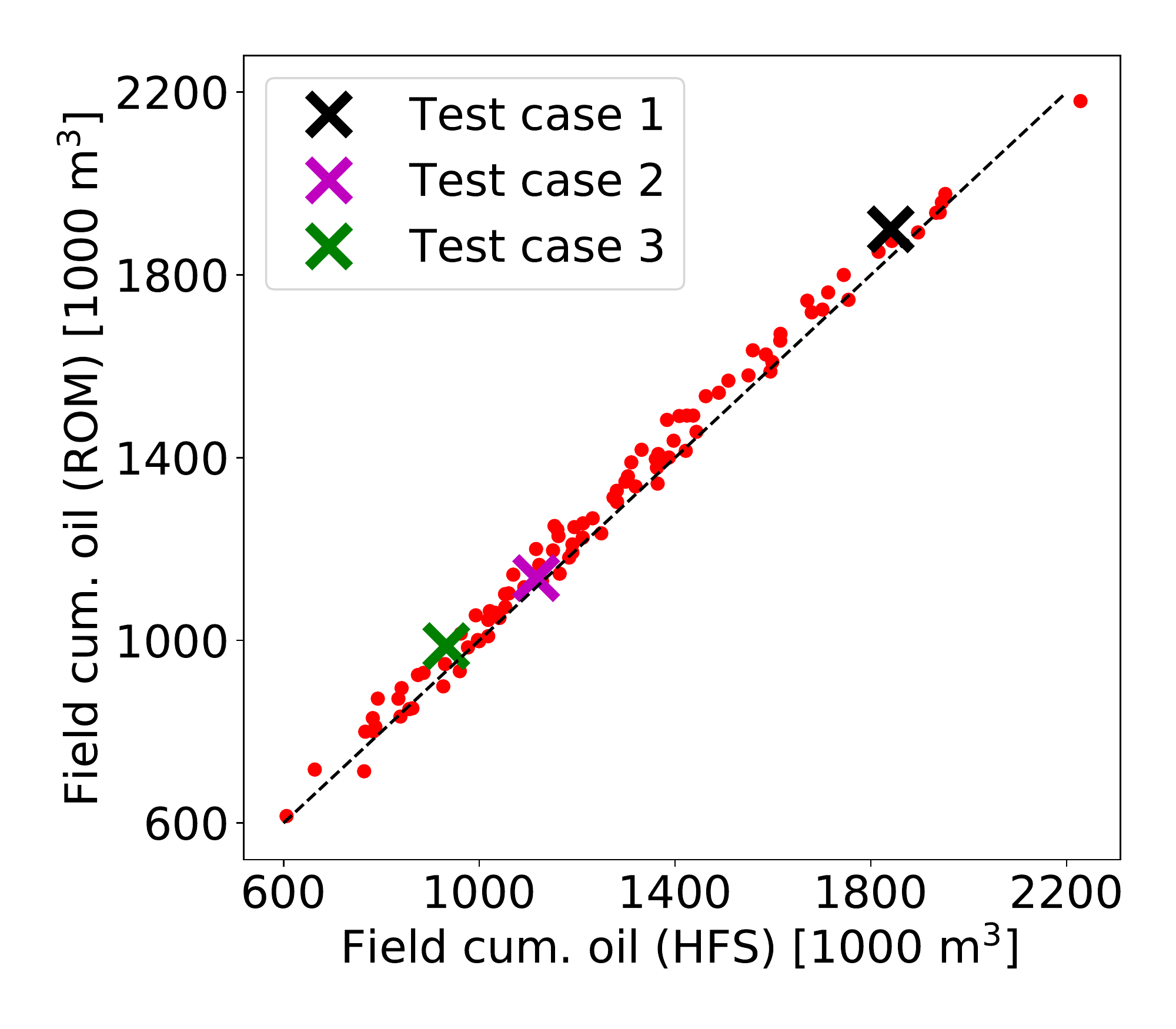}
    \caption{Cumulative oil production}
  \end{subfigure}\hfill
  \begin{subfigure}{.45\textwidth}
  	\centering
  	\includegraphics[width=\linewidth]{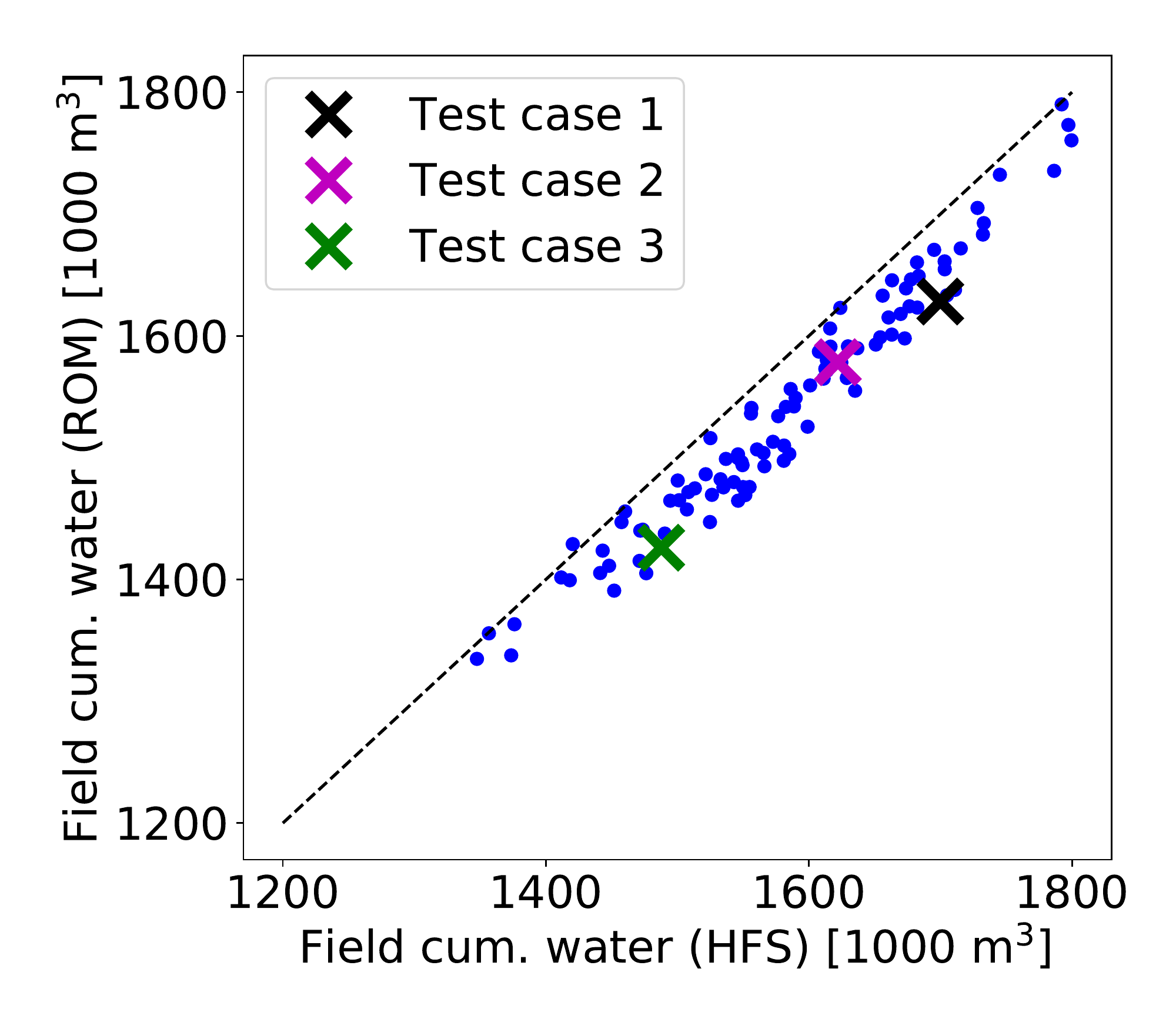}
    \caption{Cumulative water production}
  \end{subfigure} 
	\caption{Cumulative oil and water production for all 100 test cases}
	\label{fig::cum_rates}
\end{figure}

In Fig.~\ref{fig::cum_rates} we present crossplots of $Q_o$ and $Q_w$ for the HFS and ROM solutions for the 100 test cases. The three $\times$'s on each plot indicate the results for the Test Cases~1, 2 and 3. It is evident that these cases are quite different in terms of $Q_o$ and $Q_w$, and in this sense span the range of the 100 test cases. We see that the points in both plots fall near the 45$^{\circ}$ line, which demonstrates that our ROM solutions are in close agreement with the HFS. The results for $Q_w$ in Fig.~\ref{fig::cum_rates}(b) indicate that the ROM consistently under-predicts cumulative water production. The under-prediction is relatively small, however, as the range covered in this plot is narrow. Note finally that a slight over-prediction for cumulative oil production is evident in Fig.~\ref{fig::cum_rates}(a).

%%%%%%%%%%%%%%%%%%%%%%%%%%
%%%%%%% cum rates %%%%%%%%
%%%%%%%%%%%%%%%%%%%%%%%%%%

We now introduce a number of error measures, which will be used to assess the general performance of the E2C ROM. These error metrics follow those used in \citep{jin2019reduced}. The relative error for oil or water production rate, for a single production well $p$, is defined as:
\begin{equation}\label{equ::3.2.1}
e_{j}^{p} = \frac{\int_{0}^{T}\big| q_{\text{ROM}}^{j, p}(t) - q_{\text{HFS}}^{j, p}(t)\big| \text{d}t}{\int_{0}^{T}\big| q_{\text{HFS}}^{j, p}(t)\big|\text{d}t},
\end{equation}
where $j = o, w$ is the fluid phase, $q^{j, p}(t)$ is the oil or water production rate at time $t$ for production well $p$, the subscripts $\text{HFS}$ and $\text{ROM}$ denote the high-fidelity and ROM results, and $T$ is the total simulation time. We define the error for overall production rate, $E_r$, in terms of $e_o$ and $e_w$ for all production wells, as:
\begin{equation}\label{equ::3.2.2}
E_{r} = \frac{1}{n_p}\sum_{p=1}^{n_p}(e_{o}^{p} + e_{w}^{p}),
\end{equation}
where $n_p$ is the total number of production wells. Similarly, the relative error in injection BHP for a single injection well $i$ is defined as:
\begin{equation}\label{equ::3.2.3}
e_{\text{BHP}}^{i} = \frac{\int_{0}^{T}\big| p_{\text{ROM}}^{w,i}(t) - p_{\text{HFS}}^{w,i}(t)\big| \text{d}t}{\int_{0}^{T}\big| p_{\text{HFS}}^{w,i}(t)\big|\text{d}t},
\end{equation}
where $p^{w,i}(t)$ denotes the injection BHP at time $t$ for injection well $i$. The overall injection well BHP error $E_{\text{BHP}}$ is then given by:
\begin{equation}\label{equ::3.2.4}
E_{\text{BHP}} = \frac{1}{n_i}\sum_{i=1}^{n_i} e_{\text{BHP}}^{i},
\end{equation}
where $n_i$ is the total number of injection wells.

Error in global quantities is also of interest. We define global pressure and saturation error as:
\begin{equation}\label{equ::3.2.6}
E_{v} = \frac{\sum_{k=1}^{n_b}\int_{0}^{T}\big| v_{\text{ROM}}^{k} - v_{\text{HFS}}^{k}\big| \text{d}t}{\sum_{k=1}^{n_b}\int_{0}^{T}\big| v_{\text{HFS}}^{k}\big|\text{d}t},
\end{equation}
where $v^{k}$ denotes the global variable of interest in grid block $k$ (pressure $p^{k}$ or saturation $S^{k}$), and $n_b$ is the total number of grid blocks in the model. 

%Note that the definitions of error measurement closely resemble the ones in our earlier work \citep{jin2019reduced}, which should be consulted for further detail.

%%%%%%%%%%%%%%%%%%%%%%%%%%
%%%%% rel. errors %%%%%%%%
%%%%%%%%%%%%%%%%%%%%%%%%%%

\begin{figure}[htbp]
  \centering
  \begin{subfigure}{.45\textwidth}
  	\centering
  	\includegraphics[width=\linewidth]{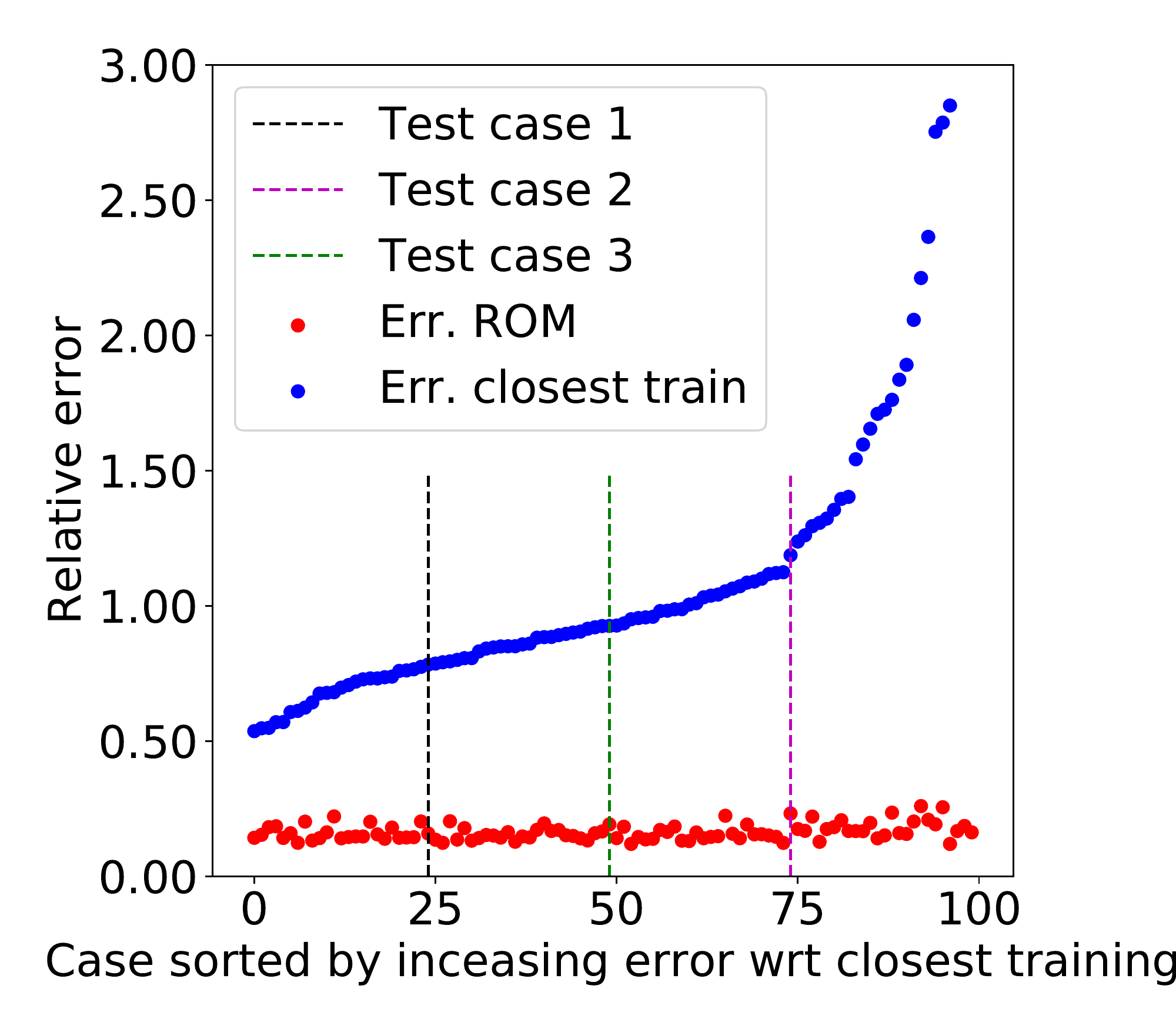}
    \caption{Production rates ($E_{r}$)}
  \end{subfigure}\hfill
  \begin{subfigure}{.45\textwidth}
  	\centering
  	\includegraphics[width=\linewidth]{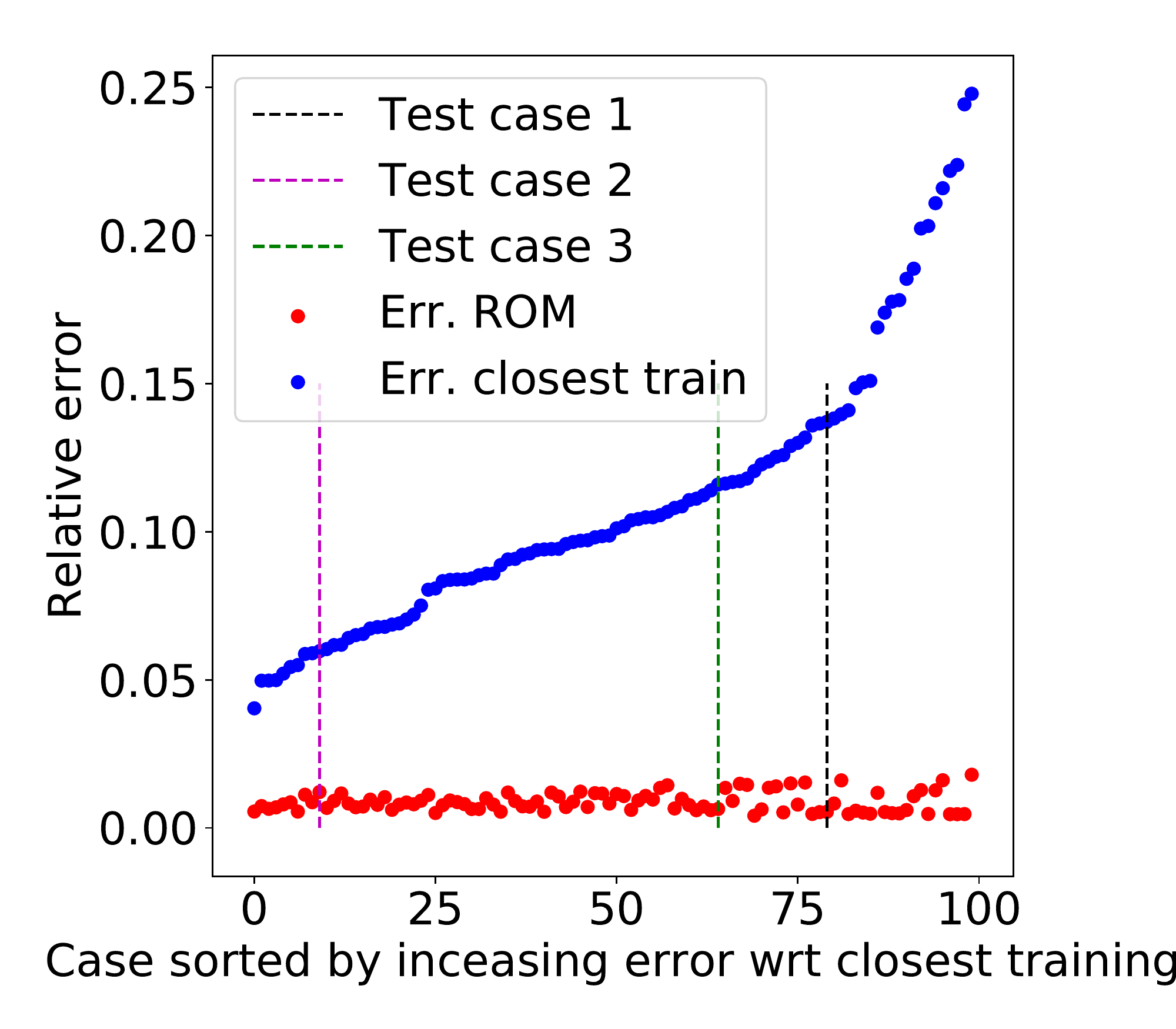}
    \caption{Injection BHPs ($E_{\text{BHP}}$)}
  \end{subfigure} \\
  \begin{subfigure}{.45\textwidth}
  	\centering
  	\includegraphics[width=\linewidth]{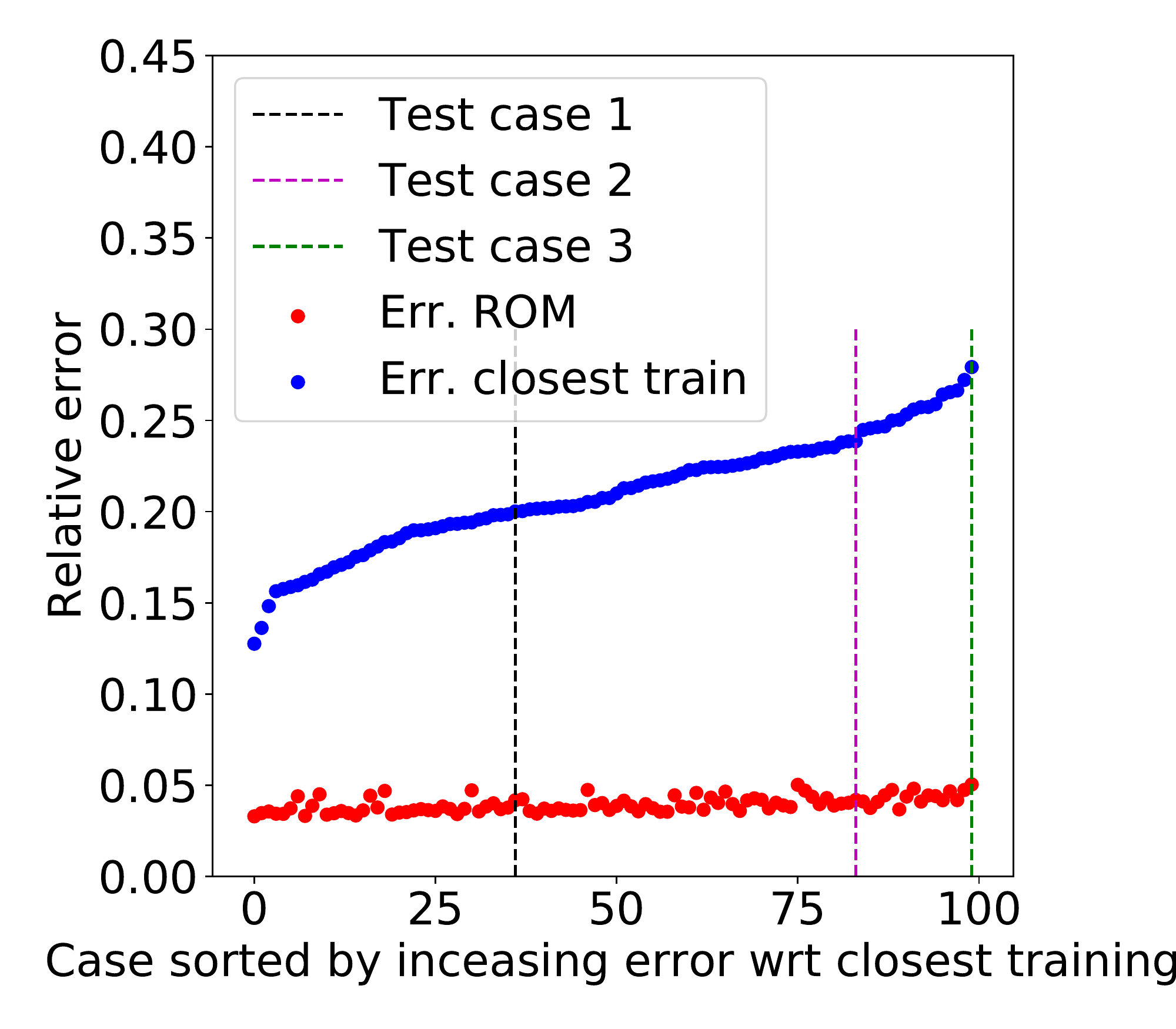}
    \caption{Saturation field ($E_{S}$)}
  \end{subfigure}\hfill
    \begin{subfigure}{.45\textwidth}
    \centering
    \includegraphics[width=\linewidth]{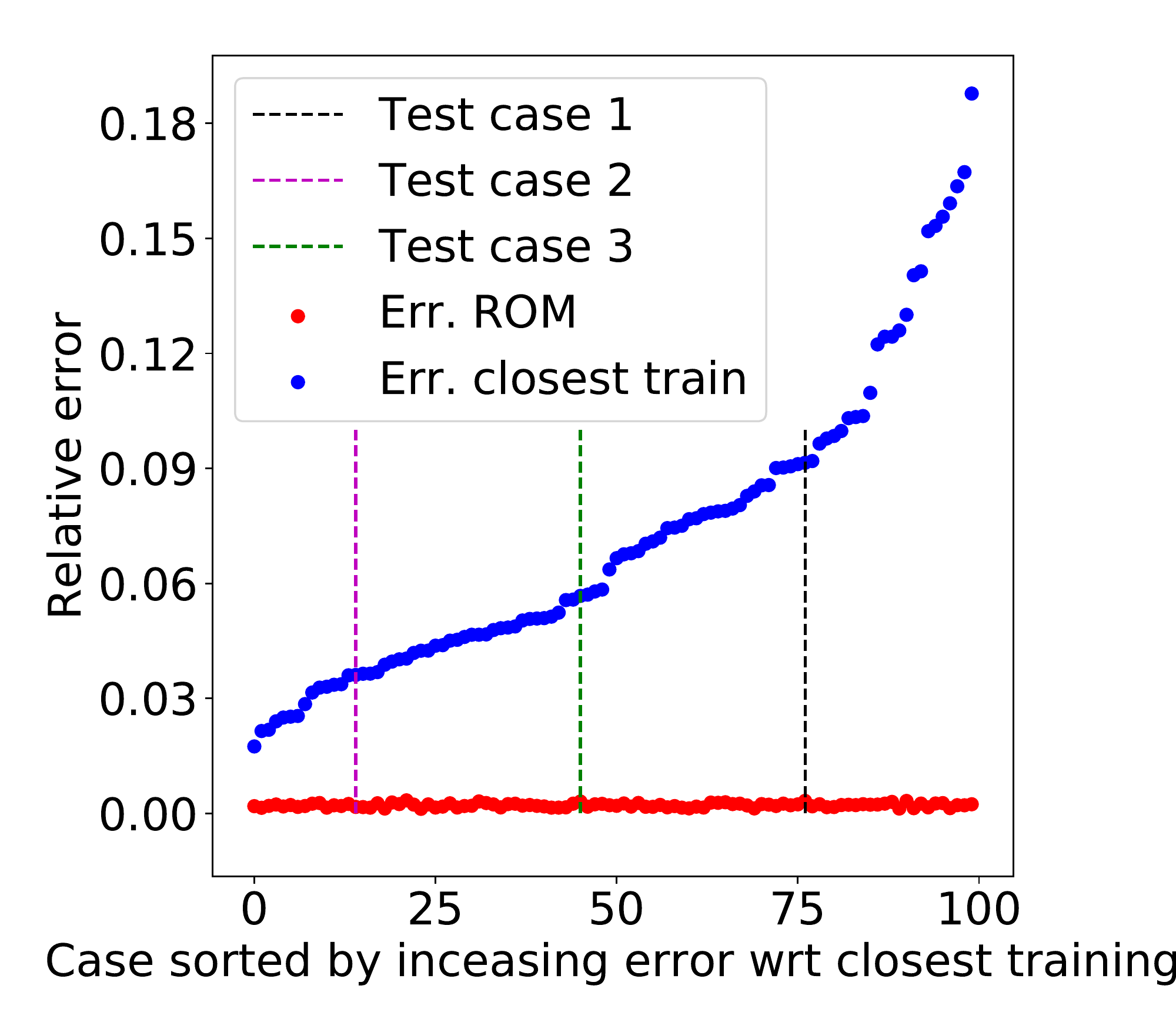}
    \caption{Pressure field ($E_{p}$)}
  \end{subfigure}
	\caption{Errors for quantities for interest}
	\label{fig::rel_err}
\end{figure}

These four error quantities are displayed as the red points in Fig.~\ref{fig::rel_err}. We also evaluate these errors for the `closest training run' for all test cases. In the plots, the points are ordered by increasing error for the `closest training run' (blue points). Results for Test Cases~1, 2 and 3 are indicated in each plot. We see that the ROM errors are consistently very small, while the errors for the `closest training run' are large in many cases. Interestingly, the ROM errors do not appear to depend on the error associated with the `closest training run.' This is a desirable feature as it suggests a high degree of robustness in the E2C ROM.

%%%%%%%%%%%%%%%%%%%%%%%%%%%%%%%%%%%%
%%%%% err vs. num. training %%%%%%%%
%%%%%%%%%%%%%%%%%%%%%%%%%%%%%%%%%%%%

\begin{figure}[htbp]
  \centering
  \begin{subfigure}{.45\textwidth}
    \centering
    \includegraphics[width=\linewidth]{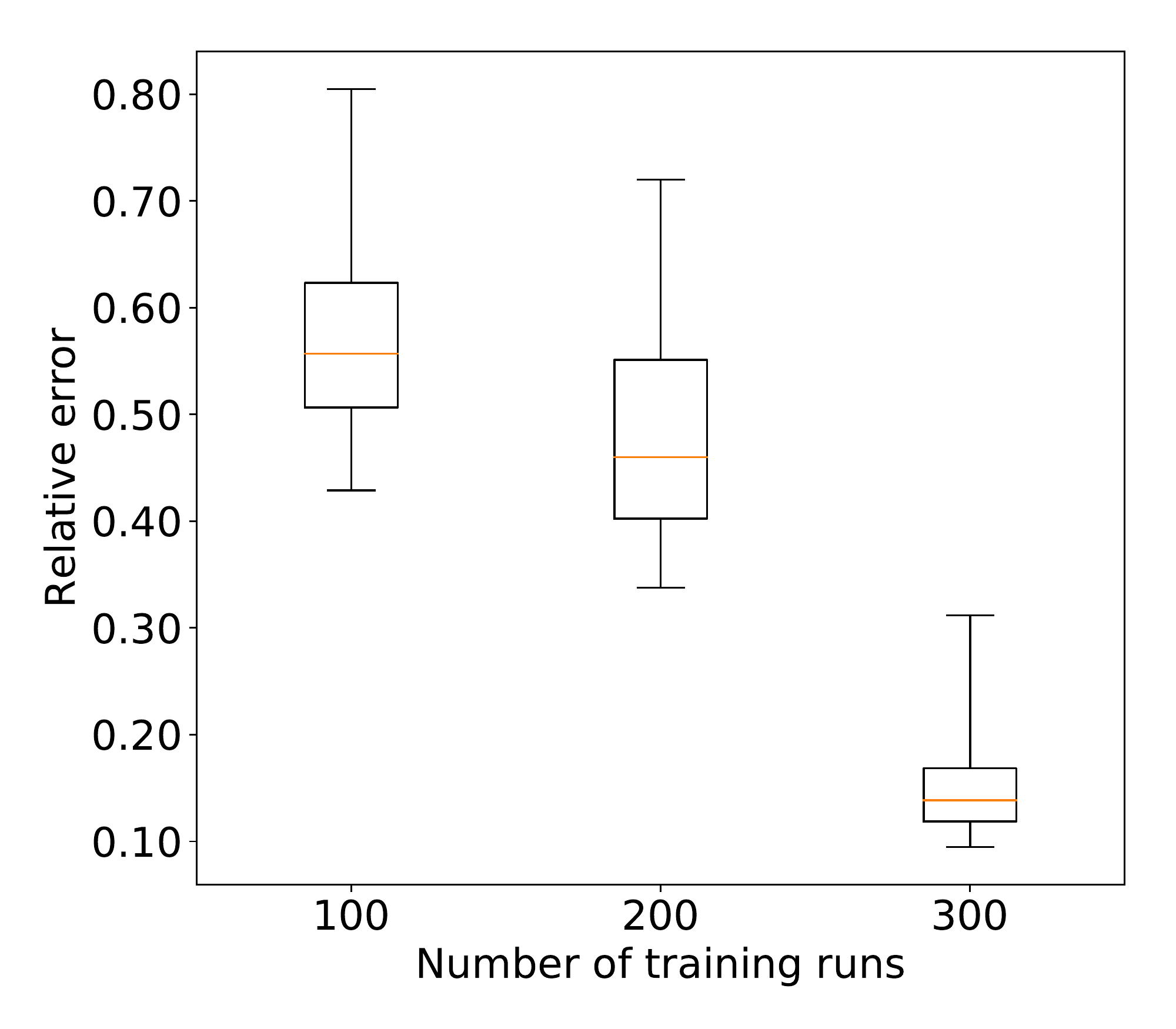}
    \caption{Error in production rates ($E_{r}$)}
  \end{subfigure}\hfill
  \begin{subfigure}{.45\textwidth}
    \centering
    \includegraphics[width=\linewidth]{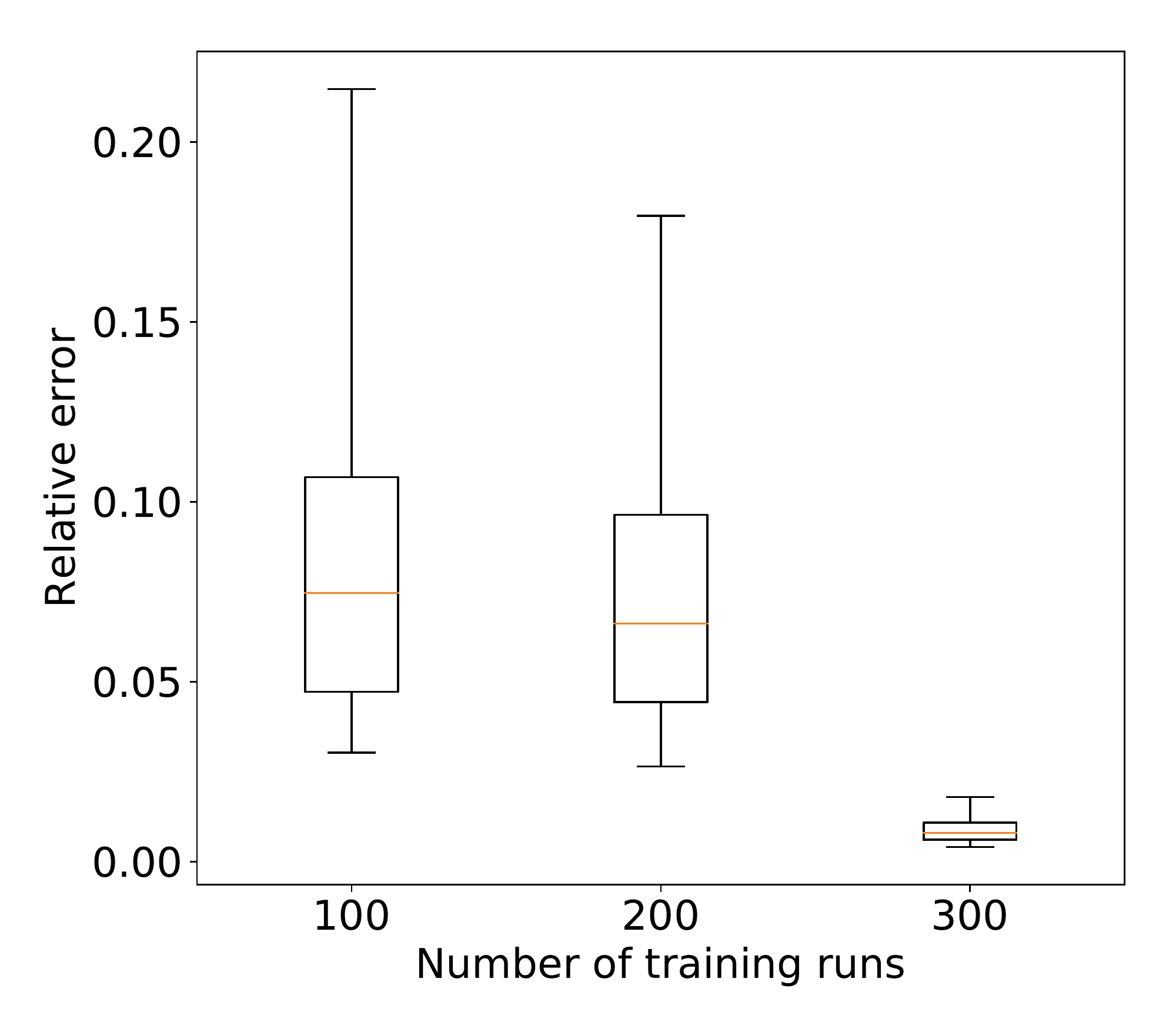}
    \caption{Error in injection BHPs ($E_{\text{BHP}}$)}
  \end{subfigure} \\
  \begin{subfigure}{.45\textwidth}
    \centering
    \includegraphics[width=\linewidth]{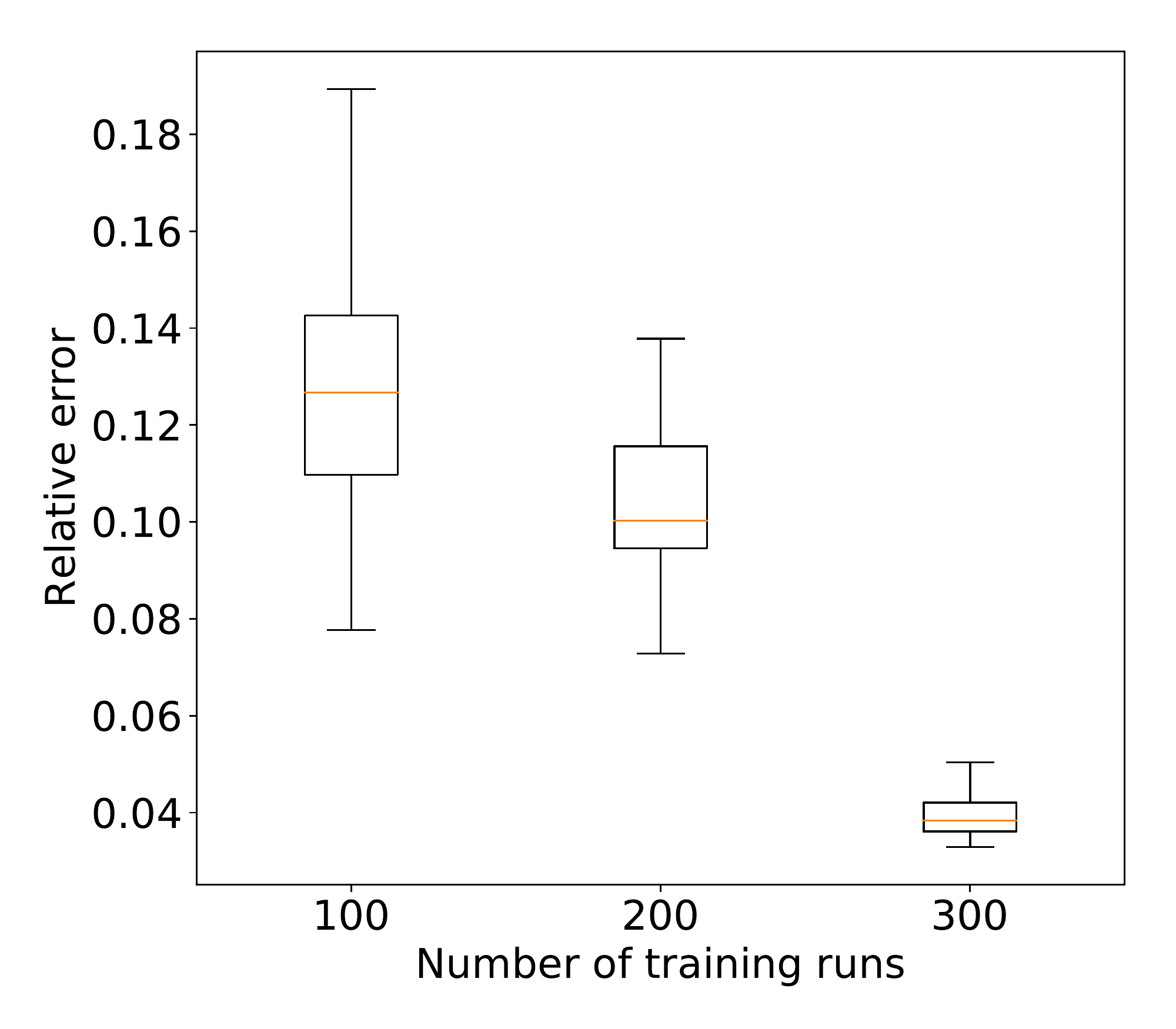}
    \caption{Error in saturation field ($E_{S}$)}
  \end{subfigure}\hfill
    \begin{subfigure}{.45\textwidth}
    \centering
    \includegraphics[width=\linewidth]{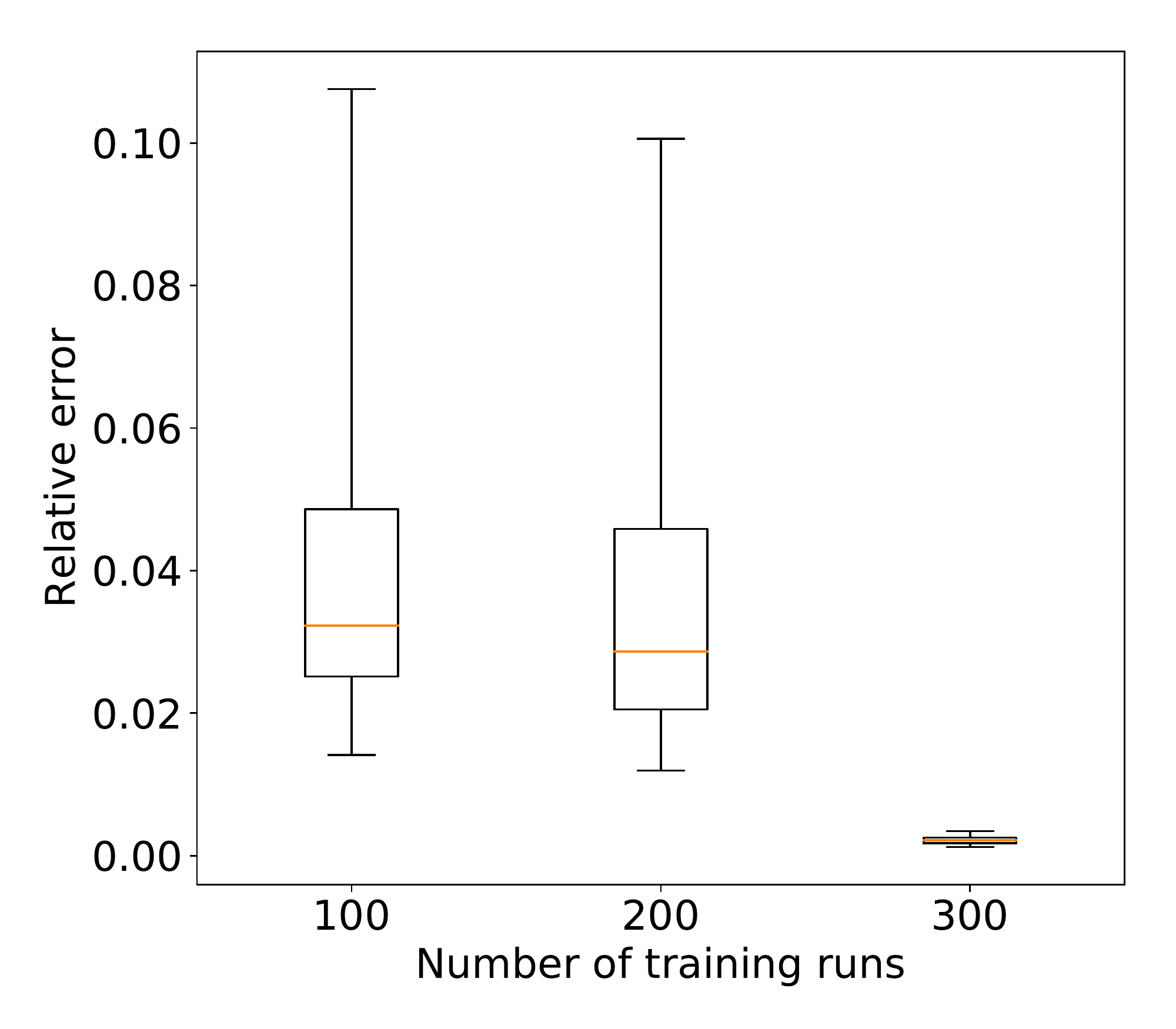}
    \caption{Error in pressure field ($E_{p}$)}
  \end{subfigure}
  \caption{ROM error with different numbers of training runs}
  \label{fig::err_cmp}
\end{figure}

We now briefly consider the use of smaller numbers of training runs in the construction of the E2C ROM. For these cases we present only summary error results. Fig.~\ref{fig::err_cmp} displays the four relative errors considered above, in terms of box plots, for 100, 200 and 300 training runs. In each box, the central orange line indicates the median error, and the bottom and top edges of the box show the 25th and 75th percentile errors. The `whiskers' extending out from the boxes indicate the minimum and maximum errors. There is significant improvement in ROM accuracy as we proceed from 200 to 300 training runs. In future work it will be useful to establish approaches to determine the required number of training runs.

Because it is difficult to display the errors for Test Cases~1, 2 and 3 in the box plots in Fig.~\ref{fig::err_cmp}, we present them in Table~\ref{tab::rel_err}. These results are with 300 training runs.
%Finally, the errors for the three selected test cases and their corresponding percentiles within the 100 test cases are shown in Table~\ref{tab::rel_err}. 
Note that the average values for $E_{r}$, $E_{\text{BHP}}$, $E_{S}$ and $E_{p}$ across all 100 test cases are about 0.14, 0.02, 0.04 and 0.002, respectively. The error values for the three test cases shown in the table can be seen to represent a reasonable spread among the full set of test cases. It is of interest to observe that the four errors do not appear to be closely correlated within a particular test case. For example, in Test Case~1, $E_p$ is in the 98th percentile, while $E_{\text{BHP}}$ is in the 15th percentile.

%could have large value for one error while small values for other errors. For instance, Test Case~1 has below-average value for $E_{r}$ and $E_{\text{BHP}}$, while its $E_{p}$ is almost the worst among all 100 test cases. On the other hand, Test Case~2 has near-worst error values for $E_{r}$, $E_{S}$ and $E_{p}$, while its $E_{\text{BHP}}$ is below the average.

% \begin{table}[!htb]
% \centering
% \begin{tabular}{l@{\hskip .5in} l@{\hskip .25in} l@{\hskip .25in} l@{\hskip .25in} l}
%     \hline
%     \textbf{Model/Avg. Err.}      &$E_{r}$        &$E_{\text{BHP}}$       &$E_{S}$        &$E_{p}$ \\
%         \\[-1em]
%     \hline
%     ROM                 &14\%           &2\%                    &4\%            &0.2\%    \\
%         \\[-1em]
%     Closest train       &105\%          &12\%                   &22\%           &9\%      \\
%     \hline  
% \end{tabular}
% \caption{Average relative error for the closest training run and the ROM on 100 test runs}
% \label{tab::rel_err}
% \end{table}

\begin{table}[!htb]
\centering
\begin{tabular}{l@{\hskip .5in} l@{\hskip .25in} l@{\hskip .25in} l@{\hskip .25in} l@{\hskip .25in} l}
    \hline
    \textbf{Test cases} &   &$E_{r}$        &$E_{\text{BHP}}$   &$E_{S}$        &$E_{p}$ \\
        \\[-1em]
    \hline
    Test Case~1             &Error      &0.11       &0.0054     &0.042      &0.0033    \\
        \\[-1em]
                           &Percentile     &37      &15         &70         &98      \\
    \hline
    Test Case~2             &Error      &0.19       &0.0064     &0.050      &0.0031    \\
        \\[-1em]
                           &Percentile     &96      &29         &99         &95      \\
   \hline
    Test Case~3             &Error      &0.16       &0.012      &0.042      &0.0017    \\
        \\[-1em]
                           &Percentile     &77      &84         &73         &24      \\
    \hline  
\end{tabular}
\caption{Average error and percentile for test cases}
\label{tab::rel_err}
\end{table}
% -------------------------------------------------------------
% -------------------------------------------------------------
\section{Concluding remarks}\label{conclusion}
% -------------------------------------------------------------
% -------------------------------------------------------------

In this work, we introduced a deep-learning-based reduced-order modeling procedure for subsurface flow simulation. The procedure was adapted from the existing embed-to-control (E2C) procedure, though we introduced some key modifications relative to the formulation in \citep{watter2015embed}. Essentially, the ROM consists of an auto-encoder (AE) and a linear transition model. In our E2C formulation, an additional physics-based loss function was combined with the data-mismatch loss function to enhance consistency with the governing flow equations. Although it is based on deep-learning concepts and methods, the various E2C ROM steps were shown to be very analogous to those used in the well-developed physics/numerics-based POD-TPWL ROM. 

%Generally accurate ROM results for key quantities of interest, including well injection BHPs, phase production rates, global pressure and saturation fields, were achieved.

In most of our evaluations, we performed 300 training runs in the offline step. Excluding the run time for the training simulations, the offline model construction required 10-12~minutes for ROM training using a Tesla V100 GPU. Online (runtime) speedups of $\mathcal{O}$(1000), relative to AD-GPRS full-order simulations, were observed for the case considered. Given the offline costs and online speedup, the use of this ROM is appropriate when many (related) simulation runs are required. This is the case in production optimization computations, data assimilation and uncertainty assessments (though in this work only a single geological model was considered).

The deep-learning-based ROM was tested on two-dimensional oil-water reservoir simulation problems involving a heterogeneous permeability field. Large variations (relative to training runs) in injection and production well control settings were prescribed in the test cases. A total of 100 test cases were considered. ROM accuracy was assessed for key quantities of interest, including well injection BHPs, phase production rates, and global pressure and saturation fields. The E2C ROM was shown to be consistently accurate over the full set of test runs. ROM error was seen to be much lower than that for the `closest training run' (appropriately defined). Error was found to increase, however, if 100 or 200 training runs were used instead of 300. 

In future work, the E2C ROM should be extended to more complicated three-dimensional problems and tested on realistic cases. Extension to three dimensions can be approached by replacing conv2D layers with conv3D layers. The ROM can be readily used with various optimization algorithms for production optimization, and its performance in this setting should be evaluated. Importantly, the E2C ROM should be applicable for use with global as well as local optimization algorithms. This is in contrast to existing POD-based ROMs, which can only be expected to be accurate in more limited neighborhoods and are thus most suitable for local-search methods. It is also of interest to explore the potential of predicting flow responses with changing well locations. If this is successful, the ROM could be applied for well location optimization, or combined well location and control optimization problems. %In addition, a new speedup measurements need to be established for more objective evaluation of our ROM performances (i.e, comparing CPU time for full-order simulation to GPU time for ROM). 
To improve the accuracy and robustness of the framework for these more challenging applications, more complicated encoder and decoder structures, such as denseNet \citep{jegou2017one}, could be evaluated. Finally, the auto-encoder used here could be extended to a VAE or the uncertainty auto-encoder \citep{grover2018uncertainty} to enable system and control uncertainties to be taken into account.


\section*{Acknowledgements}
We are grateful to the Stanford University Smart Fields Consortium (SFC) for partial funding of this work. We thank the Stanford Center for Computational Earth \& Environmental Science (CEES) for providing the computational resources used in this study. We also thank Yuke Zhu and Aditya Grover for useful discussions, and Oleg Volkov for help with the AD-GPRS software.

% -------------------------------------------------------------
% -------------------------------------------------------------
\appendix\label{append}
% -------------------------------------------------------------
% -------------------------------------------------------------

% -------------------------------------------------------------
% -------------------------------------------------------------
\section{Embed-to-control network architecture}\label{appendix-e2c}
% -------------------------------------------------------------
% -------------------------------------------------------------

\begin{figure}[htbp]
  \centering
  \begin{subfigure}{.25\textwidth}
    \centering
    \includegraphics[width=\linewidth]{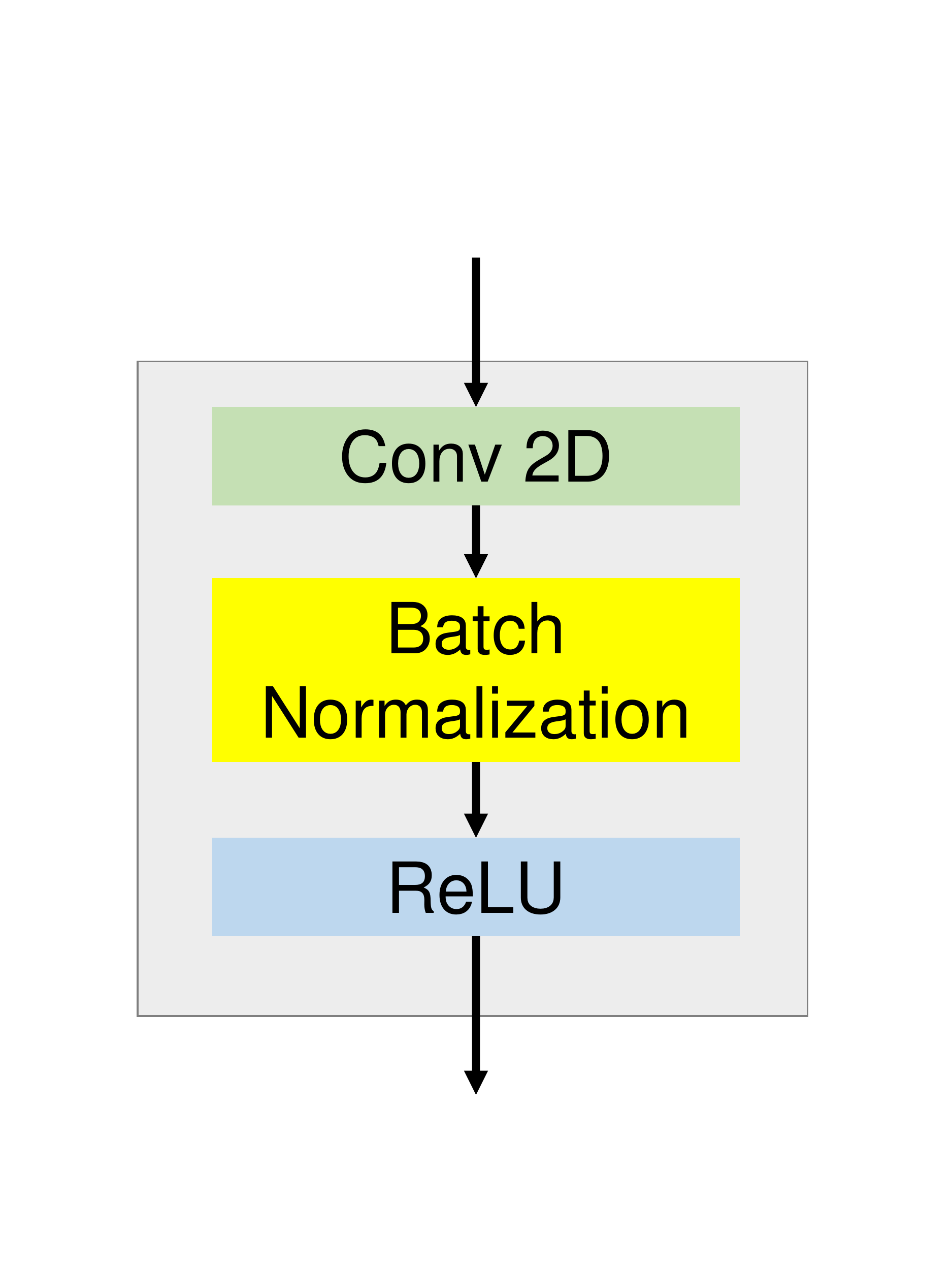}
    \caption{Encoding block}
  \end{subfigure}\hfill
  \begin{subfigure}{.25\textwidth}
    \centering
    \includegraphics[width=\linewidth]{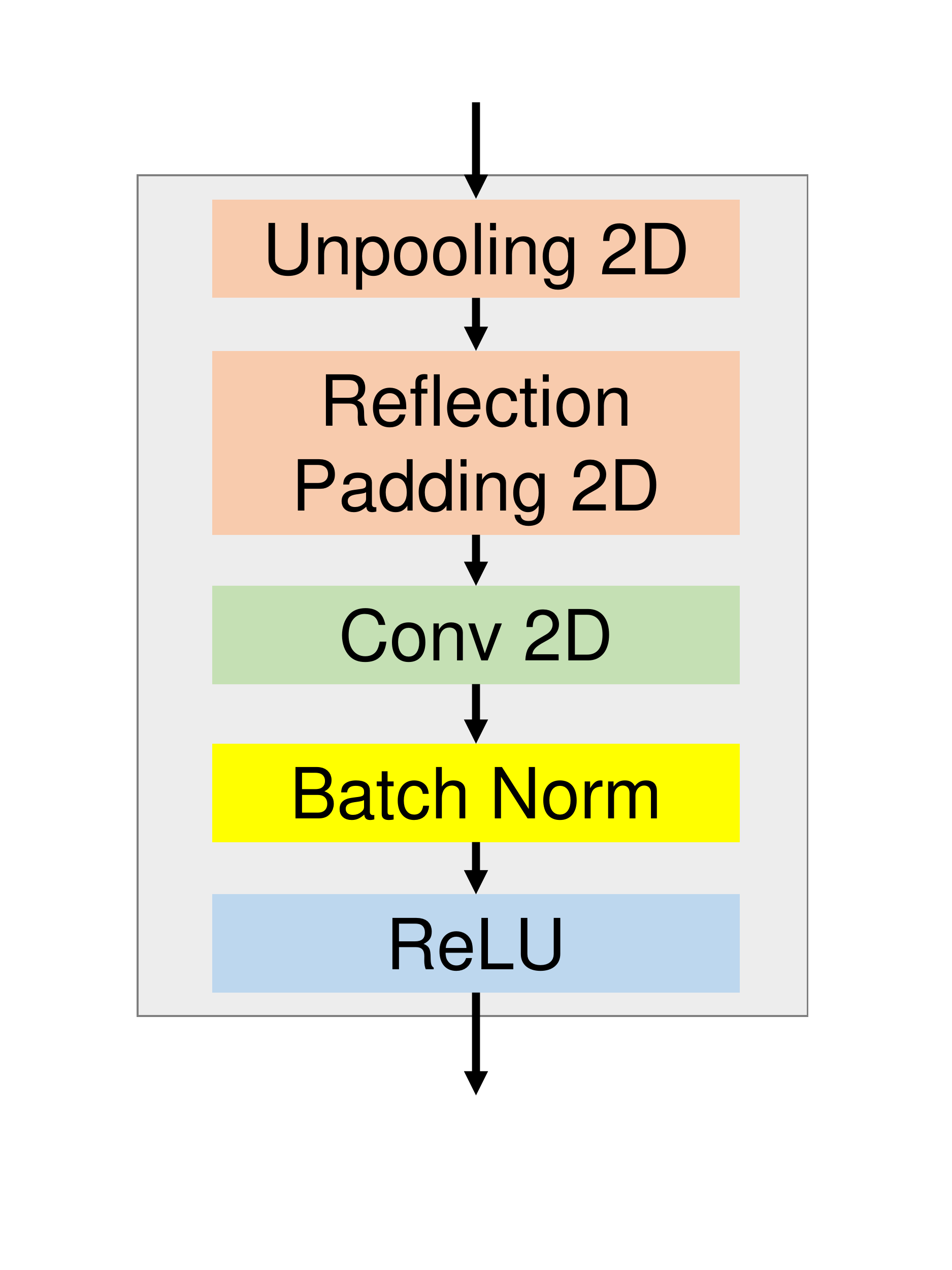}
    \caption{Decoding block}
  \end{subfigure}\hfill
  \begin{subfigure}{.25\textwidth}
    \centering
    \includegraphics[width=\linewidth]{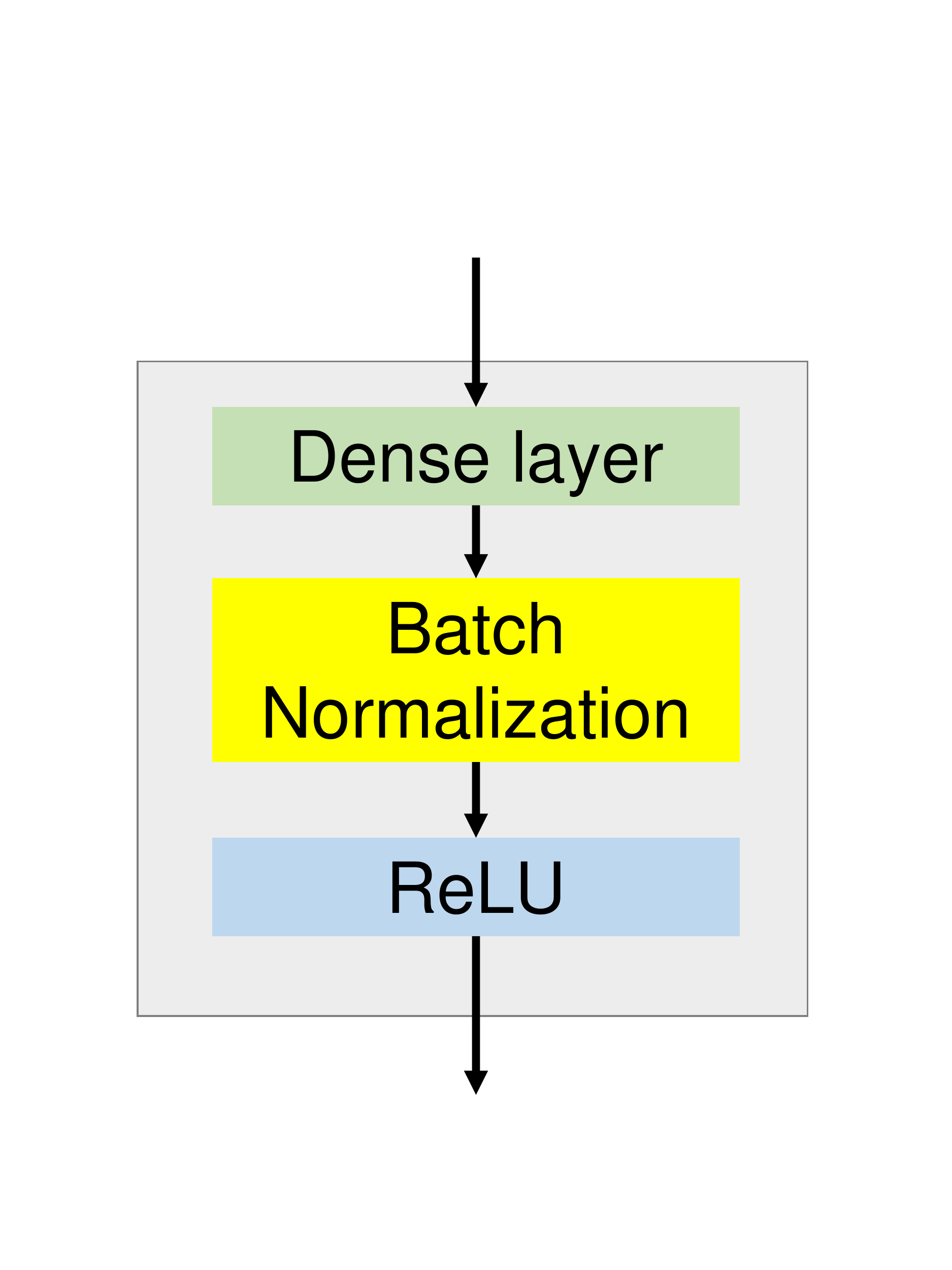}
    \caption{Trans. block}
  \end{subfigure}\hfill
  \begin{subfigure}{.25\textwidth}
    \centering
    \includegraphics[width=\linewidth]{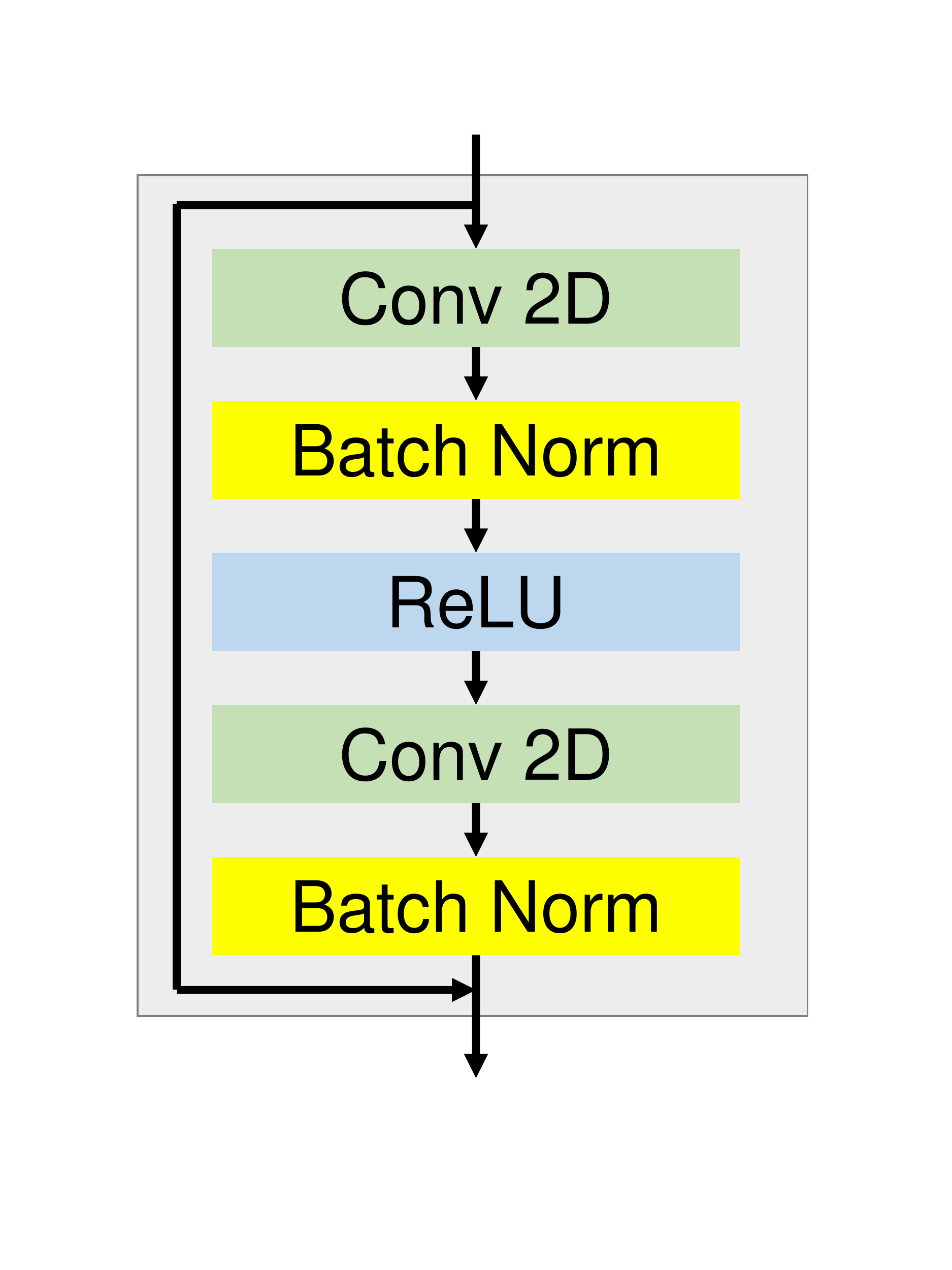}
    \caption{Res conv block}
  \end{subfigure}
  \caption{Blocks in embed-to-control models}
  \label{fig::e2c_blocks}
\end{figure}

The architecture of the encoder is summarized in Table~\ref{tab::enc}. The encoder has a stack of four encoding blocks, with 16, 32, 64 and 128 filters, respectively. The detailed structure of the encoding block is shown in Fig.~\ref{fig::e2c_blocks}(a). The size of the filter refers to that for the conv2D layer within the encoding block. The dimensions for the batchNorm layer and the ReLU are consistent with the size of the output from the conv2D layer. The encoder also has a stack of three residual convolutional (resConv) blocks, each with 128 filters of size $3\times3\times128$ and stride~1. The filter size again refers to those for the conv2D layers within the resConv block. The detailed structure of the resConv block is shown in Fig.~\ref{fig::e2c_blocks}(d). Note that before being fed into the dense layer, the output from the resConv block is reshaped from a 3D matrix of dimension $(N_x/4, N_y/4, 128)$ into a long vector. The dense layer at the end of the encoder can be treated as a linear transform with input size $N_x/4\times N_y/4\times128$, and output size of latent dimension $l_z$ ($l_z=50$ in the cases considered here). Here $N_x$ and $N_y$ are the height and width of the input images, which are both 60 in our cases.

\begin{table}[!htb]
\centering
\begin{tabular}{l@{\hskip .75in} l@{\hskip .75in} l}
    \hline
    \textbf{Layer}  & \textbf{Filter number, size and stride}                       &\textbf{Output size} \\
    \hline
    Input           &                                                       & ($N_{x}$, $N_{y}$, $2$)  \\
        \\[-1em]
    Encoding block  & 16 of $3 \times 3 \times 2$, stride 2                 & ($N_{x}/2$, $N_{y}/2$, $16$)  \\
        \\[-1em]
    Encoding block  & 32 of $3 \times 3 \times 16$, stride 1                & ($N_{x}/2$, $N_{y}/2$, $64$)  \\
        \\[-1em]
    Encoding block  & 64 of $3 \times 3 \times 32$, stride 2                & ($N_{x}/4$, $N_{y}/4$, $128$)  \\
        \\[-1em]
    Encoding block  & 128 of $3 \times 3 \times 64$, stride 1               & ($N_{x}/4$, $N_{y}/4$, $128$)  \\
        \\[-1em]
    ResConv block   & 128 of $3 \times 3 \times 128$, stride 1              & ($N_{x}/4$, $N_{y}/4$, $128$)  \\
        \\[-1em]
    ResConv block   & 128 of $3 \times 3 \times 128$, stride 1              & ($N_{x}/4$, $N_{y}/4$, $128$)  \\
        \\[-1em]
    ResConv block   & 128 of $3 \times 3 \times 128$, stride 1              & ($N_{x}/4$, $N_{y}/4$, $128$)  \\
        \\[-1em]
    Dense
                    &                                                       & ($l_{z}$, 1)  \\
    \hline  
\end{tabular}
\caption{Network architecture for encoder}
\label{tab::enc}
\end{table}

The architecture of the decoder is summarized in Table~\ref{tab::dec}. The decoder structure is analogous to that of the encoder, but with the components in reversed order. The decoder is comprised of a dense layer, a stack of three resConv blocks, a stack of four decoding blocks, and a conv2D layer. Note that the output of the dense layer is a long vector of dimension $N_x/4\times N_y/4\times128$, which is reshaped into a 3D matrix of size $(N_x/4, N_y/4, 128)$ before fed into the resConv block. The detailed structure of the decoding block is shown in Fig.~\ref{fig::e2c_blocks}(b).

\begin{table}[!htb]
\centering
\begin{tabular}{l@{\hskip .75in} l@{\hskip .75in} l}
    \hline
    \textbf{Layer}  & \textbf{Filter number, size and stride}                       &\textbf{Output size} \\
    \hline
    Input           &                                                       & ($l_{z}$, 1) \\
    Dense           &                                                       & ($N_{x}/4\times N_{y}/4\times128$, 1)  \\
        \\[-1em]
    ResConv block   & 128 of $3 \times 3 \times 128$, stride 1              & ($N_{x}/4$, $N_{y}/4$, $128$)  \\
        \\[-1em]
    ResConv block   & 128 of $3 \times 3 \times 128$, stride 1              & ($N_{x}/4$, $N_{y}/4$, $128$)  \\
        \\[-1em]
    ResConv block   & 128 of $3 \times 3 \times 128$, stride 1              & ($N_{x}/4$, $N_{y}/4$, $128$)  \\
        \\[-1em]
    Decoding block  & 128 of $3 \times 3 \times 128$, stride 1              & ($N_{x}/4$, $N_{y}/4$, $128$)  \\
        \\[-1em]
    Decoding block  & 64 of $3 \times 3 \times 128$, stride 2                & ($N_{x}/2$, $N_{y}/2$, $64$)  \\
        \\[-1em]
    Decoding block  & 32 of $3 \times 3 \times 64$, stride 1                & ($N_{x}/2$, $N_{y}/2$, $32$)  \\
        \\[-1em]
    Decoding block  & 16 of $3 \times 3 \times 32$, stride 2                & ($N_{x}$, $N_{y}$, $16$)  \\
        \\[-1em]
    Conv 2D         & 2 of $3 \times 3 \times 16$, stride 1                 & ($N_{x}$, $N_{y}$, $2$)  \\
    \hline  
\end{tabular}
\caption{Network architecture for decoder}
\label{tab::dec}
\end{table}

Fig.~\ref{fig::e2c_blocks} shows the detailed structure for the encoding blocks, the decoding blocks, the transformation blocks and the resConv blocks. An encoding block is a sequential combination of a 2D convolutional layer (conv2D), a batch-normalization (bathNorm) layer, and a rectified linear unit (ReLU). A reduction of spatial dimension (down-sampling) is achieved through tuning stride size for the conv2D layer (e.g., stride size of two for the encoding blocks in Table~\ref{tab::enc}). The decoding block has a 2D unpooling layer, which increases the size of the input by repeating the rows and columns of the data, a 2D reflection padding layer, which also increases the size of input by padding the boundary on two of the dimensions of the data, a conv2D layer, a batchNorm layer, and a ReLU. The increase of spatial dimension (up-sampling) is achieved through the 2D unpooling and 2D reflection padding layers. 

The structure of the resConv block closely resembles resNet, where an identity mapping is created to bypass the nonlinear layers. For a resConv layer, instead of the direct mapping of input $x$ to the target function $\mathcal{F}(x)$, the nonlinear layer only needs to learn a residual mapping of $\mathcal{H}(x):=\mathcal{F}(x)-x$. This will be zero in the extreme (worst) case ($\mathcal{F}(x)=x$) and guarantees that a deeper neural network will achieve higher (or at least equal) accuracy relative to its shallower counterpart. The nonlinear layer follows the standard structure of conv2D-batchNorm-ReLU. The transformation block has an architecture of dense-batchNorm-ReLU, where the dimension of the dense layer is set to 200 for the cases tested.

% -------------------------------------------------------------
% -------------------------------------------------------------
\section{Additional test case results}\label{appendix-case}
% -------------------------------------------------------------
% -------------------------------------------------------------

We now present E2C ROM results for Test Cases~2 and 3. The errors for quantities of interest for these cases are shown in Figs.~\ref{fig::cum_rates} and \ref{fig::rel_err} and in Table~\ref{tab::rel_err}. Our descriptions here are brief since these results are very comparable to those discussed in Section~\ref{results}.

%%%%%%%%%%%%%%%%%%%%%%%%%%%%%%%%
%%%%%%%%%%% Case 2 %%%%%%%%%%%%%
%%%%%%%%%%%%%%%%%%%%%%%%%%%%%%%%
\subsection{Results for Test Case~2}

\begin{figure}[htbp]
  \centering
  \begin{subfigure}{.45\textwidth}
    \centering
    \includegraphics[width=\linewidth]{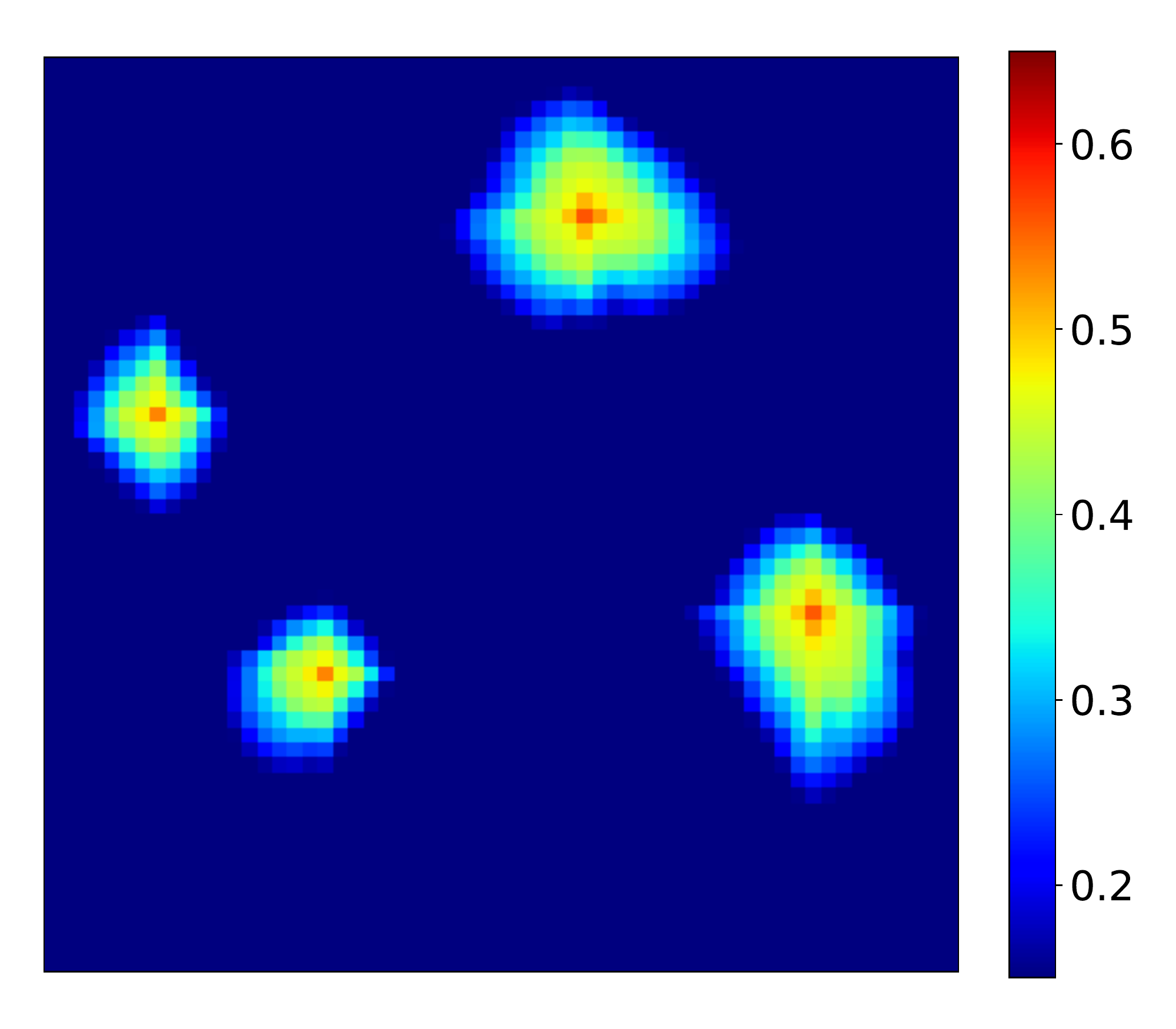}
    \caption{High-fidelity solution ($\text{HFS}_{\text{test}}$)}
  \end{subfigure}\hfill
  \begin{subfigure}{.45\textwidth}
    \centering
    \includegraphics[width=\linewidth]{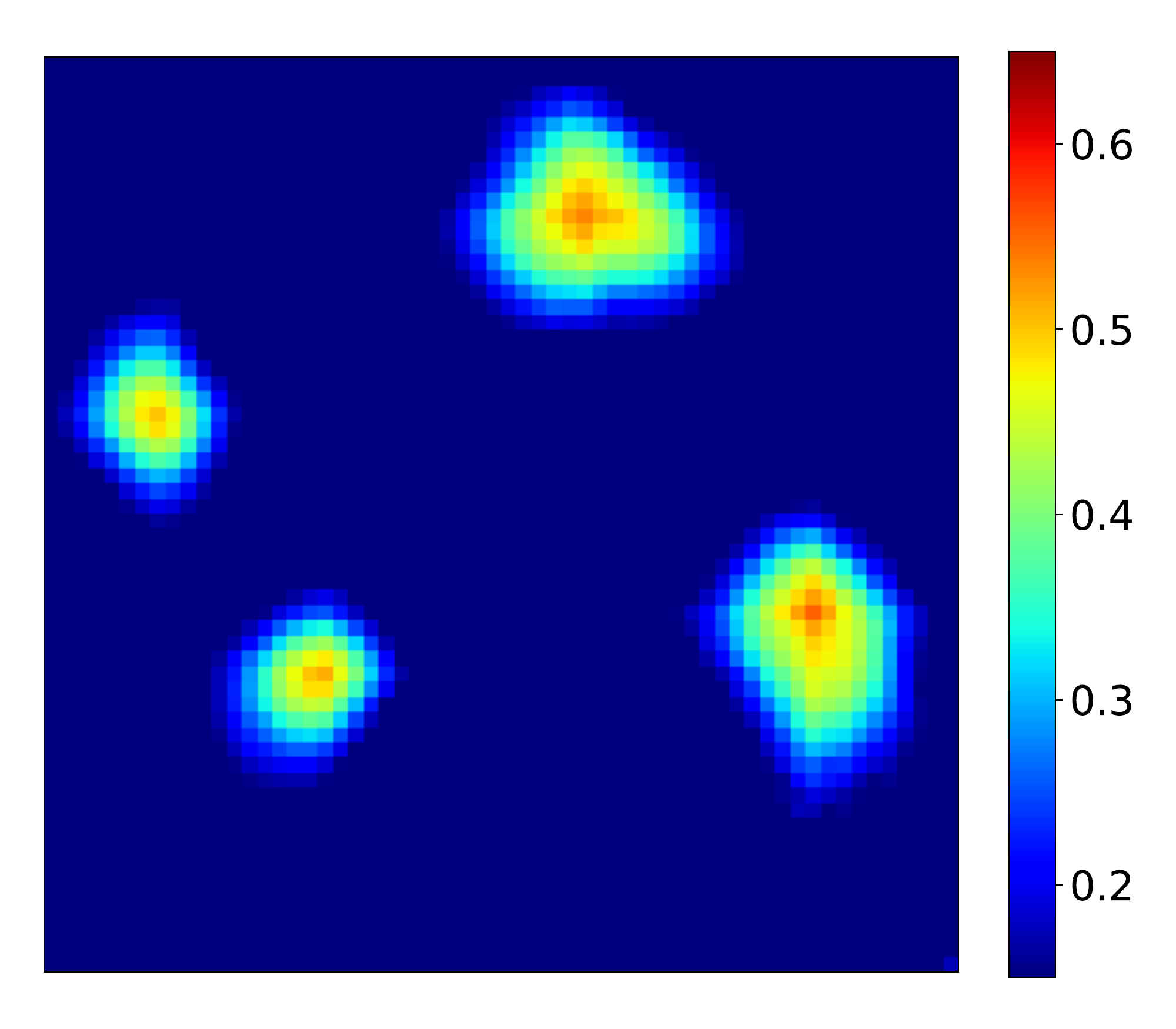}
    \caption{ROM solution ($\text{ROM}_{\text{test}}$)}
  \end{subfigure} \\
  \begin{subfigure}{.45\textwidth}
    \centering
    \includegraphics[width=\linewidth]{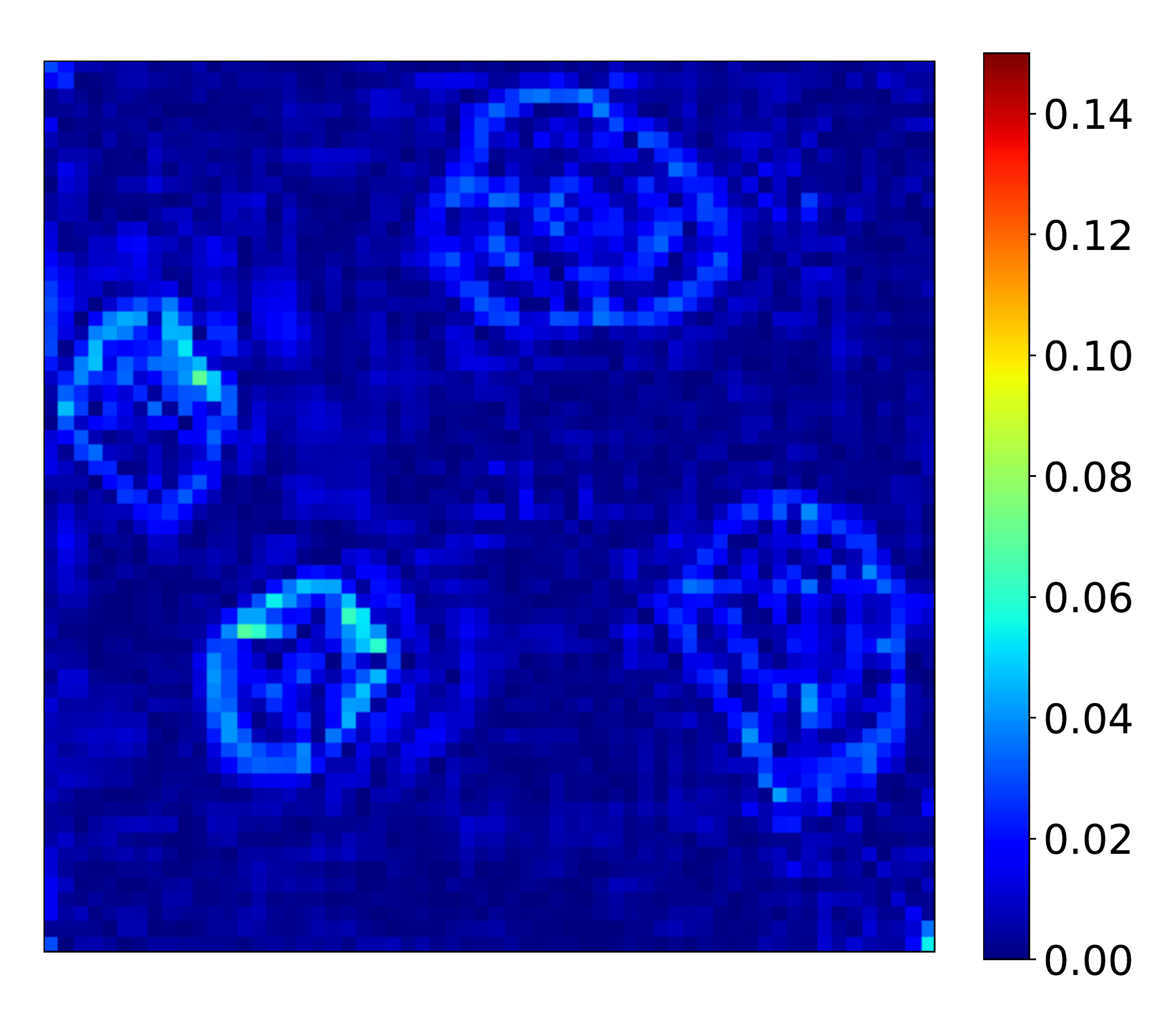}
    \caption{$|\text{HFS}_{\text{test}} - \text{ROM}_{\text{test}}|$}
  \end{subfigure}\hfill
    \begin{subfigure}{.45\textwidth}
    \centering
    \includegraphics[width=\linewidth]{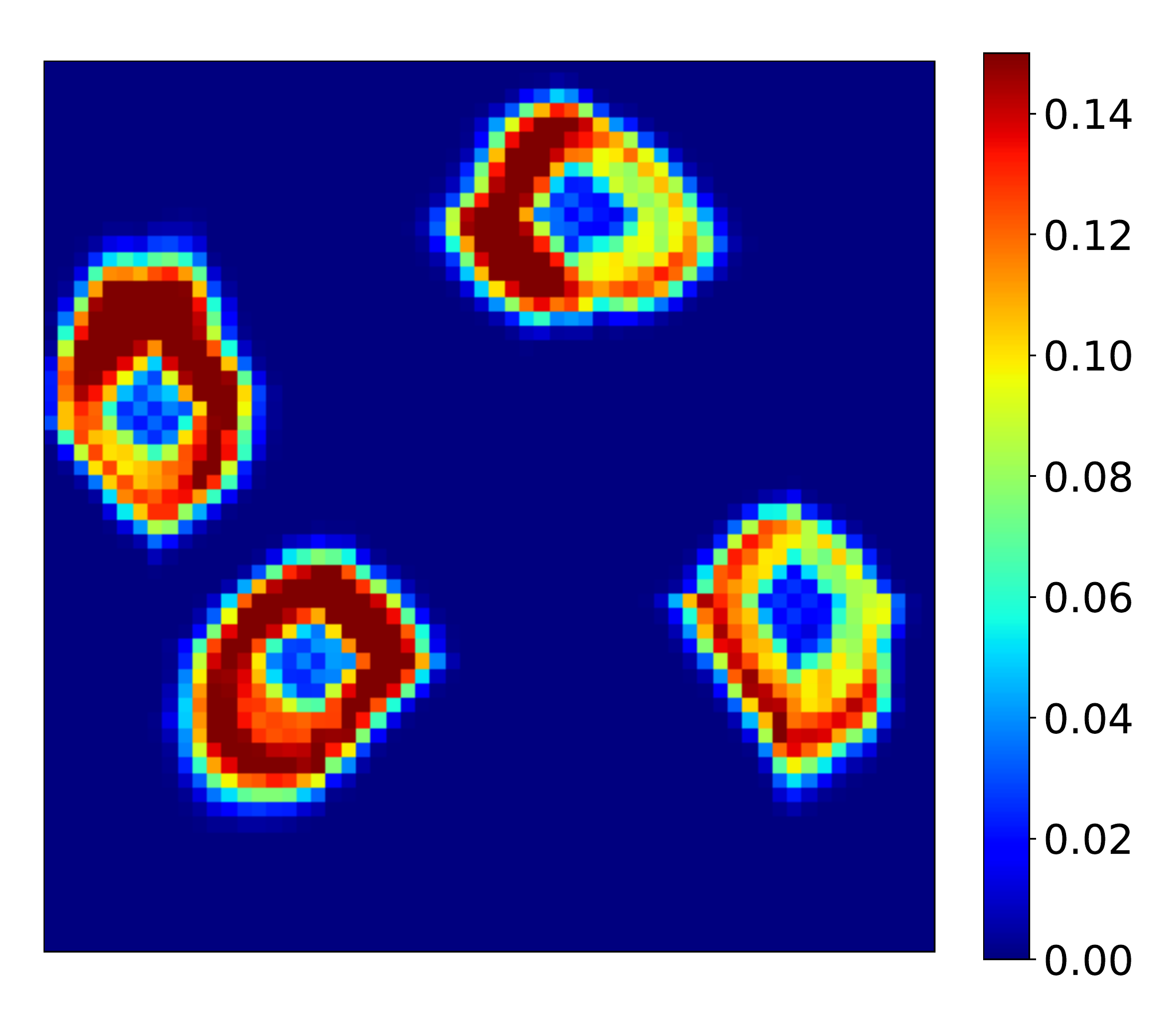}
    \caption{$|\text{HFS}_{\text{test}} - \text{HFS}_{\text{train}}|$}
  \end{subfigure}
  \caption{Test Case~2: saturation field at 200 days}
  \label{fig::test_2_sat_200}
\end{figure}

\begin{figure}[htbp]
  \centering
  \begin{subfigure}{.45\textwidth}
    \centering
    \includegraphics[width=\linewidth]{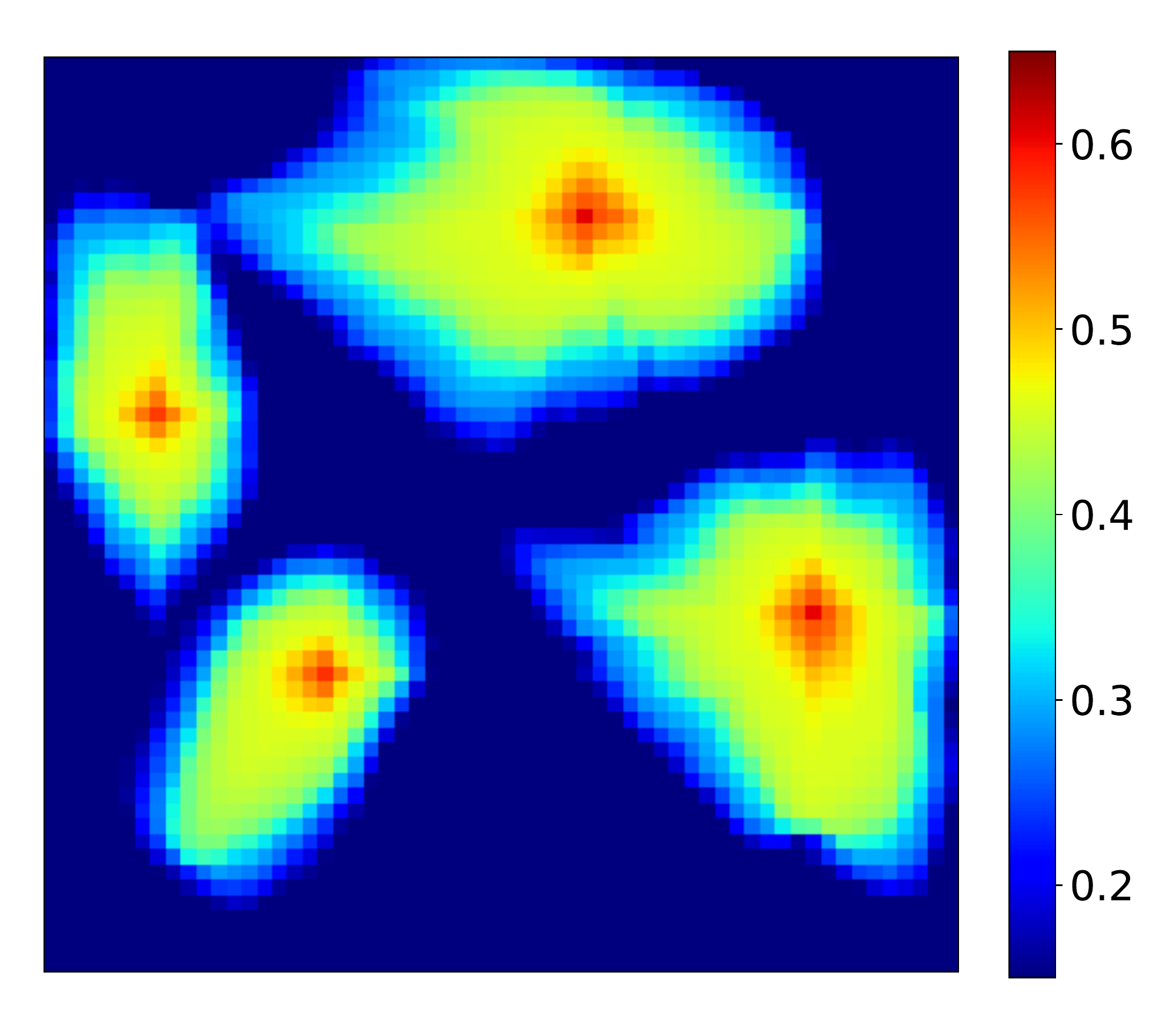}
    \caption{High-fidelity solution ($\text{HFS}_{\text{test}}$)}
  \end{subfigure}\hfill
  \begin{subfigure}{.45\textwidth}
    \centering
    \includegraphics[width=\linewidth]{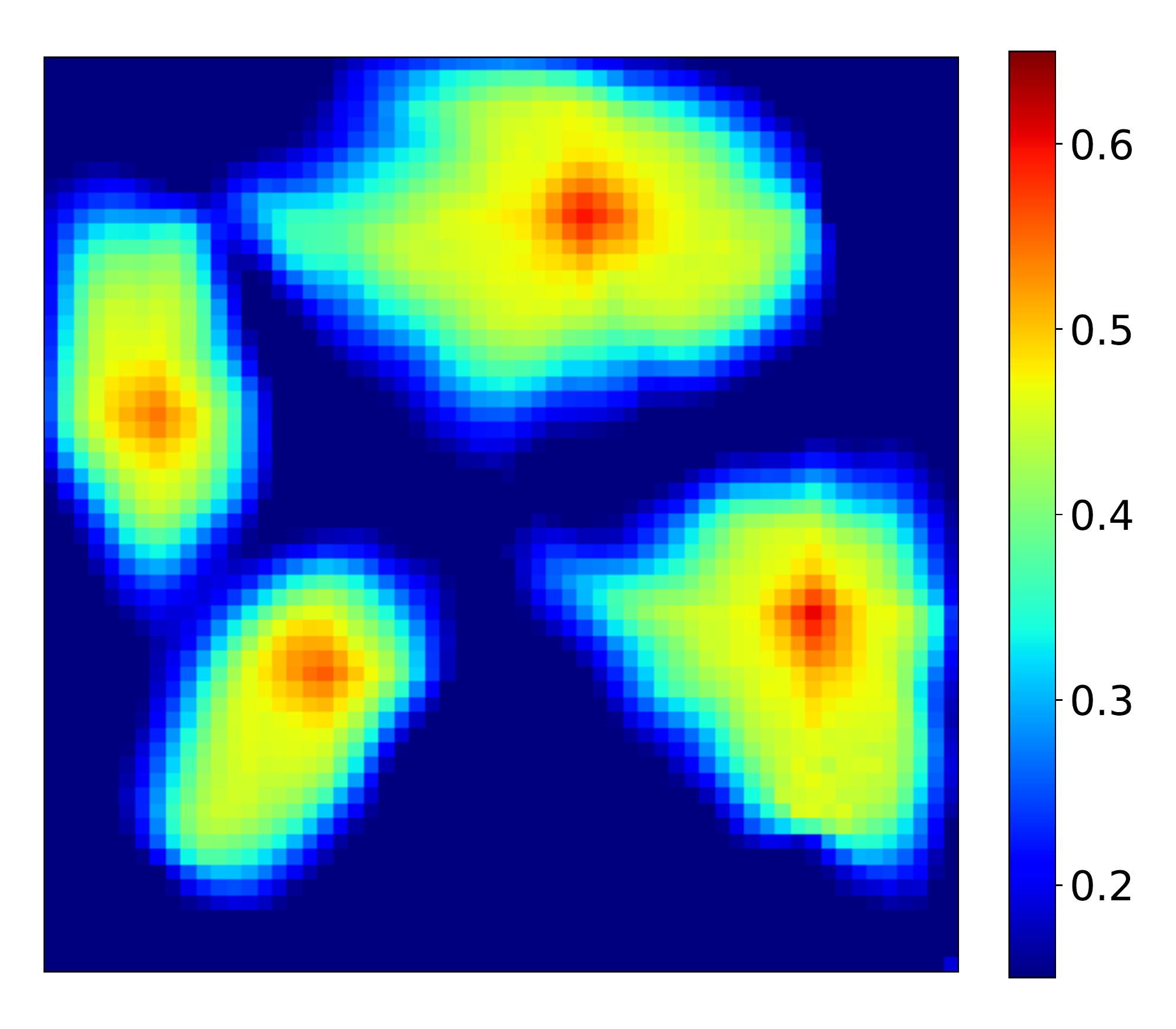}
    \caption{ROM solution ($\text{ROM}_{\text{test}}$)}
  \end{subfigure} \\
  \begin{subfigure}{.45\textwidth}
    \centering
    \includegraphics[width=\linewidth]{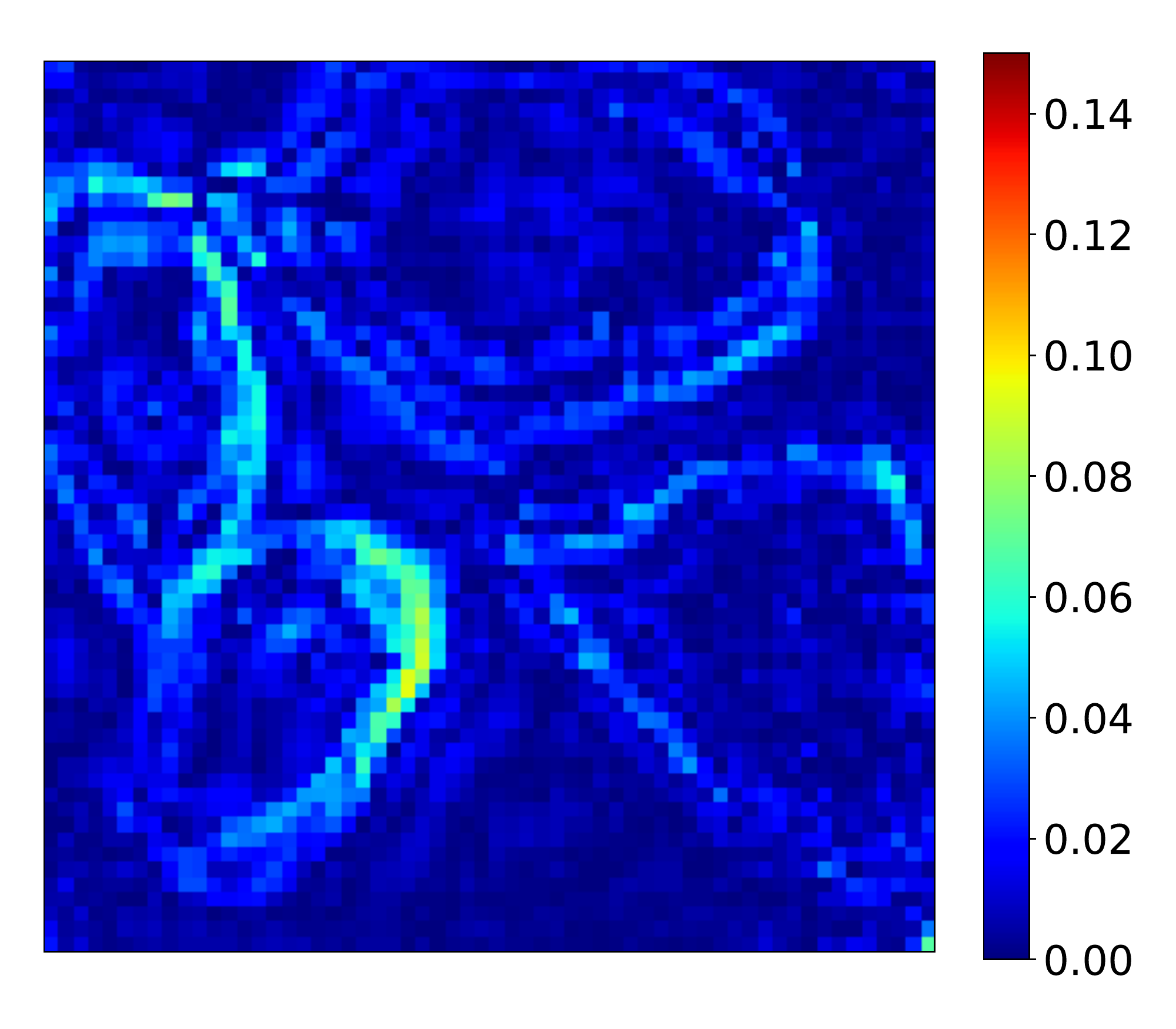}
    \caption{$|\text{HFS}_{\text{test}} - \text{ROM}_{\text{test}}|$}
  \end{subfigure}\hfill
    \begin{subfigure}{.45\textwidth}
    \centering
    \includegraphics[width=\linewidth]{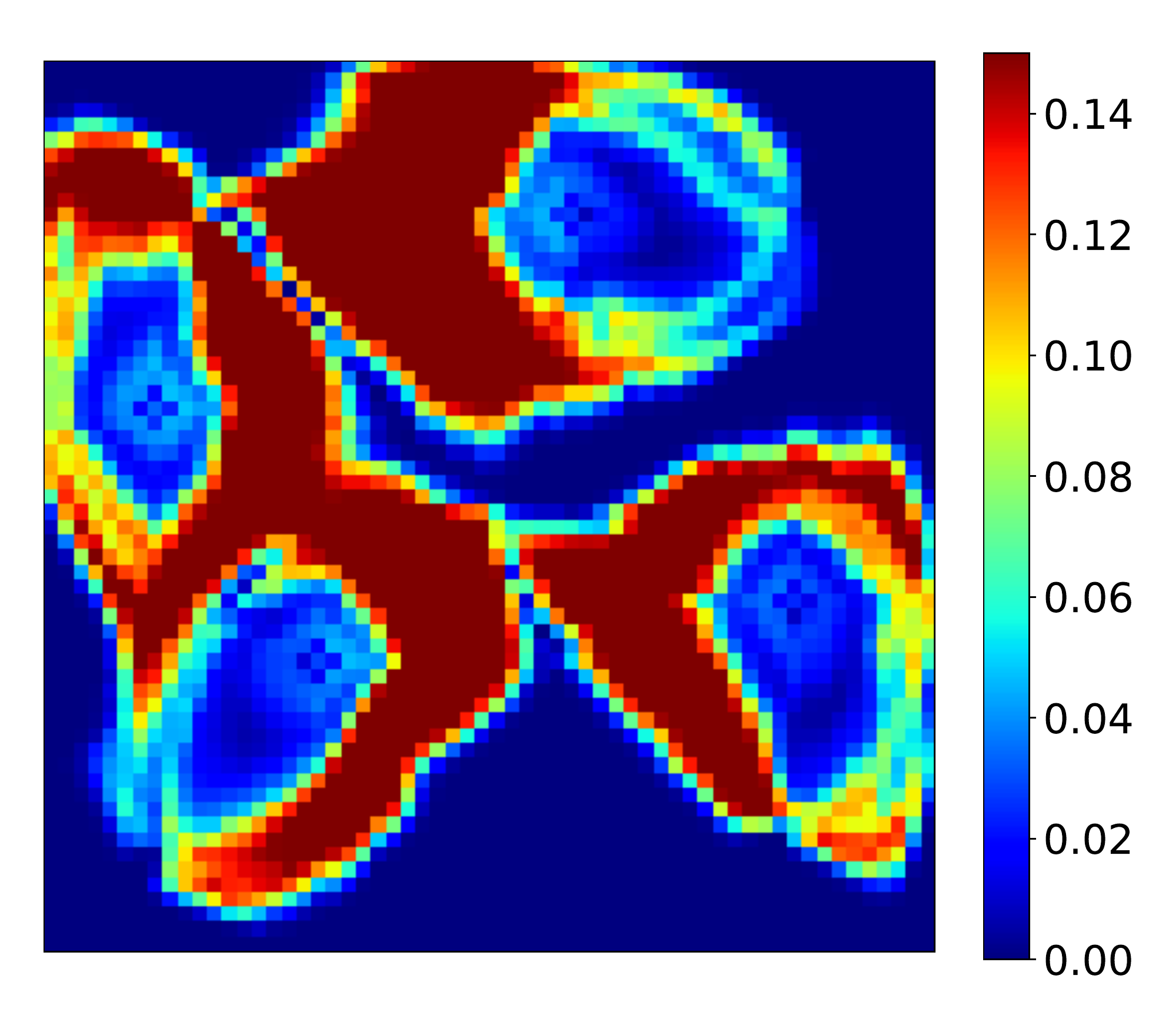}
    \caption{$|\text{HFS}_{\text{test}} - \text{HFS}_{\text{train}}|$}
  \end{subfigure}
  \caption{Test Case~2: saturation field at 1000 days}
  \label{fig::test_2_sat_1000}
\end{figure}

\begin{figure}[htbp]
  \centering
  \begin{subfigure}{.45\textwidth}
    \centering
    \includegraphics[width=\linewidth]{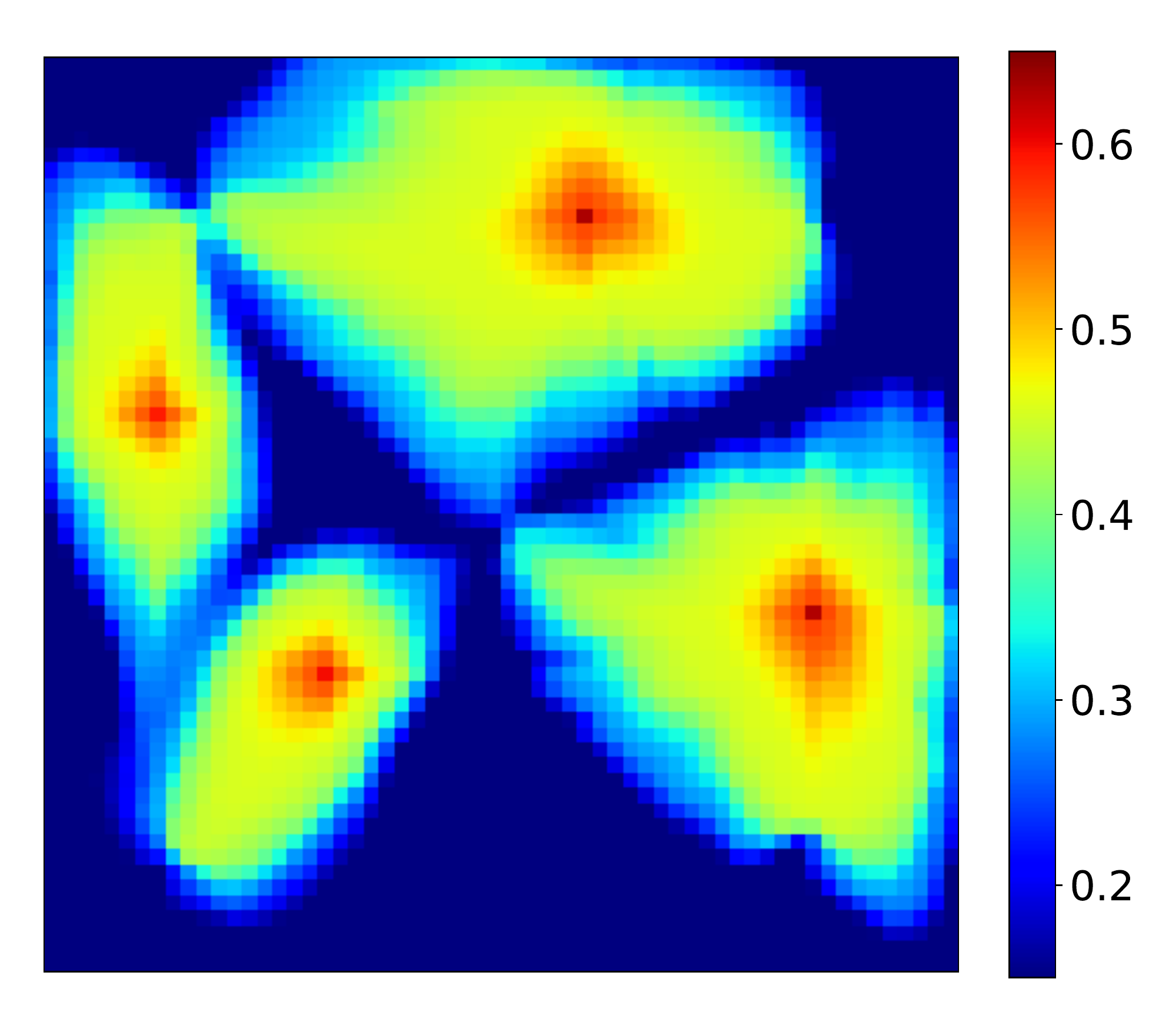}
    \caption{High-fidelity solution ($\text{HFS}_{\text{test}}$)}
  \end{subfigure}\hfill
  \begin{subfigure}{.45\textwidth}
    \centering
    \includegraphics[width=\linewidth]{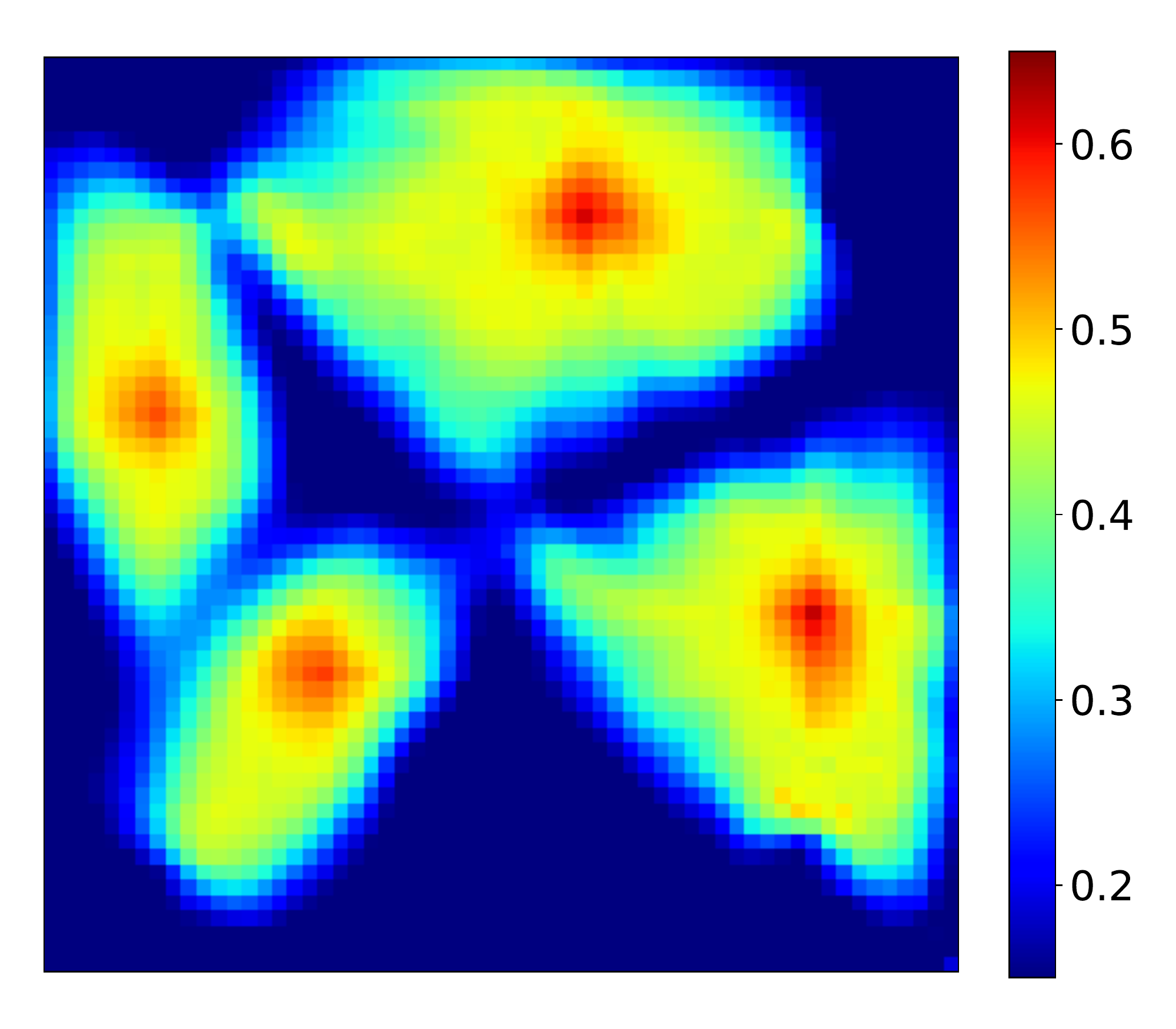}
    \caption{ROM solution ($\text{ROM}_{\text{test}}$)}
  \end{subfigure} \\
  \begin{subfigure}{.45\textwidth}
    \centering
    \includegraphics[width=\linewidth]{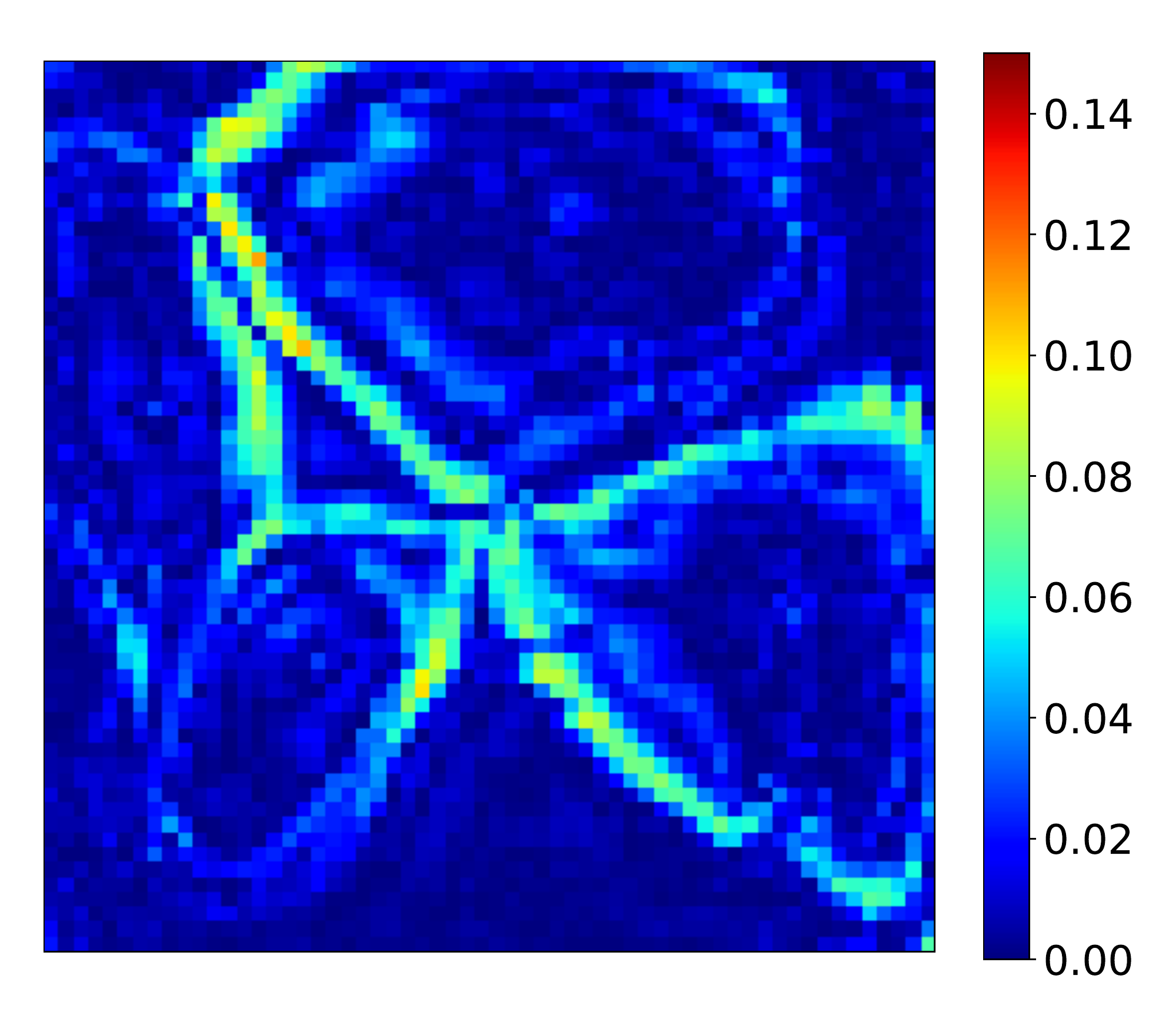}
    \caption{$|\text{HFS}_{\text{test}} - \text{ROM}_{\text{test}}|$}
  \end{subfigure}\hfill
    \begin{subfigure}{.45\textwidth}
    \centering
    \includegraphics[width=\linewidth]{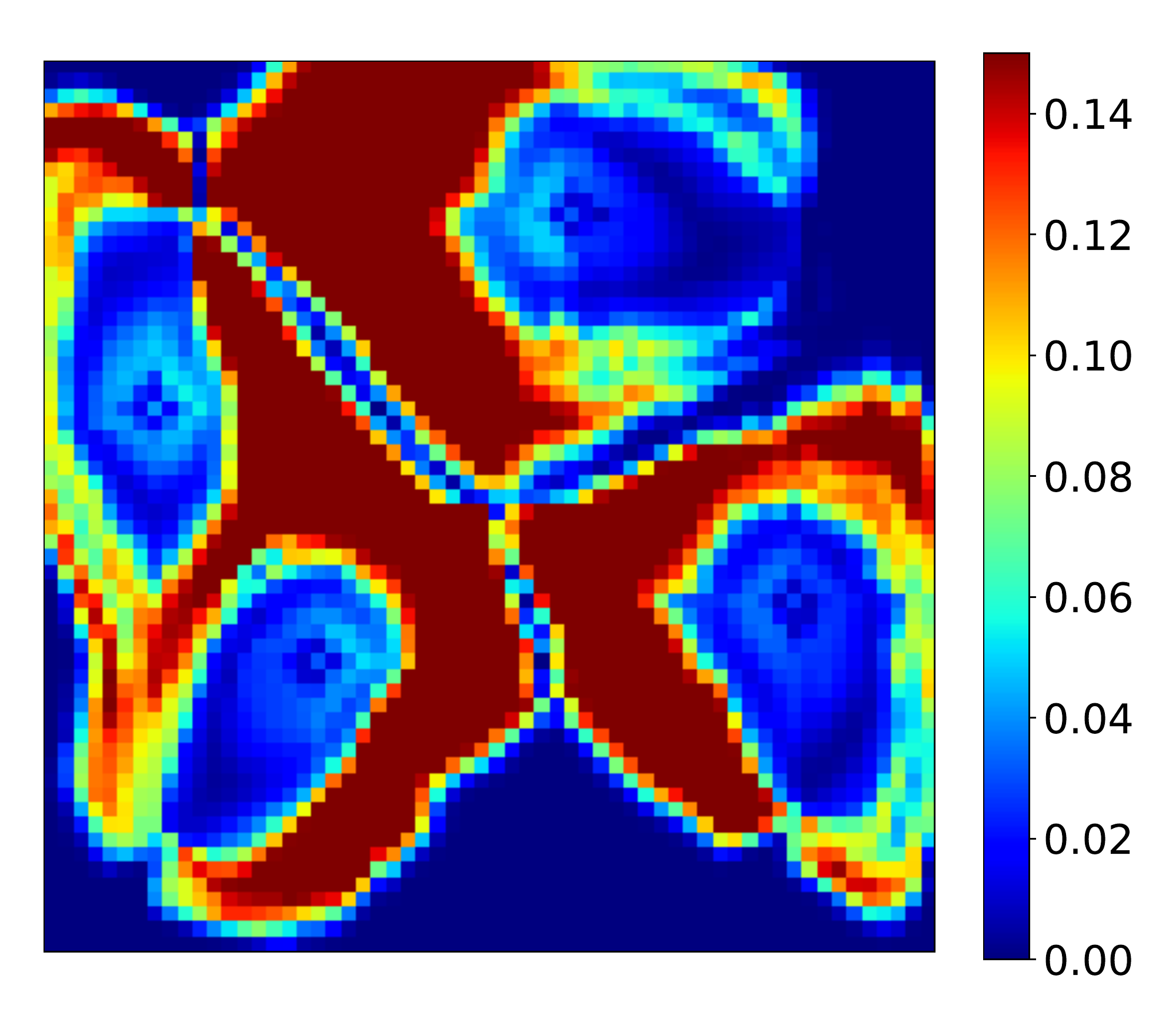}
    \caption{$|\text{HFS}_{\text{test}} - \text{HFS}_{\text{train}}|$}
  \end{subfigure}
  \caption{Test Case~2: saturation field at 1800 days}
  \label{fig::test_2_sat_1800}
\end{figure}

\begin{figure}[htbp]
  \centering
  \begin{subfigure}{.45\textwidth}
    \centering
    \includegraphics[width=\linewidth]{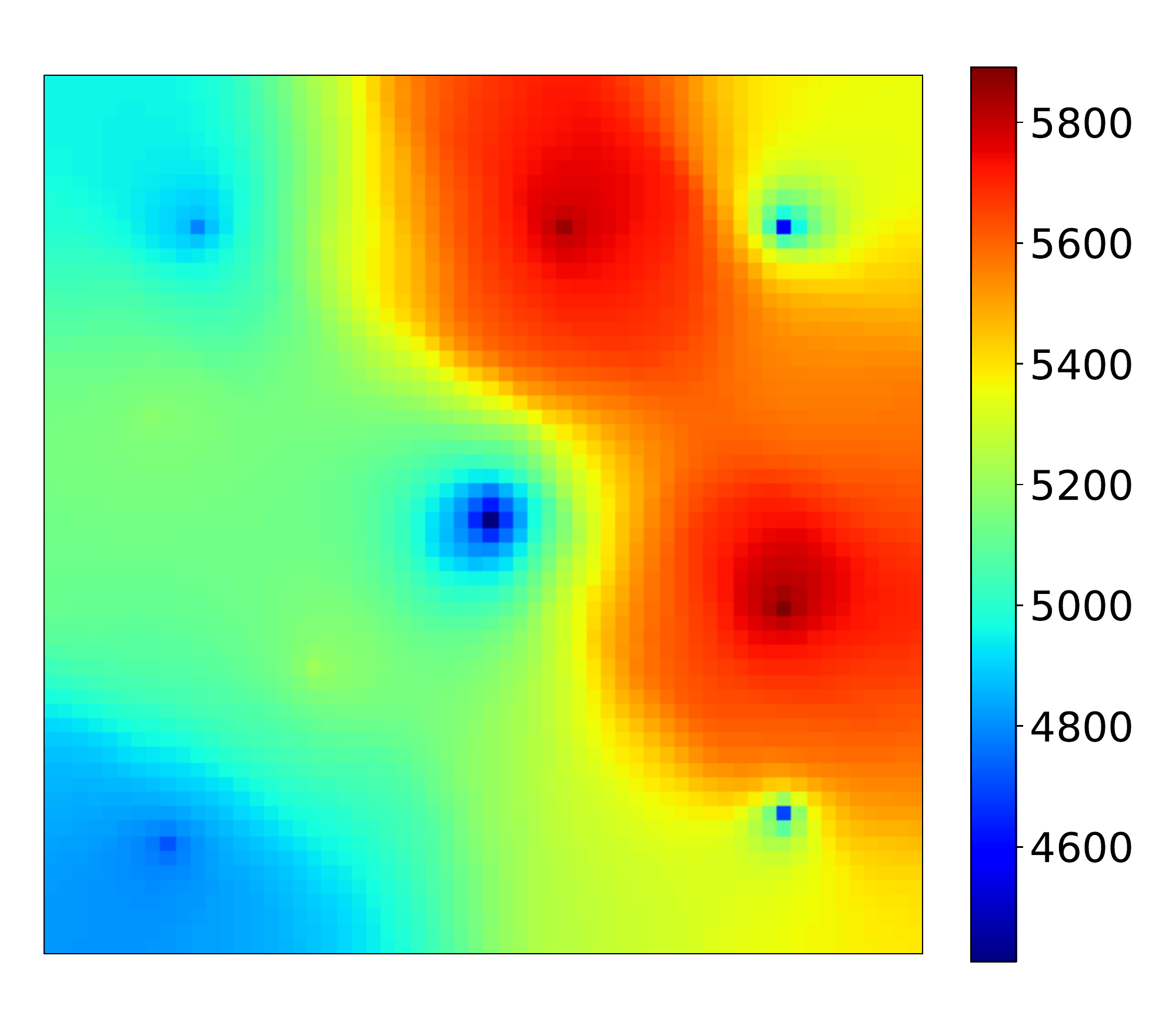}
    \caption{High-fidelity solution ($\text{HFS}_{\text{test}}$)}
  \end{subfigure}\hfill
  \begin{subfigure}{.45\textwidth}
    \centering
    \includegraphics[width=\linewidth]{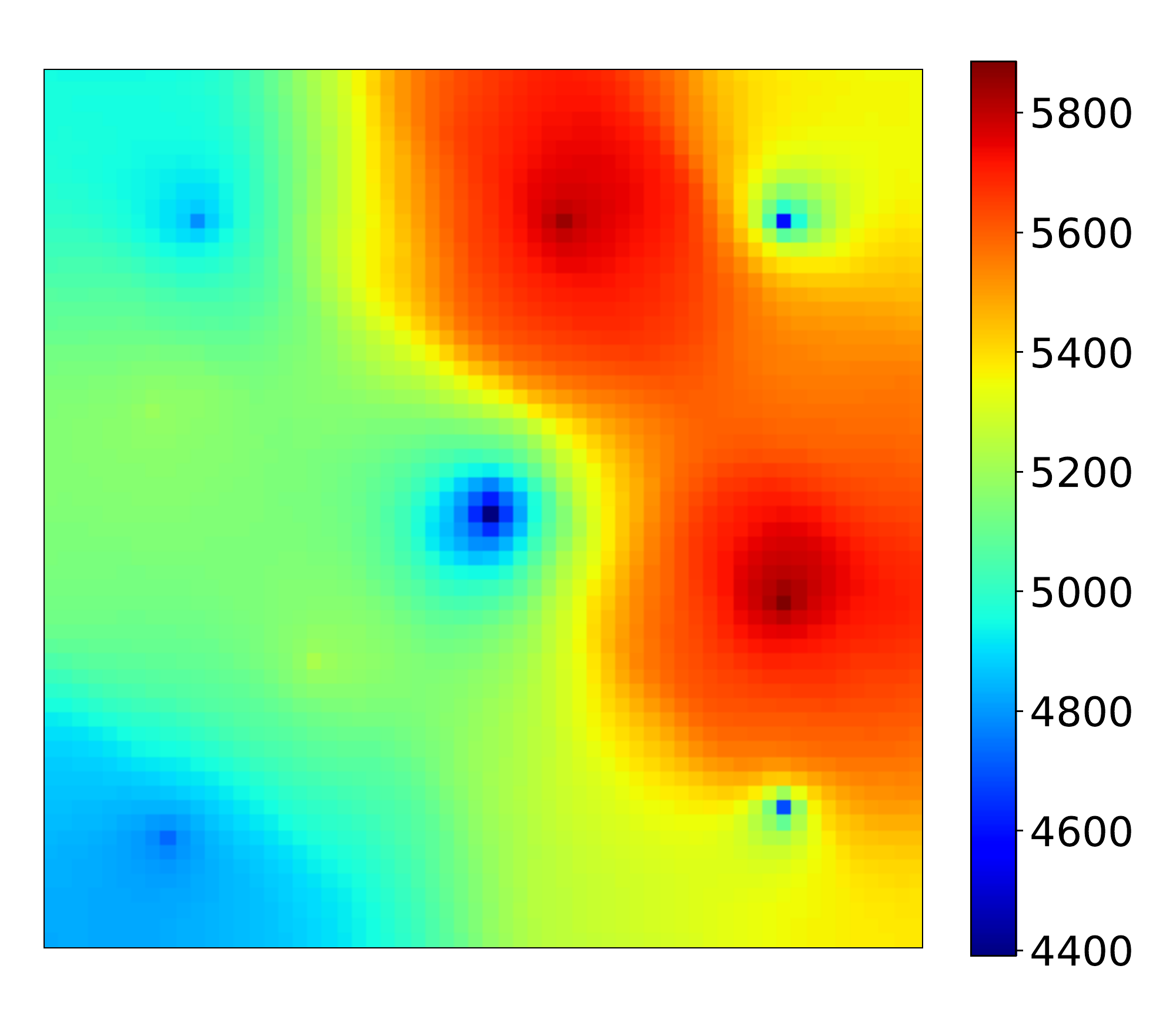}
    \caption{ROM solution ($\text{ROM}_{\text{test}}$)}
  \end{subfigure} \\
  \begin{subfigure}{.45\textwidth}
    \centering
    \includegraphics[width=\linewidth]{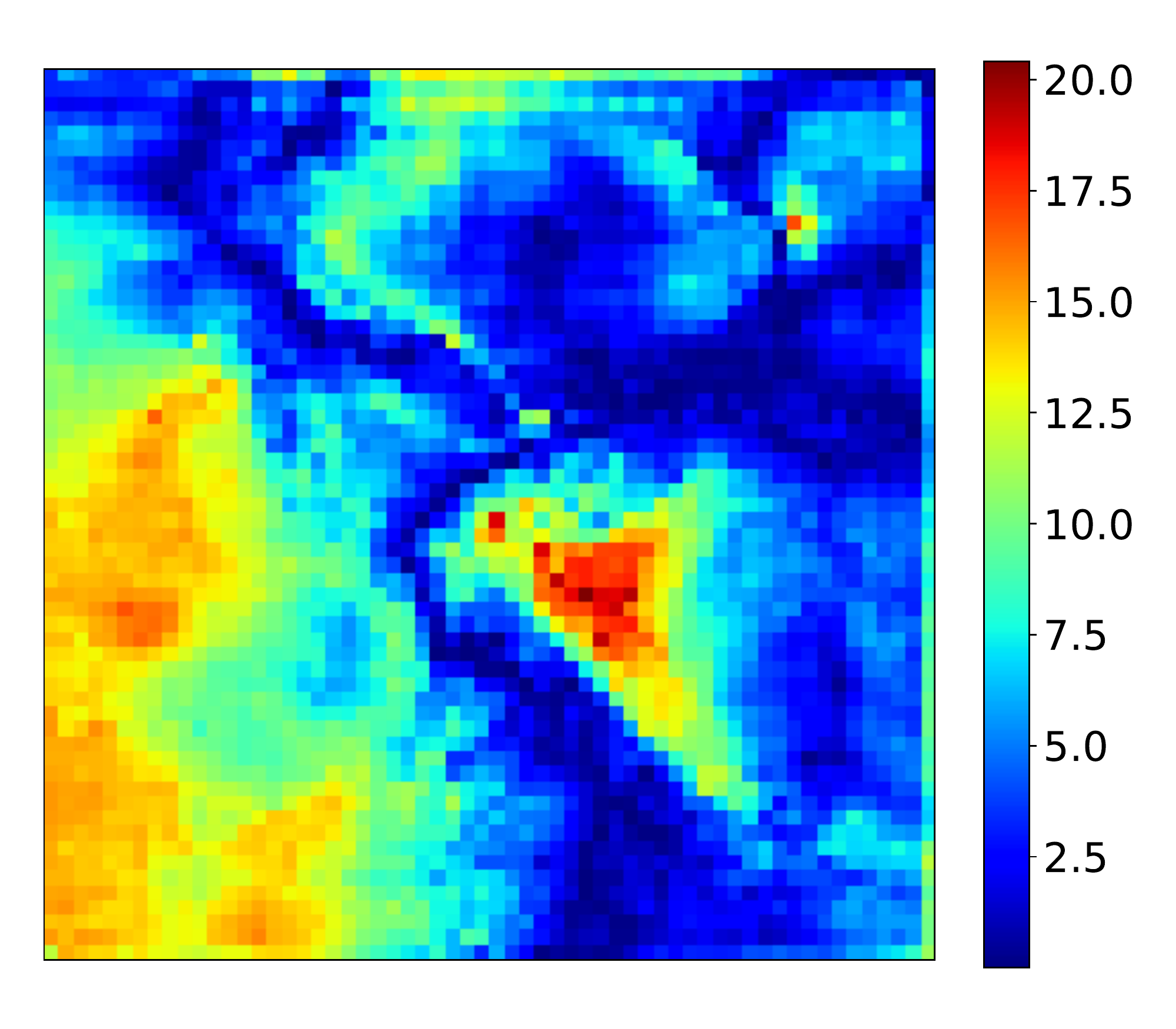}
    \caption{$|\text{HFS}_{\text{test}} - \text{ROM}_{\text{test}}|$}
  \end{subfigure}\hfill
    \begin{subfigure}{.45\textwidth}
    \centering
    \includegraphics[width=\linewidth]{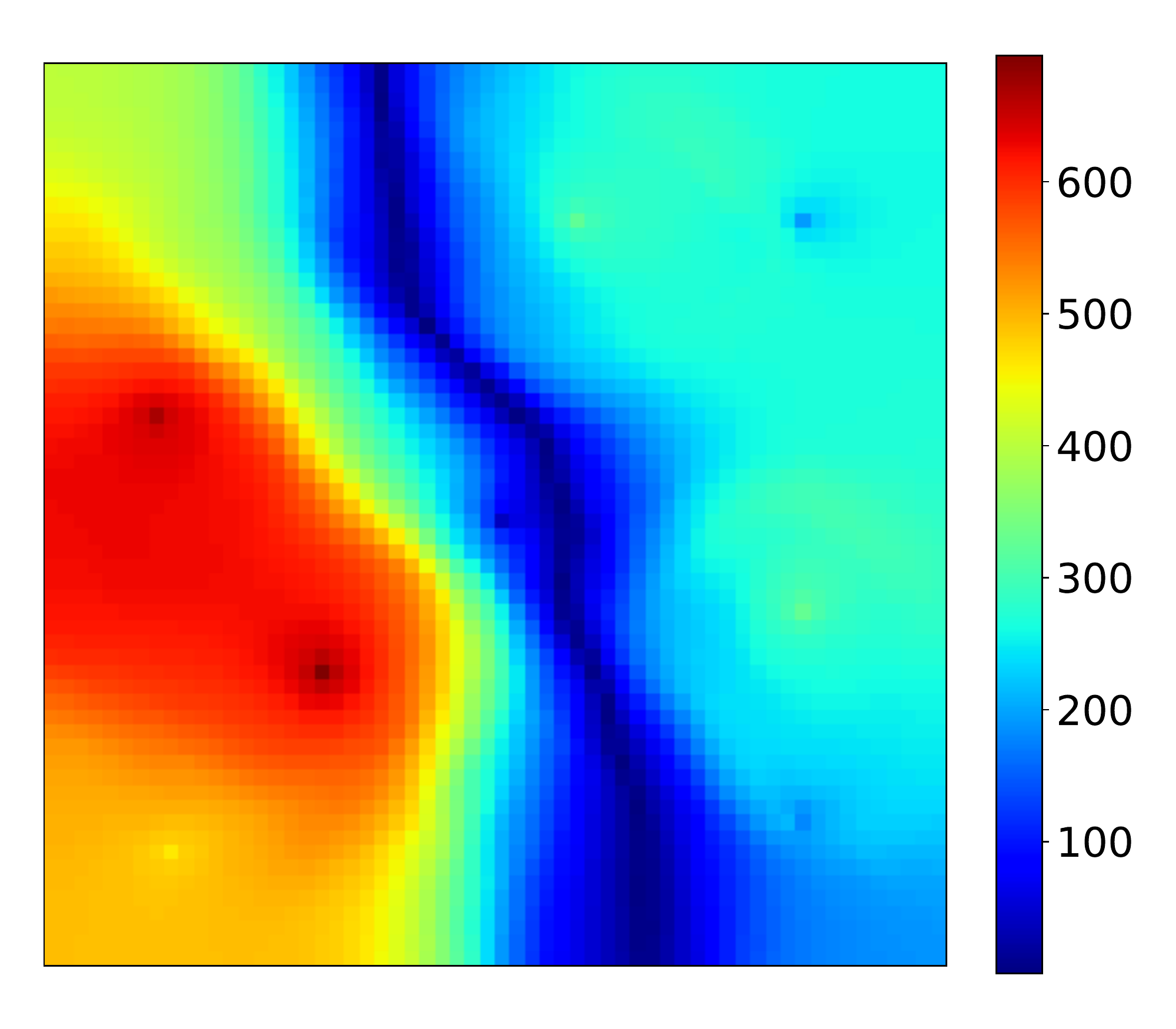}
    \caption{$|\text{HFS}_{\text{test}} - \text{HFS}_{\text{train}}|$}
  \end{subfigure}
  \caption{Test Case~2: pressure field at 1000 days (all colorbars in units of psi)}
  \label{fig::test_2_pres_1000}
\end{figure}

Figs.~\ref{fig::test_2_sat_200}, \ref{fig::test_2_sat_1000} and \ref{fig::test_2_sat_1800} display the saturation field at 200, 1000 and 1800~days, and Fig.~\ref{fig::test_2_pres_1000} shows the pressure field at 1000~days, for this case. The evolution of saturation indicates a somewhat different sweep here compared to that of Test Case~1. Specifically, comparing Fig.~\ref{fig::test_2_sat_1800}(b) to Fig.~\ref{fig::test_1_sat_1800}(b), we see that the water plume around Well~I2 (upper right) is larger, and the water plumes around I1 (upper left) and I3 (lower left) are smaller in this case. Please refer to Fig.~\ref{fig::perm} for the exact well locations. The pressure map for Test Case~2 also shows a very different pattern compared to that of Test Case~1 at 1000~days. The difference maps shown in Figs.~\ref{fig::test_2_sat_200}(c), \ref{fig::test_2_sat_1000}(c), \ref{fig::test_2_sat_1800}(c) and \ref{fig::test_2_pres_1000}(c) indicate that the ROM predictions are again accurate for these global quantities.

\begin{figure}[htbp]
  \centering
  \begin{subfigure}{.5\textwidth}
    \centering
    \includegraphics[width=\linewidth]{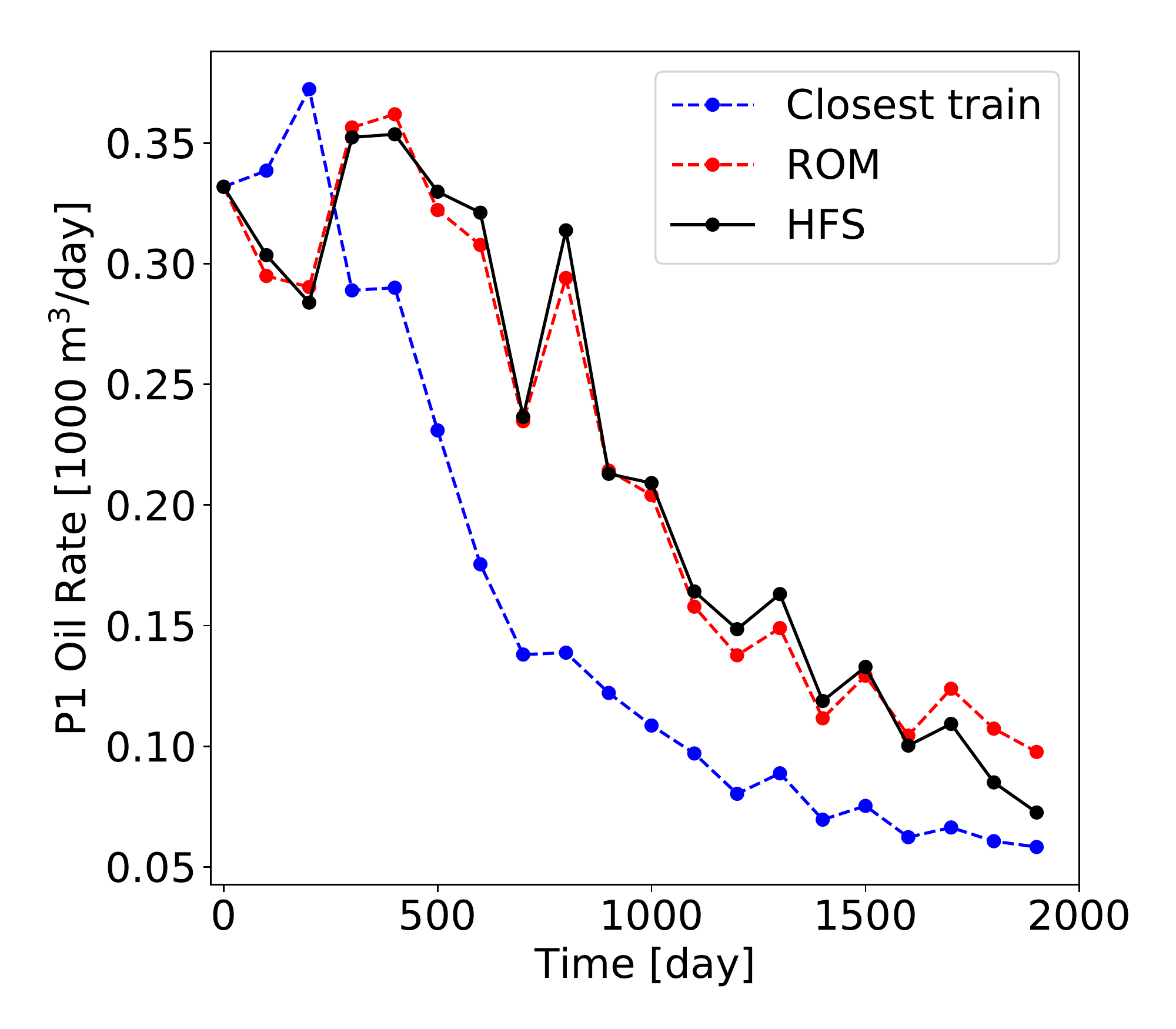}
    \caption{Oil rate}
  \end{subfigure}\hfill
  \begin{subfigure}{.5\textwidth}
    \centering
    \includegraphics[width=\linewidth]{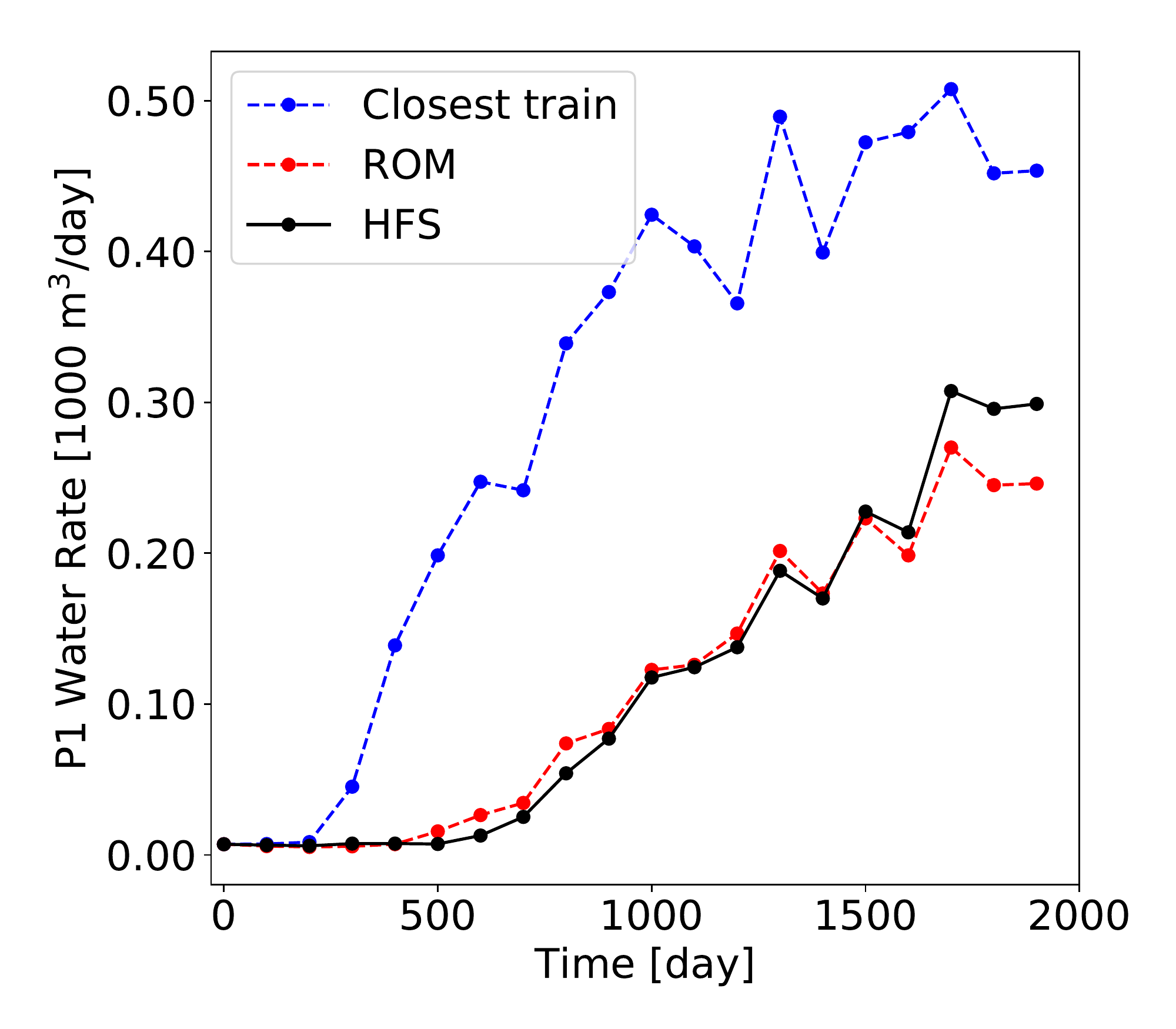}
    \caption{Water rate}
  \end{subfigure}
  \caption{Test Case~2: production rates for Well P1}
  \label{fig::test_2_rate_p1}
\end{figure}

\begin{figure}[htbp]
  \centering
  \begin{subfigure}{.5\textwidth}
    \centering
    \includegraphics[width=\linewidth]{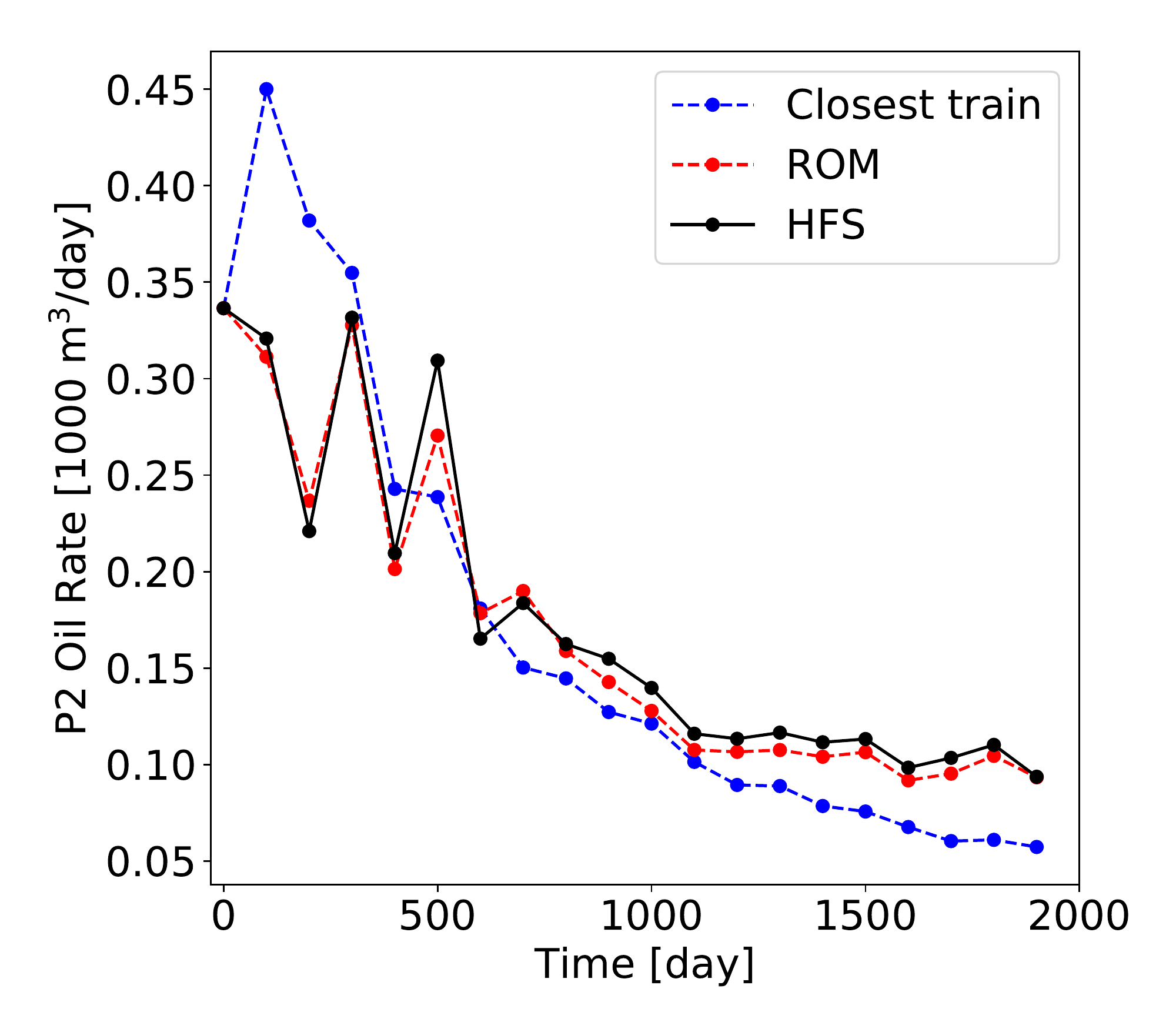}
    \caption{Oil rate}
  \end{subfigure}\hfill
  \begin{subfigure}{.5\textwidth}
    \centering
    \includegraphics[width=\linewidth]{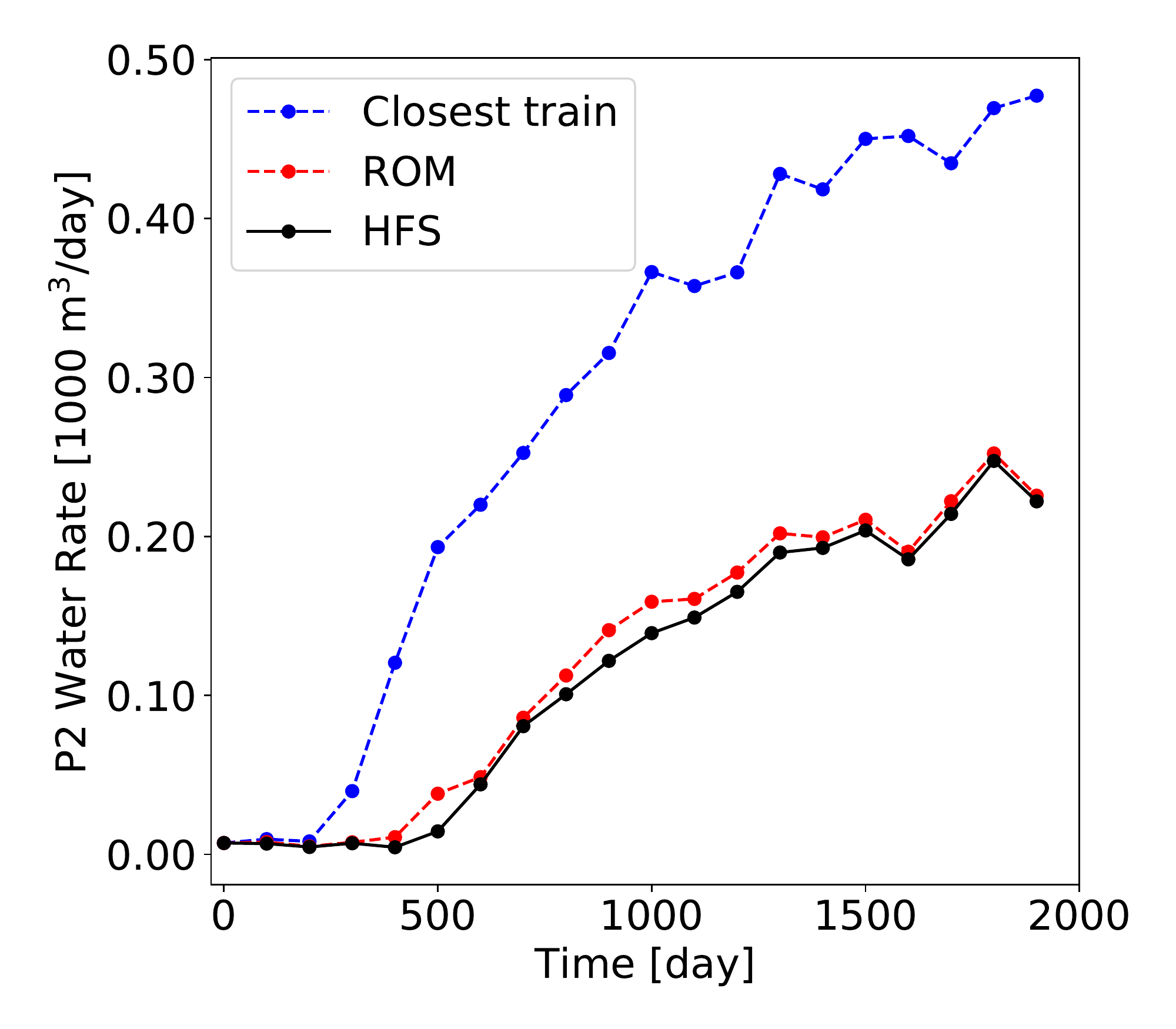}
    \caption{Water rate}
  \end{subfigure}
  \caption{Test Case~2: production rates for Well P2}
  \label{fig::test_2_rate_p2}
\end{figure}

\begin{figure}[htbp]
  \centering
  \begin{subfigure}{.45\textwidth}
    \centering
    \includegraphics[width=\linewidth]{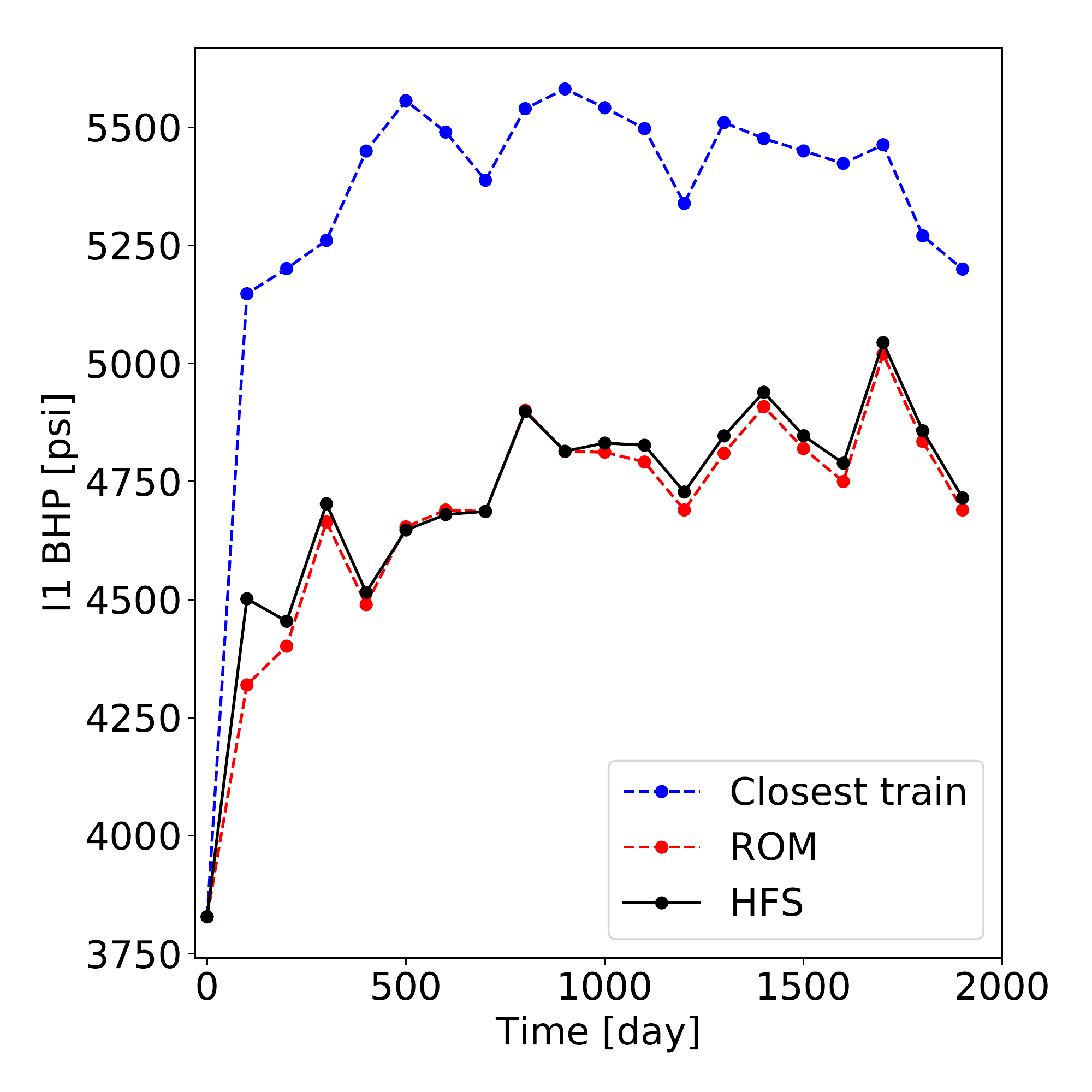}
    \caption{Well~I1}
  \end{subfigure}\hfill
  \begin{subfigure}{.45\textwidth}
    \centering
    \includegraphics[width=\linewidth]{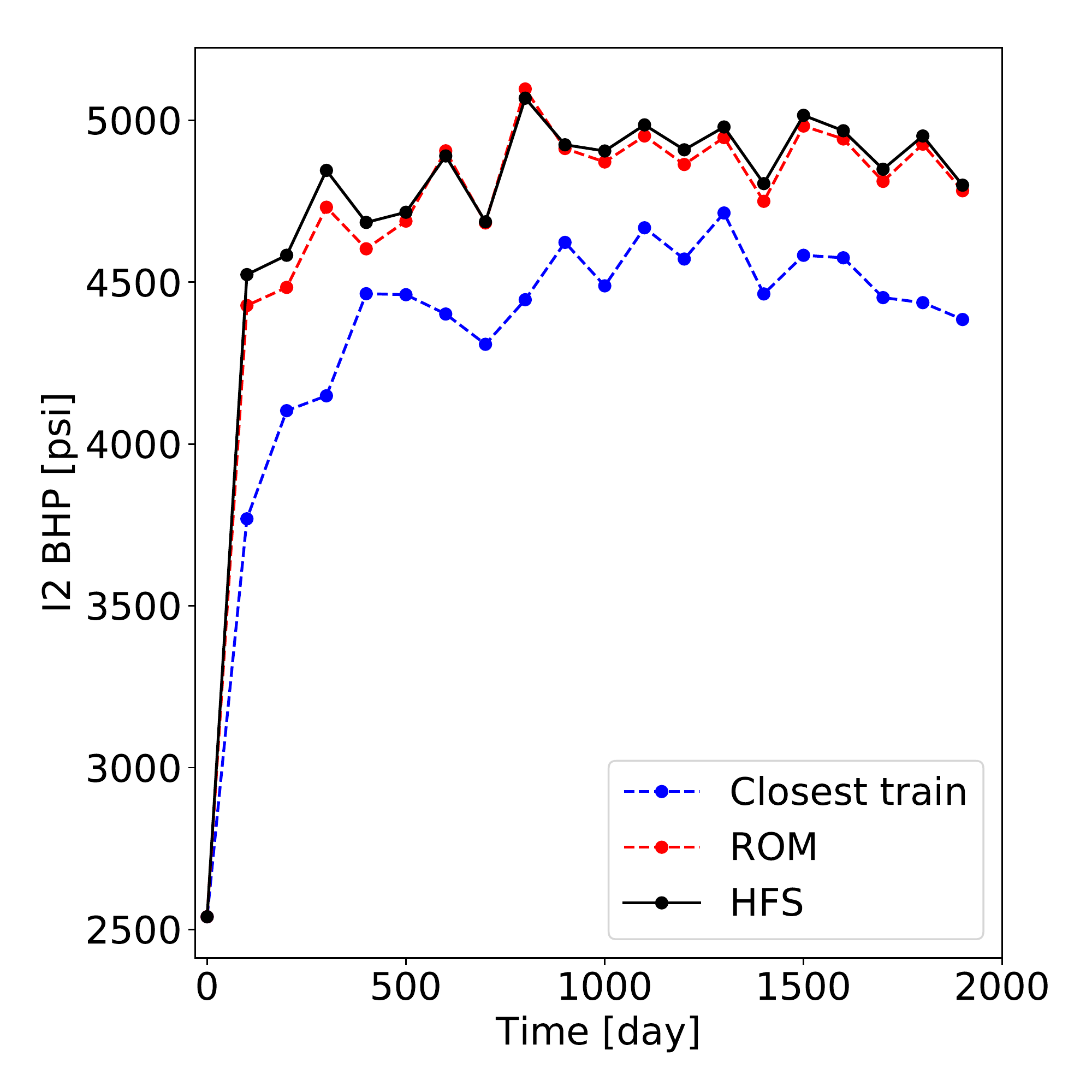}
    \caption{Well~I2}
  \end{subfigure} \\
  \begin{subfigure}{.45\textwidth}
    \centering
    \includegraphics[width=\linewidth]{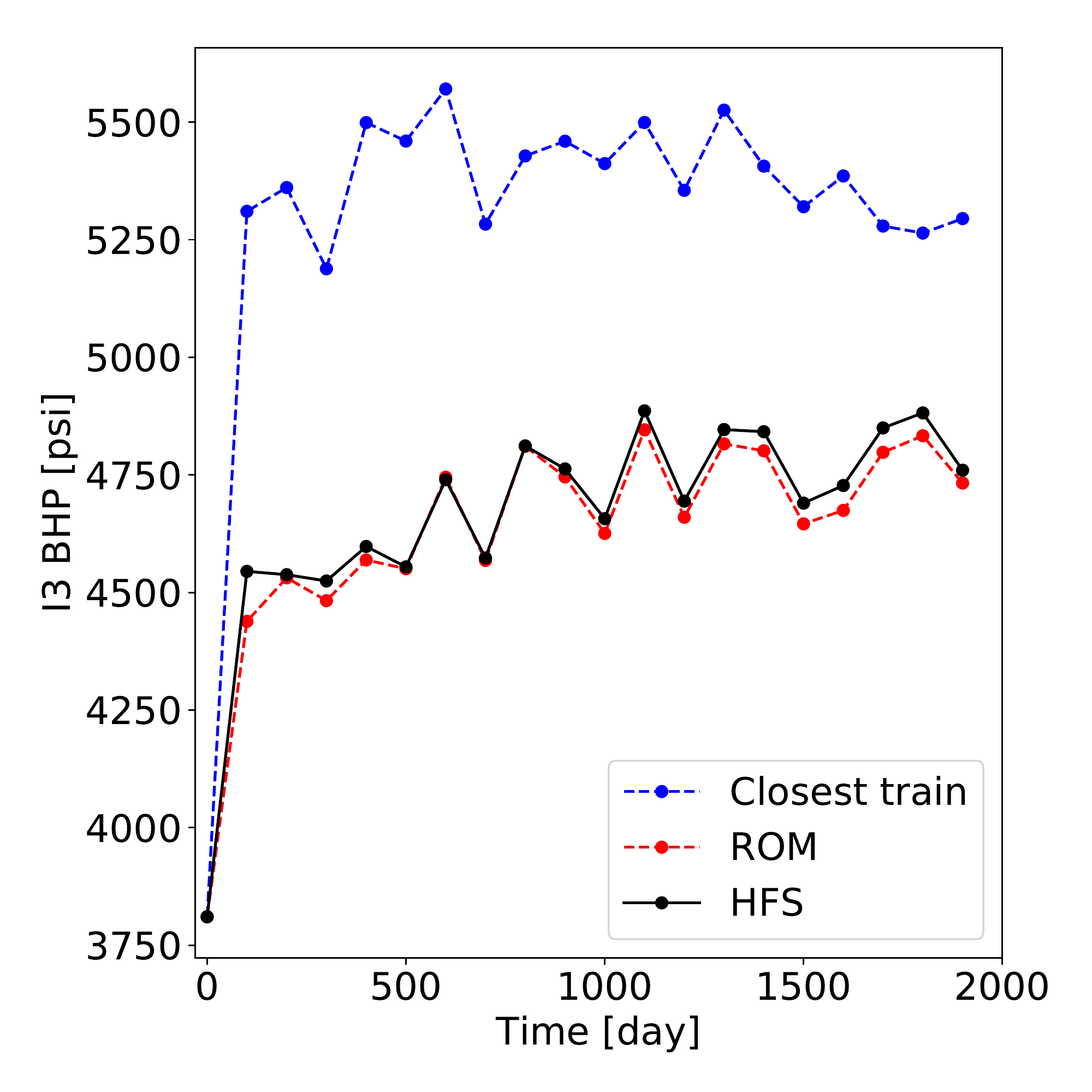}
    \caption{Well~I3}
  \end{subfigure}\hfill
    \begin{subfigure}{.45\textwidth}
    \centering
    \includegraphics[width=\linewidth]{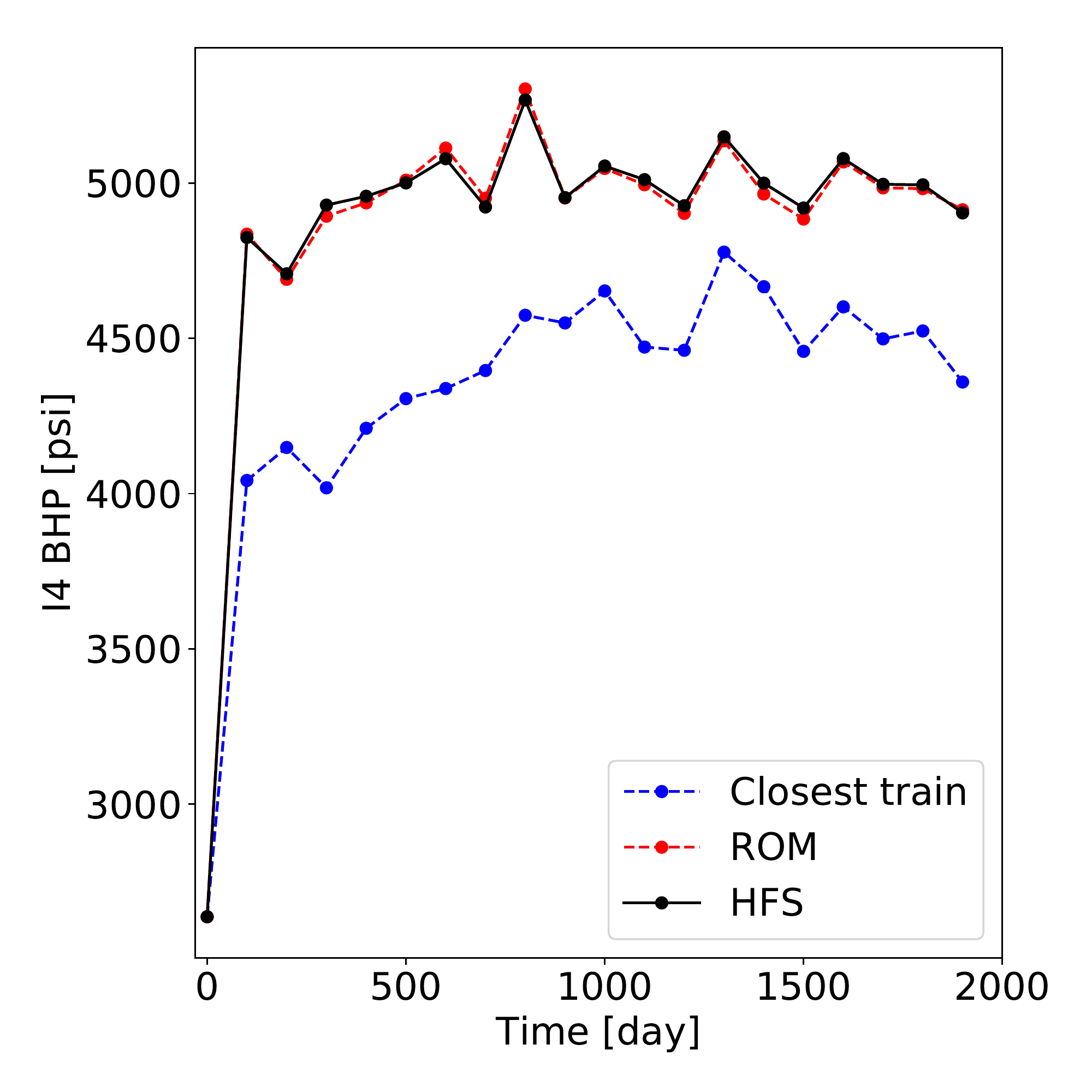}
    \caption{Well~I4}
  \end{subfigure}
  \caption{Test Case~2: injection BHPs}
  \label{fig::test_2_inj}
\end{figure}

Figs.~\ref{fig::test_2_rate_p1} and \ref{fig::test_2_rate_p2} display oil and water production rates for Wells P1 and P2, which are the major contributors to total field production in Test Case~2. Fig.~\ref{fig::test_2_inj} presents the injection BHPs for the four injection wells. The ROM solutions are again in close agreement with the high-fidelity solution, and significant discrepancy between these solutions and the `closest training run' is observed for most well quantities.

%%%%% case 2, saturation %%%%%%%%

%%%%% case 2, pressure %%%%%%%%

%%%%% case 2, prod rates %%%%%%%%

%%%%% case 2, inj bhps %%%%%%%%

%%%%%%%%%%%%%%%%%%%%%%%%%%%%%%%%
%%%%%%%%%%% Case 3 %%%%%%%%%%%%%
%%%%%%%%%%%%%%%%%%%%%%%%%%%%%%%%
\subsection{Results for Test Case~3}

\begin{figure}[htbp]
  \centering
  \begin{subfigure}{.45\textwidth}
    \centering
    \includegraphics[width=\linewidth]{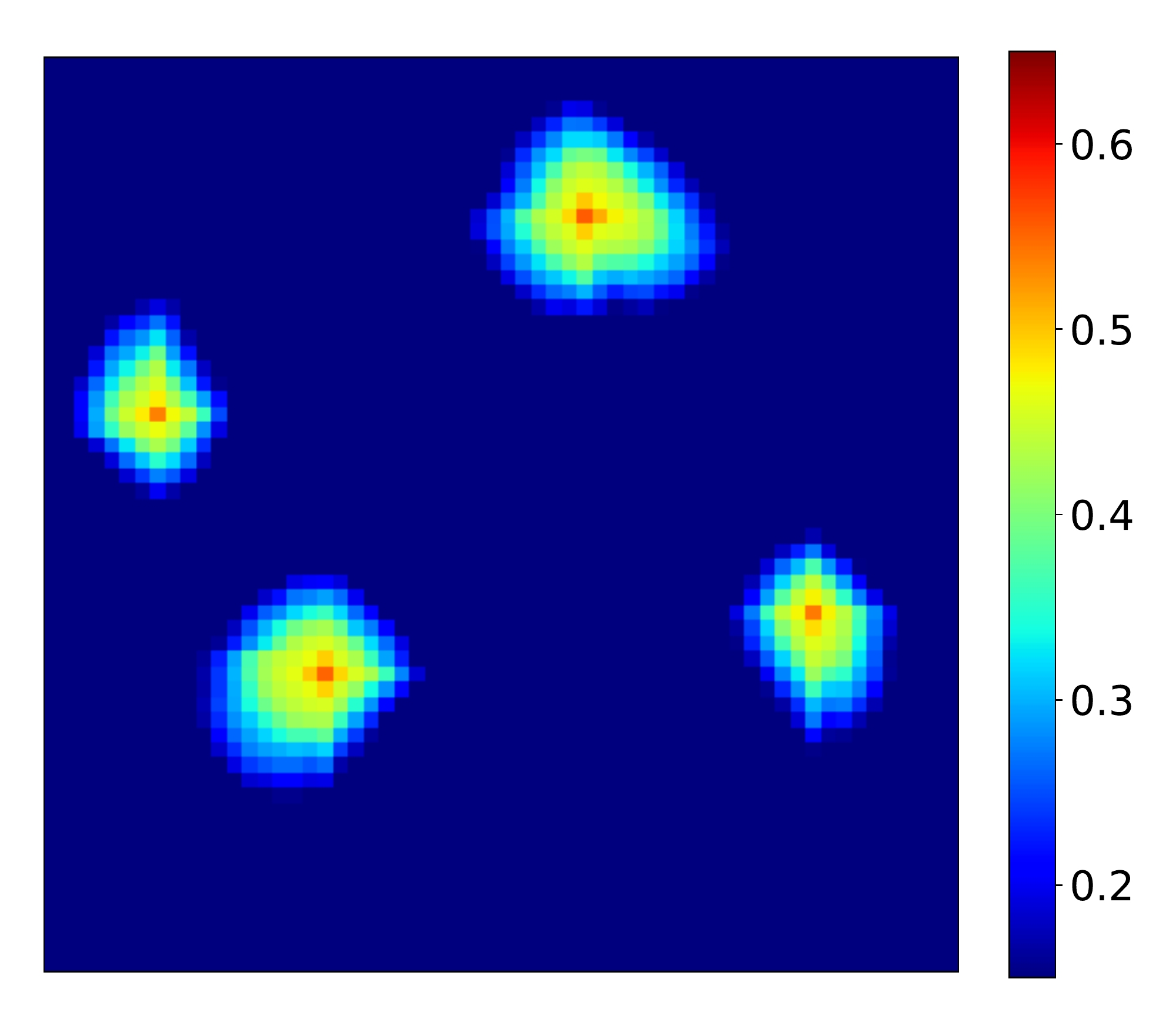}
    \caption{High-fidelity solution ($\text{HFS}_{\text{test}}$)}
  \end{subfigure}\hfill
  \begin{subfigure}{.45\textwidth}
    \centering
    \includegraphics[width=\linewidth]{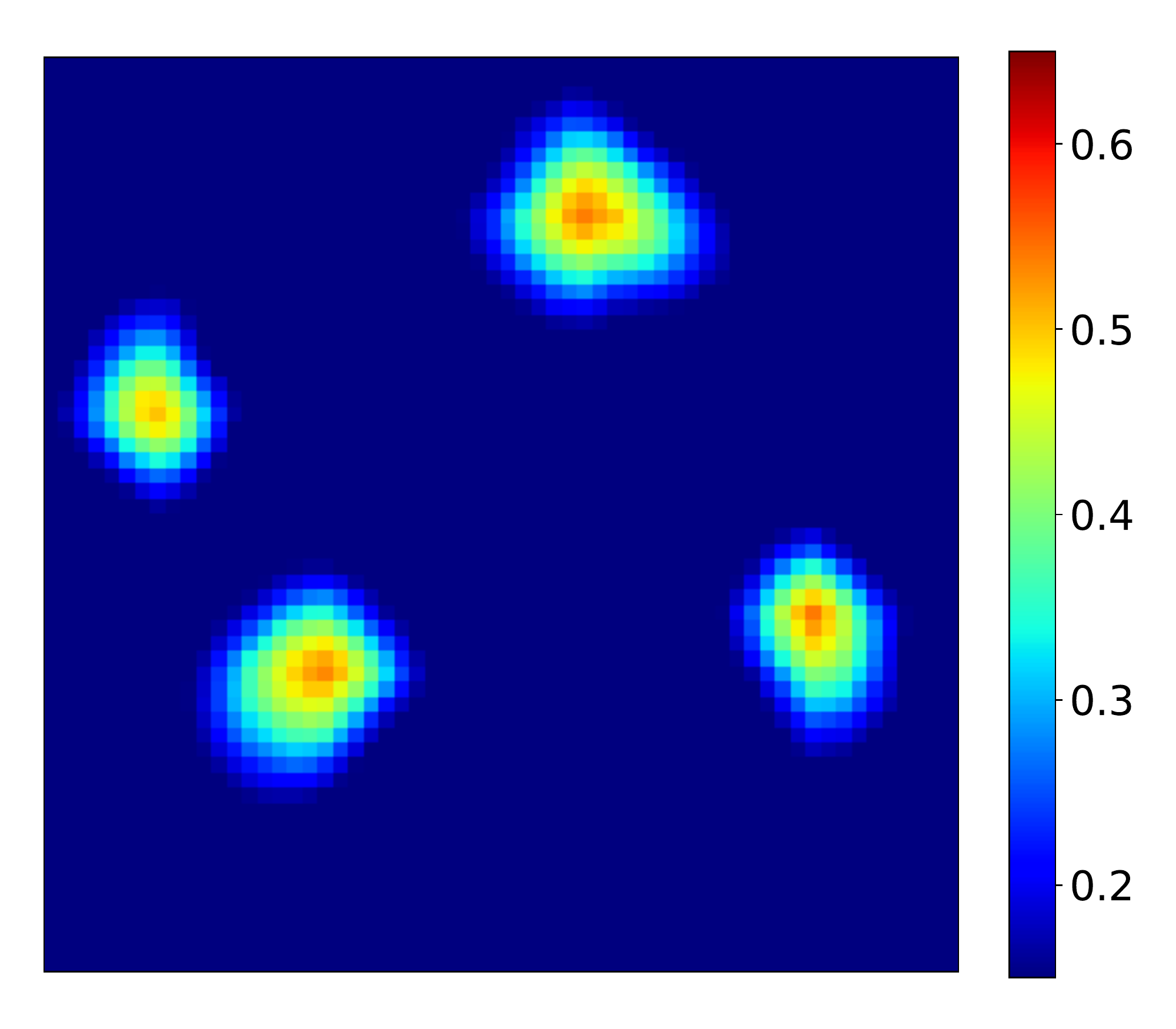}
    \caption{ROM solution ($\text{ROM}_{\text{test}}$)}
  \end{subfigure} \\
  \begin{subfigure}{.45\textwidth}
    \centering
    \includegraphics[width=\linewidth]{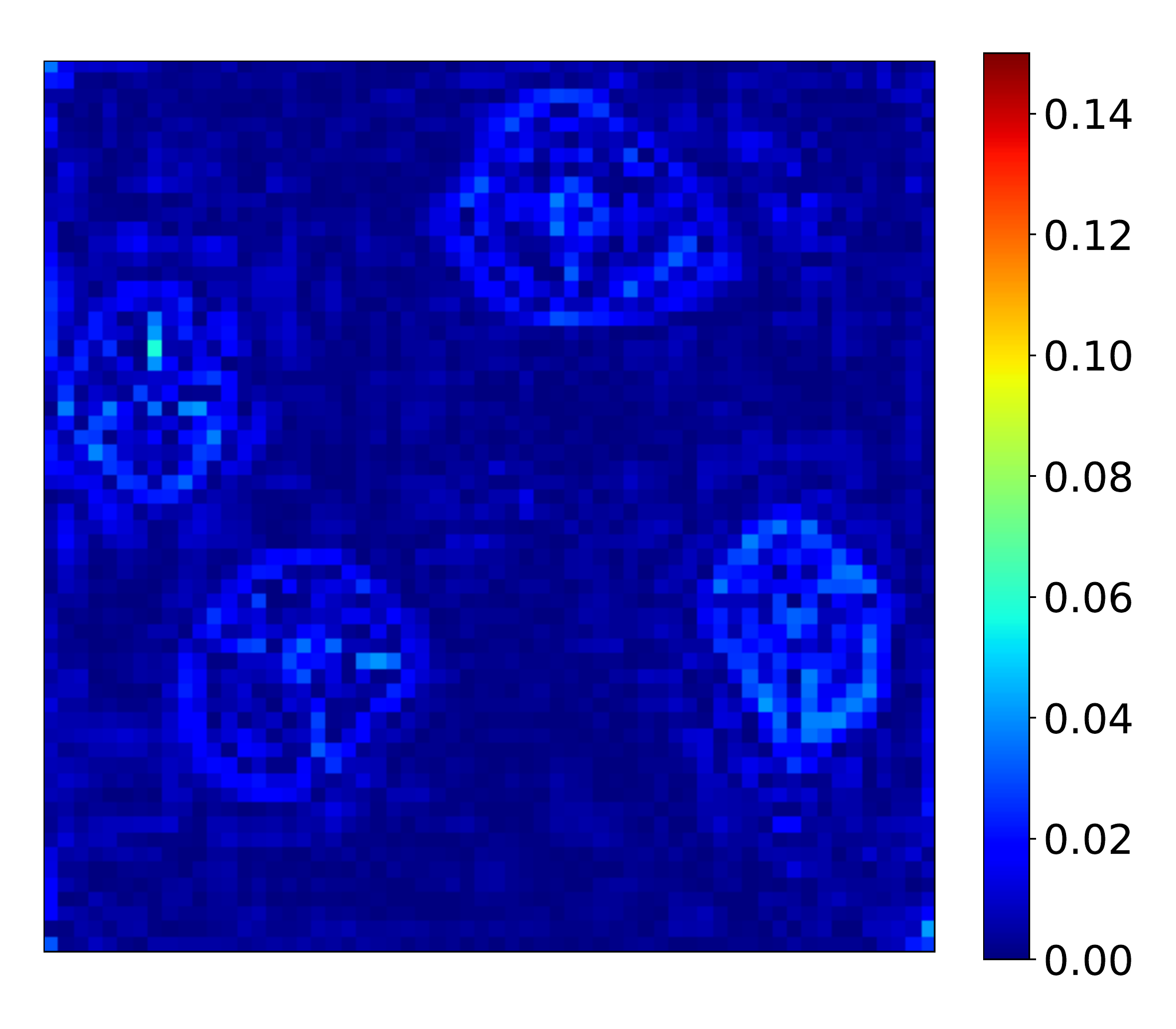}
    \caption{$|\text{HFS}_{\text{test}} - \text{ROM}_{\text{test}}|$}
  \end{subfigure}\hfill
    \begin{subfigure}{.45\textwidth}
    \centering
    \includegraphics[width=\linewidth]{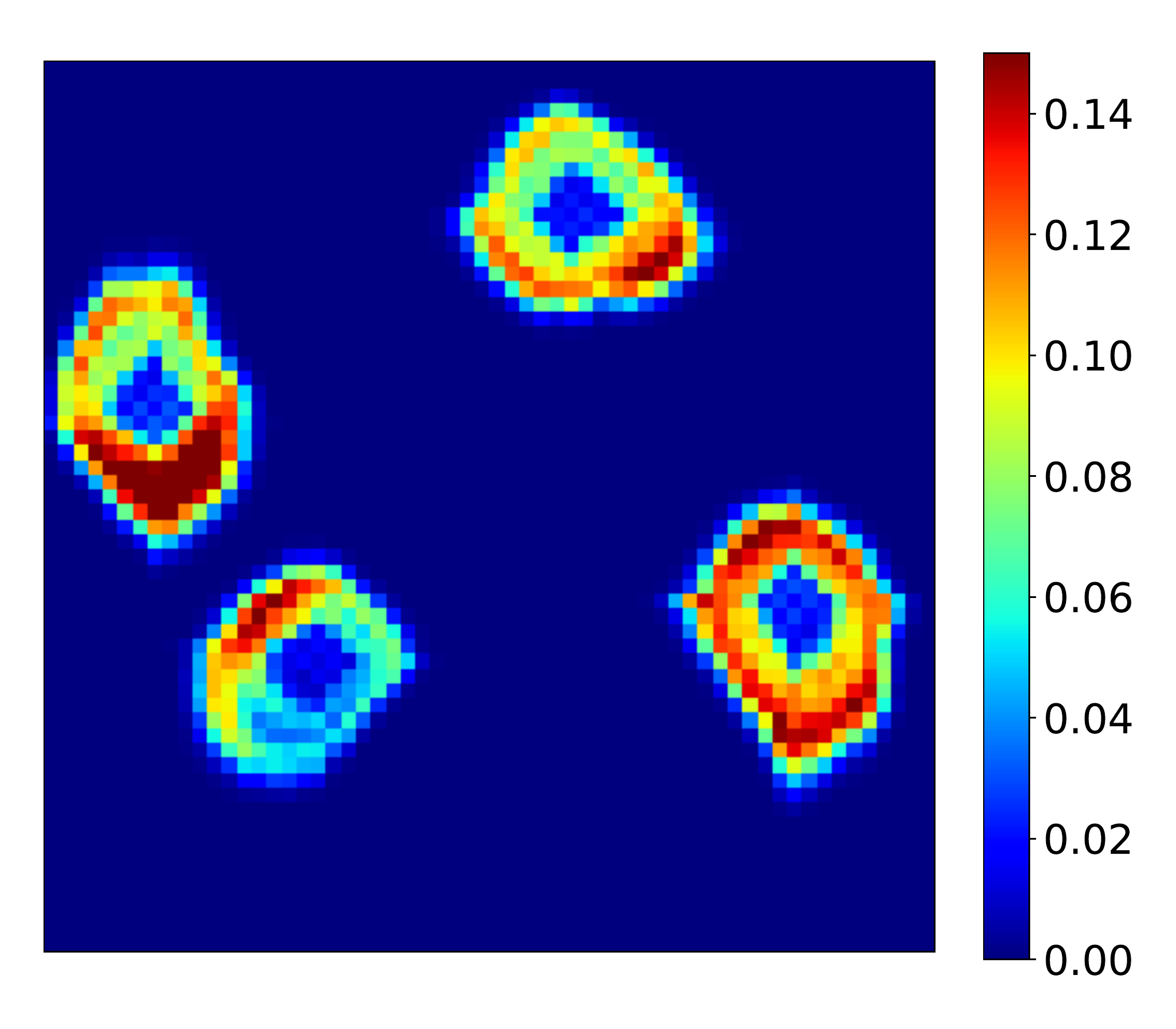}
    \caption{$|\text{HFS}_{\text{test}} - \text{HFS}_{\text{train}}|$}
  \end{subfigure}
  \caption{Test Case~3: saturation field at 200 days}
  \label{fig::test_3_sat_200}
\end{figure}

\begin{figure}[htbp]
  \centering
  \begin{subfigure}{.45\textwidth}
    \centering
    \includegraphics[width=\linewidth]{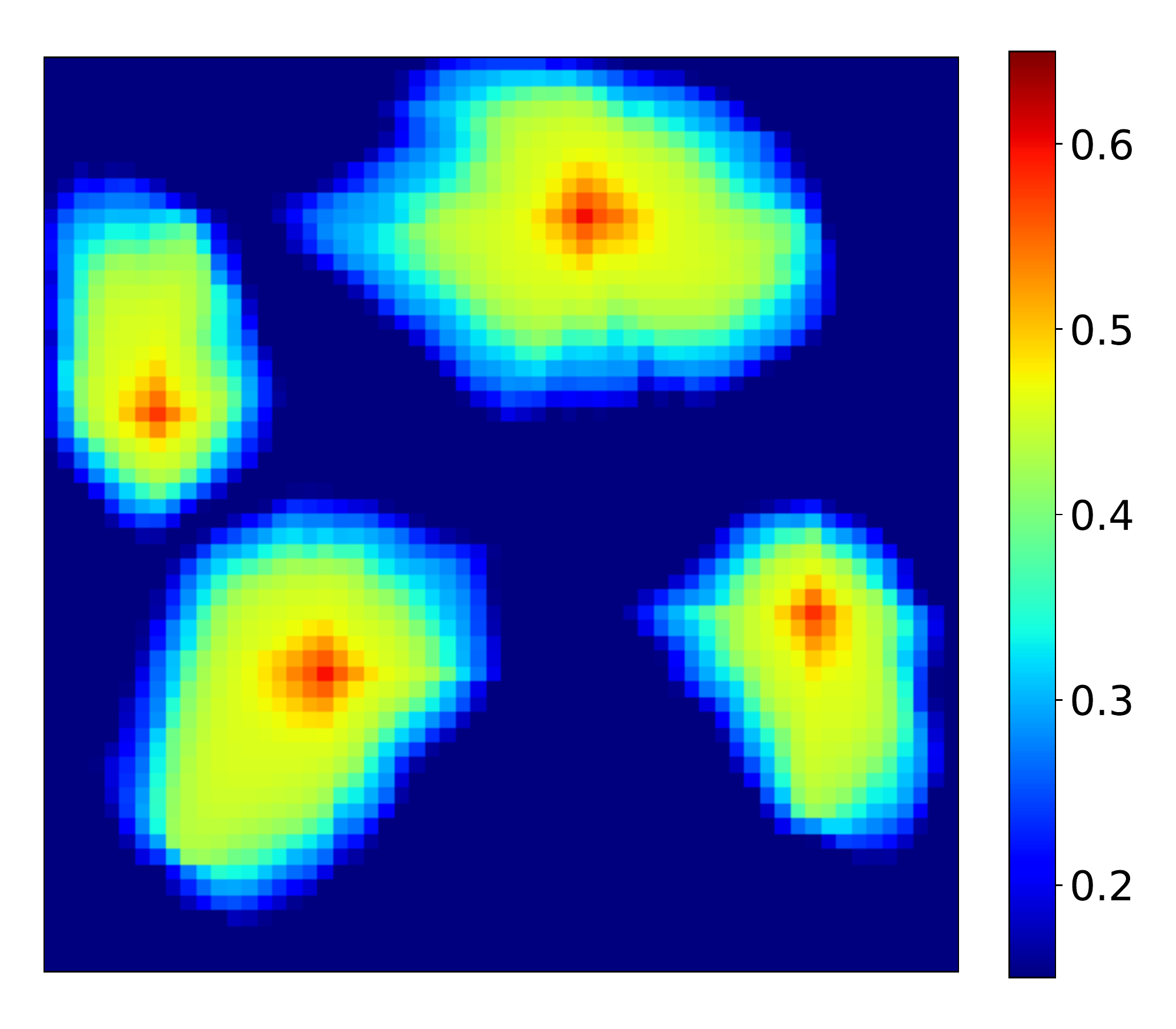}
    \caption{High-fidelity solution ($\text{HFS}_{\text{test}}$)}
  \end{subfigure}\hfill
  \begin{subfigure}{.45\textwidth}
    \centering
    \includegraphics[width=\linewidth]{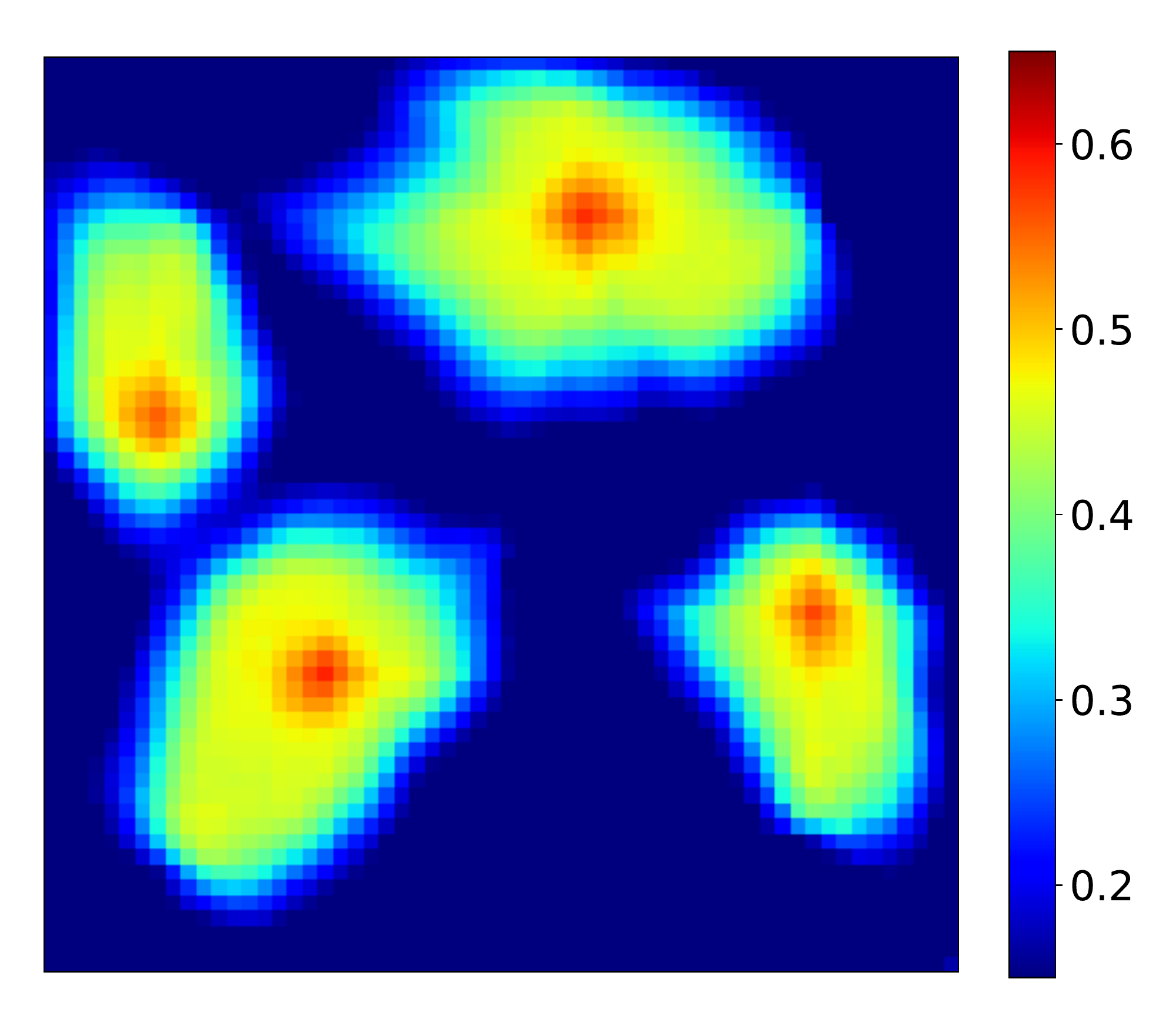}
    \caption{ROM solution ($\text{ROM}_{\text{test}}$)}
  \end{subfigure} \\
  \begin{subfigure}{.45\textwidth}
    \centering
    \includegraphics[width=\linewidth]{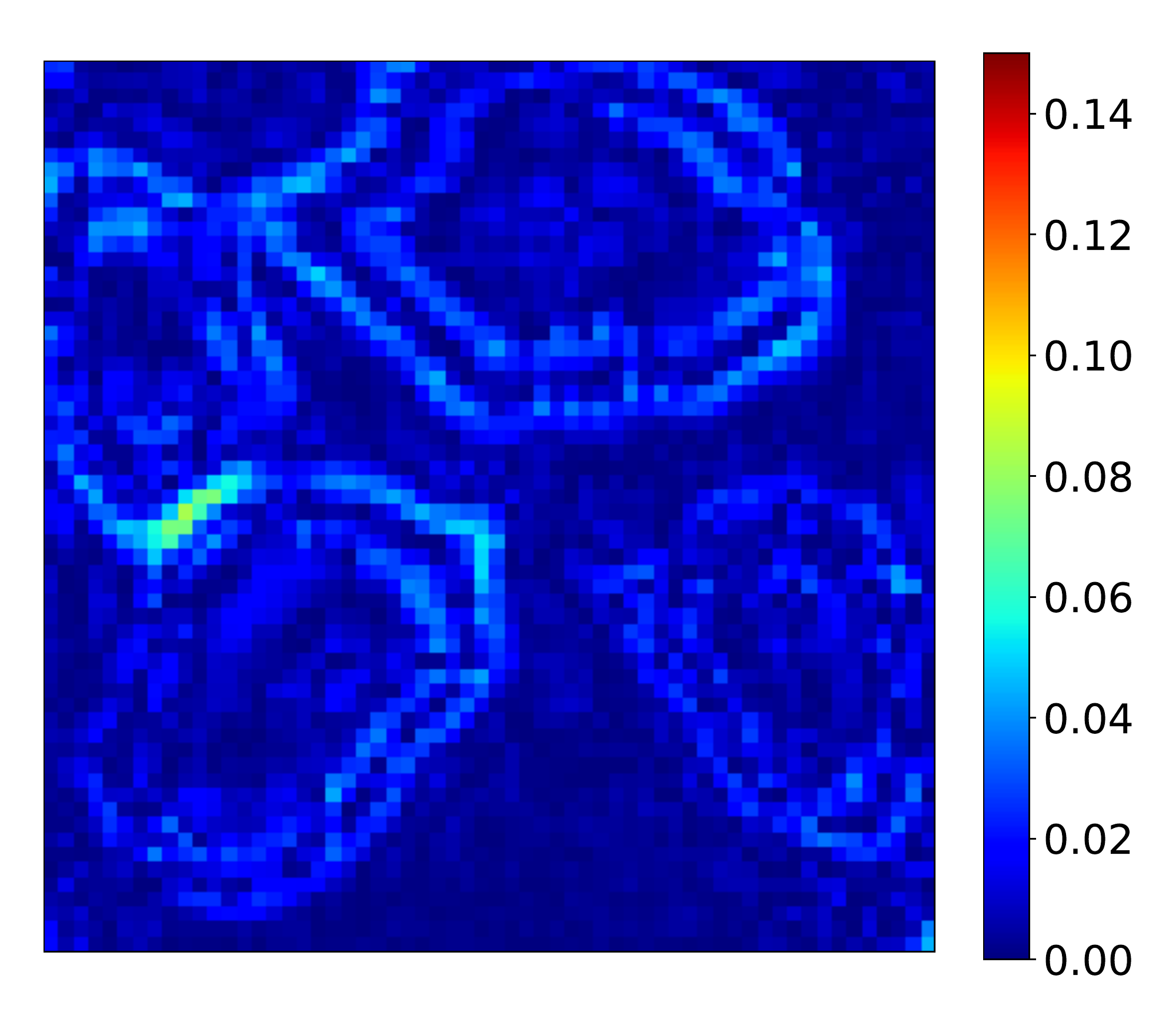}
    \caption{$|\text{HFS}_{\text{test}} - \text{ROM}_{\text{test}}|$}
  \end{subfigure}\hfill
    \begin{subfigure}{.45\textwidth}
    \centering
    \includegraphics[width=\linewidth]{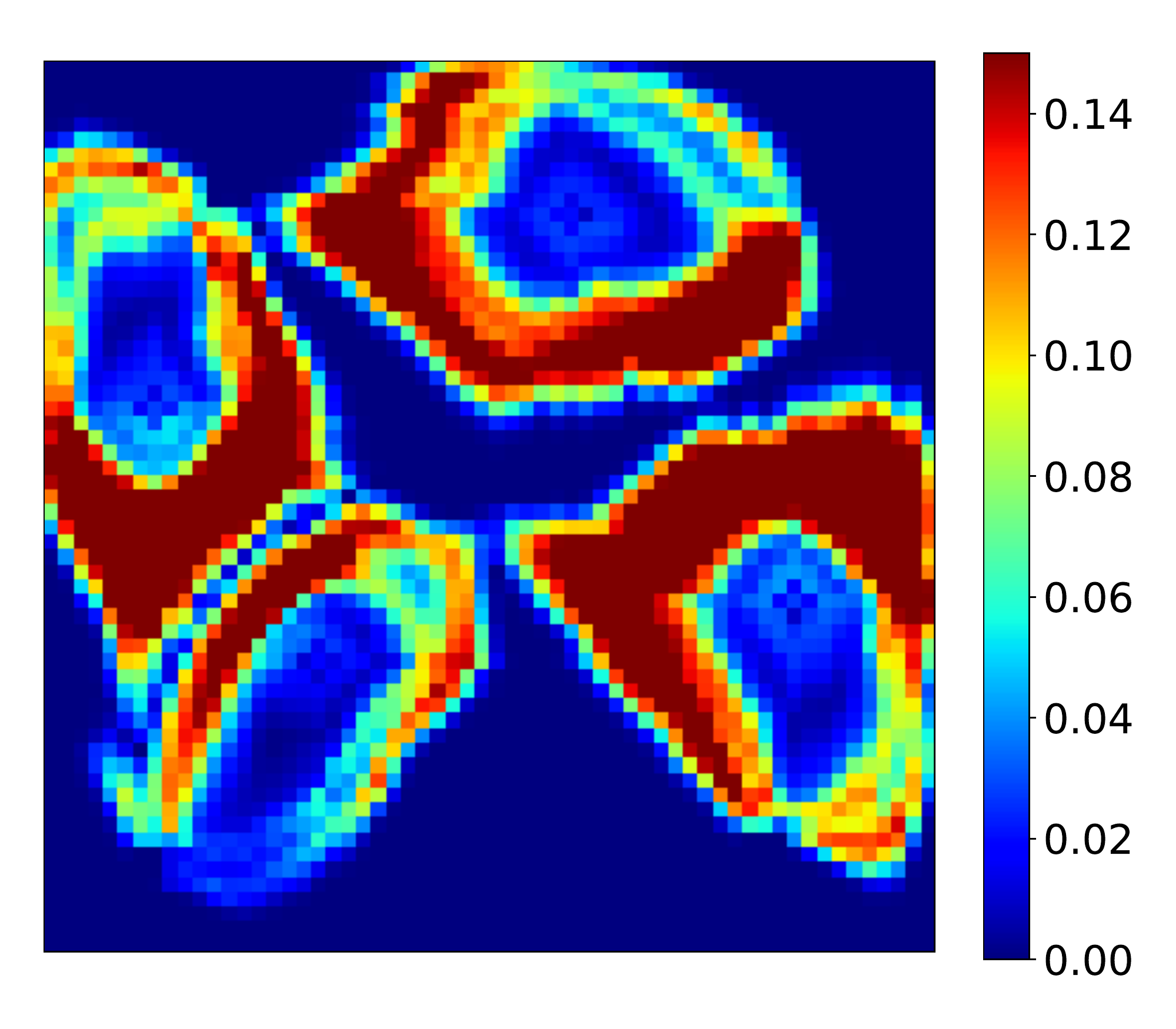}
    \caption{$|\text{HFS}_{\text{test}} - \text{HFS}_{\text{train}}|$}
  \end{subfigure}
  \caption{Test Case~3: saturation field at 1000 days}
  \label{fig::test_3_sat_1000}
\end{figure}

\begin{figure}[htbp]
  \centering
  \begin{subfigure}{.45\textwidth}
    \centering
    \includegraphics[width=\linewidth]{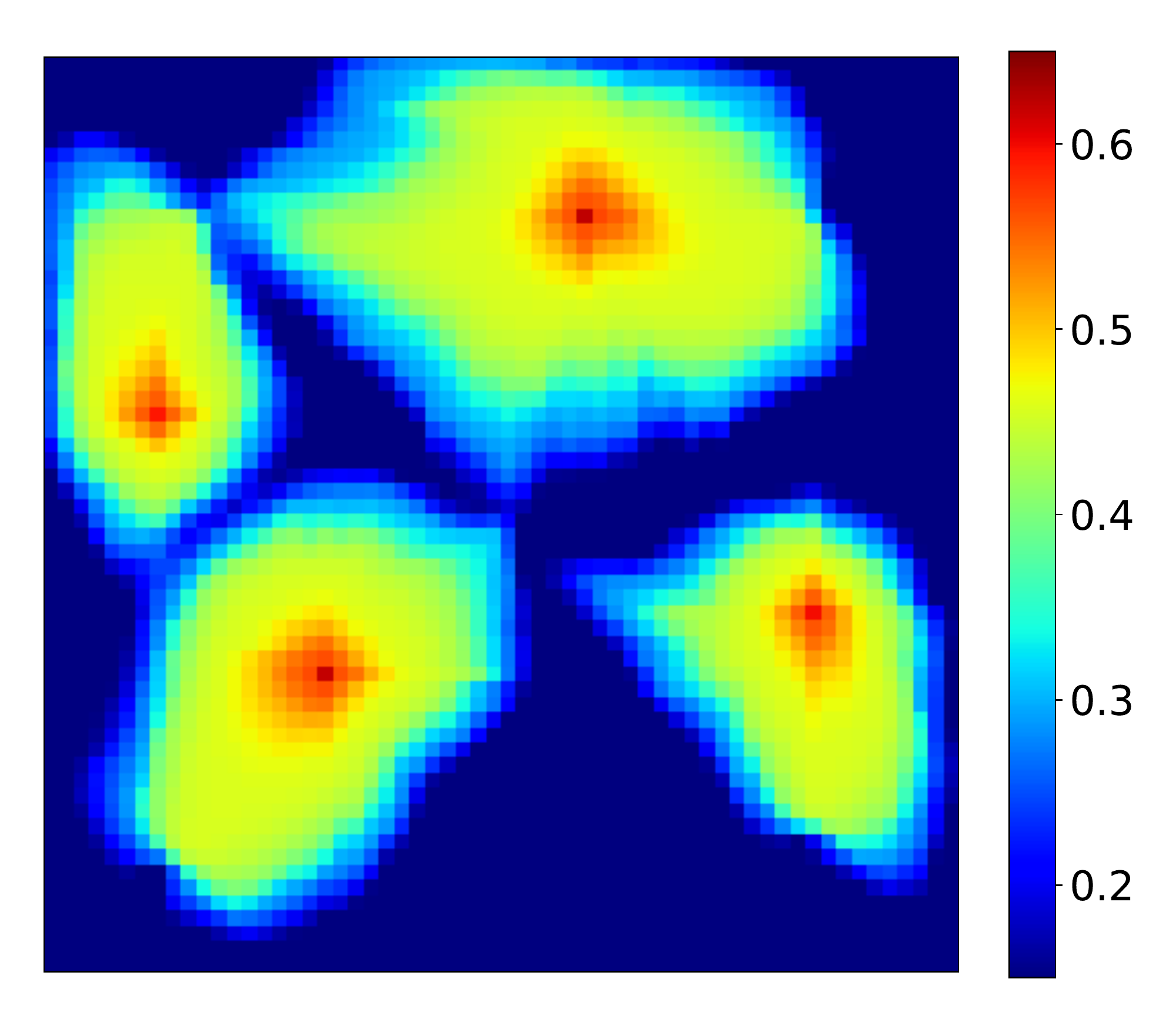}
    \caption{High-fidelity solution ($\text{HFS}_{\text{test}}$)}
  \end{subfigure}\hfill
  \begin{subfigure}{.45\textwidth}
    \centering
    \includegraphics[width=\linewidth]{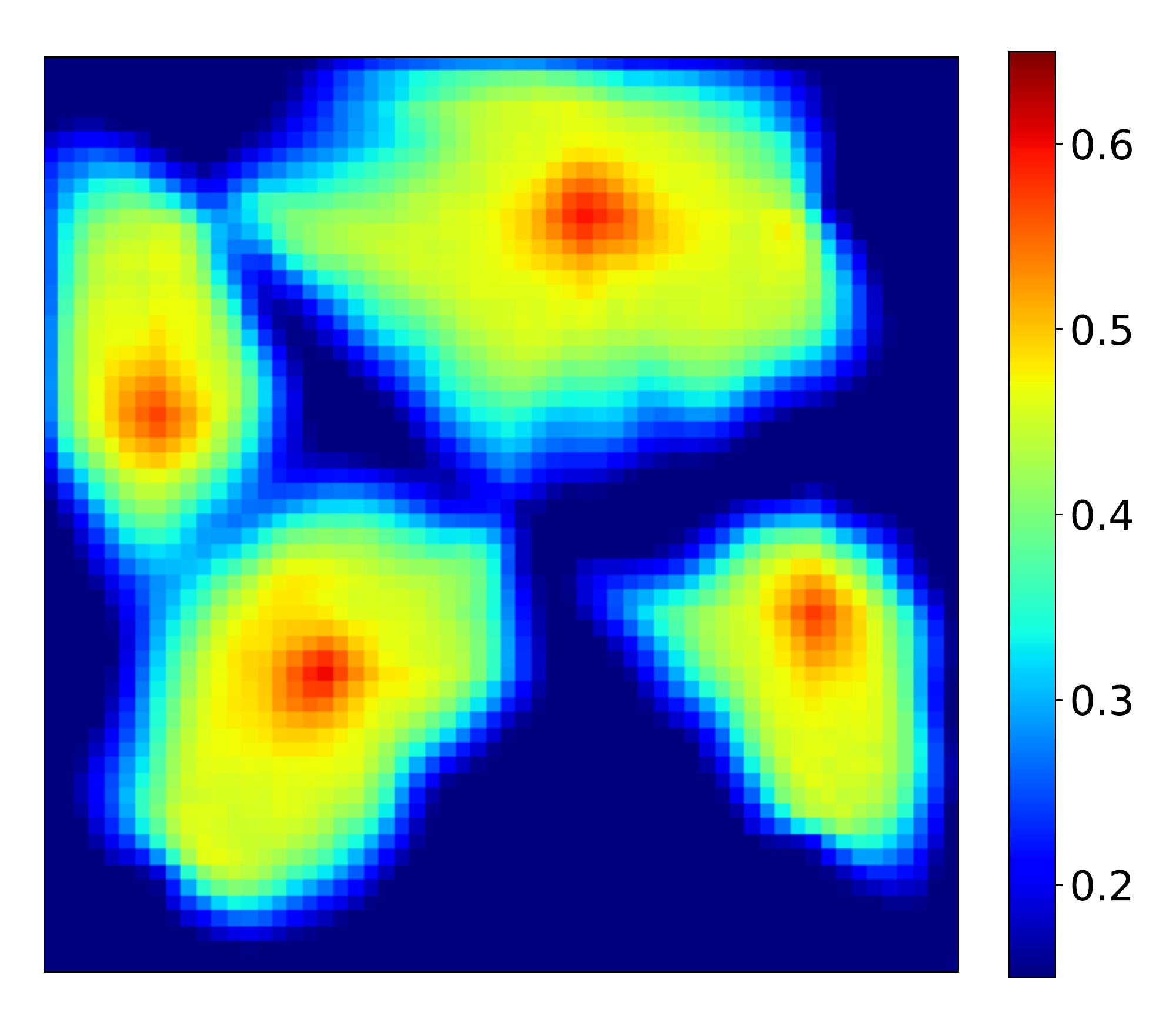}
    \caption{ROM solution ($\text{ROM}_{\text{test}}$)}
  \end{subfigure} \\
  \begin{subfigure}{.45\textwidth}
    \centering
    \includegraphics[width=\linewidth]{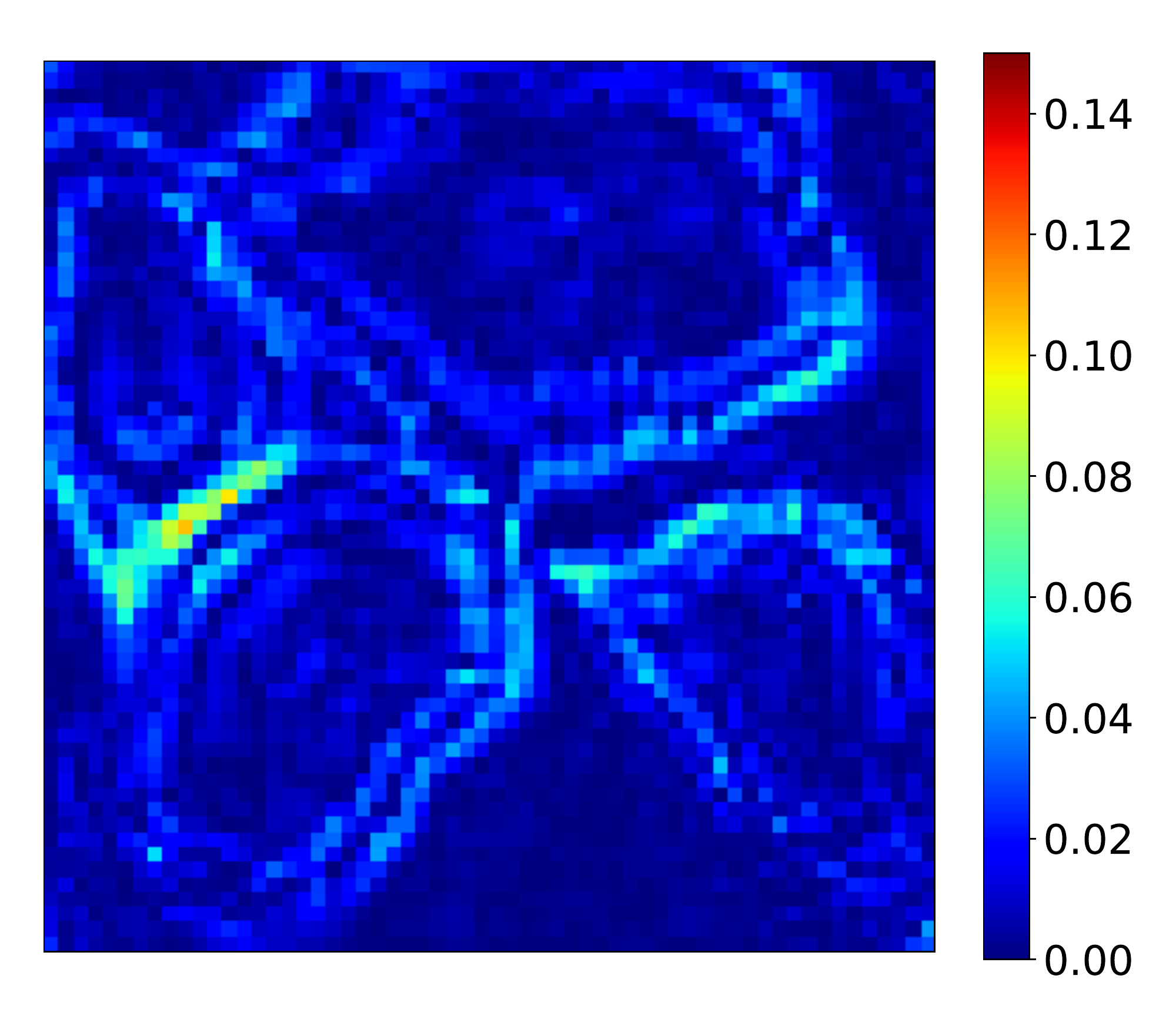}
    \caption{$|\text{HFS}_{\text{test}} - \text{ROM}_{\text{test}}|$}
  \end{subfigure}\hfill
    \begin{subfigure}{.45\textwidth}
    \centering
    \includegraphics[width=\linewidth]{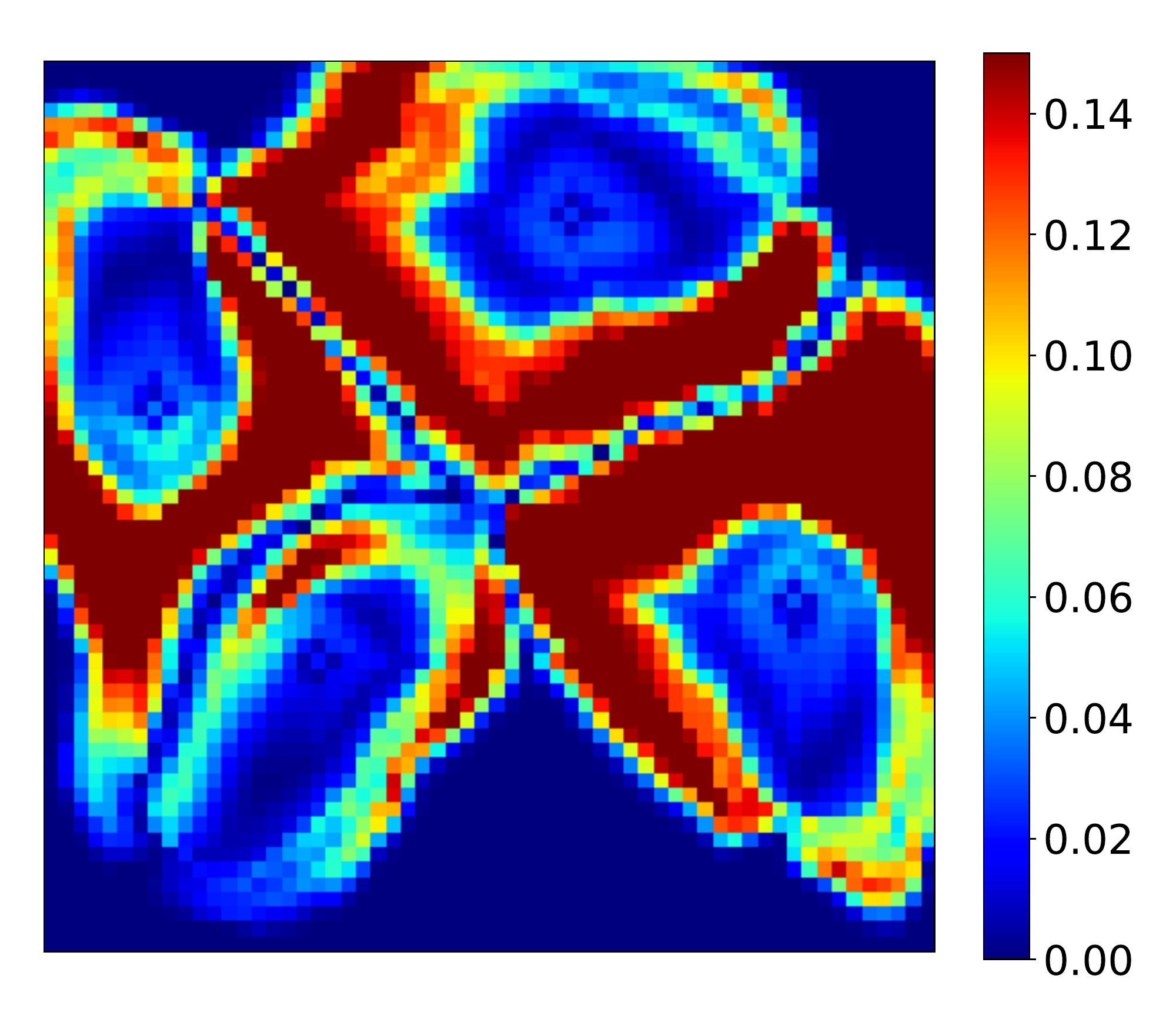}
    \caption{$|\text{HFS}_{\text{test}} - \text{HFS}_{\text{train}}|$}
  \end{subfigure}
  \caption{Test Case~3: saturation field at 1800 days}
  \label{fig::test_3_sat_1800}
\end{figure}

\begin{figure}[htbp]
  \centering
  \begin{subfigure}{.45\textwidth}
    \centering
    \includegraphics[width=\linewidth]{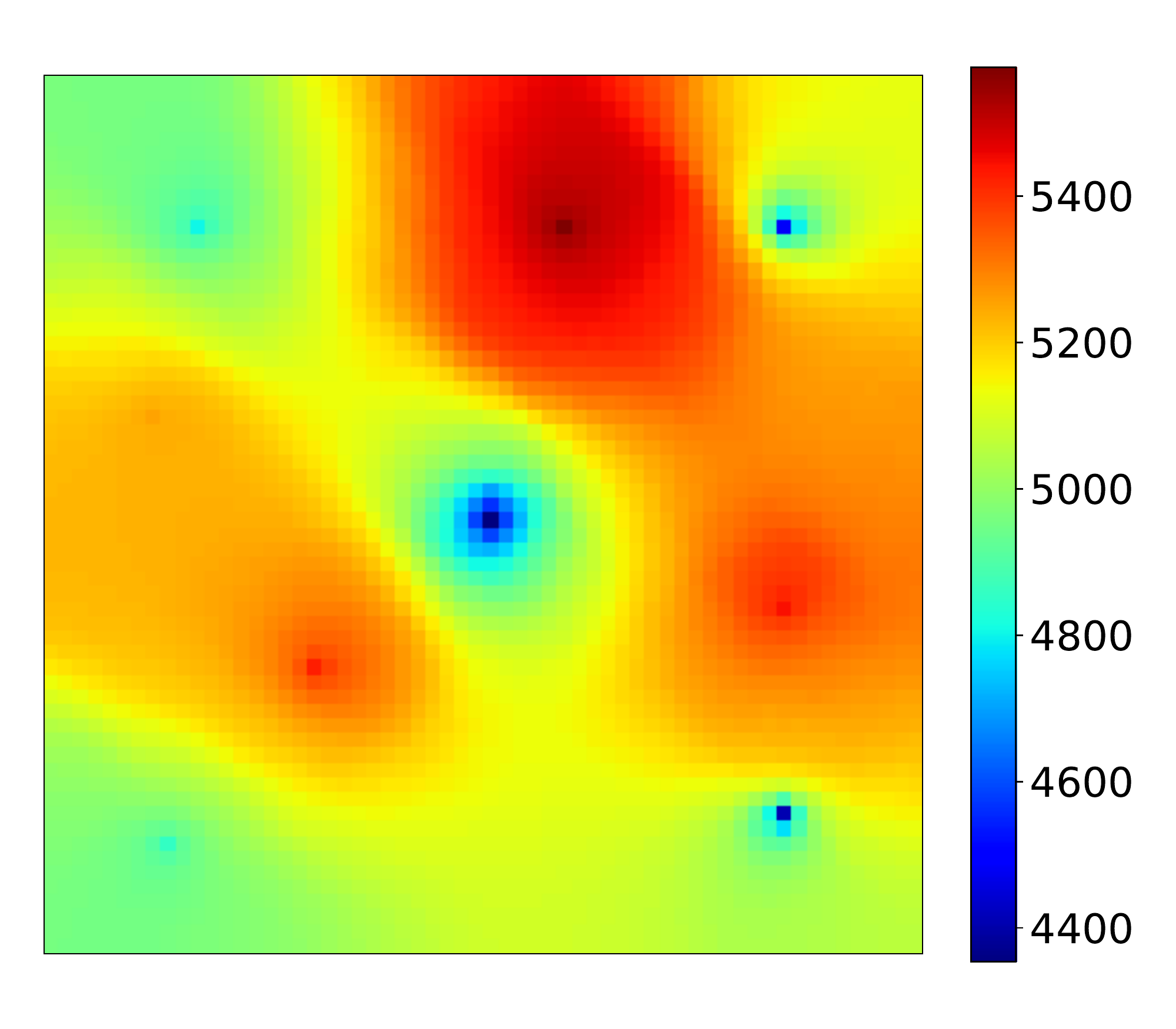}
    \caption{High-fidelity solution ($\text{HFS}_{\text{test}}$)}
  \end{subfigure}\hfill
  \begin{subfigure}{.45\textwidth}
    \centering
    \includegraphics[width=\linewidth]{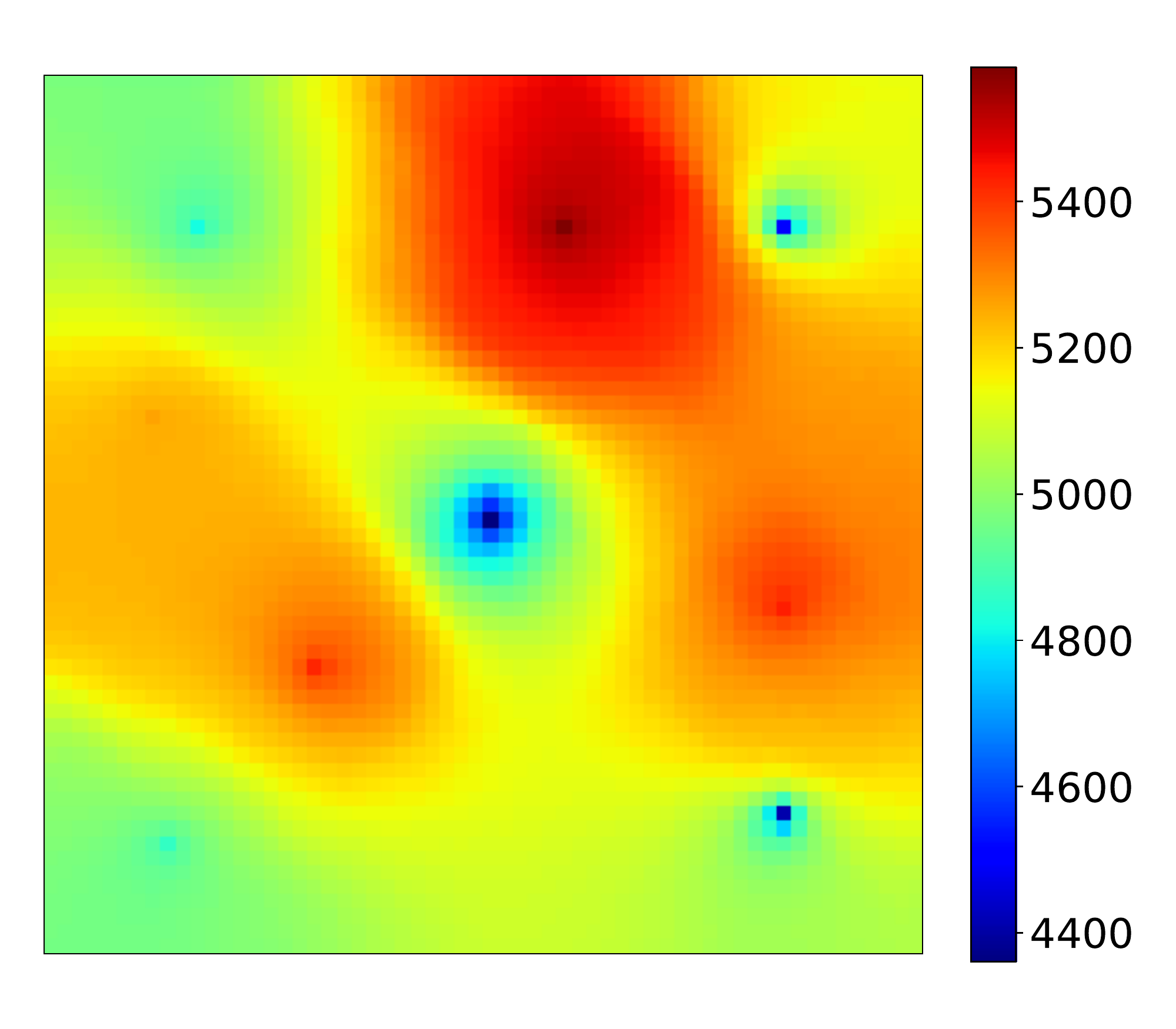}
    \caption{ROM solution ($\text{ROM}_{\text{test}}$)}
  \end{subfigure} \\
  \begin{subfigure}{.45\textwidth}
    \centering
    \includegraphics[width=\linewidth]{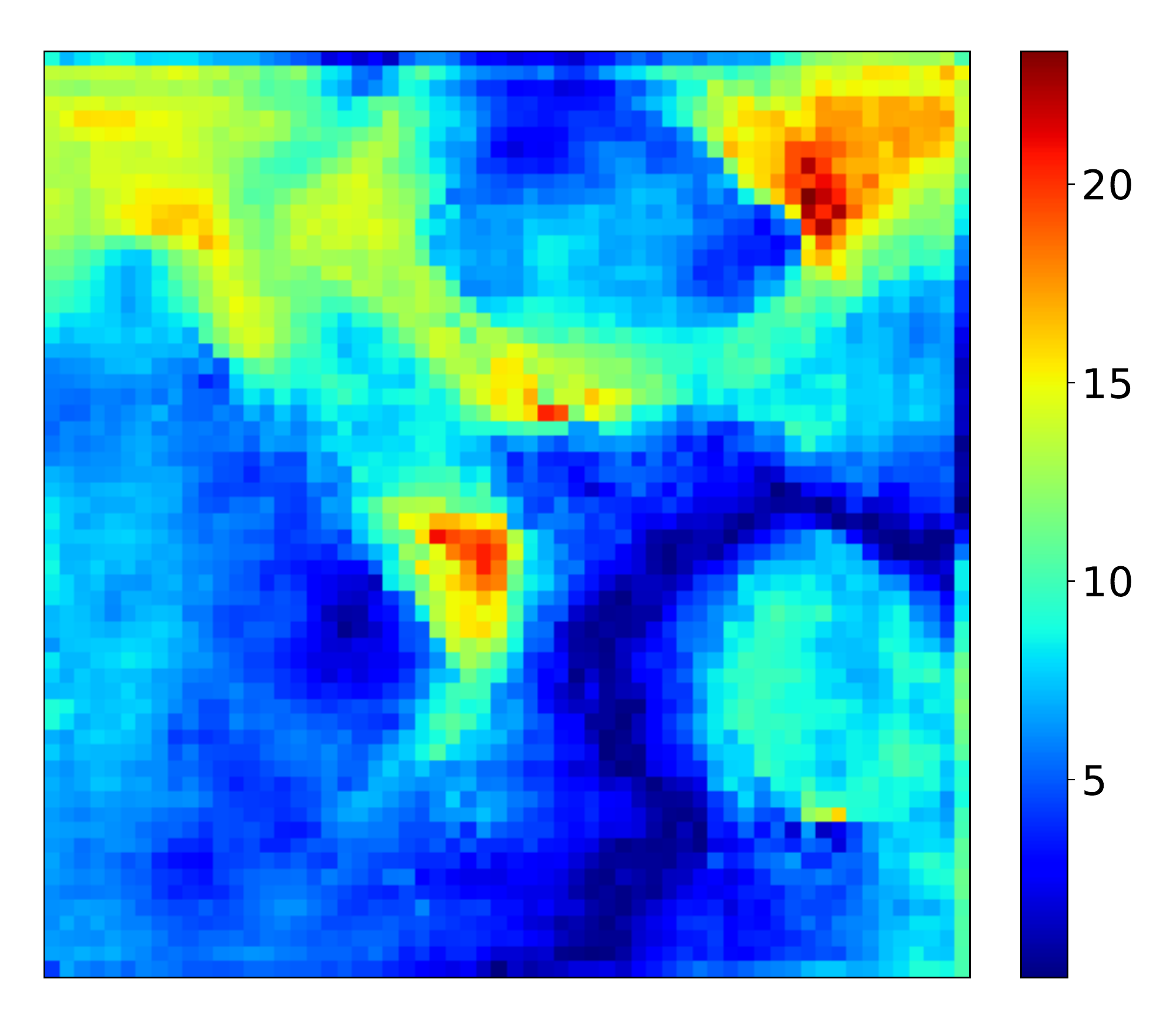}
    \caption{$|\text{HFS}_{\text{test}} - \text{ROM}_{\text{test}}|$}
  \end{subfigure}\hfill
    \begin{subfigure}{.45\textwidth}
    \centering
    \includegraphics[width=\linewidth]{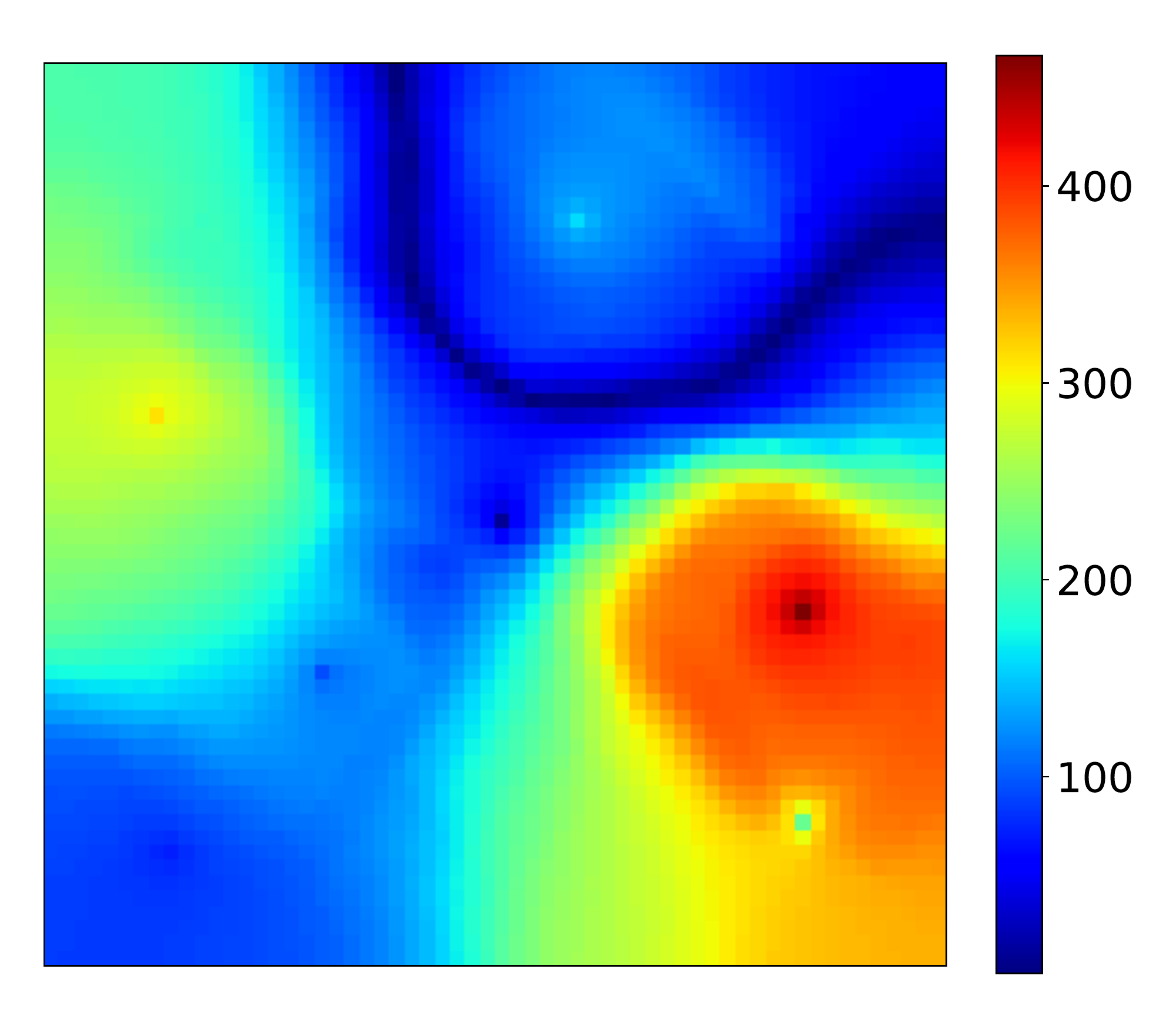}
    \caption{$|\text{HFS}_{\text{test}} - \text{HFS}_{\text{train}}|$}
  \end{subfigure}
  \caption{Test Case~3: pressure field at 1000 days (all colorbars in units of psi)}
  \label{fig::test_3_pres_1000}
\end{figure}

\begin{figure}[htbp]
  \centering
  \begin{subfigure}{.5\textwidth}
    \centering
    \includegraphics[width=\linewidth]{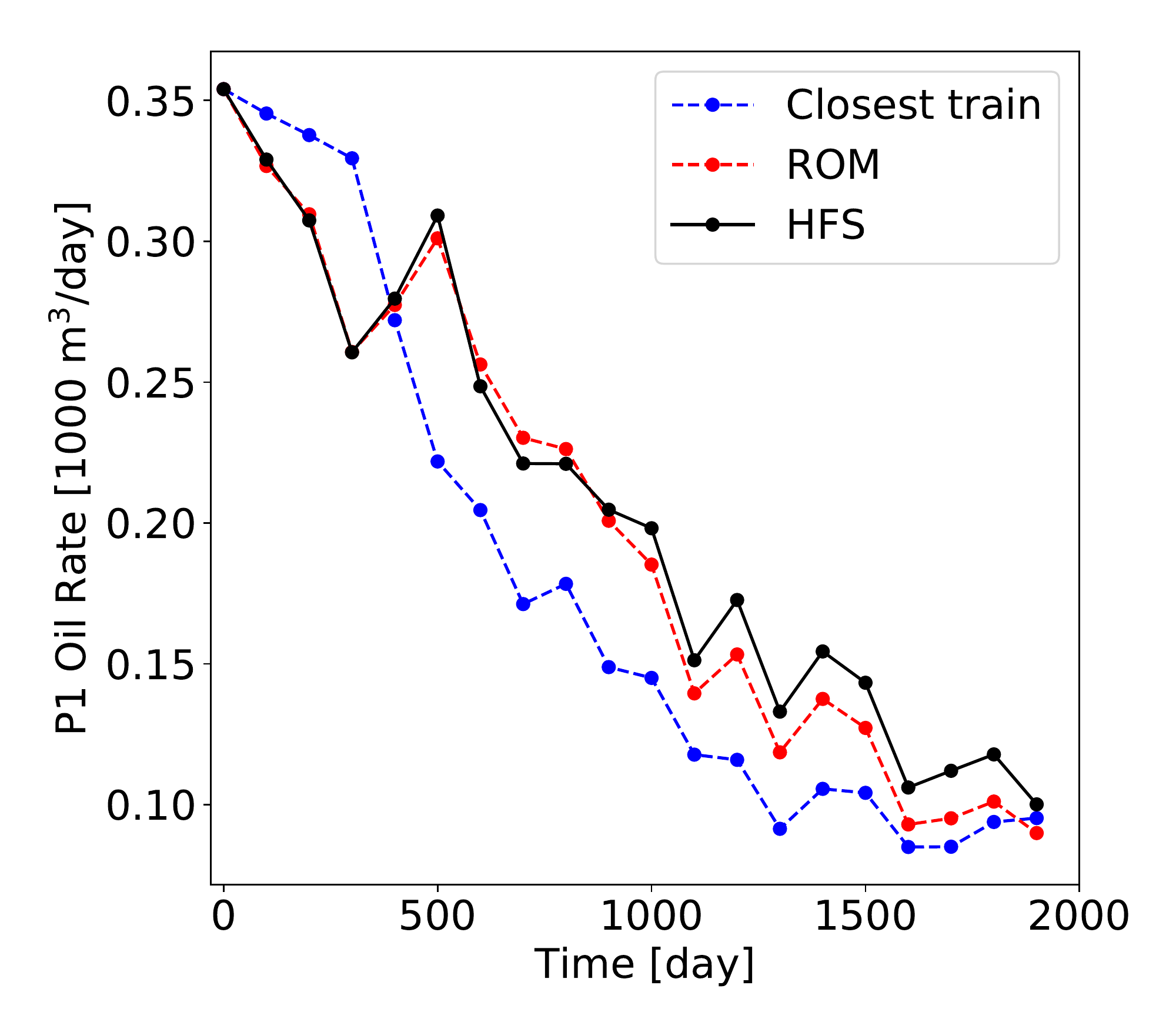}
    \caption{Oil rate}
  \end{subfigure}\hfill
  \begin{subfigure}{.5\textwidth}
    \centering
    \includegraphics[width=\linewidth]{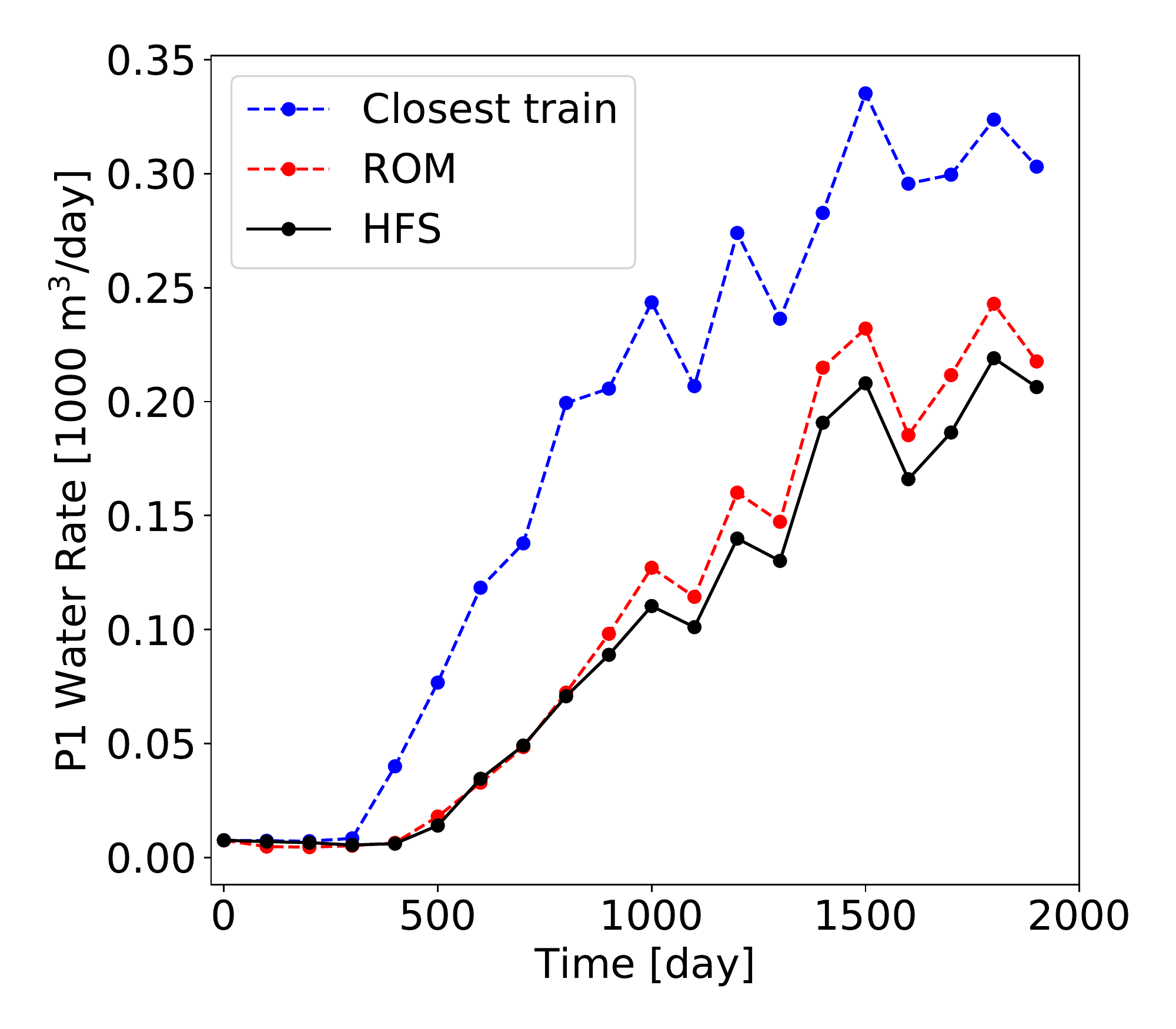}
    \caption{Water rate}
  \end{subfigure}
  \caption{Test Case~3: production rates for Well P1}
  \label{fig::test_3_rate_p1}
\end{figure}

\begin{figure}[htbp]
  \centering
  \begin{subfigure}{.5\textwidth}
    \centering
    \includegraphics[width=\linewidth]{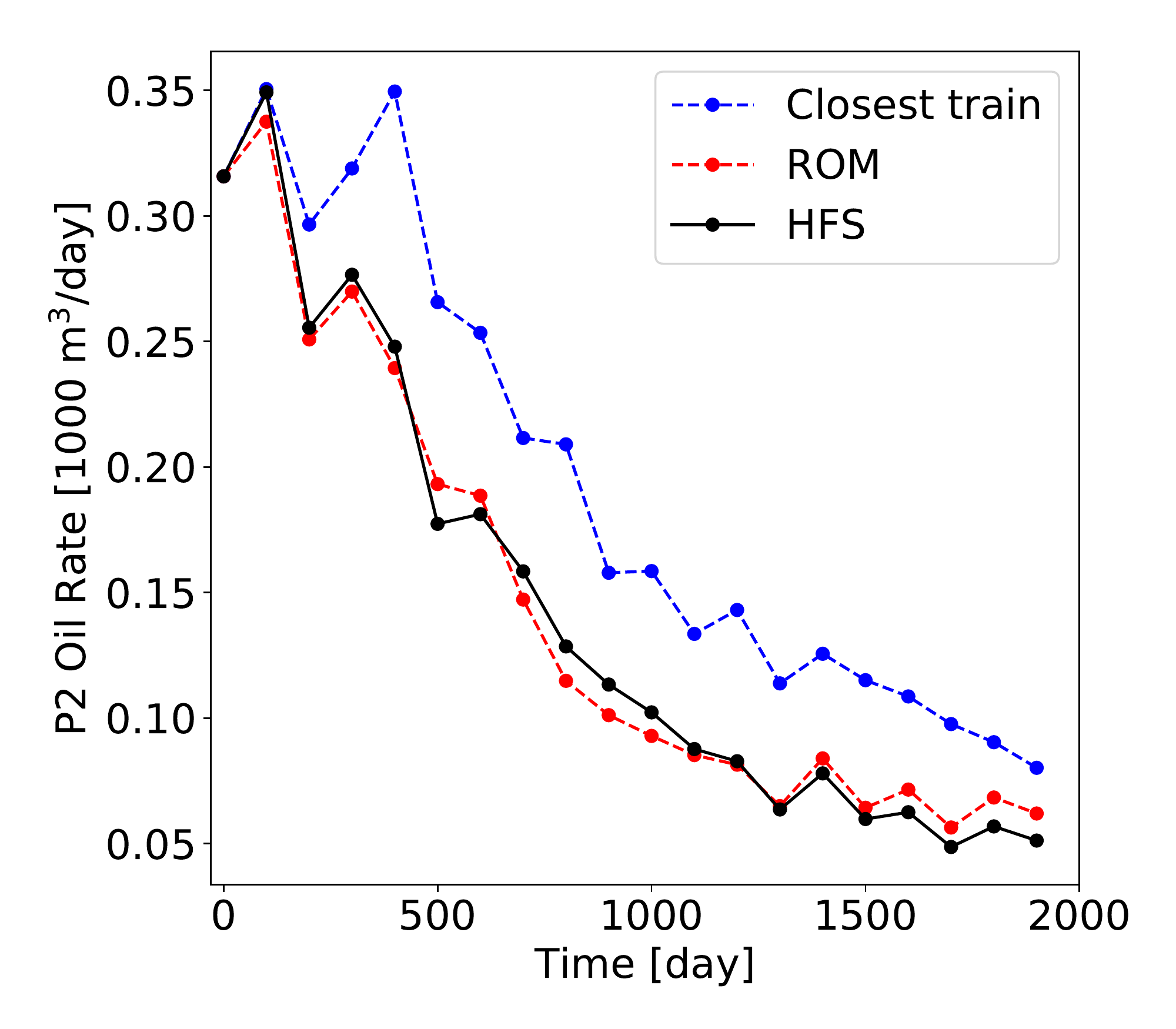}
    \caption{Oil rate}
  \end{subfigure}\hfill
  \begin{subfigure}{.5\textwidth}
    \centering
    \includegraphics[width=\linewidth]{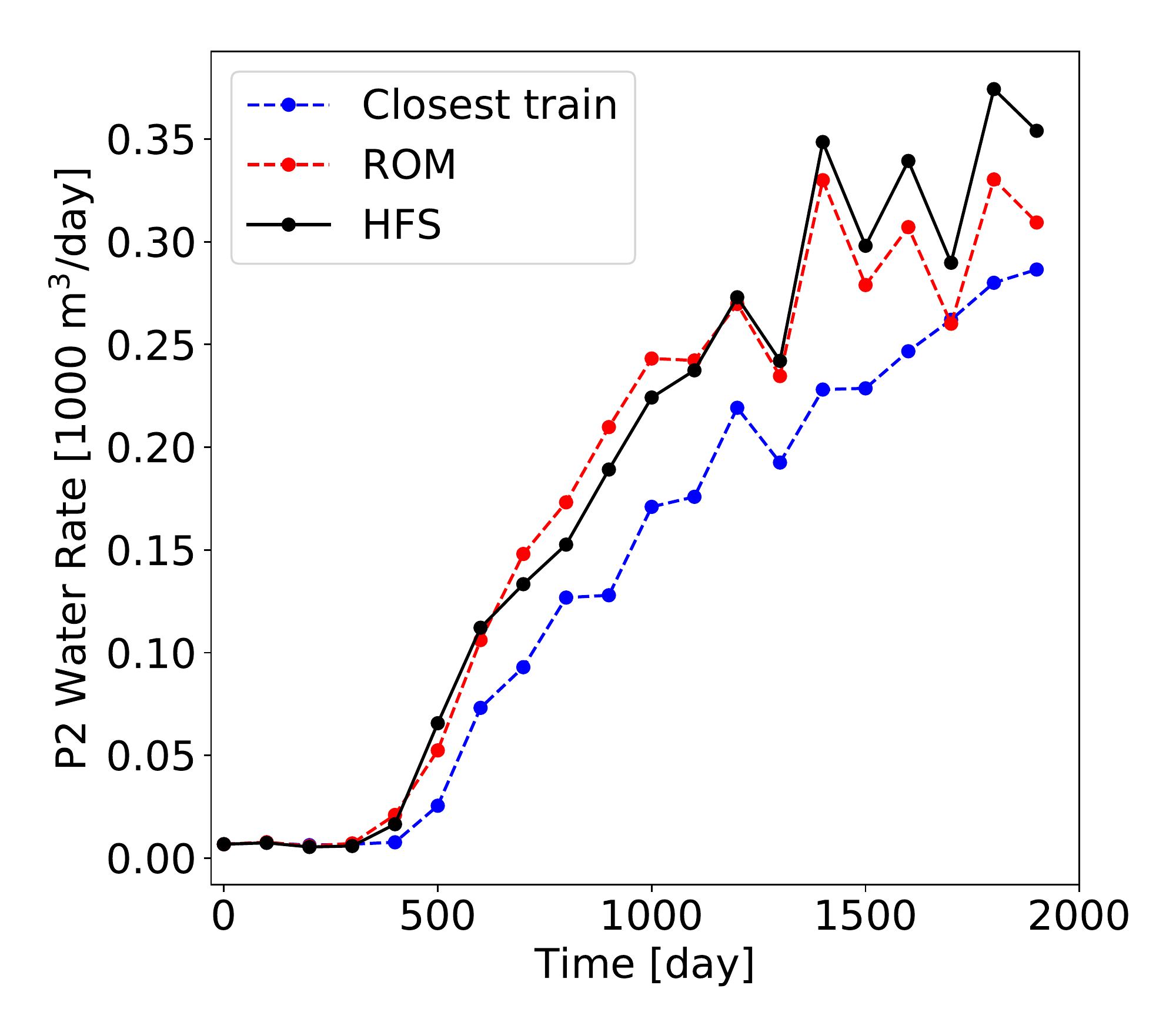}
    \caption{Water rate}
  \end{subfigure}
  \caption{Test Case~3: production rates for Well P2}
  \label{fig::test_3_rate_p2}
\end{figure}

\begin{figure}[htbp]
  \centering
  \begin{subfigure}{.45\textwidth}
    \centering
    \includegraphics[width=\linewidth]{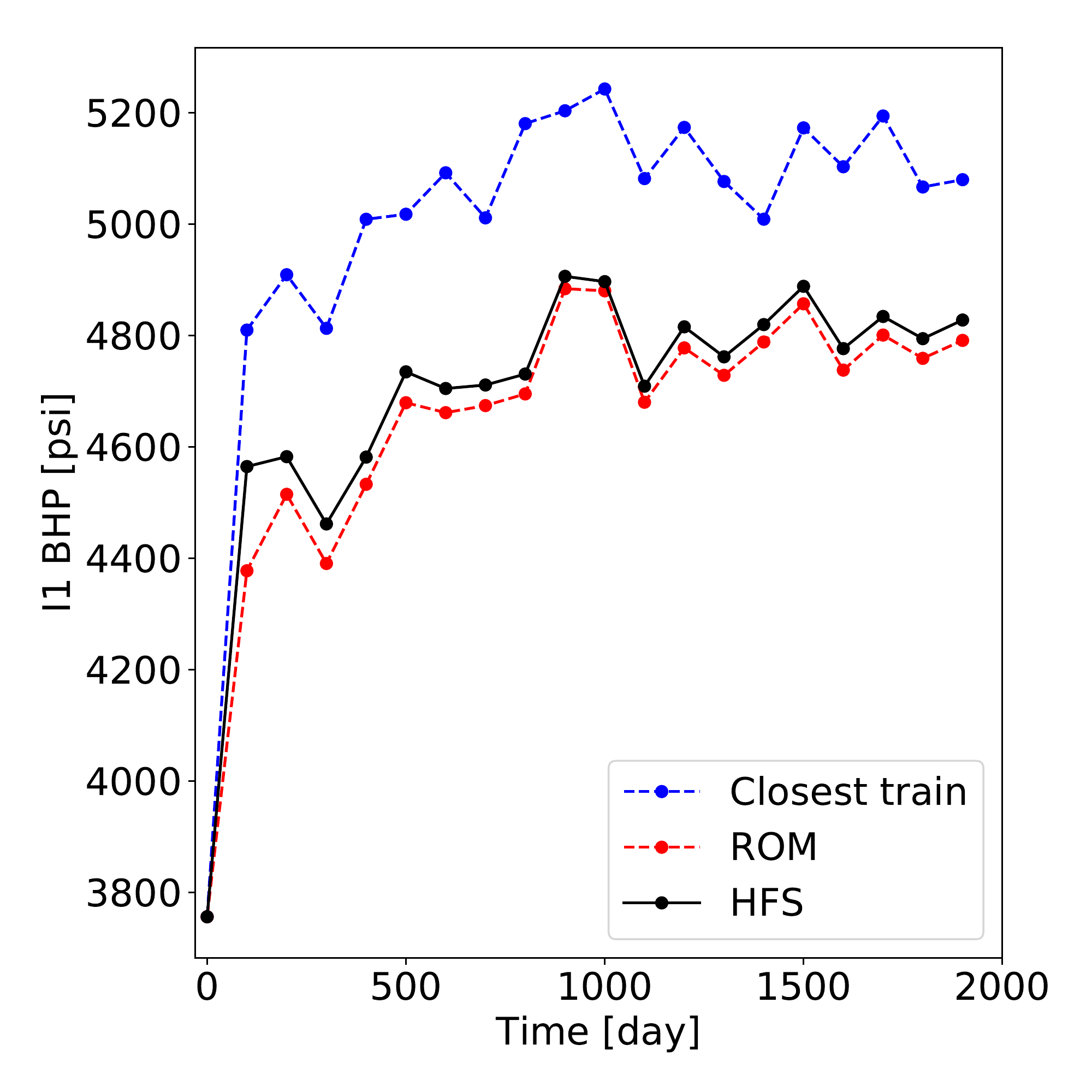}
    \caption{Well~I1}
  \end{subfigure}\hfill
  \begin{subfigure}{.45\textwidth}
    \centering
    \includegraphics[width=\linewidth]{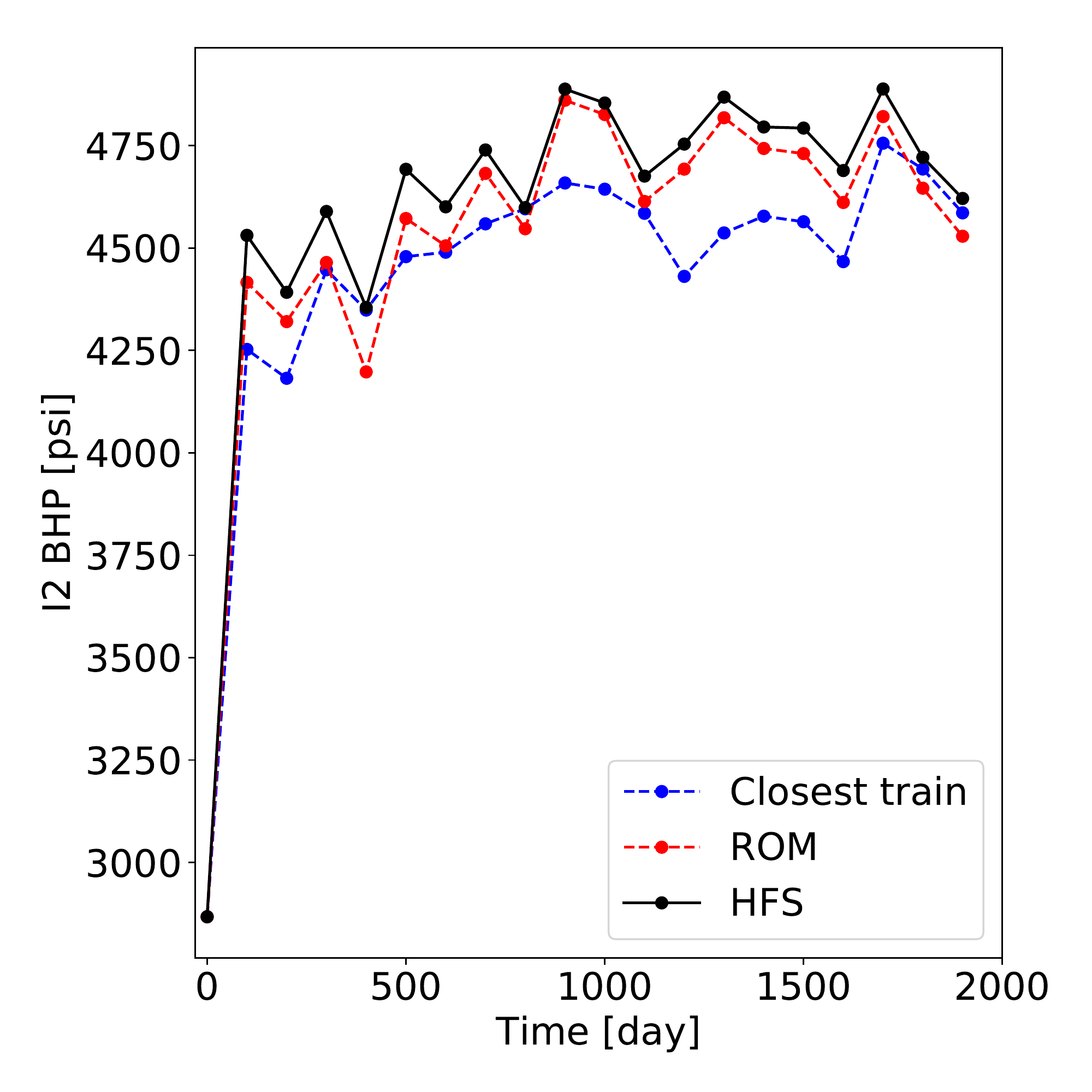}
    \caption{Well~I2}
  \end{subfigure} \\
  \begin{subfigure}{.45\textwidth}
    \centering
    \includegraphics[width=\linewidth]{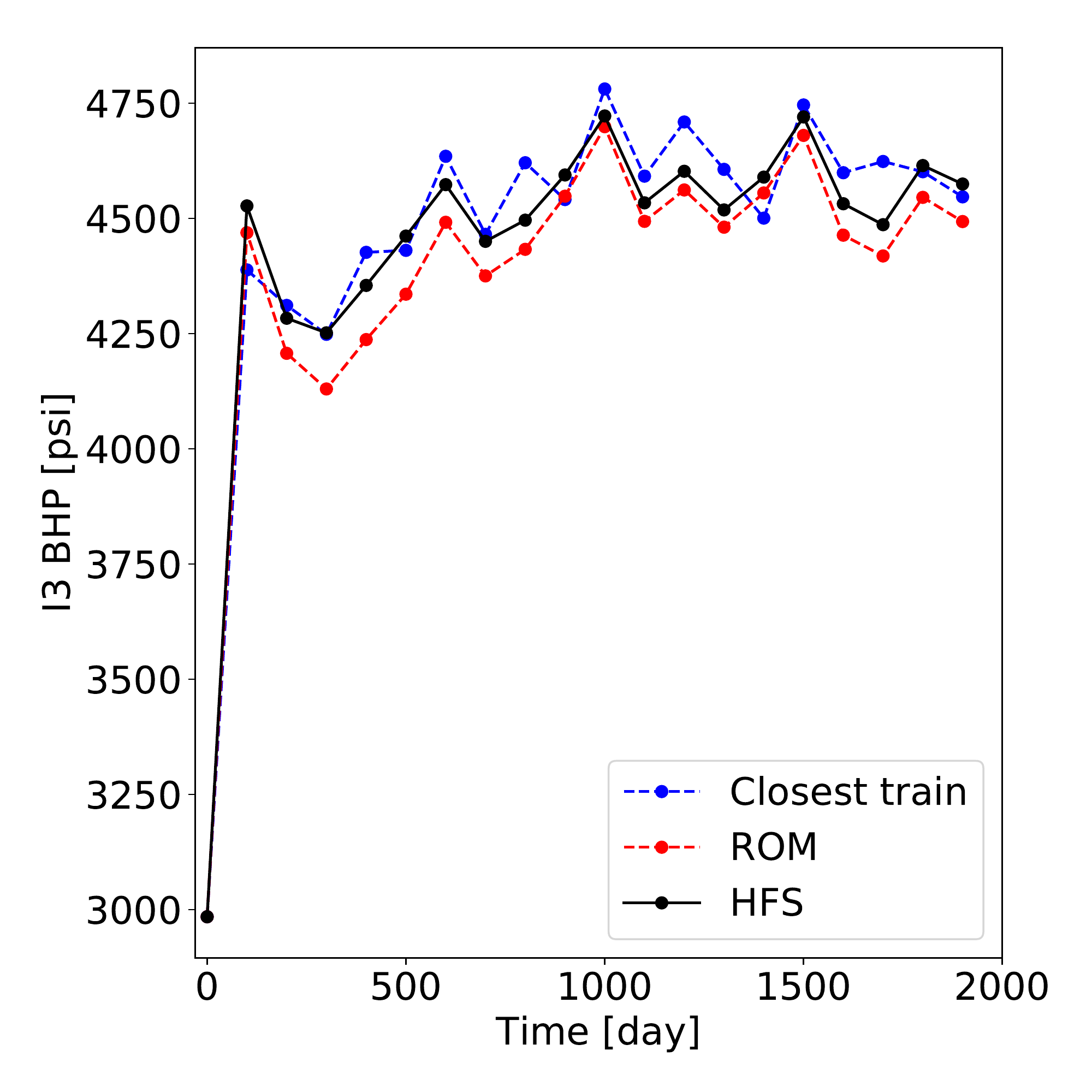}
    \caption{Well~I3}
  \end{subfigure}\hfill
    \begin{subfigure}{.45\textwidth}
    \centering
    \includegraphics[width=\linewidth]{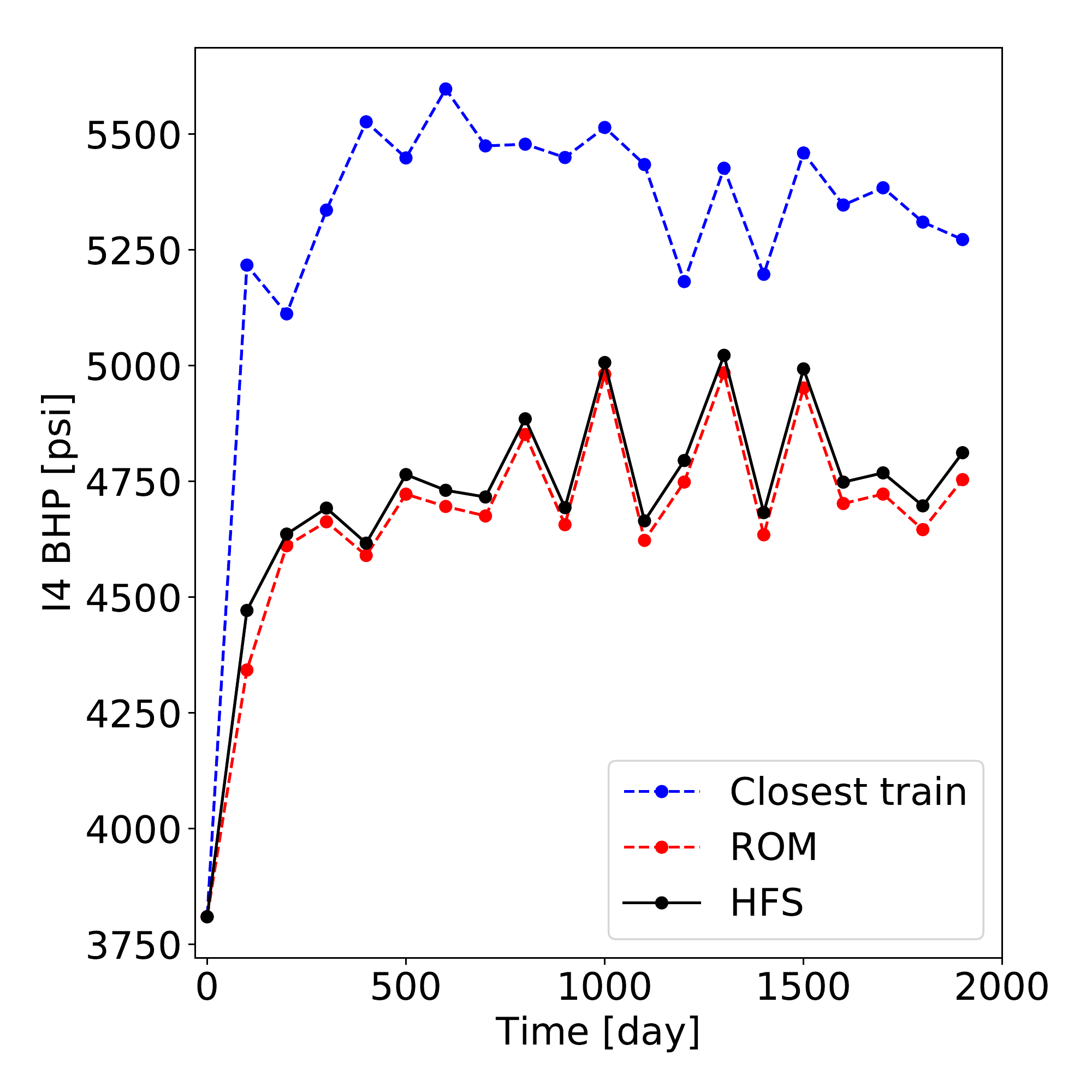}
    \caption{Well~I4}
  \end{subfigure}
  \caption{Test Case~3: injection BHPs}
  \label{fig::test_3_inj}
\end{figure}

Analogous results for the evolution of saturation for Test Case~3 are shown in Figs.~\ref{fig::test_3_sat_200}, \ref{fig::test_3_sat_1000} and \ref{fig::test_3_sat_1800}, and the pressure field at 1000~days is presented in Fig.~\ref{fig::test_3_pres_1000}. The reservoir sweep is somewhat different here than in the other two cases, with the water plume around Well I4 (lower right) clearly smaller in this case. The difference maps again indicate a high level of accuracy for the E2C ROM. Production and injection well predictions appear in Figs.~\ref{fig::test_3_rate_p1}, \ref{fig::test_3_rate_p2} and \ref{fig::test_3_inj}. We again obtain accurate ROM results for these important quantities.

%%%%% case 3, saturation %%%%%%%%

%%%%% case 3, pressure %%%%%%%%

%%%%% case 3, prod rates %%%%%%%%

%%%%% case 3, inj bhps %%%%%%%%

%% References


% \bibliographystyle{elsarticle-num}
% \bibliographystyle{elsarticle-harv}
\bibliographystyle{elsarticle-num-names}
\clearpage
%\bibliography{ref}

\end{document}